\documentclass[11pt,a4paper]{article}
\pdfoutput=1

\usepackage[colorlinks=true, linkcolor=black!50!blue, urlcolor=blue, citecolor=blue, anchorcolor=blue]{hyperref}
\usepackage[font=small,labelfont=bf,margin=0mm,labelsep=period,tableposition=top]{caption}
\usepackage[a4paper,top=3cm,bottom=2.5cm,left=2.5cm,right=2.5cm,bindingoffset=0mm]{geometry}

\usepackage{graphicx,placeins}
\usepackage{float}
\usepackage{afterpage}
\usepackage{epsfig,cite}
\usepackage{amssymb}
\usepackage{amsmath}
\usepackage{dsfont}
\usepackage{multirow}
\usepackage{url}
\usepackage{xcolor,colortbl}
\usepackage{float}
\usepackage{afterpage}
\usepackage{url}
\usepackage{hyperref}
\usepackage{booktabs}
\usepackage{mathrsfs}
\usepackage[shortlabels]{enumitem}

\usepackage{tikz}
\usepackage{tikz-3dplot}
\usepackage[compat=1.0.0]{tikz-feynman}

\usepackage{enumitem}
\usepackage{hyperref}
\usepackage{cite}

\usepackage{pifont}

\usetikzlibrary{shapes, arrows}
\usetikzlibrary{decorations.pathreplacing}
\usetikzlibrary{positioning, calc}
\tikzstyle{fitted} = [rectangle, minimum width=5cm, minimum height=1cm, text centered, draw=black, fill=red!30]
\tikzstyle{operations} = [rectangle, rounded corners, minimum width=2cm,text centered, draw=black, fill=red!30]
\tikzstyle{roundtext} = [rectangle, rounded corners, minimum width=2cm, minimum height=0.8cm, text centered, draw=black, fill=red!30]
\tikzstyle{n3py} = [rectangle, rounded corners, minimum width=3cm, minimum height=1cm, text centered, draw=black, fill=green!30]
\tikzstyle{myarrow} = [thick,->,>=stealth]
\tikzstyle{line} =[draw, -latex']
\tikzstyle{decision} = [diamond, draw, fill=red!20, text width=7.5em, text centered,  inner sep=0pt, minimum height=2em, aspect=4]
\tikzstyle{cloud} = [draw, ellipse,fill=green!20, minimum height=2em]
\tikzstyle{inout} = [rectangle, draw, fill=green!20, text width=9.5em, text centered, rounded corners, minimum height=2em, minimum width=10em]
\tikzstyle{block}=[rectangle, draw, fill=blue!20, text width=9.5em,
                   text centered, rounded corners, minimum height=2em,
                   minimum width=10em]

\definecolor{darkgreen}{rgb}{0.0, 0.5, 0.13}

\def\smallfrac#1#2{\hbox{$\frac{#1}{#2}$}}

\newcommand{\be}{\begin{equation}}
\newcommand{\ee}{\end{equation}}
\newcommand{\bea}{\begin{eqnarray}}
\newcommand{\eea}{\end{eqnarray}}
\newcommand{\bi}{\begin{itemize}}
\newcommand{\ei}{\end{itemize}}
\newcommand{\ben}{\begin{enumerate}}
\newcommand{\een}{\end{enumerate}}
\newcommand{\la}{\left\langle}
\newcommand{\ra}{\right\rangle}

\newcommand{\lp}{\left(}
\newcommand{\rp}{\right)}

\def\gsim{\mathrel{\rlap{\lower4pt\hbox{\hskip1pt$\sim$}}
    \raise1pt\hbox{$>$}}}         
\def\lsim{\mathrel{\rlap{\lower4pt\hbox{\hskip1pt$\sim$}}
    \raise1pt\hbox{$<$}}}         

\newcommand{\draft}[1]{}

\def\beq{\begin{equation}}
\def\eeq{\end{equation}}

\def\lapprox{\lower .7ex\hbox{$\;\stackrel{\textstyle <}{\sim}\;$}}
\def\gapprox{\lower .7ex\hbox{$\;\stackrel{\textstyle >}{\sim}\;$}}

\numberwithin{equation}{section}
\numberwithin{figure}{section}
\numberwithin{table}{section}

\usepackage{tabularx}
\newcolumntype{C}[1]{>{\centering\arraybackslash}p{#1}}

\usepackage{amsmath}
\usepackage{amsfonts}
\usepackage{amssymb}
\usepackage{dsfont}
\usepackage{pifont}
\usepackage{booktabs}
\usepackage{graphicx}
\usepackage{epstopdf}
\usepackage{epsfig}
\usepackage{framed}
\usepackage{makeidx}
\usepackage{siunitx}
\usepackage[capitalise]{cleveref}
\usepackage{hyperref}
\usepackage{placeins}
\usepackage[font=small,labelfont=bf]{caption}

\begin{document}
\newgeometry{top=1.5cm,bottom=1.5cm,left=1.5cm,right=1.5cm,bindingoffset=0mm}

\vspace{-2.0cm}
\begin{flushright}
Nikhef 2025-002\\
Edinburgh 2024/10\\
CERN-TH-2024-169
\end{flushright}
\vspace{0.3cm}

\begin{center}
  {\Large \bf Parton distributions confront LHC Run II data:\\[0.3cm] a quantitative appraisal }
  \vspace{1.1cm}

   Amedeo Chiefa$^1$, Mark N. Costantini$^2$,
   Juan Cruz-Martinez$^3$,
   Emanuele R. Nocera$^4$,\\[0.15cm]
   Tanjona R. Rabemananjara$^{5,6}$,
   Juan Rojo$^{5,6}$,
   Tanishq Sharma$^{4,5,6}$,
   Roy Stegeman$^1$, and
    Maria Ubiali$^2$

 {\it \small

    \vspace{0.7cm}

    ~$^1$The Higgs Centre for Theoretical Physics, University of Edinburgh,\\
    JCMB, KB, Mayfield Rd, Edinburgh EH9 3JZ, Scotland\\[0.1cm]
     ~$^2$ DAMTP, University of Cambridge, Wilberforce Road, Cambridge, CB3 0WA, United Kingdom\\[0.1cm]
      ~$^3$CERN, Theoretical Physics Department, CH-1211 Geneva 23, Switzerland\\[0.1cm]
            ~$^4$ Dipartimento di Fisica, Universit\`a degli Studi di Torino and\\
      INFN, Sezione di Torino, Via Pietro Giuria 1, 10125 Torino, Italy\\[0.1cm]
       ~$^5$Department of Physics and Astronomy, Vrije Universiteit, NL-1081 HV Amsterdam\\[0.1cm]
      ~$^6$Nikhef Theory Group, Science Park 105, 1098 XG Amsterdam, The Netherlands\\[0.1cm]

   }

 \vspace{0.7cm}

{\bf \large Abstract}

\end{center}

We present a systematic comparison of theoretical predictions and various
high-precision experimental measurements, specifically of differential cross
sections performed by the LHC run II for Drell-Yan gauge boson, top-quark pair,
single-inclusive jet and di-jet production, and by HERA for single-inclusive
jet and di-jet production. Theoretical predictions are computed at
next-to-next-to-leading order (NNLO) accuracy in perturbative Quantum
Chromodynamics. The most widely employed sets of Parton Distribution
Functions (PDFs) are used, and PDF, strong coupling, and missing higher order
uncertainties are taken into account. We quantitatively assess the predictive
power of each PDF set and the contribution of the different sources of
experimental and theoretical uncertainty to the agreement between data and
predictions. We show that control over all of these aspects is crucial to
precision physics studies, such as the determination of Standard Model
parameters at the LHC.

\clearpage

\tableofcontents

\section{Introduction}
\label{sec:intro}

The remarkable progress witnessed by the determination of the parton
distribution functions (PDFs) of the proton~\cite{Kovarik:2019xvh,
  Ethier:2020way,Amoroso:2022eow,Ubiali:2024pyg} in recent years has been
driven by three factors: the extension of the input dataset, in particular
thanks to high-precision Large Hadron Collider (LHC) measurements; the
improvement of the accuracy of theoretical computations, now reaching
the approximate next-to-next-to-next-to-leading order (aN$^3$LO) in the strong
coupling; and the investigation of several methodological
aspects, specifically with respect to the quantification of PDF uncertainties.
Some groups~\cite{Alekhin:2017kpj,Hou:2019efy,Bailey:2020ooq,NNPDF:2021njg}
provide regular updates of their PDF determinations based on a broad input
dataset, while other groups focus on the interpretation of restricted subsets
of data~\cite{Moffat:2021dji,ATLAS:2021vod,Accardi:2023gyr}.
All of these PDF sets differ, sometimes by amounts that are larger than their
nominal uncertainties, which can have very disparate size in different PDF
sets. In order to understand the origin of these differences, several benchmark
studies have been performed over the years~\cite{Alekhin:2011sk,
  Botje:2011sn,Ball:2012wy,Rojo:2015acz,Accardi:2016ndt,
  Andersen:2016qtm,Andersen:2014efa,Butterworth:2015oua,Watt:2012tq,
  Hou:2016sho,Gao:2013bia,Carrazza:2015aoa,Carrazza:2016htc,Carrazza:2015hva,
  PDF4LHCWorkingGroup:2022cjn,Jing:2023isu}. In particular, it was
shown~\cite{PDF4LHCWorkingGroup:2022cjn} that PDFs determined from an identical
set of experimental data and theoretical predictions with three different
methodological frameworks~\cite{Hou:2019efy,Bailey:2020ooq,NNPDF:2017mvq}
displayed similar central values but somewhat different uncertainties. These
differences should likely be ascribed to the different methodological framework.

These differences challenge the theoretical interpretation of the outcome of PDF
determinations in terms of the underlying proton structure, as highlighted by
the ongoing discussion concerning the possible existence of intrinsic charm
quarks in the proton~\cite{Ball:2022qks,NNPDF:2023tyk,Guzzi:2022rca}.
Furthermore, they degrade the physics reach of core LHC analyses sensitive to
PDFs, concerning both the measurements of fundamental Standard Model (SM)
parameters --- like the strong coupling $\alpha_s(m_Z)$, the $W$-boson mass
$m_W$, and the effective leptonic mixing angle $\sin^2 \theta_{\rm eff}^\ell$ ---
and direct (resonance-like) and indirect (effective-field-theory (EFT)-like)
searches for physics beyond the SM. The first aspect is illustrated, for
instance, by the large PDF dependence of the
high-mass forward-backward asymmetry in Drell-Yan gauge boson
production~\cite{Ball:2022qtp,Fu:2023rrs,Fiaschi:2022wgl}. The second aspect
is illustrated by the interplay between PDFs and possible EFT contamination in
high-$p_T$ top-quark pair and Drell-Yan cross section
measurements~\cite{Greljo:2021kvv,Iranipour:2022iak,Kassabov:2023hbm,
  Hammou:2023heg,Costantini:2024xae,Hammou:2024xuj,Gao:2022srd}.

Recent LHC analyses have highlighted this far-from-ideal state of affairs.
Here are three examples. First, the ATLAS determination of the strong
coupling $\alpha_s(m_Z)$ from neutral-current Drell-Yan differential
measurements  in the transverse momentum of the $Z$ boson~\cite{ATLAS:2023lsr}.
This is the most precise $\alpha_s$ determination ever performed from a
single experiment, with a quoted uncertainty $\delta=9\cdot 10^{-4}$. Of this
value, the uncertainty due to the PDF is estimated to be the dominant
component, $\delta_{\rm pdf}=5\cdot 10^{-4}$, using the MSHT20 aN$^3$LO
fit~\cite{McGowan:2022nag}. However, if the PDF uncertainly is defined as the
difference between central predictions obtained with the
CT18A~\cite{Hou:2019efy} and NNPDF4.0~\cite{NNPDF:2021njg} PDF sets, one gets
an uncertainty which is four times larger, $\delta_{\rm pdf}=2\times 10^{-3}$.
Second, the CMS measurement of the effective leptonic mixing angle
$\sin^2\theta^\ell_{\rm eff}$~\cite{CMS:2024ony}. In this case, the PDF
uncertainty is estimated to be $\delta_{\rm pdf}=0.14\%$ using the
CT18Z PDF set~\cite{Hou:2019efy}, while the spread of the 
central results obtained with CT18~\cite{Hou:2019efy} and
MSHT~\cite{Bailey:2020ooq} (before profiling) is around a factor of
five larger, $\delta_{\rm pdf}=0.7\%$. Third, the updated ATLAS measurement
of the $W$ mass at 7 TeV~\cite{ATLAS:2023fsi}. In this case, the PDF
uncertainty is estimated to be $\delta_{\rm pdf}=7.7~(14.6)$~MeV in the lepton
transverse momentum $p_T^\ell$ (transverse mass $m_T$) channel, to be
compared with the spread between NNPDF4.0 and MSHT20 which gives twice that
estimate, $\delta_{\rm pdf}=17~(21)$~MeV. Similar considerations apply to the
precise $m_W$ measurement performed by the CMS collaboration~\cite{CMS:2024nau}.
Each of these analyses selects a different baseline PDF set, with other
possible choices of PDFs yielding a central value potentially outside the
quoted PDF uncertainties. Finally --- and crucially --- these analyses do not
satisfactorily consider the back-reaction of the precision measurement under
scrutiny on all other datasets entering a PDF fit~\cite{Forte:2020pyp}, which
is especially relevant when profiling/reweighting techniques for {\it in situ}
calibration are used.

These considerations highlight the importance of understanding the origin of the
differences observed when computing theoretical predictions with different PDF
sets. Complementing existing benchmark studies that tackle this question,
here we investigate whether existing PDF sets can be discriminated according to
their predictive power of high-precision measurements not included in their
determination. We will specifically consider cross sections measured by the LHC
run II for Drell-Yan gauge boson, top-quark pair, single-inclusive jet and
di-jet production, and by HERA for single-inclusive jet and di-jet production.
We will compare these experimental data to theoretical predictions computed at
next-to-next-to-leading order (NNLO) accuracy in perturbative Quantum
Chromodynamics (QCD) and quantitatively assess their mutual agreement.
We will take into account all sources of theoretical uncertainty in this
assessment, namely PDF, $\alpha_s$, and missing higher order (MHO)
uncertainties. We will study the dependence of this assessment on the input PDF
set. This exercise is an extension of the future test introduced
in~\cite{Cruz-Martinez:2021rgy}.

The outline of this paper is as follows. In Sect.~\ref{sec:data} we present
the considered LHC and HERA measurements and the computation of the
corresponding theoretical predictions. In Sect.~\ref{sec:approach} we
describe how we quantitatively assess the agreement between experimental data
and theoretical predictions, and in particular how we account for PDF,
$\alpha_s$, and MHO uncertainties in this assessment. In Sect.~\ref{sec:results}
we present a selection of representative results for each class of measurements,
highlighting the relative contribution of the various sources of theoretical
uncertainty in the description of the data, and commenting on features that are
common to or different from various PDF sets. We summarise our findings in
Sect.~\ref{sec:summary}. Two appendices complete the paper.
Appendix~\ref{app:unreg} quantifies the impact of regularizing
ill-conditioned experimental covariance matrices in the assessment of the
data-theory comparison. Appendix~\ref{app:extra_results} collects
the complete set of results not shown in Sect.~\ref{sec:results}.

\section{Experimental data and theoretical predictions}
\label{sec:data}

In this section, we present the experimental data considered in this work and
the details of the corresponding theoretical computations. The data has been
selected according to the following criteria.

\begin{itemize}

\item We consider datasets for scattering processes that provide information
  on PDFs of different partons (quarks, antiquarks, gluon) in a broad kinematic
  region of $x$ and $Q^2$. For a given process, we select the dataset based on
  the largest integrated luminosity available.

\item We avoid datasets that are already included in PDF determinations used
  to compute theoretical predictions, to avoid double-counting. The only
  exception is  the recent re-analysis of the $Z$ data at a centre-of-mass
  energy of 8~TeV by ATLAS~\cite{ATLAS:2023lsr}.
 
\item We consider datasets for which the corresponding theoretical predictions
  can be computed at NNLO in perturbative QCD using event generators interfaced
  to fast interpolation grids. This avoids reliance on $K$-factors and
  allows one to readily evaluate predictions upon changes of input PDF set and
  factorisation and renormalisation scales.

\item  We only consider datasets for which the corresponding experimental
  information is publicly available, in particular through the
  {\sc HEPdata} repository~\cite{Maguire:2017ypu}.

\end{itemize}

Taking into account these requirements, the ATLAS, CMS, LHCb, H1, and ZEUS
datasets that are considered in this study are summarised in
Table~\ref{tab:input_datasets}, classified by process type. For each dataset
we indicate the experiment, the final-state channel, the measured differential
distribution(s), the centre-of-mass energy, the integrated luminosity, the
number of data points (after kinematic cuts), and the corresponding publication
reference. For the ATLAS and CMS top-quark pair production and for the CMS
single-inclusive jet production datasets, we list all the separate
distributions provided by the corresponding analyses. In this work, we select
a subset of these distributions, which we deem most representative as
explained in Sect.~\ref{subsec:compatibility}. In the following, we discuss
the main features of these datasets and describe the associated theoretical
calculations.

\begin{table}[!t]
  \centering
  \scriptsize
  \renewcommand{\arraystretch}{2.0}
  \begin{tabularx}{\textwidth}{XXXcccrc}
  \toprule
  Process
  & Experiment
  & Final State
  & Observable
  & $\sqrt{s}$~(TeV)
  & $\mathcal{L}$~(fb$^{-1}$)
  & $n_{\rm dat}$
  & Ref.
  \\
  \midrule
  \multirow{4}{*}{LHC $W,Z$}
  & ATLAS
  & Z $p_T$ spectrum
  & $\left(\frac{1}{\sigma}\right)\frac{d\sigma}{dp_T^{\ell\ell}}$
  & 13
  & 36.1
  & 38
  & \cite{ATLAS:2019zci}
  \\
  & CMS
  & W incl. prod. 
  & $\frac{d\sigma}{d|\eta|}$
  & 13
  & 35.9
  & 36
  & \cite{CMS:2020cph}
  \\
  & LHCb
  & Z incl. forward prod. 
  & $\frac{d\sigma}{dy^{Z}}$
  & 13
  & 5.1
  & 18
  &\cite{LHCb:2021huf}
  \\
  & ATLAS
  & Z incl. prod. 
  & $\frac{d\sigma}{d|y|}$
  & 8
  & 20.2
  & 7
  & \cite{ATLAS:2023lsr}
  \\
  \midrule
  \multirow{12}{*}{top-pair}
  & \multirow{3}{*}{ATLAS}
  & \multirow{3}{*}{all-hadronic}
  & $\left(\frac{1}{\sigma}\right)\frac{d\sigma}{dm_{t\bar{t}}}$
  & 13
  & 36.1
  & 9
  & \cite{ATLAS:2020ccu}
  \\
  &
  &
  & $\left(\frac{1}{\sigma}\right)\frac{d\sigma}{d|y_{t\bar{t}}|}$
  & 13
  & 36.1
  & 12
  & \cite{ATLAS:2020ccu} 
  \\
  &
  &
  & $\left(\frac{1}{\sigma}\right)\frac{d^2\sigma}{d|y_{t\bar{t}}|dm_{t\bar{t}}}$
  & 13
  & 36.1
  & 11
  & \cite{ATLAS:2020ccu}   
  \\
  \cline{2-8}
  & \multirow{5}{*}{ATLAS}
  & \multirow{5}{*}{$\ell+$jets}
  & $\left(\frac{1}{\sigma}\right)\frac{d\sigma}{dm_{t\bar{t}}}$
  & 13
  & $36.1$
  & 9
  & \cite{ATLAS:2019hxz}
  \\
  &
  &
  & $\left(\frac{1}{\sigma}\right)\frac{d\sigma}{dp_T^t}$
  & 13
  & $36.1$
  & 8
  & \cite{ATLAS:2019hxz}
  \\
  &
  &
  & $\left(\frac{1}{\sigma}\right)\frac{d\sigma}{d|y_t|}$
  & 13
  & $36.1$
  & 5
  & \cite{ATLAS:2019hxz}
  \\
  &
  &
  & $\left(\frac{1}{\sigma}\right)\frac{d\sigma}{d|y_{t\bar{t}}|}$
  & 13
  & $36.1$
  & 7
  & \cite{ATLAS:2019hxz}
  \\
  \cline{2-8}
  & \multirow{5}{*}{CMS}
  & \multirow{5}{*}{$\ell+$jets}
  & $\left(\frac{1}{\sigma}\right)\frac{d\sigma}{dm_{t\bar{t}}}$
  & 13
  & $137$
  & 15
  & \cite{CMS:2021vhb}
  \\ 
  &
  &
  & $\left(\frac{1}{\sigma}\right)\frac{d\sigma}{dp_T^t}$
  & 13
  & $137$
  & 16
  & \cite{CMS:2021vhb}
  \\  
  &
  &
  & $\left(\frac{1}{\sigma}\right)\frac{d\sigma}{d|y_{t\bar{t}}|}$
  & 13
  & $137$
  & 10
  & \cite{CMS:2021vhb}
  \\  
  &
  &
  & $\left(\frac{1}{\sigma}\right)\frac{d\sigma}{d|y_t|}$
  & 13
  & $137$
  & 11
  & \cite{CMS:2021vhb}
  \\  
  &
  &
  & $\left(\frac{1}{\sigma}\right)\frac{d^2\sigma}{d|y_{t\bar{t}}|dm_{t\bar{t}}}$
  & 13
  & $137$
  & 35
  & \cite{CMS:2021vhb}
  \\   
  \midrule
  \multirow{3}{*}{LHC jets}
  & ATLAS
  & incl. jet $R=0.4$
  & $\frac{d^2\sigma}{dp_Td|y|}$
  & 13
  & 3.2
  & 177
  & \cite{ATLAS:2017ble}
  \\
  & CMS
  & incl. jets $R=0.4~(0.7)$
  & $\frac{d^2\sigma}{dp_Td|y|}$
  & 13
  & 36.3 (33.5)
  & 78
  & \cite{CMS:2021yzl}
  \\
  & ATLAS
  & di-jets $R=0.4$
  & $\frac{d^2\sigma}{dm_{jj}d|y^*|}$
  & 13
  & 3.2
  & 136
  & \cite{ATLAS:2017ble}
  \\
  \midrule
  \multirow{7}{*}{HERA jets}
  & H1
  & incl. jet (low $Q^2$)
  & $\frac{d^2\sigma}{dQ^{2}dp_T}$
  & 0.319
  & 0.29
  & 37
  & \cite{H1:2016goa}
  \\
  & H1
  & incl. jet (high $Q^2$)
  & $\frac{d^2\sigma}{dQ^{2}dp_T}$
  & 0.319
  & 0.351
  & 24
  & \cite{H1:2014cbm}
  \\
  & ZEUS
  & incl. jet
  & $\frac{d^2\sigma}{dQ^2dE_T}$
  & 0.300
  & 0.038
  & 30
  & \cite{ZEUS:2002nms}
  \\
  & ZEUS
  & incl. jet
  & $\frac{d^2\sigma}{dQ^2dE_T}$
  & 0.319
  & 0.082
  & 30
  & \cite{ZEUS:2006xvn}
  \\
  & H1
  & di-jets (low $Q^2$)
  & $\frac{d^2\sigma}{dQ^{2}d\langle p_T \rangle}$
  & 0.319
  & 0.29
  & 37
  & \cite{H1:2016goa}
  \\
  & H1
  & di-jets (high $Q^2$)
  & $\frac{d^2\sigma}{dQ^{2}d\langle p_T \rangle}$
  & 0.319
  & 0.351
  & 24
  & \cite{H1:2014cbm}
  \\
  & ZEUS
  & di-jets
  & $\frac{d^2\sigma}{dQ^2d\la E_T\ra}$
  & 0.319
  & 0.374
  & 22
  & \cite{ZEUS:2010vyw}
  \\
  \bottomrule
\end{tabularx}

  \vspace{0.3cm}
  \caption{The ATLAS, CMS, LHCb, H1, and ZEUS datasets considered in
    this work, classified by process type. For each dataset we indicate
    the experiment, the final-state channel, the measured differential
    distribution(s), the centre-of-mass energy $\sqrt{s}$, the integrated
    luminosity $\mathcal{L}$, the number of data points $n_{\rm dat}$ (after
    kinematic cuts), and the corresponding publication reference. For the CMS
    single-inclusive jet production and for the ATLAS and CMS top-quark pair
    production datasets, we list the separate distributions provided by the
    corresponding analyses.}
  \label{tab:input_datasets}
\end{table}

\subsection{Drell-Yan weak boson production at the LHC}
\label{subsec:LHC_DY}

Neutral- and charged-current Drell-Yan production is used to probe
quark-flavour PDF separation, through rapidity distributions in the central
(ATLAS and CMS) and forward (LHCb) regions~\cite{Rojo:2017xpe,Thorne:2008am},
and the gluon PDF, through transverse momentum
distributions~\cite{Boughezal:2017nla}. In the former case, the leading
partonic channel is initiated by quarks and antiquarks; in the latter case,
a non-zero $p_T$ distribution arises from the $gq(\bar{q})$ partonic initial
state followed by a hard $g\to q\bar{q}$ splitting. Here we consider three
LHC Run II representative measurements for each of these categories: one by
ATLAS~\cite{ATLAS:2019zci}, one by CMS~\cite{CMS:2020cph}, and one by
LHCb~\cite{LHCb:2016fbk}. All these measurements correspond to a
centre-of-mass energy of 13~TeV. We also consider the recent re-analysis of the
inclusive $Z$ boson production measurement at a centre-of-mass energy of 8~TeV
by ATLAS extrapolated to the full leptonic phase space~\cite{ATLAS:2023lsr}.

The ATLAS Run II measurement~\cite{ATLAS:2019zci} corresponds to an integrated
luminosity of 36.1~fb$^{-1}$. It consists of the $Z$-boson production
cross section, reconstructed from the combination of events resulting from 
electron and muon decays, differential in the transverse momentum of the
dilepton pair $p_T^{\ell\ell}$. The measurement is performed in a fiducial
phase space, defined by the lepton transverse momentum $p_T^\ell>27$~GeV, the
absolute lepton pseudorapidity $|\eta_\ell|<2.5$, and the dilepton invariant
mass $66<m_{\ell\ell}<116$~GeV. Cross sections are provided for both the
absolute distribution and the distribution normalised to the fiducial cross
section. The full breakdown of correlated systematic uncertainties is
available and taken into account.

The CMS Run II measurement~\cite{CMS:2020cph} corresponds to an integrated
luminosity of 35.9~fb$^{-1}$. It consists of the $W^\pm$ boson production
cross section, reconstructed from the combination of events resulting from
electron and muon decays. This measurement is presented as a double-differential
distribution in the absolute lepton rapidity $|\eta|$, with
$|\eta|<2.4$, and in the lepton transverse momentum $p_T^\ell$, with
$26<p_T^\ell<56$~GeV. It is available for each $W$ polarisation state and
averaged over polarisations. For each boson, 18 equally large bins in
$|\eta|$ and a single bin in $p_T^\ell$ are provided. The full breakdown of
correlated systematic uncertainties is available and taken into account.

The LHCb Run II measurement~\cite{LHCb:2021huf} corresponds to an integrated
luminosity of 5.1~fb$^{-1}$. It consists of the $Z$-boson production cross
section, reconstructed only from muon decays, in the fiducial phase space
defined by the muon transverse momentum $p_T^\mu>20$~GeV, the dimuon invariant
mass $60<m_{\mu\mu}<120$~GeV, and the muon rapidity $2.0<\eta_{\mu}<4.5$. The
presented cross section is differential in the rapidity of the $Z$ boson $y^Z$.
The full breakdown of correlated systematic uncertainties is
available and taken into account.

Finally, we consider the recent ATLAS measurement of $Z$ boson production
based on the 2012 dataset at a centre-of-mass energy of 8~TeV, which
corresponds to an integrated luminosity of 20.2~fb$^{-1}$~\cite{ATLAS:2023lsr}.
The measurement is extrapolated to the full phase space of the decay electrons
and muons in the dilepton rapidity range $|y|< 3.6$, and covers
the $Z$ pole invariant mass region, $80\le m_{\ell\ell} \le$100~GeV.
We specifically consider the cross section differential in $|y|$.
The dependence on the transverse momentum of the dilepton pair is integrated
over. The precision of this measurement, excluding the luminosity uncertainty,
ranges from 0.2\%, for $|y|\le 2.0$, to 0.9\% at more forward
rapidities. This measurement is based on a re-analysis of events that were
previously used in another two measurements~\cite{ATLAS:2016gic,ATLAS:2017rue}
from which double- and triple-differential distributions in the fiducial region
for the final-state leptons were reconstructed. The distributions are
differential, respectively, in the invariant mass $m_{\ell\ell}$ and absolute
rapidity $|y|$ of the dilepton pair, and in $m_{\ell\ell}$,
$|y|$, and the cosine of the Collins-Soper angle, $\cos\theta^*$. The
covered invariant mass region extends below, across and above the $Z$ peak.
The double differential measurement was included in the
MSHT20~\cite{Bailey:2020ooq} and NNPDF4.0~\cite{NNPDF:2021njg} PDF fits.
For this reason, the new measurement~\cite{ATLAS:2023lsr} does not fulfil the
second selection criterion established at the beginning of this section.
We make an exception for this measurement because, first, it exhibits a
significant PDF dependence, and, second, it underlies the most precise
determination of the strong coupling ever performed at a hadron collider,
in which PDF uncertainties are the leading uncertainties.
In Sect.~\ref{subsec:ATLAS_8TEV} we will discuss the interplay between the
original~\cite{ATLAS:2016gic,ATLAS:2017rue} and new~\cite{ATLAS:2023lsr}
measurements.  

For all these measurements, theoretical predictions accurate to NNLO QCD are
computed in the form of {\sc PineAPPL} interpolation
grids~\cite{Carrazza:2020gss}  with
{\sc NNLOjet}~\cite{NNLOJET:2025xyz,Cruz-Martinez:2025ffa}. The computation
incorporates in particular the NNLO QCD corrections to the transverse momentum
distributions of the $Z$ boson from
Refs.~\cite{Gehrmann-DeRidder:2015wbt,Gehrmann-DeRidder:2016cdi}.
The central renormalisation and factorisation scales are set to
\begin{equation}
  \mu_F=\mu_R=\sqrt{m_{\ell\ell}^2+(p_T^{\ell\ell})^2}\,,\qquad\mu_F=\mu_R=M_V\,,
\end{equation}
respectively for the $Z$ transverse momentum distribution and the gauge boson rapidity distributions (with $M_V$ the gauge boson mass, $Z$ or $W$). In the former case,
we also apply a kinematic cut $p_T^{\ell\ell}>30$~GeV to remove the region where
resummation corrections, not accounted for by our fixed-order computation,
may be relevant~\cite{Boughezal:2017nla,Wiesemann:2020gbm}.
Electroweak, QED, and photon-induced corrections, though known, are not
considered here.

\subsection{Top quark pair production at the LHC}
\label{subsec:LHC_toppair}

Top-quark pair production at the LHC, which is initiated by gluon-gluon
scattering, primarily probes the gluon PDF at large
$x$~\cite{Czakon:2013tha,Czakon:2016olj,Bailey:2019yze,Catani:2019iny,Catani:2019hip,
Catani:2020tko,Kassabov:2023hbm,Alekhin:2024bhs}. In addition to their
PDF sensitivity, top-quark pair cross sections also constrain the top-quark
mass $m_t$ and the strong coupling
$\alpha_s(m_Z)$~\cite{CMS:2014rml,ATLAS:2022aof}. Here we
consider the ATLAS measurement reconstructed from the
all-hadronic~\cite{ATLAS:2020ccu} and lepton+jets~\cite{ATLAS:2019hxz} final
states, and the CMS measurement reconstructed from the lepton+jets final
state~\cite{CMS:2021vhb}. We consider only parton-level measurements
presented in terms of the kinematic variables of the final-state top quarks.
The reason being that theoretical computations accurate to NNLO QCD for
particle-level measurements~\cite{Czakon:2020qbd} are not available in a
numerical format suitable for this analysis. All these measurements were taken
during the LHC Run II, at a centre-of-mass energy of 13 TeV.

The ATLAS measurements correspond to an integrated luminosity of 36~fb$^{-1}$.
Cross sections are provided, absolute and normalised to the total cross section,
as single- and double-differential distributions in various kinematic variables.
For the sake of this work, we consider a subset of them, either the normalised
or the absolute differential distributions. The choice depends on the stability
of the experimental covariance matrix, as we will explain in
Sect.~\ref{subsec:compatibility}. For the all-hadronic measurement, we choose
the single-differential absolute (normalised) distributions in the invariant
mass (absolute rapidity) of the top-quark pair, $m_{t\bar t}$ ($y_{t\bar t}$), and
the double-differential absolute distribution in these two variables. For the
lepton+jets measurement, we choose the single-differential normalised
distributions in the invariant mass of the top quark pair, $m_{t\bar t}$, in the
transverse momentum of the top quark, $p_T^t$, in the absolute rapidity of the
top-quark pair, $y_{t\bar t}$, and in the absolute rapidity of the top-quark,
$y_t$.

The CMS measurement corresponds to an integrated luminosity of 137~fb$^{-1}$,
that is, all the events recorded during the LHC Run II.
Absolute and normalised cross sections are provided as single- and
double-differential distributions in various kinematic variables. We
select a subset of them, specifically the single-differential normalised
distributions in $m_{t\bar{t}}$, $|y_{t\bar{t}}|$ , $|y_{t}|$, and $p_{T}^t$, and the
double-differential normalised distribution in $(m_{t\bar{t}},|y_{t\bar{t}}|)$.

Theoretical predictions accurate to NNLO QCD are computed using
{\sc MATRIX}~\cite{Grazzini:2017mhc}, which has been interfaced to 
{\sc PineAPPL}~\cite{Carrazza:2020gss}. The central factorisation and
renormalisation scales are set to 
\begin{equation}
  \mu_R =\mu_F=H_T/2=\sqrt{m_t^2+\left( p_T^t\right)^2}\Big/2\,,
\end{equation}
where $m_t$ and $p_T^t$ are the mass and the transverse momentum of the top
quark. This choice follows the recommendation of~\cite{Czakon:2016dgf}. A value
of $m_t^{\rm pole}=172.5$~GeV has been used for the top-quark pole mass,
consistently with the latest PDG average~\cite{ParticleDataGroup:2024cfk}.
These computations have been benchmarked, when possible, against {\sc FastNLO}
tables~\cite{Czakon:2017dip} generated with the code presented
in~\cite{Czakon:2015owf}. Electroweak, QED, and photon-induced cross sections
are not included.

\subsection{Jet production at the LHC}
\label{subsec:LHC_jets}

Collider jet production at the LHC is a traditional probe of the gluon PDF in
the large-$x$ region~\cite{Rojo:2014kta,AbdulKhalek:2020jut,Cridge:2023ozx,
  Ablat:2024uvg}, though it provides also information on the large-$x$ quark
PDFs. Here we consider the ATLAS measurement of single-inclusive jet and di-jet
production~\cite{ATLAS:2017ble}, and the CMS measurement of single-inclusive jet
production~\cite{CMS:2021yzl}. Both of them  were taken during the LHC
Run II, at a centre-of-mass energy of 13~TeV.

The ATLAS measurements correspond to an integrated luminosity
of 3.2~fb$^{-1}$. Whereas this amounts to only a small fraction
of the events recorded during Run II, no other unfolded measurements of
single-inclusive jet or di-jet production based on a larger sample have been
presented by ATLAS to date. The single-inclusive jet measurement is presented
as a set of double differential cross sections in the jet transverse momentum
$p_T$, with 100 GeV $\leq p_T\leq 3.5$~TeV, and the jet absolute rapidity
$|y|$, with $|y|<3.0$. The di-jet measurement is presented as
a set of double differential cross sections in the di-jet invariant mass
$m_{jj}$, with 300 GeV $\leq m_{jj}\leq 9$~TeV, and the half absolute rapidity
separation between the two leading jets, $|y^*|$, with $|y^*|<3.0$.
Single-inclusive jets and di-jets are reconstructed by means of the anti-$k_T$
clustering algorithm~\cite{Cacciari:2008gp} for a jet radius of $R=0.4$.
The full breakdown of correlated systematic uncertainties is available,
separately for the single-inclusive jet and di-jet measurements, and taken into
account.

The CMS measurement corresponds to an integrated luminosity of 
36.3 (33.5) fb$^{-1}$ and a jet radius of $R=0.4$ ($R=0.7$).
We consider the measurement with $R=0.4$.
Cross sections are double differential in the jet transverse momentum $p_T$,
with $97 \leq p_T\leq 3.1$~TeV, and in the jet absolute rapidity $|y|$,
with $|y|<2.0$. The experimental covariance matrix of the measurement is
provided and taken into account.

For all the aforementioned measurements, theoretical predictions, accurate to
NNLO QCD in the leading color approximation, were computed with the
{\sc NNLOjet} code~\cite{Gehrmann-DeRidder:2019ibf}. The central
factorisation and renormalisation scales where chosen as
\begin{equation}
  \mu_F=\mu_R=p_T\,,\qquad \mu_F=\mu_R=m_{jj}\,,
\end{equation}
respectively for single-inclusive jets and di-jets. These predictions were
released~\cite{Britzger:2022lbf} as interpolation grids in the
{\sc APPLfast} format through the {\sc Ploughshare}
website~\cite{Ploughshareurl}. For the sake of this work, these
grids have been converted to the {\sc PineAPPL}
format~\cite{Carrazza:2020gss}. We do not account for NLO electroweak
corrections or photon-initiated contributions, neither for single-inclusive
jets nor for di-jets. Monte Carlo uncertainties due to the generation of a
finite number of events are generally smaller than MHO and $\alpha_s$
uncertainties, and are thus ignored.

\subsection{Jet production at HERA}
\label{subsec:HERA_jets}

Jet production in DIS can probe the gluon PDF at large $x$ as well. This
process was measured at HERA by the H1 and ZEUS experiments and demonstrated
to constrain the gluon PDF and the strong
coupling~\cite{ZEUS:2023zie,H1:2021xxi,H1:2017bml} in comparison to fits based
on inclusive DIS measurements only. Here we consider four H1
measurements~\cite{H1:2010mgp,H1:2016goa,H1:2014cbm} and 
three ZEUS measurements~\cite{ZEUS:2002nms,ZEUS:2006xvn,ZEUS:2010vyw} for
single-inclusive jet and di-jet cross sections, as indicated in
Table~\ref{tab:input_datasets}.

The H1 measurements correspond to the HERA-II data-taking period with a
centre-of-mass energy of 319~GeV. Two pairs of single-inclusive jet
and di-jet measurements are available, which focus on different regions of the
exchanged boson virtuality $Q^2$: a low-$Q^2$ pair,
$5.5\leq Q^2\leq 80$~GeV$^2$, and a high-$Q^2$ pair,
$150\leq Q^2\leq 15000$~GeV$^2$. The integrated luminosity is, respectively,
290~pb$^{-1}$ and 351~pb$^{-1}$. On top of the virtuality $Q^2$, cross sections
are differential in the jet transverse momentum $p_T$ or the di-jet average
momentum $\la p_T\ra$, respectively for the single-inclusive jet and the
di-jet measurements. Massless jets are identified using the $k_T$ algorithm 
with the $R$ parameter set to $R=1$. Experimental correlations are available for
all measurements, including for points in different single-inclusive jet and
di-jet bins at different $Q^2$.

The ZEUS measurements correspond to the HERA-I data-taking period, with a
centre-of-mass energy of 300~GeV and an integrated luminosity of 38~pb$^{-1}$,
and to the HERA-II data-taking period, with a centre-of-mass energy of
319~GeV and an integrated luminosity of 82~pb$^{-1}$ and 374~pb$^{-1}$.
Cross sections are presented as differential distributions in the
vector boson virtuality $Q^2$, with $Q^2\geq 125$~GeV$^2$, and the jet
transverse energy $E_T$ or the di-jet average transverse energy
$\la E_T\ra$, respectively for the single-inclusive jet and the di-jet
measurements. Experimental correlations are available across bins within the
same set, but not across bins in different datasets. Unlike inclusive DIS
structure functions~\cite{Abramowicz:2015mha}, no combination between the H1
and ZEUS results exists to date. A final ZEUS measurement of single-inclusive
jet production cross has been published recently~\cite{ZEUS:2023zie}, however
we do not consider it because NNLO QCD corrections to matrix elements are not
readily available in a format suitable for fast convolution with PDFs. 

Theoretical predictions accurate to NNLO QCD were computed with the
{\sc NNLOjet} code~\cite{Currie:2016ytq,Currie:2017tpe} in the
zero-mass variable-flavour-number scheme. The central factorisation and
renormalisation scales are
\begin{equation}
  \mu=\mu_F=\mu_R=Q^2+(p_T)^2\,,\qquad \mu=\mu_F=\mu_R=Q^2+\la p_T\ra^2\,,
\end{equation}
respectively for single-inclusive jets and di-jets.
Data points for which $\mu\leq 10$~GeV were excluded to ensure that the
scale is larger than the $b$-quark mass. This is necessary because jets
are built from massless partons. As in the case of LHC jets, theoretical
predictions were released as interpolation grids through
the {\sc Ploughshare} website~\cite{Ploughshareurl}.
We convert these grids to the {\sc PineAPPL} format~\cite{Carrazza:2020gss}.

\section{Methodological framework}
\label{sec:approach}

In this section, we describe how we quantitatively assess the agreement between
the measurements presented in Sect.~\ref{sec:data} and the corresponding
theoretical predictions for different PDF sets. We first define the figure of
merit that we use, and specifically explain how we take into account
experimental and theoretical uncertainties in it. We then discuss how this
figure of merit may become misleading if the experimental covariance matrix is
ill-conditioned, and illustrate how we identify and handle such cases.
We finally review the PDF sets that we consider as input for the
computation of the theoretical predictions.

\subsection{Figure of merit}
\label{subsec:def}

We quantify the agreement between experimental data and theoretical predictions
by computing the (reduced) $\chi^2$ for each dataset
\begin{equation}
  \chi^{2}
  =
  \frac{1}{n_{\rm dat}}
  \sum_{i, j=1}^{n_{\mathrm{dat}}}
  \left(T^{(0)}_i-D_i\right)
  \left(\operatorname{cov}^{-1}\right)_{i j}
  \left(T^{(0)}_j-D_j\right)\,,
  \label{eq:chi2}
\end{equation}
where $n_{\mathrm{dat}}$ is the number of data points in the considered dataset,
$\{ D_i\}$ are the central values of the experimental data, $\{ T_i^{(0)}\}$
are the corresponding theoretical predictions, and $\operatorname{cov}_{i j}$
is the covariance matrix. Theoretical predictions are evaluated, for both Monte
Carlo and Hessian PDF sets, as the prediction obtained with the central PDF
$f^{(0)}$, $T_i^{(0)}=T_i(f^{(0)})$. Note that the $\chi^2$ in Eq.~\eqref{eq:chi2}
is normalised to the number of data points. Therefore, in case of perfect
agreement between data and theory, one expects $\chi^{2}\sim 1$,
with statistical fluctuations of the order of the standard deviation of the
$\chi^2$ distribution, $\sigma_{\chi^2}=\sqrt{2/n_{\mathrm{dat}}}$. 

To evaluate Eq.~\eqref{eq:chi2}, one needs to compute the covariance
matrix $\operatorname{cov}_{i j}$. In addition to experimental uncertainties,
one should consider all the relevant sources of theoretical uncertainties,
in particular, those associated to missing higher orders (MHO), to PDFs, and
to the value of the strong coupling $\alpha_s(m_Z)$. Assuming that all of these
theoretical uncertainties follow a Gaussian distribution and that they are
mutually independent, they can be incorporated in the covariance matrix
following the formalism developed in~\cite{NNPDF:2019ubu,NNPDF:2024dpb}.
Specifically, in this formalism the covariance matrix in Eq.~\eqref{eq:chi2}
reads as
\begin{equation}
  \operatorname{cov}_{ij}
  =
  \left(\operatorname{cov}_{\rm exp}\right)_{ij}
  +
  \left(\operatorname{cov}_{\rm th}\right)_{ij}\,,
\end{equation}
where the theory covariance matrix is in turn the sum of a MHO, PDF, and
$\alpha_s$ contributions
\begin{equation}
\label{eq:total_th_covmat}
  \left(\operatorname{cov}_{\rm th}\right)_{ij}
  =
  \left(\operatorname{cov}_{\rm mho}\right)_{ij}
  +
  \left(\operatorname{cov}_{\rm pdf}\right)_{ij}
  +
  \left(\operatorname{cov}_{\rm as}\right)_{ij}\,.
\end{equation}

The experimental covariance matrix is sometimes provided together with the
experimental measurements, otherwise, in most cases, we reconstruct it from 
knowledge of experimental uncertainties as
\begin{equation}
  \left(\operatorname{cov}_{\exp }\right)_{i j}
  =
  \delta_{i j}\sigma_i^{\text {(uncorr) }}\sigma_j^{\text{(uncorr)}}
  +
  \sum_{\ell=1}^{n_{\rm corr}}\sigma_{i, \ell}^{(\rm corr)}\sigma_{j, \ell}^{(\rm corr)}\,.
  \label{eq:covexp}
\end{equation}
Here $\sigma_i^{\text {(uncorr)}}$ is the sum in quadrature of all the 
uncorrelated uncertainties, and $\sigma_{i, \ell}^{(\rm corr)}$ is the
set of $n_{\rm corr}$ correlated uncertainties. These could be
additive or multiplicative, however this distinction is not relevant here,
given that Eq.~\eqref{eq:chi2} is only used to quantify the agreement
between data and theory. In a fit of PDFs, this distinction is instead relevant
because multiplicative uncertainties may bias the determination of the PDF
central value and variance. Therefore they would require a specific treatment,
by re-defining either the experimental covariance matrix with the $t_0$
prescription~\cite{Ball:2009qv} or the figure of merit with 
additional nuisance parameters~\cite{Pumplin:2002vw}. Note that whenever we
reconstruct the experimental covariance matrix using Eq.~\eqref{eq:covexp},
we implicitly assume that correlated uncertainties are
100\% correlated, given that no specific correlation model is provided for the
considered datasets, see Sect.~\ref{sec:data}. Decorrelation remains however
possible, using the procedure summarised in Sect.~\ref{subsec:compatibility},
and we will actually make use of it, as discussed further below.

The contribution to the covariance matrix due to MHO is evaluated
following~\cite{NNPDF:2019ubu,NNPDF:2024dpb}. Specifically, MHO are estimated
as the difference between theoretical predictions computed with fixed and varied
renormalisation and factorisation scales, $\mu_R$ and $\mu_F$.
Several prescriptions defining scale variations are possible. As is common
practice in LHC analyses, we adopt the 7-point variation prescription,
which gives the MHO covariance matrix
\be
\left(\operatorname{cov}_{\rm mho}\right)_{i j}
=
\smallfrac{1}{3}
\big\{
\Delta_i^{+0} \Delta_j^{+0} + \Delta_i^{-0}\Delta_j^{-0}
+ \Delta_i^{0+} \Delta_j^{0+} + \Delta_i^{0-}\Delta_j^{0-}+ \Delta_i^{++}\Delta_j^{++}
+ \Delta_i^{--} \Delta_j^{--}
\big\}\,,
\label{eq:cov_mat}
\ee
where, for each data point $i,j$, the shifts are defined as
\be
\Delta_i\lp \kappa_R,\kappa_F\rp
=
T_i\lp \mu_R=\kappa_R\mu^{(0)}_R,\mu_F=\kappa_R\mu^{(0)}_F\rp
-
T_i\lp\mu_R^{(0)},\mu_F^{(0)}\rp\,,
\label{eq:shift_theory_predictions}
\ee
with $\lp\mu_R^{(0)},\mu_F^{(0)}\rp$ the central renormalisation and factorisation
scales and
\be
\begin{array}{lll}
\Delta_i^{+0}
=
\Delta_i\lp 2,1\rp \,,
&
\Delta_i^{-0}
=
\Delta_i\lp 1/2,1\rp \,,
&
\Delta_i^{0+}
=
\Delta_i\lp 1,1/2\rp \,,\\
\Delta_i^{0-}
=
\Delta_i\lp 1,1/2\rp\,,
&
\Delta_i^{++}
=
\Delta_i\lp 2,2\rp\,,
&
\Delta_i^{--}
=
\Delta_i\lp 1/2,1/2\rp\,.
\end{array}
\ee
The shifts in the NNLO theory predictions associated to the scale variations,
Eq.~\eqref{eq:shift_theory_predictions}, are directly evaluated from the
{\sc PineAPPL} grids~\cite{Carrazza:2020gss}. In general, the 7-point MHO
theory covariance matrix defined by Eq.~\eqref{eq:cov_mat} differs from the
envelope prescription to estimate MHO uncertainties frequently used in LHC
studies.

The contribution to the covariance matrix due to PDF uncertainties is
determined, for each of the PDF sets considered (see Sect.~\ref{subsec:pdfs}),
using the definition of covariance between the random variables $T_i$ and $T_j$
\begin{equation}
  \left(\operatorname{cov}_{\rm pdf}\right)_{i j}
  =
  \mathbb{E}\left[(T_i-\mathbb{E}[T_i])(T_j-\mathbb{E}[T_j]) \right]\,,
  \label{eq:pdf_covariance}
\end{equation}
where $\mathbb{E}[X]$ denotes the expectation value of the random variable $X$.
For Hessian PDF sets, Eq.~\eqref{eq:pdf_covariance} reads
\begin{equation}
  \left(\operatorname{cov}_{\rm pdf}\right)_{i j}
  =
  \sum_{k=1}^{n_{\text{eig}}} \left(T_i^{(k)}
  -
  T_i^{(0)} \right) \left(T_j^{(k)} - T_j^{(0)} \right)\,,
  \label{eq:covpdf_hessian}
\end{equation}
where $T_{i}^{(k)}=T_{i}(f^{(k)})$ is the theoretical prediction computed with
the PDF associated to the $k$-th eigenvalue $f^{(k)}$, and
$T_{i}^{(0)}=T_{i}(f^{(0)})$ is the theoretical prediction computed with the
central PDF $f^{(0)}$. We use the symmetric definition of Hessian PDF
uncertainties, since we assume that PDF uncertainties are Gaussian. In case of
asymmetric Hessian PDF sets, we replace the difference between $T_{i,j}^{(k)}$
and $T_{i,j}^{(0)}$ in Eq.~(3.9) with the difference between predictions obtained
with positive and negative eigenvectors. For Monte Carlo PDF sets,
Eq.~\eqref{eq:pdf_covariance} reads
\begin{equation}
  \left(\operatorname{cov}_{\rm pdf}\right)_{i j}
  =
  \frac{1}{n_{\rm rep}}
  \sum_{k=1}^{n_{\rm rep}}
  \left(T_i^{(k)}-\left\langle T_i \right\rangle_{\rm rep} \right)
  \left(T_j^{(k)}-\left\langle T_j \right\rangle_{\rm rep} \right)\,,
  \label{eq:covpdf_mc}
\end{equation}
where $T_{i}^{(k)}=T_{i}(f^{(k)})$ is the theoretical prediction computed with
the PDF associated to the $k$-th replica $f^{(k)}$, and
$\left\langle T_{i}\right\rangle_{\rm rep}=\frac{1}{n_{\rm rep}}\sum_{k=1}^{n_{\rm rep}}T_{i}^{(k)}$ is the average over replicas.

The contribution to the covariance matrix due to the uncertainty of the value
of $\alpha_s(m_Z)$ is determined as follows. We take
$\alpha_s(m_Z)=0.118\pm 0.001$ for all PDF sets considered, consistently with
the latest PDG average~\cite{ParticleDataGroup:2024cfk}, and we construct
\be
  \left(\operatorname{cov}_{\rm as}\right)_{i j}
  =
  \smallfrac{1}{2}
  \Big\{
  \Delta_{i,\alpha_s}^+\Delta_{j,\alpha_s}^+
  +
  \Delta_{i,\alpha_s}^-\Delta_{j,\alpha_s}^-
  \Big\}\,,
  \label{eq:covmat_th_alphas}
\ee
where, for each data point $i,j$,
\begin{equation}
  \begin{array}{lcl}
    \Delta_{i,\alpha_s}^+
    & \equiv
    & T_i(\alpha_s=0.119) - T_i(\alpha_s=0.118) \,, \\
    \Delta_{i,\alpha_s}^-
    & \equiv
    & T_i(\alpha_s=0.118) - T_i(\alpha_s=0.117) \,.
  \end{array}
\end{equation}
The value of $\alpha_s$ in the theory predictions is
varied consistently both in the matrix element and in the PDFs, a fact that
is streamlined thanks to the usage of {\sc PineAPPL} grids.
The combination of Eq.~\eqref{eq:covmat_th_alphas} with
Eq.~\eqref{eq:covpdf_hessian} (for a Hessian set) or
Eq.~\eqref{eq:covpdf_mc} (for a Monte Carlo set) reproduces the prescription
of~\cite{PDF4LHCWorkingGroup:2022cjn}, according to which PDF and 
$\alpha_s$ uncertainties are added in quadrature.

In Sect.~\ref{sec:results} we will quantify the agreement between experimental
data and theory predictions, obtained with different PDF sets, in terms of the
figure of merit given in Eq.~(\ref{eq:chi2}). When accounting for all sources
of experimental and theoretical uncertainties, we have
\begin{equation}
  \label{eq:chi2_v2}
  \chi^{2}_{\rm exp+th}
  =
 \frac{1}{n_{\rm dat}} \sum_{i, j=1}^{n_{\mathrm{dat}}}\left(T^{(0)}_i-D_i\right) \lp \lp  
  \operatorname{cov}_{\exp}
  + \operatorname{cov}_{\rm mho}
  + \operatorname{cov}_{\rm pdf}
  + \operatorname{cov}_{\rm as}\rp^{-1}\right)_{i j}\left(T^{(0)}_j-D_j\right)\, ,
\end{equation}
with the individual contributions to the covariance matrix combined in
quadrature. In order to understand the impact of the various sources of
uncertainties entering Eq.~\eqref{eq:chi2_v2}, we will also present results
for variants of this figure of merit restricted to a subset of the
uncertainties, in particular
\begin{equation}
  \label{eq:chi2_v3}
  \chi^{2}_{\rm exp}
  =
\frac{1}{n_{\rm dat}}  \sum_{i, j=1}^{n_{\mathrm{dat}}}\left(T^{(0)}_i-D_i\right) \lp \lp  
  \operatorname{cov}_{\exp}\rp^{-1}\right)_{i j}\left(T^{(0)}_j-D_j\right) \,,
\end{equation}
which contains only the experimental uncertainties, and
\begin{equation}
  \label{eq:chi2_v4}
  \chi^{2}_{\rm exp+mho}
  =
 \frac{1}{n_{\rm dat}} \sum_{i, j=1}^{n_{\mathrm{dat}}}\left(T^{(0)}_i-D_i\right) \lp \lp  
  \operatorname{cov}_{\exp}
  + \operatorname{cov}_{\rm mho}\rp^{-1}\right)_{i j}\left(T^{(0)}_j-D_j\right)\, ,
\end{equation}
defined without the contribution of the PDF and $\alpha_s$ uncertainties. 
In all cases, the figures of merit are presented normalised to the number of
data points of each dataset considered. We emphasise that, when evaluating
Eq.~\eqref{eq:chi2_v2}, PDFs enter in two different places: through the theory
predictions $T_i$ and through the PDF contribution to the total
covariance matrix in Eq.~\eqref{eq:total_th_covmat}.

To further assess the significance of $\chi_{\rm exp+th}^2$,
Eq.~\eqref{eq:chi2_v2}, as a measure of the agreement between experimental
data and theoretical predictions, we will make use of two additional estimators
in Sect.~\ref{sec:results}. The first estimator is the relative change in the
total $\chi^2$ due to the change of input PDF set for a given dataset
\begin{equation}
  \label{eq:delta_chi2}
  \Delta \chi^{2(i)}
  =
  \frac{\chi^{2(i)}_{\rm exp+th}-\left\langle\chi^{2}_{\rm exp+th}\right\rangle_{\rm pdfs}}{\left\langle \chi^{2}_{\rm exp+th} \right\rangle_{\rm pdfs}}\,,
\end{equation}
where the index $i$ runs over the $n_{\rm pdfs}$ input PDF sets considered in the
analysis (see Sect.~\ref{subsec:pdfs}), and the average over PDF sets is
evaluated as
\be
\label{eq:delta_chi2_average}
\left\langle \chi^{2}_{\rm exp+th} \right\rangle_{\rm pdfs}
=
\frac{1}{n_{\rm pdfs}}\sum_{i=1}^{n_{\rm pdfs}} \chi^{2(i)}_{\rm exp+th} \, .
\ee
By construction, $\sum_i \Delta \chi^{2(i)}=0$. This estimator gauges the
relative change in the value of the $\chi^2$ for a given PDF set with respect
to the average evaluated over all PDF sets considered. It therefore allows one
to disentangle PDF-related effects in the $\chi^2$ from other effects.

The second estimator quantifies the difference of the $\chi^2$, computed
with a given PDF, with respect to the $\chi^2$ averaged over all PDF sets
in terms of the number of standard deviations of the $\chi^2$ distribution
\begin{equation}
  \label{eq:delta_nsigma}
  \Delta n_{\sigma}^{(i)}
  = \frac{ \chi^{2(i)}_{\rm exp+th} - \left\langle \chi^{2}_{\rm exp+th} \right\rangle_{\rm pdfs}}{\sqrt{2/n_{\rm data}}} \,.
\end{equation}
This estimator allows one to compare the $\chi^2$ variation due to the choice
of PDF to the expected statistical fluctuations of the $\chi^2$, and therefore
check if this is significant or not. Note indeed that several of the datasets
considered contain a relatively small number of data points, so that a large
relative change of the $\chi^2$ in Eq.~\eqref{eq:delta_chi2} may be simply
explained by large fluctuations due to the small data sample.

\subsection{Stability of the experimental covariance matrix}
\label{subsec:compatibility}

The interpretation of the agreement of theoretical predictions with
experimental data, as quantified by the value of the $\chi^2$, requires
some care. As discussed in Ref.~\cite{Kassabov:2022pps}, an inaccurate
determination of experimental uncertainty correlations, in otherwise very
precise data, may result in an ill-conditioned experimental covariance matrix,
which leads in turn to anomalously large values of the $\chi^2$.

A metric to measure the conditioning of an experimental covariance matrix was
introduced in Ref.~\cite{Kassabov:2022pps}, see, in particular, Eq.~(26).
This was defined as the inverse of the smallest singular value of the
experimental correlation matrix, and called condition number $Z$. The value
$(\sqrt{2}Z)^{-1}$ was then demonstrated to be related to the amount by which
experimental correlations need to be determined to ensure that the $\chi^2$
remains stable, namely that it does not vary by more than one standard
deviation, $\sigma_{\chi^2}=\sqrt{2/n_{\rm dat}}$. A large value of $Z$ indicates a
dataset for which small variations of the correlation model can potentially
lead to large $\chi^2$ variations for unchanged data and theory
and vice-versa. In Ref.~\cite{Kassabov:2022pps} a reasonable threshold was
defined to be $Z=4$. This value corresponds to assuming that correlations on
uncertainties of the order of a few percent, such as those that
affect the LHC measurements considered in this work, be estimated with an
absolute uncertainty of less than $0.18$. This means that if the correlation
between two bins is estimated to be $1.00$, while its real value is $0.82$, one
can expect that the $\chi^2$ deviates from one by more than $1\sigma$ even if
experimental data and theoretical predictions are perfectly consistent. Note
that the smaller the uncertainty, the higher the required precision with
which correlations need be known to be within a $1\sigma$ variation of the
$\chi^2$. As explicitly shown in Ref.~\cite{Kassabov:2022pps} as part of a toy
model (see in particular Eq.~(29) and Fig.~3), for a 1\% (2.5\%) uncertainty,
correlations ought to be known with a precision of 2\% (12\%). The choice
$Z=4$ (the same for all experiments) should therefore be seen as a practical
diagnostic choice.

In some cases, a large value of $Z$ may not imply a pathological behaviour of
the experimental data. A typical case is the one in which the luminosity
uncertainty, which by definition is 100\% correlated across all bins
of a given dataset, is the largest of all uncertainties. In this case, we
expect the condition number $Z$ to be large. The same would happen with any
experimental uncertainty that is fully correlated across bins for specific
experimental reasons. In these cases, we should compute $Z$ upon excluding these
uncertainties from the reconstruction of the experimental covariance matrix.
For the sake of this work, we single out only the luminosity uncertainty as
100\% correlated, given that we do not have complete knowledge of which
other uncertainties are also undoubtedly 100\% correlated. We then split the
experimental covariance matrix into two components, one that contains only the
luminosity uncertainty, and one that contains all of the other uncertainties.
We compute $Z$, which we call $Z_{\mathcal{L}}$, for the latter covariance matrix
and regularise it, if necessary; we then compute the $\chi^2$ using the sum of
the original luminosity covariance matrix and the regularised covariance matrix.
Clearly, this procedure might decorrelate systematic uncertainties that owe to
be 100\% correlated, or decorrelate too much some other systematic
uncertainties. Nevertheless, we consider that the procedure remains a useful
diagnosis tool when information on specific decorrelation models, provided
under the experimental guidance, are not available. In this respect, we shall
also note that here our aim is not to characterise a dataset for inclusion (or
not) in a PDF determination, but rather to comparatively assess the ability of
various PDF sets to describe the data. We consider that the application of our
regularisation procedure does not alter our judgement on such an ability (see
Appendix~\ref{app:unreg}).

An alternative estimator to assess the conditioning
of the experimental correlation matrix, sometimes used in experimental
analyses, is $\lambda_\rho$, defined as the ratio of the smallest to the
largest eigenvalues of the experimental correlation matrix. A small value of
$\lambda_\rho$ indicates a large spread of eigenvalues, with the directions
associated to the smallest ones almost degenerate. These degeneracies are
those that lead to a ill-conditioned matrix.

In Table~\ref{tab:Z} we display, for each dataset listed in
Table~\ref{tab:input_datasets} and separately for each observable,
the number of data points, $n_{\rm dat}$, and the condition numbers
$\lambda_\rho$, $Z$, and $Z_{\mathcal{L}}$. For normalised
distributions $Z=Z_{\mathcal{L}}$ by construction. For datasets which do not
provide the breakdown of systematic uncertainties but instead only the overall
covariance matrix, $Z_{\mathcal{L}}$ is computed by subtracting from this
covariance matrix a covariance matrix constructed only from the
100\%-correlated luminosity uncertainty. In the case of the CMS top-quark
pair distribution, this procedure is however not applied, given that the
measurement is the combination of events recorded with different luminosities. 
We therefore leave the corresponding entry blank in Table~\ref{tab:Z}.
Whenever a dataset is presented in different variants, for example as absolute
or normalised distributions or for two different values of the jet radius $R$,
we indicate with a (*) the one used in Sect.~\ref{sec:results}. We select
the distributions that feature the lowest value of $Z_{\mathcal{L}}$. 

\begin{table}[!t]
  \centering
  \scriptsize
  \renewcommand{\arraystretch}{1.43}
  \begin{tabularx}{\textwidth}{Xrccc}
\toprule
Process
& $n_{\rm dat}$
& $\lambda_{\rho}$
& $Z$
& $Z_{\mathcal{L}}$
\\
\midrule
ATLAS 13~TeV $Z$ $1/\sigma d\sigma/dp_{T}^{\ell\ell}$
& 38
& $1.9\times 10^{-1}$
& 1.10
& 1.10
\\
CMS 13~TeV $W^+$ $d\sigma/d|\eta|$
& 18
& $8.3\times 10^{-5}$
& 25.1
& 19.0
\\
CMS 13~TeV $W^-$ $d\sigma/d|\eta|$
& 18
& $8.9\times 10^{-5}$
& 26.0
& 18.0
\\
LHCb 13~TeV $Z$ $d\sigma/dy^Z$
& 17
& $1.9\times 10^{-3}$
& 5.92
& 2.09
\\
ATLAS 8~TeV $Z$ $d\sigma/d|y|$
& 7
& $3.2\times 10^{-4}$
& 21.6
& 2.10
\\
\midrule
ATLAS 13~TeV $t\bar{t}$ all hadr. $d\sigma/dm_{t\bar{t}}$~~(*)
&  9
& $2.4\times 10^{-3}$
& 7.27
& 7.24
\\
ATLAS 13~TeV $t\bar{t}$ all hadr. $1/\sigma\ d\sigma/dm_{t\bar{t}}$
&  9
& $3.9\times 10^{-5}$
& 64.7
& 64.7
\\
ATLAS 13~TeV $t\bar{t}$ all hadr. $d\sigma/d|y_{t\bar{t}}|$
& 12
& $3.3\times 10^{-3}$
& 5.27
& 5.25
\\
ATLAS 13~TeV $t\bar{t}$ all hadr. $1/\sigma\ d\sigma/d|y_{t\bar{t}}|$~~(*)
& 12
& $8.9\times 10^{-2}$
& 1.77
& 1.77
\\
ATLAS 13~TeV $t\bar{t}$ all hadr. $d^2\sigma/dm_{t\bar{t}}\ d|y_{t\bar{t}}|$~~(*)
& 11
& $4.4\times 10^{-3}$
& 4.83
& 4.81
\\
ATLAS 13~TeV $t\bar{t}$ all hadr. $1/\sigma\ d^2\sigma/dm_{t\bar{t}}\ d|y_{t\bar{t}}|$
& 11
& $9.4\times 10^{-5}$
& 52.1
& 52.1
\\
ATLAS $t\bar{t}$ $\ell+j$ $d\sigma/dm_{t\bar{t}}$
&  9
& $5.2\times 10^{-4}$
& 16.2
& 15.9
\\
ATLAS 13~TeV $t\bar{t}$ $\ell+j$  $1/\sigma\ d\sigma/dm_{t\bar{t}}$~~(*)
&  9
& $3.0\times 10^{-3}$
& 7.62
& 7.62
\\
ATLAS 13~TeV $t\bar{t}$ $\ell+j$  $d\sigma/dp_T^t$
&  8
& $5.8\times 10^{-4}$
& 16.8
& 16.6
\\
ATLAS 13~TeV $t\bar{t}$ $\ell+j$  $1/\sigma\ d\sigma/dp_T^t$~~(*)
&  8
& $2.5\times 10^{-3}$
& 8.46
& 8.46
\\
ATLAS 13~TeV $t\bar{t}$ $\ell+j$ $d\sigma/ d|y_{t}|$
&  5
& $1.5\times 10^{-3}$
& 11.7
& 11.5
\\
ATLAS 13~TeV $t\bar{t}$ $\ell+j$ $1/\sigma\ d\sigma/d|y_{t}|$~(*)
&  5
& $9.6\times 10^{-2}$
& 2.06
&  2.06
\\
ATLAS 13~TeV $t\bar{t}$ $\ell+j$ $d\sigma/d|y_{t\bar{t}}|$
&  7
& $6.2\times 10^{-4}$
& 15.7
& 15.4
\\
ATLAS 13~TeV $t\bar{t}$ $\ell+j$   $1/\sigma\ d\sigma/d|y_{t\bar{t}}|$~(*)
&  7
& $7.8\times 10^{-2}$
& 2.26
& 2.26
\\
CMS 13~TeV $t\bar{t}$ $\ell+j$   $d\sigma/dm_{t\bar{t}}$
& 15
& $1.1\times 10^{-2}$
& 3.90
&  ----
\\
CMS 13~TeV $t\bar{t}$ $\ell+j$ $1/\sigma\ d\sigma/dm_{t\bar{t}}$~(*)
& 15
& $3.0\times 10^{-2}$
& 3.51
& 3.51
\\
CMS 13~TeV $t\bar{t}$ $\ell+j$  $d\sigma/dp_T^t$
& 16
& $7.5\times 10^{-3}$
& 4.04
& ---
\\
CMS 13~TeV $t\bar{t}$ $\ell+j$   $1/\sigma\ d\sigma/dp_T^t$~(*)
& 16
& $1.3\times 10^{-1}$
& 1.78
& 1.78
\\
CMS $t\bar{t}$ $\ell+j$ $d\sigma/d|y_t|$
& 11
& $3.3\times 10^{-3}$
& 5.75
& ---
\\
CMS 13~TeV $t\bar{t}$ $\ell+j$   $1/\sigma\ d\sigma/d|y_t|$~(*)
& 11
& $2.7\times 10^{-1}$
& 1.36
& 1.36
\\
CMS 13~TeV $t\bar{t}$ $\ell+j$ $d\sigma/d|y_{t\bar{t}}|$
& 10
& $1.2\times 10^{-3}$
& 9.68
& ---
\\
CMS 13~TeV $t\bar{t}$ $\ell+j$   $1/\sigma\ d\sigma/d|y_{t\bar{t}}|$~(*)
& 10
& $1.9\times 10^{-1}$
& 1.53
& 1.53
\\
CMS 13~TeV $t\bar{t}$ $\ell+j$ $d^2\sigma/dm_{t\bar{t}}\ d|y_{t\bar{t}}|$
& 35
& $8.1\times 10^{-5}$
& 22.4
& ---
\\
CMS 13~TeV $t\bar{t}$ $\ell+j$j $1/\sigma\ d^2\sigma/dm_{t\bar{t}}\ d|y_{t\bar{t}}|$~(*)
& 35
& $1.8\times 10^{-4}$
& 17.2
& 17.2
\\
\midrule
ATLAS 13~TeV single-inclusive jets  $d^2\sigma/dp_Td|y|$
& 177
& $2.6\times 10^{-5}$
& 16.9
& 16.2
\\
CMS 13~TeV single-inclusive jets ($R=0.4$) $d^2\sigma/dp_Td|y|$~(*)
& 78
& $1.1\times 10^{-4}$
& 13.3
& 13.1
\\
CMS 13~TeV single-inclusive jets ($R=0.7$) $d^2\sigma/dp_Td|y|$
& 78
& $9.0\times 10^{-5}$
& 14.8
& 14.5
\\
ATLAS 13~TeV di-jets $d^2\sigma/dm_{jj}d|y^*|$
& 136
& $3.8\times 10^{-5}$
& 16.8
& 15.6
\\
\midrule
H1 single-inclusive-jets (low $Q^2$) $d^2\sigma/dQ^{2} dp_T$
& 48
& $7.6\times 10^{-3}$
& 6.00
& 5.91
\\
H1 single-inclusive-jets (high $Q^2$) $d^2\sigma/dQ^{2} dp_T$
& 24
& $7.0\times 10^{-3}$
& 1.46
&  1.19
\\
ZEUS single-inclusive jets (low luminosity) $d^2\sigma/dQ^2dE_{T}$
& 30
& $5.0\times 10^{-2}$
& 1.87
& 1.82
\\
ZEUS single-inclusive jets (high luminosity) $d^2\sigma/dQ^2dE_{T}$
& 30
& $1.9\times 10^{-2}$
& 2.56
& 2.43
\\
H1 di-jets (low $Q^2$) $d^2\sigma/dQ^{2} d \langle p_T \rangle$
& 48
& $9.0\times 10^{-2}$
& 7.67
& 7.42
\\
H1 di-jets (high $Q^2$) $d^2\sigma/dQ^{2}\ d \langle p_T \rangle$
& 24
& $1.0\times 10^{-1}$
& 1.60
& 1.45
\\
ZEUS di-jets $d^2\sigma/dQ^2d\langle E_{T} \rangle $
& 22
& $1.5\times 10^{-2}$
& 2.83
& 2.72
\\
\bottomrule
\end{tabularx}

  \vspace{0.3cm}
  \caption{The number of data points, $n_{\rm dat}$, the condition numbers
    $\lambda_\rho$, $Z$, and $Z_{\mathcal{L}}$ for all datasets considered, see
    the text for their definition. When the $Z_{\mathcal{L}}$ estimator cannot be
    unambiguously computed (as explained in the text) the corresponding entry is
    left blank. Whenever different variants or distributions exist for
    a dataset, we indicate with a (*) the one used in Sect.~\ref{sec:results}.}
  \label{tab:Z}
\end{table}

The values of the condition numbers $\lambda_\rho$ and $Z$ reported in
Table~\ref{tab:Z} consistently indicate that the experimental correlation and
covariance matrices are ill-conditioned for a subset of the analyzed datasets,
according to the criterion of Refs.~\cite{Kassabov:2022pps,NNPDF:2021njg}
($Z>4$). For some of them, such as the ATLAS $d\sigma^Z/d|y_{\ell\bar{\ell}}|$
measurement at 8~TeV, and to a lesser extent for LHCb
$d\sigma^Z/dy_{\ell\bar{\ell}}$, this high $Z$ value is explained by the dominance
of the luminosity uncertainty: in these cases, $Z_{\mathcal{L}}$ is indeed
significantly smaller than $Z$. For all the other datasets,
$Z\sim Z_{\mathcal{L}}$. Relatively high values of $Z$ are found for the ATLAS and
CMS single-inclusive jet and di-jet datasets, a fact that was already observed
in the case of the corresponding measurements at 8~TeV, for which various
decorrelation models have been proposed and
tested~\cite{ATLAS:2017kux,Harland-Lang:2017ytb,AbdulKhalek:2020jut,
  ATLAS:2021vod,Kassabov:2022pps}. We finally observe that the value of $Z$ can
fluctuate by a large amount across different differential
measurements in the same dataset. For instance, the 13~TeV ATLAS
$t\bar{t}$ hadronic dataset provides single-differential distributions in
$m_{t\bar t}$ and in $|y_{t\bar t}|$, associated to values of $Z$ respectively of
$64.5$ and $1.77$.

The $\chi^2$ of the datasets listed in Table~\ref{tab:Z}
will therefore need to be interpreted with care, in particular taking into
account the possibility that it be spuriously high due to a misestimate of
experimental correlations. To avoid this issue, in Sect.~\ref{sec:results}
we will compute the $\chi^2$ upon regularisation of the experimental covariance
matrix, for all the datasets with $Z_{\mathcal{L}}>4$. We use the procedure laid
out in Ref.~\cite{Kassabov:2022pps}. This procedure consists in clipping the
singular values of the correlation matrix to a constant, whenever these are
smaller than that, while leaving the rest of the singular vectors unchanged.
This way, directions that do not contribute to instability are not affected and
the alteration to the original matrix is minimal. The clipping constant is
chosen to be $\delta^{-1}=Z$, where the value of $Z=4$ was determined
empirically in Ref.~\cite{Kassabov:2022pps}. The values of the $\chi^2$
computed with the unregularised experimental covariance matrix are
collected in Appendix~\ref{app:unreg}.

\subsection{PDF sets}
\label{subsec:pdfs}

The computation of the theoretical predictions that enter the $\chi^2$ require
a choice of PDFs as input. In this work, we consider the following PDF sets:
ABMP16~\cite{Alekhin:2017kpj}, CT18, CT18A, and CT18Z~\cite{Hou:2019efy},
MSHT20~\cite{Bailey:2020ooq}, NNPDF3.1~\cite{Ball:2017nwa},
NNPDF4.0~\cite{NNPDF:2021njg}, PDF4LHC15~\cite{Butterworth:2015oua},
and PDF4LHC21~\cite{PDF4LHCWorkingGroup:2022cjn}. These PDF sets are
the most widely used by LHC experimental collaborations in their analyses.
The main features of each of them are summarised as follows.

\begin{description}

\item[ABMP16~\cite{Alekhin:2017kpj}.] This PDF determination is based on
  DIS, Drell-Yan, single top and top-quark pair production measurements.
  The underlying theory calculations are based on a Fixed Flavour Number (FFN)
  scheme, with $n_f=3,4,5$. The strong coupling constant is determined
  alongside the PDFs yielding $\alpha_s(m_Z)=0.1147 \pm 0.0008$ with $n_f=5$,
  though a variant with a fixed value $\alpha_s(m_Z)=0.118$ is also provided.
  The PDFs are parametrised at the input scale $Q_0=1$~GeV with a fixed
  functional form. The charm PDF is assumed to be purely perturbative,
  therefore it is generated by partonic DGLAP evolution above the charm quark
  mass, whose value is a parameter of the fit. Hessian symmetric PDF
  uncertainties are determined from variations $\Delta \chi^2=1$.

\item[CT18~\cite{Hou:2019efy}.] The CT18 family of PDF determinations is based
  on DIS, Drell-Yan, single-inclusive jet, and top-quark pair production
  measurements. The underlying theory calculations are based on a General
  Mass Variable Flavour Number (GM-VFN) scheme, specifically
  ACOT-$\chi$~\cite{Aivazis:1993pi,Aivazis:1993kh,Kramer:2000hn,Tung:2001mv},
  and use a fixed value of the strong coupling as input. Parton distributions
  are parametrised at the input scale $Q_0=1.3$~GeV, equal to the charm pole
  mass $m_c^{\rm pole}$, in terms of Bernstein polynomials, the charm PDF is
  purely perturbative, and Hessian symmetric PDF uncertainties are determined
  by means of a dynamical tolerance factor $\Delta \chi^2>1$. The ATLAS 7~TeV
  $W/Z$ data~\cite{Aaboud:2016btc} is not included in the default CT18 PDF set.
  Alternate sets are determined including this dataset (CT18A), a new scale
  choice for low-$x$ DIS data (CT18X), or all of the above with a slightly
  higher value of the charm mass (CT18Z).

\item[MSHT20~\cite{Bailey:2020ooq}.] This PDF determination is based on DIS,
  Drell-Yan, Drell-Yan with jet, single-inclusive jet, and top-quark
  pair production measurements. The fit is based on the Thorne-Roberts variant
  of the GM-VFN scheme~\cite{Thorne:2012az}, and uses a fixed value of the
  strong coupling as input. Parton distributions are parametrised at the input
  scale $Q_0=1$~GeV in terms of Chebyschev polynomials, the charm PDF is purely
  perturbative (with charm pole mass $m_c^{\rm pole}=1.4$~GeV), and Hessian
  symmetric uncertainties are determined by means of a dynamical tolerance
  factor $\Delta \chi^2>1$. 

\item[NNPDF3.1~\cite{NNPDF:2017mvq}.] This PDF determination is based on DIS,
  Drell-Yan, Drell-Yan with jet, single-inclusive jet, and top-quark pair
  production measurements. The fit is based on the FONLL GM-VFN
  scheme~\cite{Forte:2010ta} and uses a fixed value of the strong coupling
  constant as input. Parton distributions are parametrised at the initial scale
  $Q_0=1.65$~GeV in terms of deep neural networks, optimised by means of a
  genetic algorithm. The charm PDF is fitted on the same footing as lighter
  quark flavours (with a charm pole mass $m_c^{\rm pole}=1.51$~GeV). PDF
  uncertainties are determined from a Monte Carlo sampling of experimental 
  uncertainties.
  
\item[NNPDF4.0~\cite{NNPDF:2021njg}.] This PDF determination is based on DIS,
  Drell-Yan, Drell-Yan with jet, single-inclusive jet and di-jet, single top and
  top-quark pair, and prompt photon production measurements. The fit is based
  on the same treatment of quark masses, running coupling, charm quark PDF, and
  uncertainty representation as NNPDF3.1. In comparison to NNPDF3.1, NNPDF4.0
  is however characterised by several methodological differences: newer
  theoretical constraints, in particular on PDF positivity and integrability,
  are implemented; PDFs are parametrised with a single neural network,
  optimised by means of gradient descent; hyperparameters, such as those that
  define the architecture of the neural network, are determined by means of an
  automated scan of the space of models that selects the optimal
  one~\cite{Carrazza:2021yrg,NNPDF:2021njg}; and the
  methodology is closure tested~\cite{DelDebbio:2021whr}.
  
\item[PDF4LHC15~\cite{Butterworth:2015oua}.] This PDF set is the Monte Carlo
  combination of the CT14~\cite{Dulat:2015mca},
  MMHT2014~\cite{Harland-Lang:2014zoa}, and NNPDF3.0~\cite{Ball:2012cx} PDF
  sets. The combination is performed by first converting the CT14 and MMHT2014
  Hessian PDF sets into Monte Carlo PDF sets by means of the algorithm
  developed in~\cite{Watt:2012tq,Hou:2016sho}. For each of the three PDF sets
  300 Monte Carlo replicas are generated, that are subsequently collated in a
  single set. The number of replicas is finally reduced by means of the
  compression algorithm developed in~\cite{Carrazza:2015hva} or converted to
  a single Hessian set by means of the algorithm
  developed in~\cite{Carrazza:2015aoa}.

\item[PDF4LHC21~\cite{PDF4LHCWorkingGroup:2022cjn}.] This PDF set is the
  Monte Carlo combination of the CT18$^\prime$, MSHT20, and NNPDF3.1$^\prime$ PDF
  sets. The CT18$^\prime$ and NNPDF3.1$^\prime$ PDF sets are variants of the CT18
  and NNPDF3.1 PDF sets: both of these differ from the corresponding baseline
  sets for the values of the charm and bottom quark pole masses, which are set
  to values common to those used in MSHT20, $m_c^{\rm pole}=1.4$~GeV and
  $m_b^{\rm pole}=4.75$~GeV. The NNPDF3.1$^\prime$ PDF set differs from NNPDF3.1
  for a number of additional variations in the input dataset and in the
  details of the theoretical computations, see Sect.~2.3
  in~\cite{PDF4LHCWorkingGroup:2022cjn}. The combination is carried out as in
  PDF4LHC15.

\end{description}

\begin{figure}[!t]
  \centering
  \includegraphics[width=0.49\textwidth]{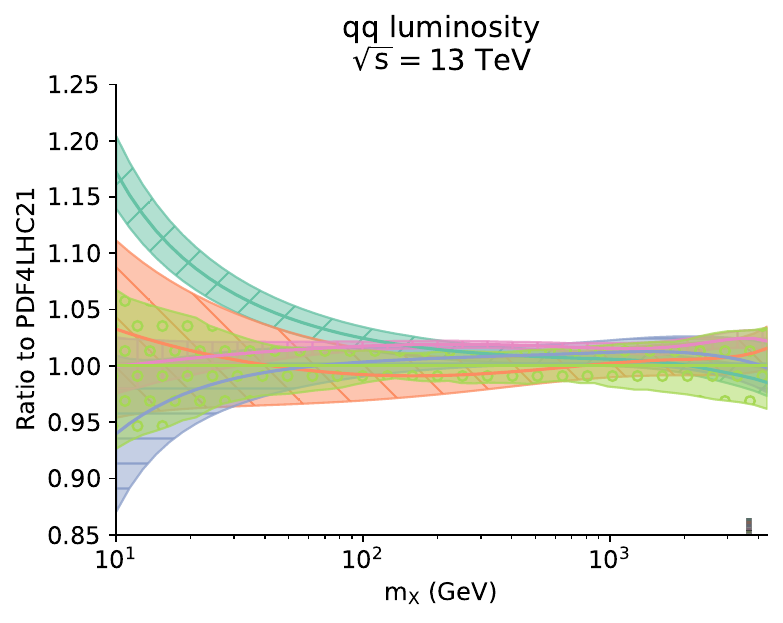}
  \includegraphics[width=0.49\textwidth]{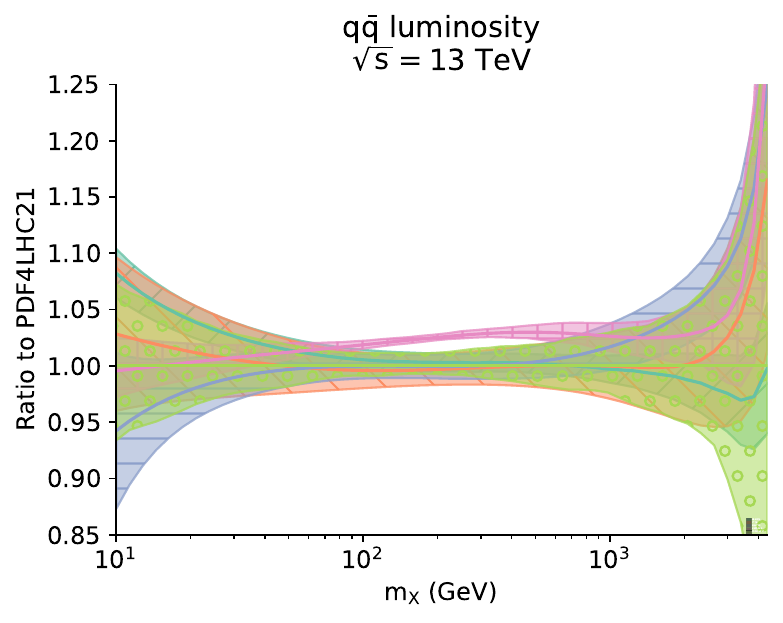}\\
  \includegraphics[width=0.49\textwidth]{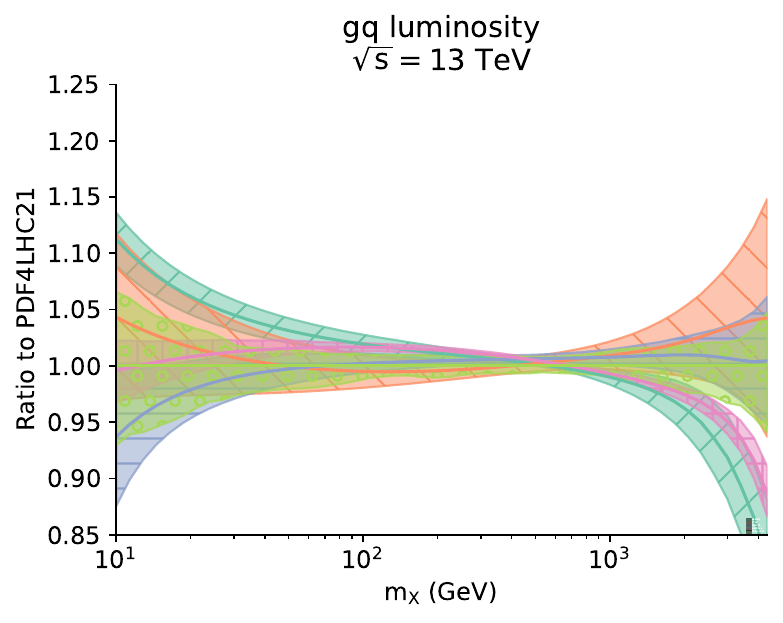}
  \includegraphics[width=0.49\textwidth]{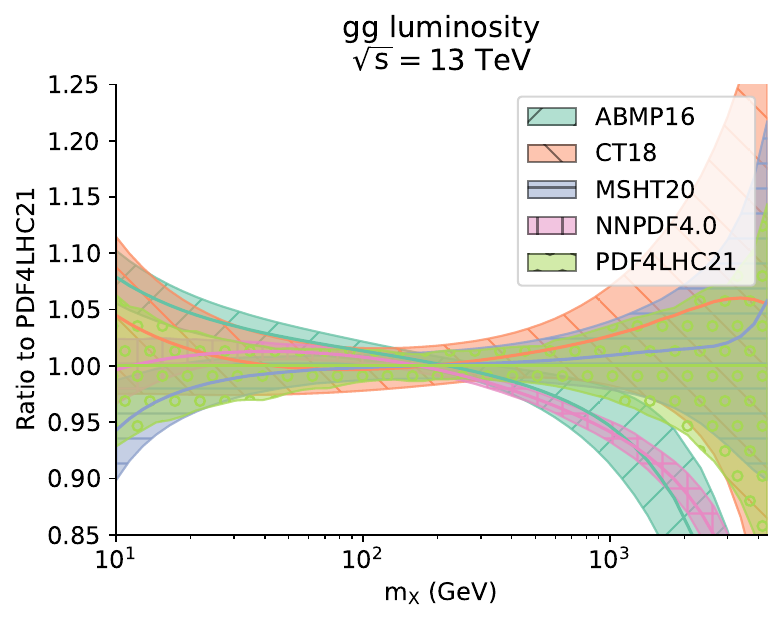}
  \caption{The quark-quark (top left), quark-antiquark (top right), gluon-quark
    (bottom left), and gluon-gluon (bottom right) partonic luminosities,
    Eqs.~(1--4) of~\cite{Mangano:2016jyj}, as a function of the invariant mass
    of the final state $m_X$ at a centre-of-mass energy $\sqrt{s}=13$~TeV
    obtained with the ABMP16, CT18, MSHT20, NNPDF4.0, and PDF4LHC21 PDF sets.
    Results are normalised to PDF4LHC21.}
  \label{fig:lumis}
\end{figure}

In all cases, we use PDF sets accurate to NNLO with a common, fixed value of
$\alpha_s(m_Z)=0.118$. Note that NNLO corrections to hadronic processes were
included in all of the aforementioned PDF sets by means of $K$-factors,
whereas here we make predictions by means of exact NNLO computations. This fact
is however immaterial, given the very weak dependence of $K$-factors on
PDFs~\cite{Sharma:2024cnc}. In the case of ABMP16, we use the set with $n_f=5$
active flavours. For ABMP16, CT18, and MSHT20, we consider Hessian sets;
for NNPDF3.1, NNPDF4.0, PDF4LHC15, and PDF4LHC21, we consider Monte Carlo sets
composed of 100 replicas. In Fig.~\ref{fig:lumis} we compare the partonic
luminosities, defined by Eqs.~(1--4) of~\cite{Mangano:2016jyj}, obtained with
the ABMP16, CT18, MSHT20, NNPDF4.0, and PDF4LHC21 PDF sets. Results are
displayed as a function of the invariant mass of the final state $m_X$ at a
centre-of-mass energy $\sqrt{s}=13$~TeV and are normalised to PDF4LHC21.
Comparison using other PDF sets can be seen in~\cite{Amoroso:2022eow}.

We do not consider PDF sets including QED
corrections~\cite{Cridge:2021pxm,Xie:2023qbn,NNPDF:2024djq},
aN$^3$LO QCD corrections~\cite{McGowan:2022nag,NNPDF:2024nan}
or MHOU~\cite{NNPDF:2024dpb}, the reason being that these are not commonly
used in LHC experimental analyses. This said, the computation of the $\chi^2$
does not change if one uses any of these PDF sets. We will study the
phenomenological implications of QED, aN$^3$LO, and of MHOU corrections to the
PDFs in the appraisal of LHC data in future work. An exception is
represented by the high-precision ATLAS 8~TeV inclusive dilepton
rapidity measurement~\cite{ATLAS:2023lsr}, for which predictions
based on the NNPDF4.0 QED~\cite{NNPDF:2024djq}, MHOU~\cite{NNPDF:2024dpb},
and aN$^3$LO~\cite{NNPDF:2024nan} PDF sets will be considered in
Sect.~\ref{subsec:ATLAS_8TEV}.

\section{Data-theory comparison appraisal}
\label{sec:results}

In this section, we quantify the agreement between the experimental data
and the corresponding theoretical predictions presented in Sect.~\ref{sec:data}
according to the estimators and upon variations of the input PDF sets discussed
in Sect.~\ref{sec:approach}. We examine datasets for each process in turn.
For each of these, we provide three complementary ways of visualizing the
data-theory agreement: a table with the values of $\chi^2_{\rm exp+th}$, and
$\chi^2_{\rm exp}$, Eqs.~\eqref{eq:chi2_v2} and~\eqref{eq:chi2_v3}, evaluated
with all the PDF sets summarised in Sect.~\ref{subsec:pdfs}\footnote{For ease
of reference, in this table we also provide the number of data points
$n_{\rm dat}$ and the standard deviation of the $\chi^2$ per point distribution
$\sqrt{2/n_{\rm dat}}$}; a set of histograms in which the total
$\chi^2_{\rm exp+th}$, Eq.~\eqref{eq:chi2_v2}, is split into the components
$\chi^2_{\rm exp+mho}$, Eq.~\eqref{eq:chi2_v4}, and $\chi^2_{\rm exp+th}$,
Eq.~\eqref{eq:chi2_v3}, albeit only for CT18, MSHT20, NNPDF4.0, and PDF4LHC21;
and a set of data-theory comparison plots, only for NNPDF4.0 and PDF4LHC21, in
which the PDF+$\alpha_s$ and MHO uncertainties are displayed separately
for selected data points. For all measurements with $Z_{\mathcal{L}}>4$ (see
Table~\ref{tab:Z}), the experimental covariance matrix is regularised
as explained in Sect.~\ref{subsec:compatibility}. We finally provide a
collective visualisation of the $\Delta\chi^{2(i)}$ and $\Delta n_\sigma^{(i)}$
estimators, Eqs.~\eqref{eq:delta_chi2_average} and \eqref{eq:delta_nsigma}, for
the CT18, MSHT20, NNPDF4.0, and PDF4LHC21 PDF sets. The values of
$\chi^2_{\rm exp+th}$ obtained without regularisation of
the experimental covariance matrix are given, for the subset of measurements
with $Z_{\mathcal{L}}>4$, in Appendix~\ref{app:unreg}. Additional histogram and
data-theory comparison plots, for the subset of measurements not highlighted
in this section, are given in Appendix~\ref{app:extra_results}.

\subsection{Drell-Yan weak boson production measurements at 13 TeV}
\label{subsec:DY_13TEV}

\begin{figure}[!t]
  \centering
  \includegraphics[width=0.49\textwidth]{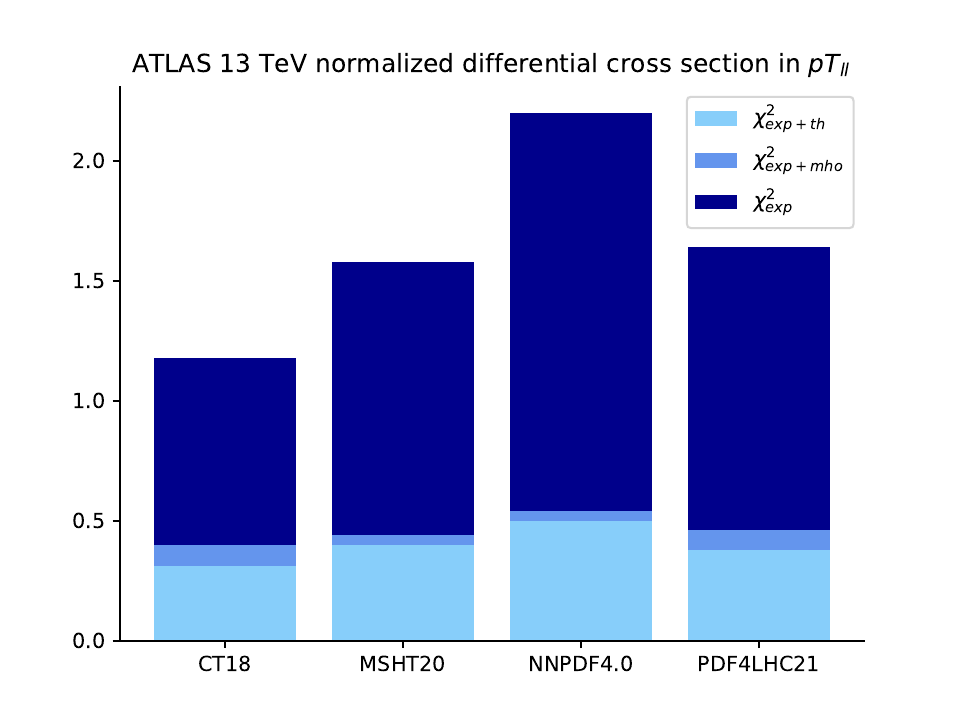}
  \includegraphics[width=0.49\textwidth]{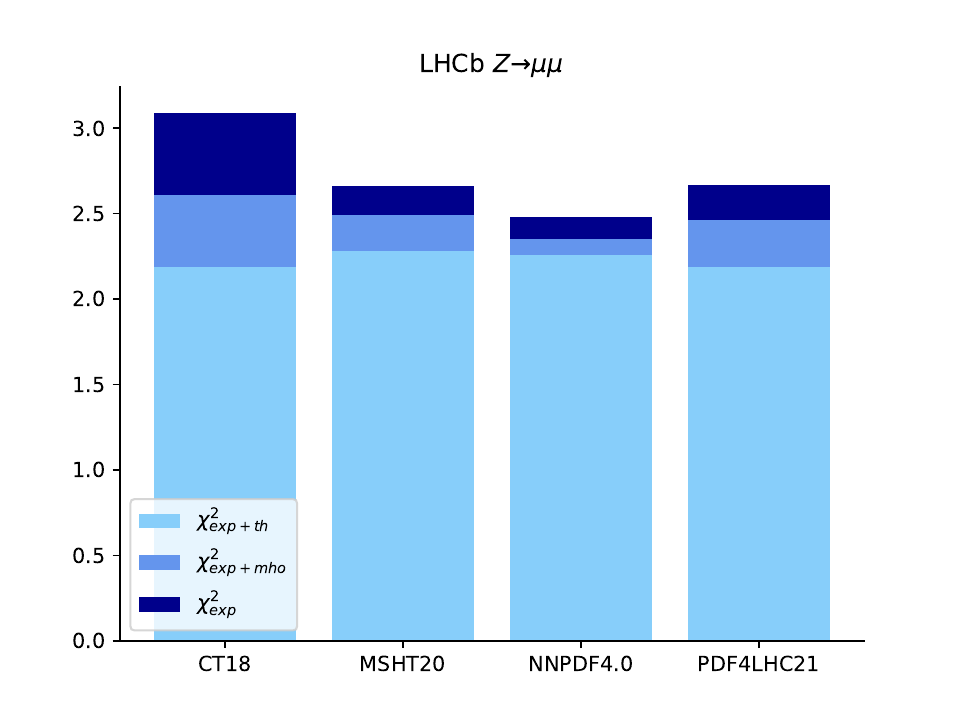}\\
  \includegraphics[width=0.49\textwidth]{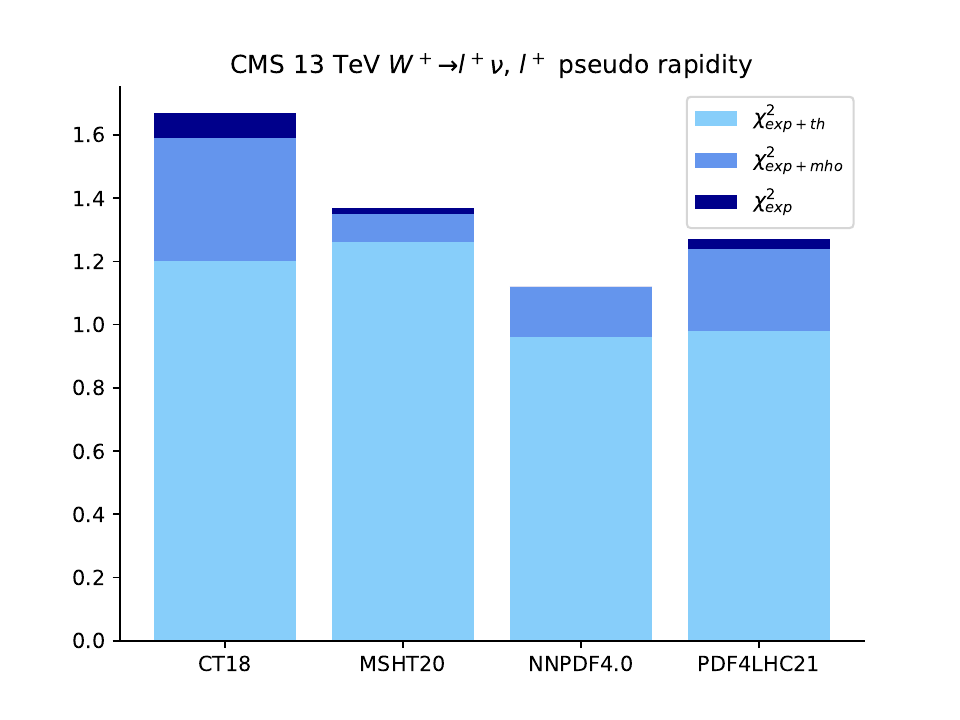} 
  \includegraphics[width=0.49\textwidth]{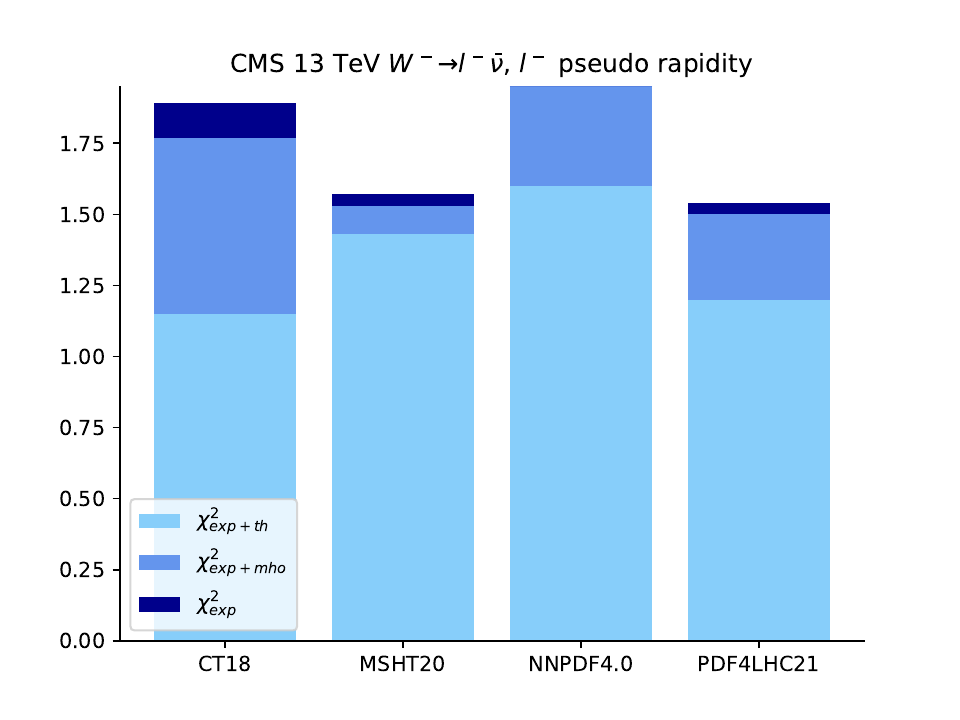}
  \caption{The breakdown of $\chi^2_{\rm exp+th}$, Eq.~\eqref{eq:chi2_v2},
    into $\chi^2_{\rm exp+mho}$, Eq.~\eqref{eq:chi2_v4}, and $\chi^2_{\rm exp}$,
    Eq.~\eqref{eq:chi2_v3}, for the ATLAS ($n_{\rm dat}=38$,
    $\sqrt{2/n_{\rm dat}} = 0.23$), CMS (each set made of $n_{\rm dat}=18$,
    $\sqrt{2/n_{\rm dat}} = 0.23$), and LHCb ($n_{\rm dat}=18$,
    $\sqrt{2/n_{\rm dat}} = 0.33$) Drell-Yan gauge boson
    production measurements at the LHC 13~TeV. Note that the inclusion of MHOUs 
    has a negligible impact on the $\chi^2$ values of the CMS 13 TeV
    $W$ distributions when theoretical predictions are computed with NNPDF4.0, 
    hence the dark blue component of the corresponding histogram is hardly
    visible.}
  \label{fig:DY-chi2}
\end{figure}

\begin{figure}[!t]
  \centering
  \includegraphics[width=0.49\textwidth]{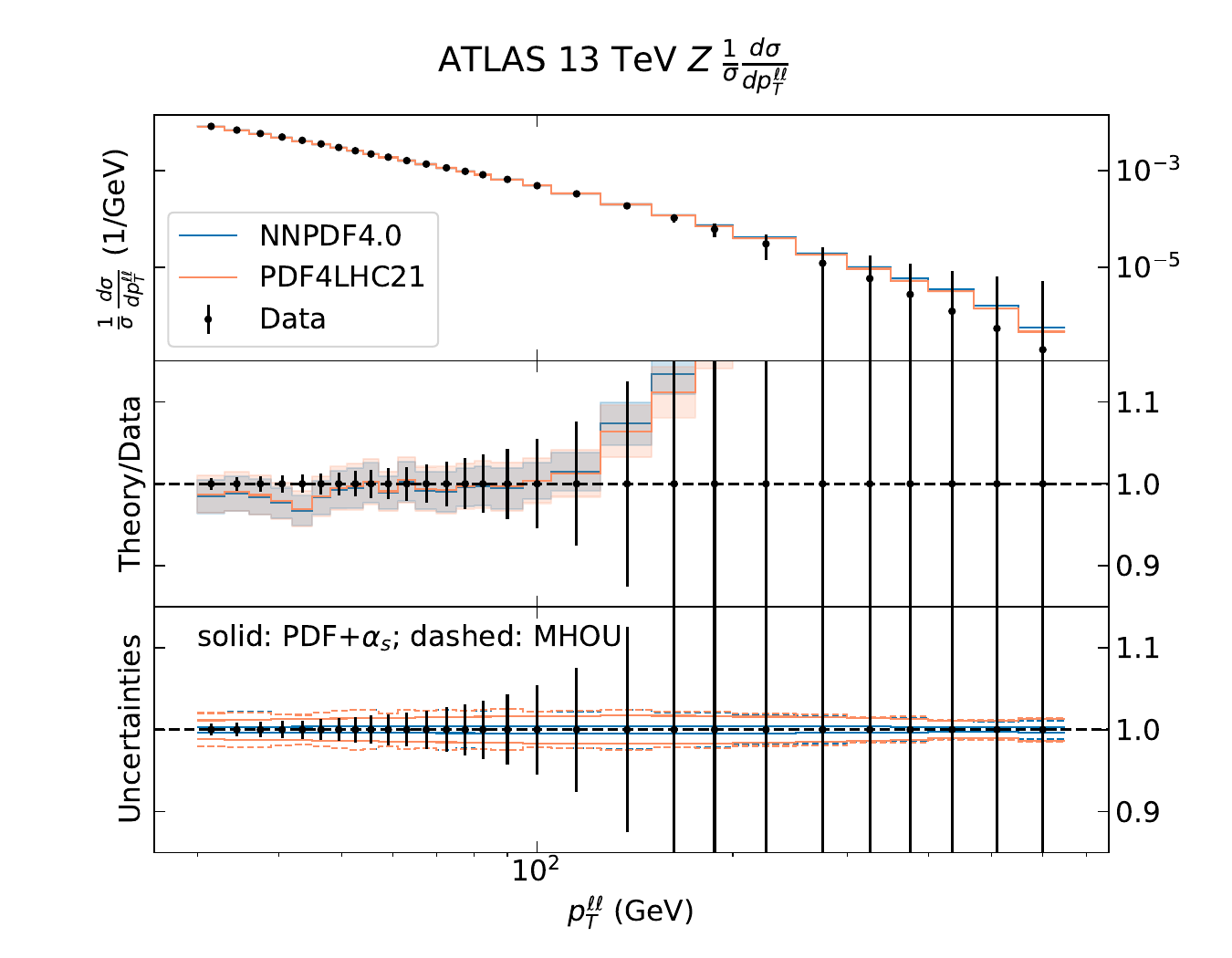}
  \includegraphics[width=0.49\textwidth]{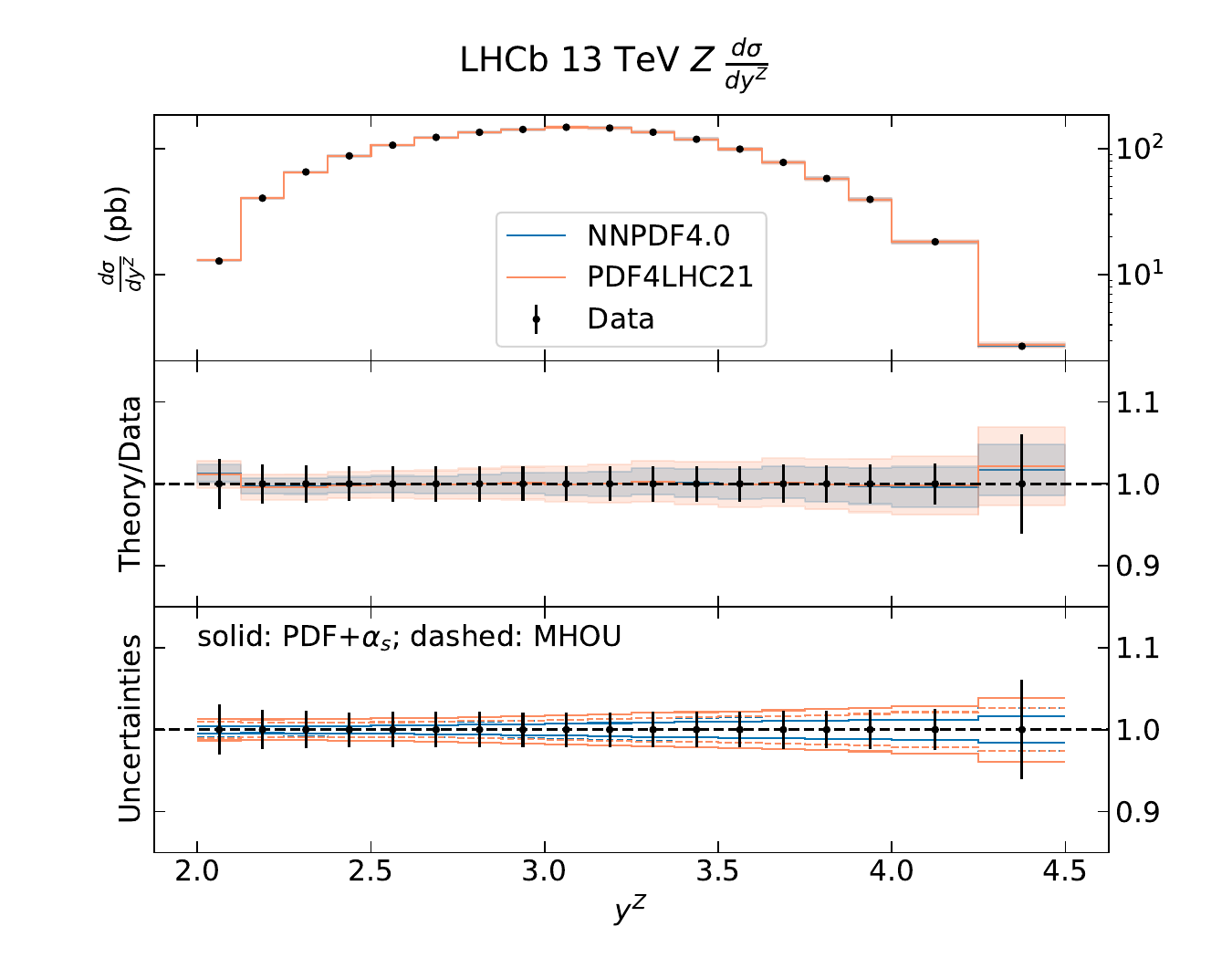}\\
  \includegraphics[width=0.49\textwidth]{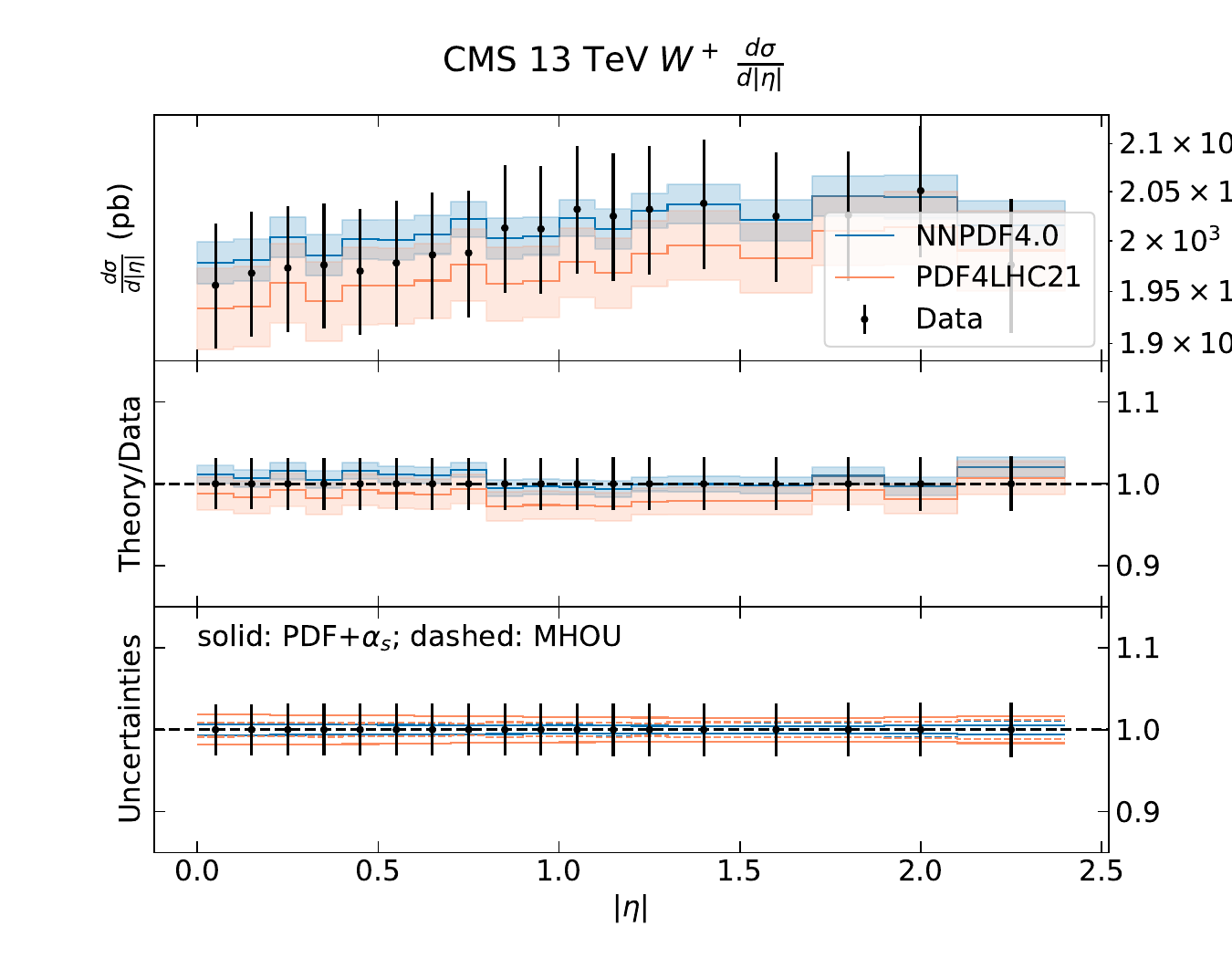}
  \includegraphics[width=0.49\textwidth]{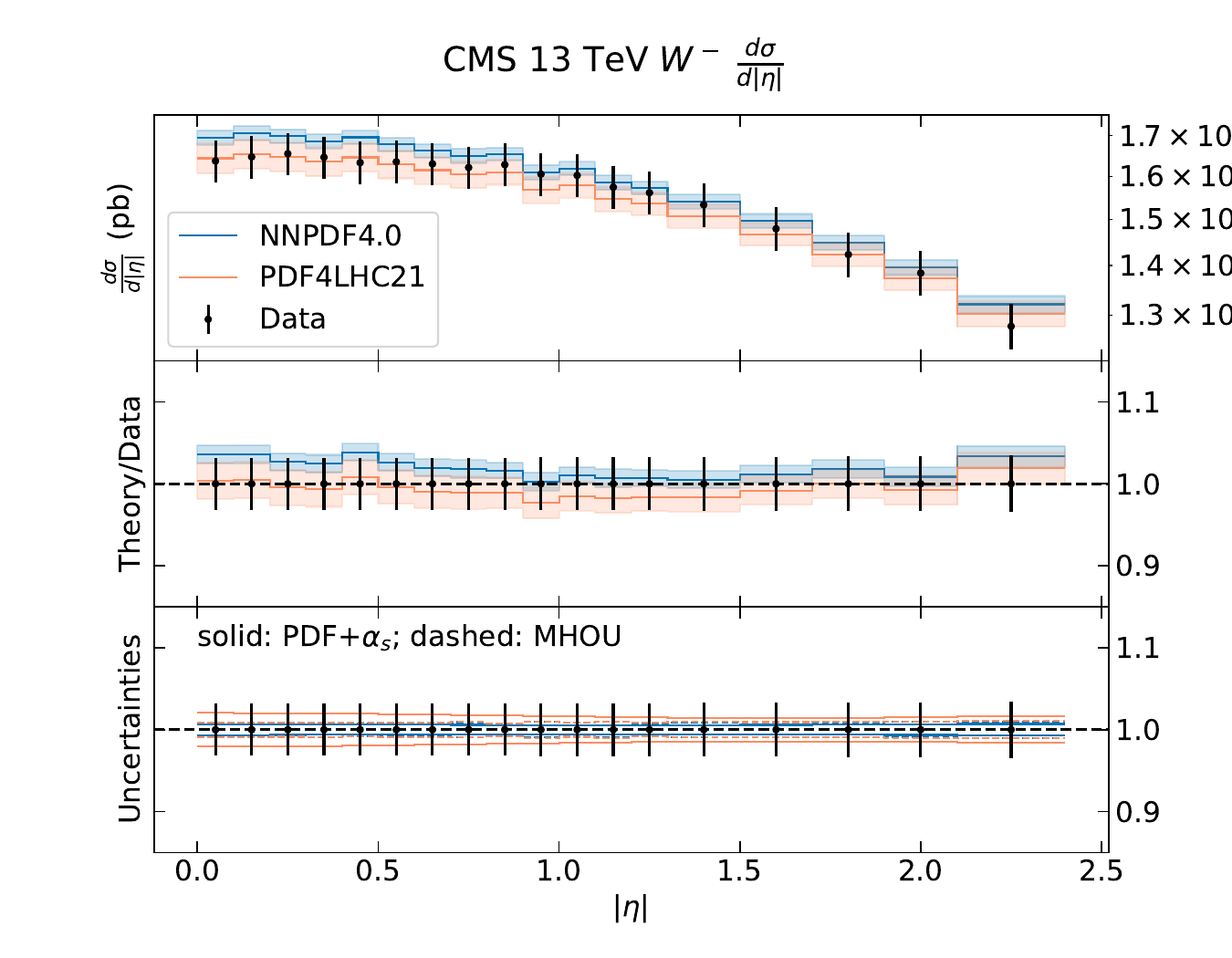}\\
  \caption{Data-theory comparison for the ATLAS, CMS, and LHCb Drell-Yan gauge
    boson production measurements at the LHC 13~TeV of
    Table~\ref{tab:input_datasets}. (Upper panels) The measured and predicted
    cross sections, with experimental and total (MHO and PDF+$\alpha_s$)
    theoretical uncertainties. (Middle panels) The same cross sections
    normalised to the experimental central value. (Lower panels) The relative
    PDF+$\alpha_s$ (dashed) and MHO (solid) uncertainties separately. In all
    panels, the experimental error bars correspond to the the total
    uncorrelated uncertainty. Correlated uncertainties are kept into account by
    shifting the central experimental value as explained
    in Appendix~B of~\cite{Pumplin:2002vw}.}
  \label{fig:DY-datatheory}
\end{figure}

We start by considering the three LHC Drell-Yan weak boson production
measurements at a centre-of-mass energy of 13~TeV outlined in
Sect.~\ref{sec:data}. The values of $\chi^2_{\rm exp}$
and $\chi^2_{\rm exp+th}$, computed with each of the PDF sets summarised in
Sect.~\ref{subsec:pdfs}, are reported in Table~\ref{tab:chi2-DY}. The
experimental covariance matrix of the CMS dataset is regularised as explained
in Sect.~\ref{subsec:compatibility}, see Appendix~\ref{app:unreg} for the
unregularised values of $\chi^2_{\rm exp+th}$. The breakdown of $\chi^2_{\rm exp+th}$
into $\chi^2_{\rm exp+mho}$ and $\chi^2_{\rm exp}$ is displayed in
Fig.~\ref{fig:DY-chi2}. The data-theory comparison is displayed in
Fig.~\ref{fig:DY-datatheory}. Each plot consists of three panels: the upper one
displays the measured and predicted cross sections, with experimental
and total (MHO and PDF+$\alpha_s$) theoretical uncertainties; the middle one
displays the same cross sections normalised to the experimental central value;
the lower one displays the relative PDF+$\alpha_s$ and MHO uncertainties
separately. Experimental error bars correspond to the total uncorrelated
uncertainty. Correlated uncertainties are included by shifting the central
experimental value, by an amount determined as explained
in Appendix~B of~\cite{Pumplin:2002vw}.

\begin{table}[!t]
  \scriptsize
  \centering
  \renewcommand{\arraystretch}{1.5}
  \begin{tabularx}{\textwidth}{Xrccccccccccc}
Dataset
& \rotatebox{0}{$n_{\rm dat}$}
& \rotatebox{0}{$\sqrt{2/n_{\rm dat}}$}
& 
& \rotatebox{80}{ABMP16}
& \rotatebox{80}{CT18}
& \rotatebox{80}{CT18A}
& \rotatebox{80}{CT18Z}
& \rotatebox{80}{MSHT20}
& \rotatebox{80}{NNPDF3.1}
& \rotatebox{80}{NNPDF4.0}
& \rotatebox{80}{PDF4LHC15}
& \rotatebox{80}{PDF4LHC21} \\
\toprule
\multirow{2}{*}{ATLAS 13~TeV $Z$ $\frac{1}{\sigma}\frac{d\sigma}{dp_{T}^{\ell\ell}}$}
& \multirow{2}{*}{38}
& \multirow{2}{*}{0.23}
& $\chi^2_{\rm exp+th}$
& 0.36 & 0.31 & 0.42 & 0.59 & 0.40 & 0.39 & 0.50 & 0.31 & 0.38 \\
&
&
& $\chi^2_{\rm exp}$
& 0.80 & 1.18 & 2.38 & 4.91 & 1.58 & 1.20 & 2.20 & 0.83 & 1.64 \\ 
\midrule
\multirow{2}{*}{CMS 13~TeV $W^+$ $\frac{d\sigma}{d|\eta|}$}
& \multirow{2}{*}{18}
& \multirow{2}{*}{0.33}
& $\chi^2_{\rm exp+th}$
& 1.31 & 1.20 & 1.11 & 1.06 & 1.26 & 0.85 & 0.96 & 1.15 & 0.98 \\ 
&
&
& $\chi^2_{\rm exp}$
& 1.41 & 1.67 & 1.30 & 1.31 & 1.37 & 0.97 & 1.12 & 1.38 & 1.27 \\
\midrule
\multirow{2}{*}{CMS 13~TeV $W^-$ $\frac{d\sigma}{d|\eta|}$}
& \multirow{2}{*}{18}
& \multirow{2}{*}{0.33}
& $\chi^2_{\rm exp+th}$
& 1.56 & 1.15 & 1.11 & 1.10 & 1.43 & 1.12 & 1.60 & 1.14 & 1.20 \\
&
&
& $\chi^2_{\rm exp}$
& 1.60 & 1.89 & 1.43 & 1.38 & 1.57 & 1.64 & 1.95 & 1.54 & 1.54 \\ 
\midrule
\multirow{2}{*}{LHCb 13~TeV $Z$ $\frac{d\sigma}{dy^Z}$}
& \multirow{2}{*}{18}
& \multirow{2}{*}{0.33}
& $\chi^2_{\rm exp+th}$
& 2.14 & 2.19 & 2.26 & 2.08 & 2.28 & 2.21 & 2.26 & 2.15 & 2.07 \\ 
&
&
& $\chi^2_{\rm exp}$
& 2.28 & 3.09 & 2.91 & 2.62 & 2.66 & 2.70 & 2.48 & 3.06 & 2.67 \\ 
\bottomrule
\end{tabularx}

  \vspace{0.3cm}
  \caption{The values of $\chi^2_{\rm exp+th}$, Eq.~\eqref{eq:chi2_v2}, and of
    $\chi^2_{\rm exp}$, Eq.~\eqref{eq:chi2_v3}, for the ATLAS, CMS, and LHCb
    Drell-Yan gauge boson production measurements at the LHC 13~TeV of
    Table~\ref{tab:input_datasets}, computed with each of the PDF sets
    summarised in Sect.~\ref{subsec:pdfs}. The experimental covariance
    matrix of the CMS dataset is regularised as explained in
    Sect.~\ref{subsec:compatibility}. The unregularised values of
    $\chi^2_{\rm exp}$ are collected in table~\ref{tab:chi2-unreg}
    of Appendix~\ref{app:unreg}.}
  \label{tab:chi2-DY} 
\end{table}

From inspection of Table~\ref{tab:chi2-DY} and of Fig.~\ref{fig:DY-chi2},
we observe that the values of $\chi^2_{\rm exp+th}$, computed with different
input PDFs, are generally closer to each other than the corresponding values
of $\chi^2_{\rm exp}$. This fact suggests that the inclusion of theory
uncertainties is essential to assess the predictive power of a given PDF
set. Moreover, the values of $\chi^2_{\rm exp+th}$ are very similar across PDF
sets: this is manifest in the case of the ATLAS and LHCb datasets, and true on
average for the CMS dataset. In this latter case, the PDF sets with larger
values of $\chi^2_{\rm exp+th}$ on the $W^+$ dataset have the smaller values of
$\chi^2_{\rm exp+th}$ on the $W^-$ dataset, and the other way around. 
For the CMS $W^+$ distribution, whereas the total $\chi^2_{\rm exp+th}$ remains
within $1\sigma$ of the $\chi^2$ distribution per unit point for all PDF sets,
the experimental $\chi^2_{\rm exp}$ is close to one only for NNPDF3.1 and
NNPDF4.0. In the case of the CMS $W^-$ distribution, the ABMP16 and NNPDF4.0, and
to a lesser extent the MSHT20 set do yield a somewhat worse description, in that the corresponding
values of $\chi^2_{\rm exp+th}$ are almost $2\sigma$ away from the unit
expectation. For the purely experimental $\chi^2_{\rm exp}$, the predictions
obtained with NNPDF4.0 and CT18 are almost $3\sigma$ away from one. 
Despite these differences, when all uncertainties are kept into account,
we cannot single out a PDF set that, overall, generalises better than another
on these datasets.

The breakdown of $\chi^2_{\rm exp+th}$ into its theoretical components depends on
the dataset and on the PDF set. The component due to MHO, gauged from the
difference between $\chi^2_{\rm exp}$ and $\chi^2_{\rm exp+mho}$, dominates the ATLAS
measurement, irrespective of the PDF set, whereas it is less prominent in the
other datasets. For CMS, this is almost immaterial, irrespective of the PDF
set. For LHCb, irrespective of the PDF set, this is typically as large as the
component due to PDF+$\alpha_s$ uncertainties, gauged from the difference
between $\chi^2_{\rm exp+mho}$ and $\chi^2_{\rm exp+th}$. This latter component may
depend on the PDF set, being usually larger for PDF sets affected by the
largest uncertainties, such as CT18 and PDF4LHC21, see Fig.~\ref{fig:lumis}.
All these facts are a consequence of how the various partonic channels
contribute to the cross sections of these processes. The ATLAS measurement
receives its leading contribution, which is $\mathcal{O}(\alpha_s)$, from the
quark-gluon partonic luminosity. The CMS and LHCb measurements receive their
leading contributions, which are $\mathcal{O}(\alpha_s^0)$, from
quark-antiquark partonic luminosities, yet in different regions of $x$, given
that they are central and forward rapidity measurements: the former at
intermediate values of $x$; the latter at large values of $x$.

The quality of the data description is generally good, being
$\chi^2_{\rm exp+th}\sim 1$, except for LHCb, for which $\chi^2_{\rm exp+th}\sim 2$,
irrespective of the PDF set. For ATLAS, $\chi^2_{\rm exp+th}\sim 0.4$, again
irrespective of the PDF set. This value is anomalously small, and may point
towards the fact that MHOU are actually overestimated by scale variations.
Discrepancies between data and theory that may
lead to these results are seen in Fig.~\ref{fig:DY-datatheory}, where the
alignment of experimental data and theoretical predictions is optimal,
within their uncertainties, for all datasets. We therefore conclude that the
somewhat high $\chi^2_{\rm exp+th}$ for LHCb is due to experimental correlations,
and will likely decrease once the dataset is included in a fit.
Note finally that the quality of the data description of the CMS measurement
would have been rather worse, at face value,
had the regularisation procedure described in Sect.~\ref{subsec:compatibility}
not been applied. The values of $\chi^2_{\rm exp+th}$ obtained without it are
reported in Appendix~\ref{app:unreg}. As we can see from
Fig.~\ref{fig:DY-datatheory}, theoretical predictions are
almost spot on experimental measurements. The otherwise very large values of
the $\chi^2$ obtained without regularisation are spurious, and denote an
ill-conditioning of their experimental covariance matrix.

\subsection{The ATLAS 8 TeV inclusive $Z$ boson production measurement}
\label{subsec:ATLAS_8TEV}

\begin{figure}[!t]
  \centering
  \includegraphics[width=0.49\textwidth]{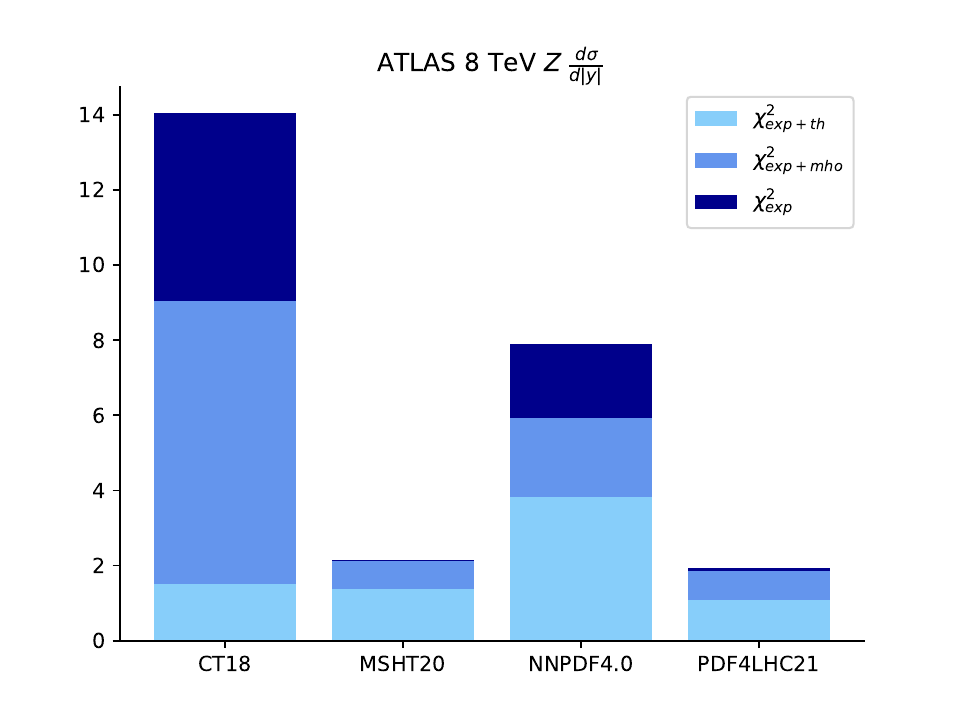}
  \includegraphics[width=0.49\textwidth]{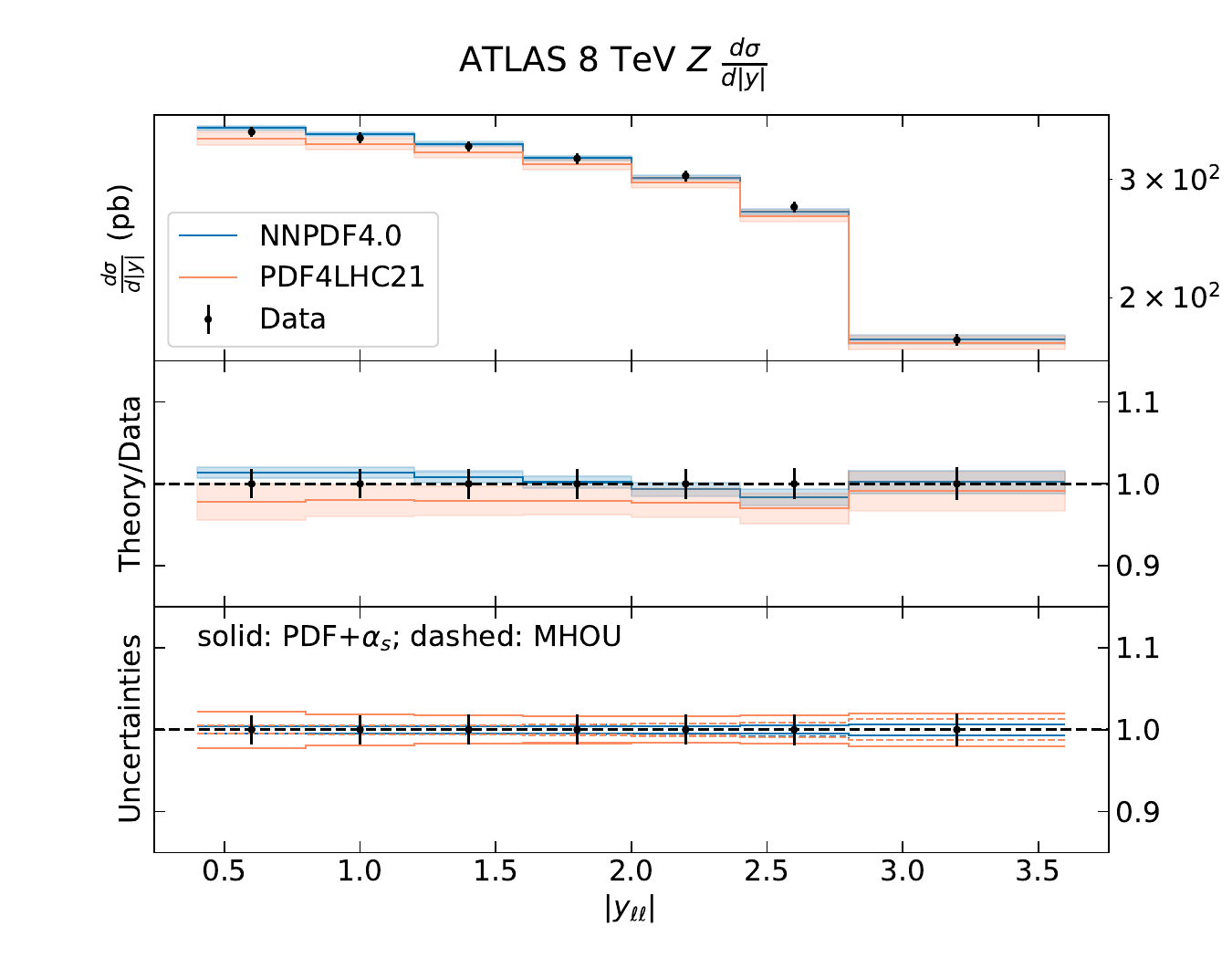}\\
  \caption{Same as Figs.~\ref{fig:DY-chi2} (left) and~\ref{fig:DY-datatheory}
    (right) for the ATLAS Drell-Yan gauge boson production measurements at the
    LHC 8~TeV~\cite{ATLAS:2023lsr} ($n_{\rm dat}=7$, $\sqrt{2/n_{\rm dat}}=0.53$).}
  \label{fig:ATLAS8TEV}
\end{figure}

We then consider the ATLAS measurement of Drell-Yan Z boson production at the
LHC 8~TeV outlined in Sect.~\ref{sec:data}. The values of $\chi^2_{\rm exp}$
and $\chi^2_{\rm exp+th}$, computed with each of the PDF sets summarised in
Sect.~\ref{subsec:pdfs}, are reported in Table~\ref{tab-chi2-atlasz-all}. 
The breakdown of $\chi^2_{\rm exp+th}$ into $\chi^2_{\rm exp+mho}$ and
$\chi^2_{\rm exp}$ and the data-theory comparison
are displayed in Fig.~\ref{fig:ATLAS8TEV}, in the same format as
Figs.~\ref{fig:DY-chi2} and \ref{fig:DY-datatheory}.

\begin{table}[!t]
  \scriptsize
  \centering
  \renewcommand{\arraystretch}{1.5}
  \begin{tabularx}{\textwidth}{Xrccccccccccc}
Dataset
& \rotatebox{0}{$n_{\rm dat}$}
& \rotatebox{0}{$\sqrt{2/n_{\rm dat}}$}
&
& \rotatebox{80}{ABMP16}
& \rotatebox{80}{CT18}
& \rotatebox{80}{CT18A}
& \rotatebox{80}{CT18Z}
& \rotatebox{80}{MSHT20}
& \rotatebox{80}{NNPDF3.1}
& \rotatebox{80}{NNPDF4.0}
& \rotatebox{80}{PDF4LHC15}
& \rotatebox{80}{PDF4LHC21} \\
\toprule
\multirow{2}{*}{ATLAS 8~TeV $Z$ $\frac{d\sigma}{d|y|}$}
& \multirow{2}{*}{7}
& \multirow{2}{*}{0.53}
& $\chi^2_{\rm exp+th}$
& 4.25 & 1.52 & 1.52 & 1.18 & 1.37 & 1.61 & 3.83 & 1.23 & 1.09 \\ 
&
&
& $\chi^2_{\rm exp}$
& 7.36 & 14.0 & 4.63 & 4.31 & 2.14 & 4.70 & 7.90 & 7.41 & 1.93 \\
\bottomrule
\end{tabularx}

  \vspace{0.3cm}
  \caption{Same as Table~\ref{tab:chi2-DY} for the ATLAS Drell-Yan gauge boson
    production measurements at the LHC 8~TeV~\cite{ATLAS:2023lsr}.}
  \label{tab-chi2-atlasz-all} 
\end{table}

From inspection of Table~\ref{tab-chi2-atlasz-all} and
Fig.~\ref{fig:ATLAS8TEV}, we observe that the values of $\chi^2_{\rm exp+th}$
decrease significantly with respect to $\chi^2_{\rm exp}$. As already remarked
for the other Drell-Yan data, this fact further indicates that a careful
account of theoretical uncertainties is crucial to assess the predictive
power of a PDF set. For CT18 and NNPDF4.0, the MHO and PDF+$\alpha_s$
contributions to the $\chi^2$ have approximately the same size, and are
relatively large. For MSHT20 and PDF4LHC21, the MHO contribution to the $\chi^2$
is essentially immaterial. This is possibly due to the fact that there is a
large variance in the quality of the description of this dataset before
including theoretical uncertainties in the computation of the $\chi^2$:
even if all PDF sets provide an unsatisfactory description of the data,
MSHT20 and PDF4LHC21 have a $\chi^2_{\rm exp}$ of order 2, whereas all of the
others have a $\chi^2_{\rm exp}$ of order 5--10. Once theoretical uncertainties
are included, one gets $\chi^2_{\rm exp+th}$ of the order of 1, except for ABMP16
and NNPDF4.0, for which $\chi^2_{\rm exp+th}$ is equal to 3.47 and 3.83.
The discrepancy between experimental data and theoretical predictions obtained
with NNPDF4.0 instead of PDF4LHC21 is visible in the right panel of
Fig.~\ref{fig:ATLAS8TEV}. The shape of the NNPDF4.0 prediction displays a
peculiar dip around a value of the dilepton rapidity of 2.7.

This is a case in which NNPDF4.0 seems to perform poorly. This fact is a little
surprising because, as discussed in Sect.~\ref{subsec:LHC_DY}, an earlier
version~\cite{ATLAS:2017rue} of the ATLAS measurement of~\cite{ATLAS:2023lsr}
was included in NNPDF4.0. This is a substantial difference with respect to all
of the other measurements examined in this work, which, albeit corresponding
to production processes already included in NNPDF4.0, are completely out of
sample. For this reason, we consider that, only in this case, some more
extensive investigations are needed. To this purpose, we recompute the values
of $\chi^2_{\rm exp}$ and $\chi^2_{\rm exp+th}$ using the NNPDF4.0 PDF sets
that include QED corrections~\cite{NNPDF:2024djq}, MHOUs~\cite{NNPDF:2024dpb},
and aN$^3$LO corrections and MHOUs~\cite{NNPDF:2024nan}. All these
PDF sets include the ATLAS Drell-Yan $Z$ boson production measurements at 8~TeV
presented in~\cite{ATLAS:2016gic,ATLAS:2017rue}. Furthermore, to understand the
interplay of these measurements with the new version considered
here~\cite{ATLAS:2023lsr} (see Sect.~\ref{sec:data} for details), we perform
the following additional fits:
\begin{enumerate}[(a)]
\item  a fit equivalent to the NNLO NNPDF4.0 baseline fit excluding the
  ATLAS measurement of~\cite{ATLAS:2017rue};
\item a fit equivalent to the NNLO NNPDF4.0 baseline fit in which the
  ATLAS measurement of~\cite{ATLAS:2017rue} is replaced with that
  of~\cite{ATLAS:2023lsr};
\item a fit equivalent to fit (b), in which the ATLAS measurement
  of~\cite{ATLAS:2023lsr} is weighted as explained in Sect.~4.2.3
  of~\cite{NNPDF:2021njg};
\item a fit equivalent to fit (b), including MHOUs.
\end{enumerate}

\begin{table}[!t]
  \centering
  \scriptsize
  \renewcommand{\arraystretch}{1.5}
  \begin{tabularx}{\textwidth}{Xrccccccccccc}
Dataset
& \rotatebox{0}{$n_{\rm dat}$}
& \rotatebox{0}{$\sqrt{2/n_{\rm dat}}$}
&
& \rotatebox{80}{NNPDF4.0}
& \rotatebox{80}{aN$^3$LO MHOU}
& \rotatebox{80}{MHOU}
& \rotatebox{80}{QED}
& \rotatebox{80}{fit (a)}
& \rotatebox{80}{fit (b)}
& \rotatebox{80}{fit (c)}
& \rotatebox{80}{fit (d)}\\
\toprule
\multirow{2}{*}{ATLAS 8~TeV $Z$ $\frac{d\sigma}{d|y|}$~\cite{ATLAS:2023lsr}}
& \multirow{2}{*}{7}
& \multirow{2}{*}{0.53}
& $\chi^2_{\rm exp+th}$
& 3.83 & 3.32 & 3.33 & 3.93 & 3.43 & 2.24 & 0.17 & 1.95 \\
&
&
& $\chi^2_{\rm exp}$
& 7.90 & 8.42 & 8.38 & 8.77 & 7.24 & 3.49 & 0.18 & 3.19 \\
\midrule
\multirow{2}{*}{ATLAS 8~TeV $Z$ $\frac{d\sigma}{d|y|}$~\cite{ATLAS:2017rue}}  
& \multirow{2}{*}{60}
& \multirow{2}{*}{0.18}
& $\chi^2_{\rm exp+th}$
& 1.08 & 1.05 & 1.01 & 1.09 & 1.08 & 1.06 & 1.02 & 1.01 \\
&
&
& $\chi^2_{\rm exp}$
& 1.23 & 1.18 & 1.11 & 1.25 & 1.24 & 1.28 & 1.41 & 1.17 \\
\midrule
\multirow{2}{*}{ATLAS 8~TeV $Z$ $\frac{d\sigma}{d|y|}$ (at $Z$-peak)~\cite{ATLAS:2017rue}}
& \multirow{2}{*}{24}
& \multirow{2}{*}{0.29}
& $\chi^2_{\rm exp+th}$
& 1.30 & 1.27 & 1.27 & 1.29 & 1.31 & 1.30 & 1.28 & 1.28 \\
&
&
& $\chi^2_{\rm exp}$
& 1.31 & 1.28 & 1.27 & 1.30 & 1.31 & 1.31 & 1.28 & 1.31 \\
\bottomrule
\end{tabularx}

  \vspace{0.3cm}
  \caption{Same as Table~\ref{tab-chi2-atlasz-all} for the baseline NNPDF4.0
    PDF set and for the additional NNPDF4.0-like PDF sets described in the text.
    Values are displayed separately for the measurement
    from~\cite{ATLAS:2023lsr}, for the measurement from~\cite{ATLAS:2017rue},
    and for the subset of the latter corresponding to the invariant mass bin
    of the $Z$ peak.}
  \label{tab:Z8tev}
  \end{table}

The values of $\chi^2_{\rm exp}$ and $\chi^2_{\rm exp+th}$ computed with the baseline
NNPDF4.0 PDF set and with all the aforementioned PDF sets are collected in
Table~\ref{tab:Z8tev}. Values are displayed for the ATLAS measurement
of~\cite{ATLAS:2023lsr}, which is included only in fits (b), (c), and (d),
for the ATLAS measurement of~\cite{ATLAS:2017rue}, which is included in the
NNPDF4.0, aN$^3$LO MHOU, MHOU, and QED fits, and for the subset of the ATLAS
measurement of~\cite{ATLAS:2017rue} corresponding to the invariant mass bin of
the $Z$ peak. This way, the kinematic coverage is the same as
in~\cite{ATLAS:2023lsr}. The results corresponding to the NNPDF4.0 baseline
fit are the same as in Table~\ref{tab-chi2-atlasz-all}.

From Table~\ref{tab:Z8tev}, we make the following conclusions. The ATLAS
dataset of~\cite{ATLAS:2017rue} is described fairly well by NNPDF4.0,
including the bin corresponding to the $Z$-peak invariant mass,
whereas the dataset of~\cite{ATLAS:2023lsr} is not, even when accounting for
theoretical uncertainties in the computation of $\chi^2_{\rm exp+th}$. This
state of affairs does not change if one considers variants of the NNPDF4.0 PDF
sets including N$^3$LO corrections, MHOUs, or QED corrections. It is therefore
unlikely that theoretical inaccuracy is a limitation in the description of the
ATLAS measurement of~\cite{ATLAS:2023lsr}. The ATLAS dataset
of~\cite{ATLAS:2017rue} is described with comparable quality by a PDF
set determined from a fit without it (fit (a)); in this case, the description
of the ATLAS dataset of~\cite{ATLAS:2023lsr} does not improve in a
significant way. If instead one tries to fit the ATLAS measurement
of~\cite{ATLAS:2023lsr} (fit (b)), the value of $\chi^2_{\rm exp+th}$
($\chi^2_{\rm exp}$) improves by about $2\sigma$ ($8\sigma$). At the same time,
the description of the ATLAS measurement of~\cite{ATLAS:2017rue} does
not change in a significant way. We therefore conclude that the old and new
measurements are not in tension between each other. The picture can be further
improved if one repeats fit (b) with inclusion of MHOUs (fit (d)): in this
case, the values of $\chi^2_{\rm exp+th}$ and $\chi^2_{\rm exp}$ reduce by about
another half of a sigma. One may finally wonder whether the ATLAS measurement
of~\cite{ATLAS:2023lsr} is in tension with other datasets included in NNPDF4.0.
In this respect, fit (c) reveals that an overly good description of the dataset
can be achieved if it is overweighted. The description of the ATLAS measurement
of~\cite{ATLAS:2017rue} is not significantly altered, in comparison to the other
fits, thus confirming that the two measurements are consistent with each other.
However, the global fit quality deteriorates significantly: the total
$\chi^2_{\rm exp}$ per point increases from 1.16 (in the default NNPDF4.0 fit) to
1.24. Because there are about 4600 fitted data points, this corresponds to a
worsening of about $4\sigma$. Inspection of individual dataset figures reveals
that this is due to a deterioration in the description of several Drell-Yan
measurements, which see the following increase in the value of $\chi^2_{\rm exp}$:
D0 $W$ muon asymmetry production~\cite{D0:2014kma} from 1.91 to 4.27
($n_{\rm dat}=9$); ATLAS $W$, $Z$ production, 7~TeV~\cite{ATLAS:2016nqi} from
1.67 to 3.07 ($n_{\rm dat}=53$); LHCb $Z$ production, 7~TeV~\cite{LHCb:2012gii}
from 1.65 to 2.48 ($n_{\rm dat}=9$); LHCb $W^\pm$ production,
7~TeV~\cite{LHCb:2015okr} from 1.97 to 4.12 ($n_{\rm dat}=29$); and LHCb $Z$
production, 8~TeV~\cite{LHCb:2015kwa} from 1.33 to 2.32 ($n_{\rm dat}=17$).
We therefore conclude that the ATLAS measurement of~\cite{ATLAS:2023lsr}
is in tension with these other datasets.

In summary, the ATLAS measurement of~\cite{ATLAS:2023lsr} is consistent with
its earlier version~\cite{ATLAS:2017rue}, but in
tension with other Drell-Yan measurements included in NNPDF4.0. An acceptable
description of it can be achieved if this is included in the fit and if
MHOU are taken into account. The value of $\chi^2_{\rm exp+th}=1.95$ is indeed
only slightly less than $2\sigma$ away from the unit expectation. This level
of disagreement does not appear to be more pathological than that observed for
few other datasets included in NNPDF4.0, and is also similar to that
observed for other datasets examined in the following. Note however that, for
all of the other datasets examined in this work, we will look at the data-theory
(dis)agreement only {\it before} inclusion in a fit. Refitting could
possibly improve the overall picture, however investigations along this
direction are beyond the scope of this work, as they will be part of the
dataset selection involved with a future NNPDF release.

\subsection{Top-quark pair production measurements}
\label{subsec:LHCtoppair}

\begin{figure}[!t]
  \centering
  \includegraphics[width=0.49\textwidth]{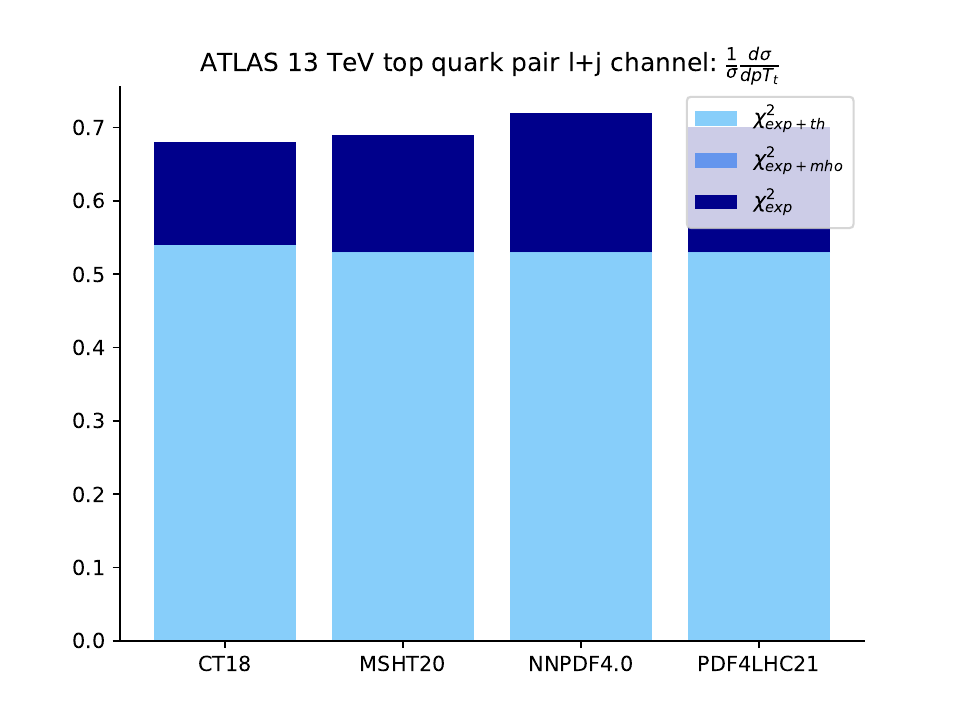}
  \includegraphics[width=0.49\textwidth]{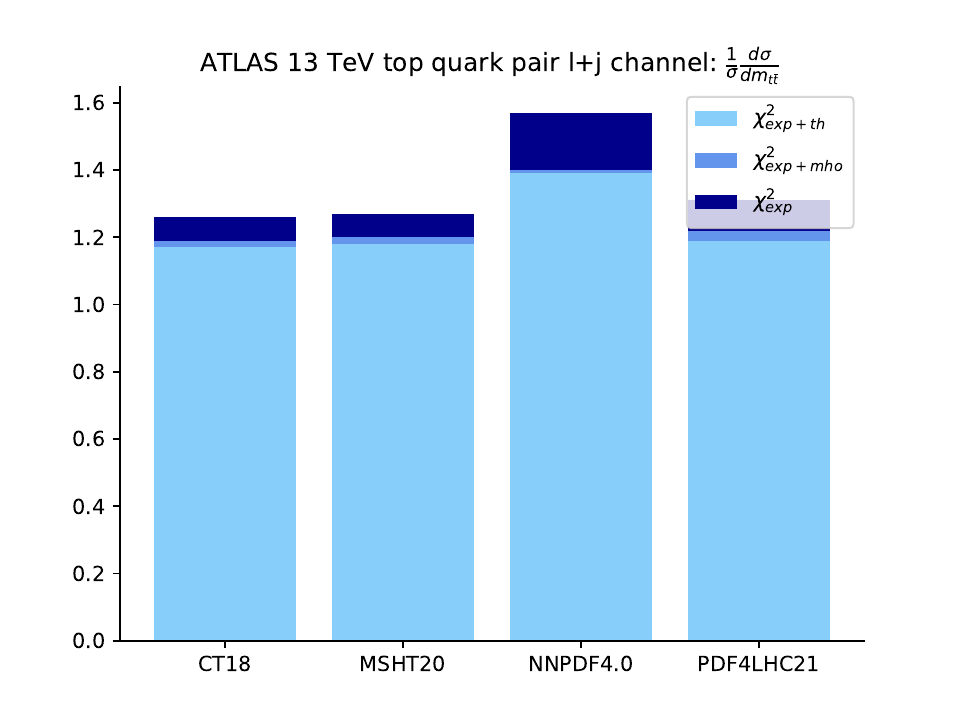}\\
  \includegraphics[width=0.49\textwidth]{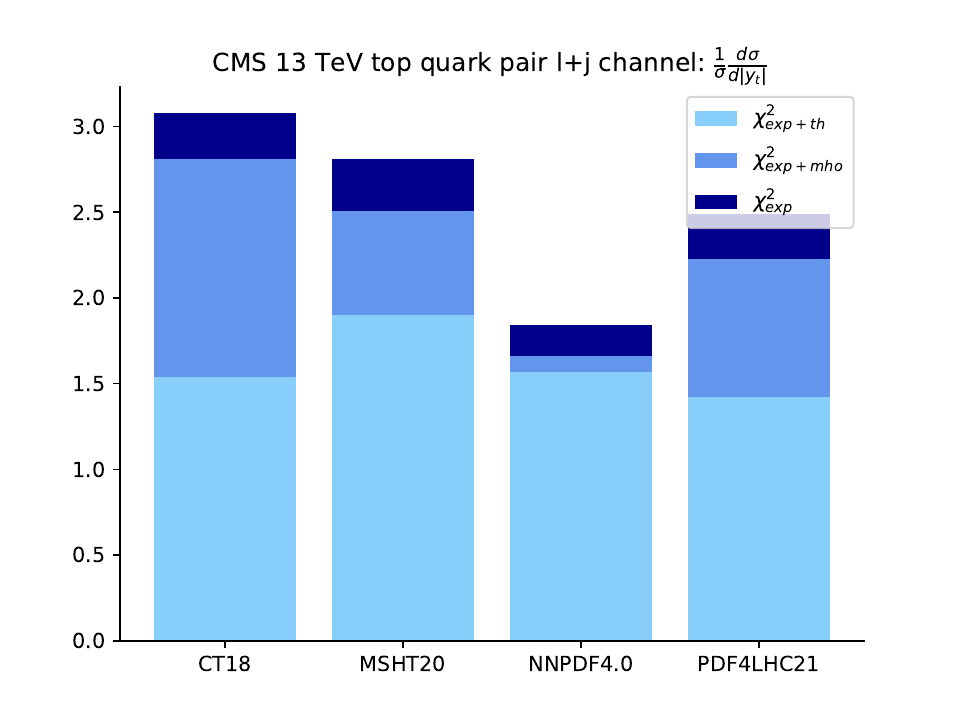}
  \includegraphics[width=0.49\textwidth]{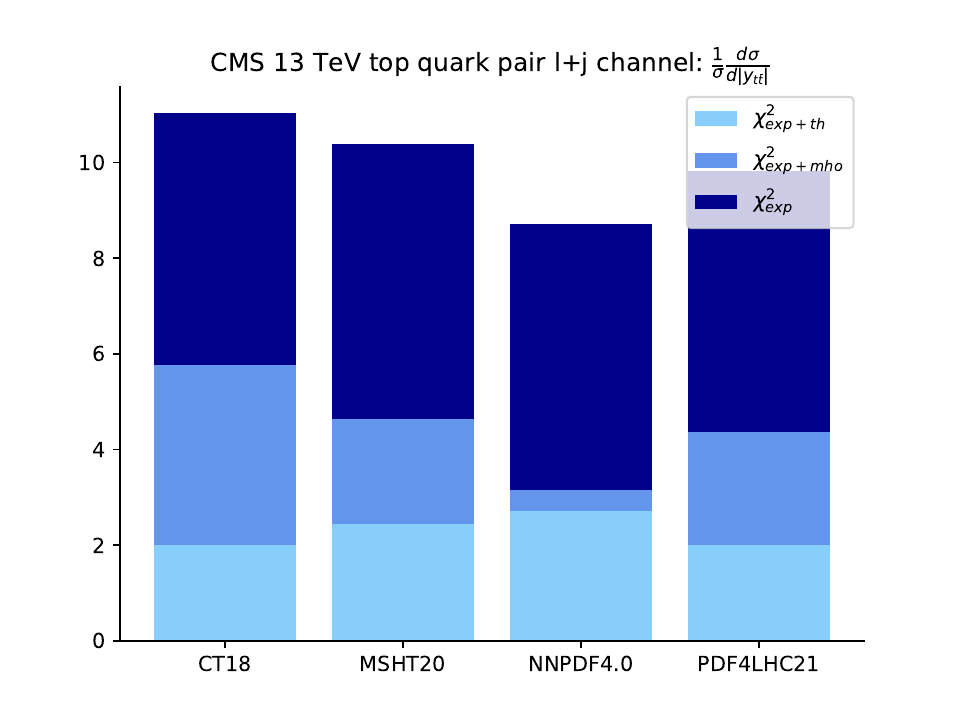}\\
  \includegraphics[width=0.49\textwidth]{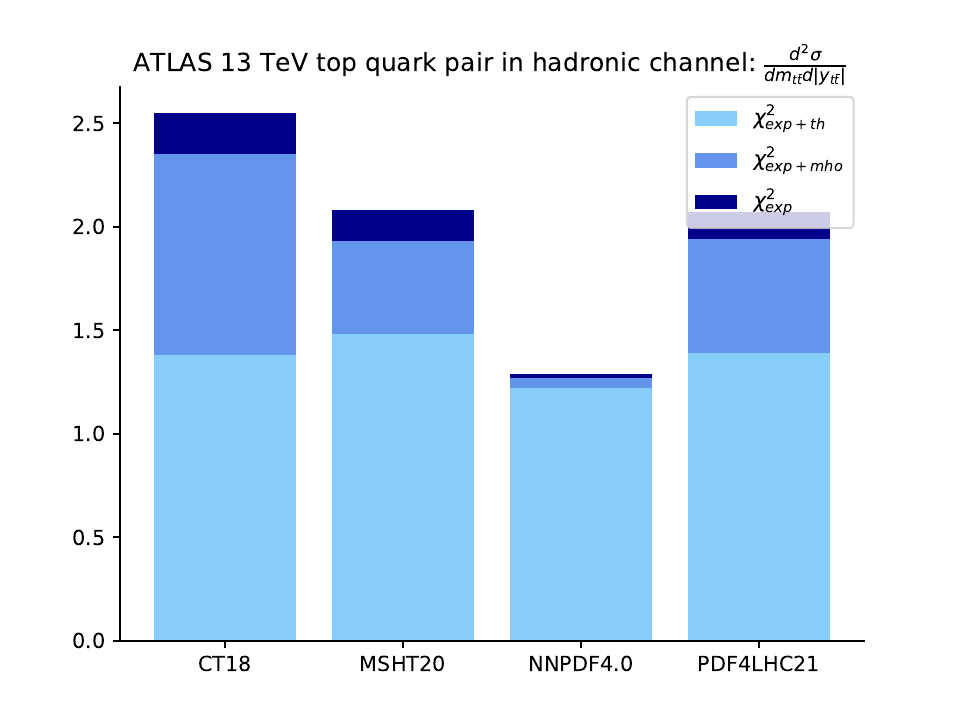}
  \includegraphics[width=0.49\textwidth]{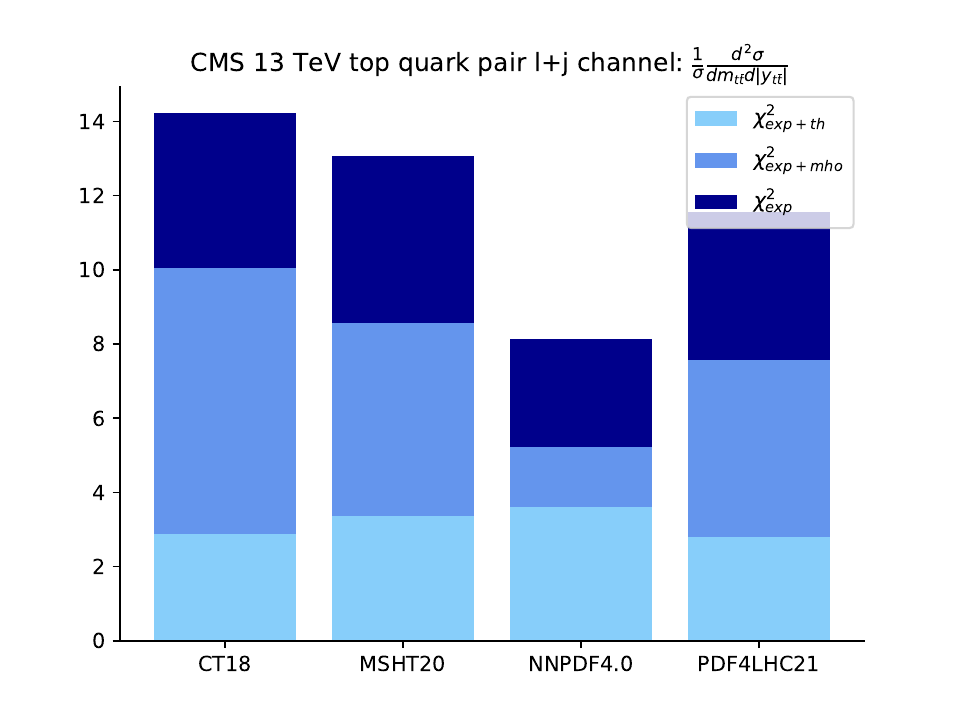}\\
  \caption{Same as Fig.~\ref{fig:DY-chi2} for a representative subset of
    top-quark pair production measurements at the LHC 13~TeV. Specifically,
    from top to bottom, left to right: the ATLAS lepton+jets normalised cross
    sections single-differential in the transverse momentum of the top quark 
    $p_T^t$  ($n_{\rm dat} = 8$, $\sqrt{2/n_{\rm dat}} = 0.50$), 
    and in the invariant mass of the top-quark pair $m_{t\bar t}$  ($n_{\rm dat} = 9$, $\sqrt{2/n_{\rm dat}} = 0.47$); the
    CMS lepton+jets normalised cross sections, single-differential in the
    absolute rapidity of the top quark and of the top-quark pair $|y_t|$ ($n_{\rm dat} = 11$, $\sqrt{2/n_{\rm dat}} = 0.43$) and
    $|y_{t\bar t}|$  ($n_{\rm dat} = 10$, $\sqrt{2/n_{\rm dat}} = 0.45$), the ATLAS all-hadronic absolute cross section,
    double-differential in the invariant mass and absolute rapidity of the
    top-quark pair  ($n_{\rm dat} = 11$, $\sqrt{2/n_{\rm dat}} = 0.43$); and the CMS lepton+jets normalised cross section,
    double-differential in the invariant mass and absolute rapidity of the
    top-quark pair ($n_{\rm dat} = 35$, $\sqrt{2/n_{\rm dat}} = 0.24$). Histogram plots for the other datasets are collected in
    Fig.\ref{fig:chi2histo_ttbar_additional} of
    Appendix~\ref{app:extra_results}.}
  \label{fig:top-ATLAS-2-chi2} 
\end{figure}

\begin{figure}[!t]
  \includegraphics[width=0.49\textwidth]{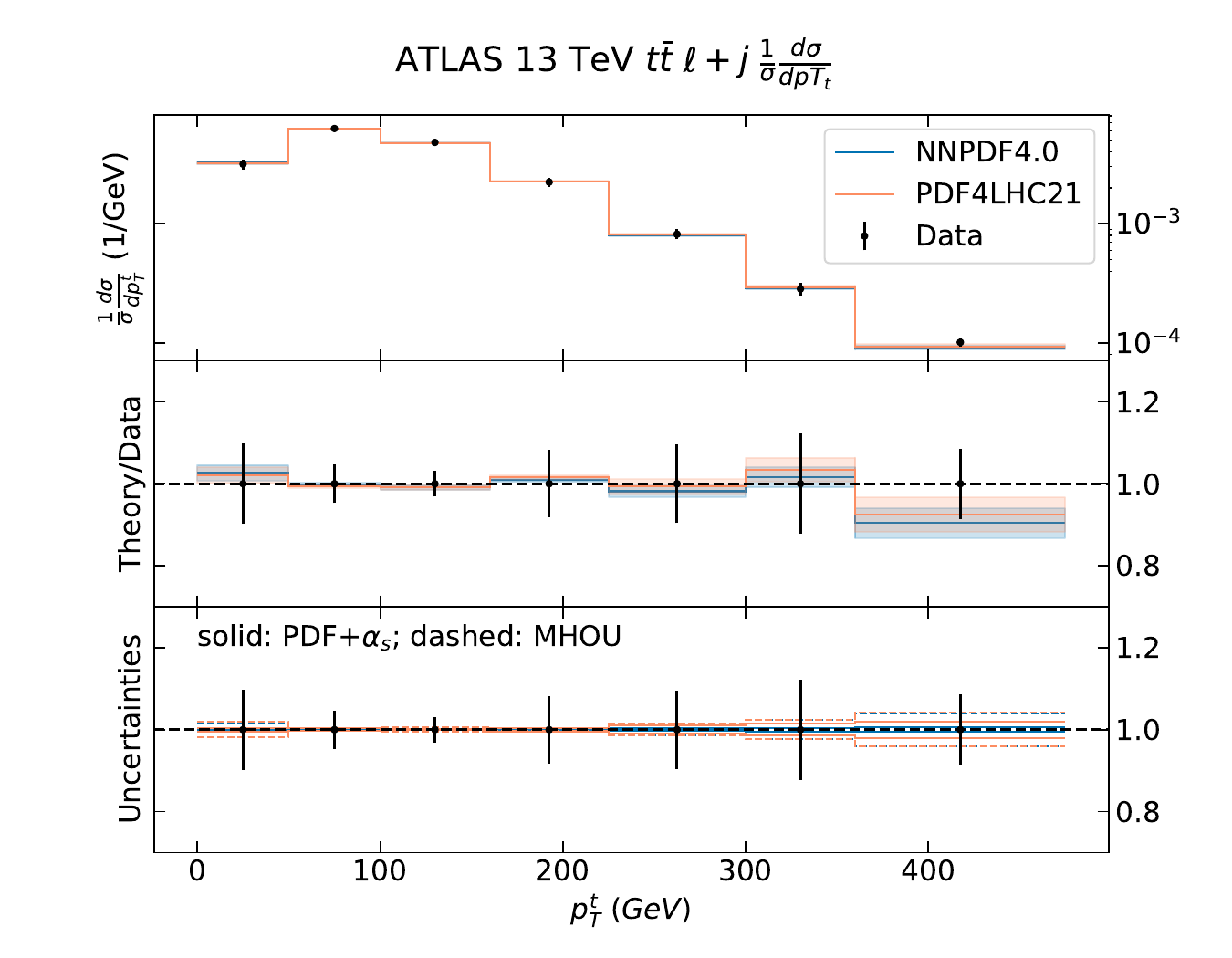}
  \includegraphics[width=0.49\textwidth]{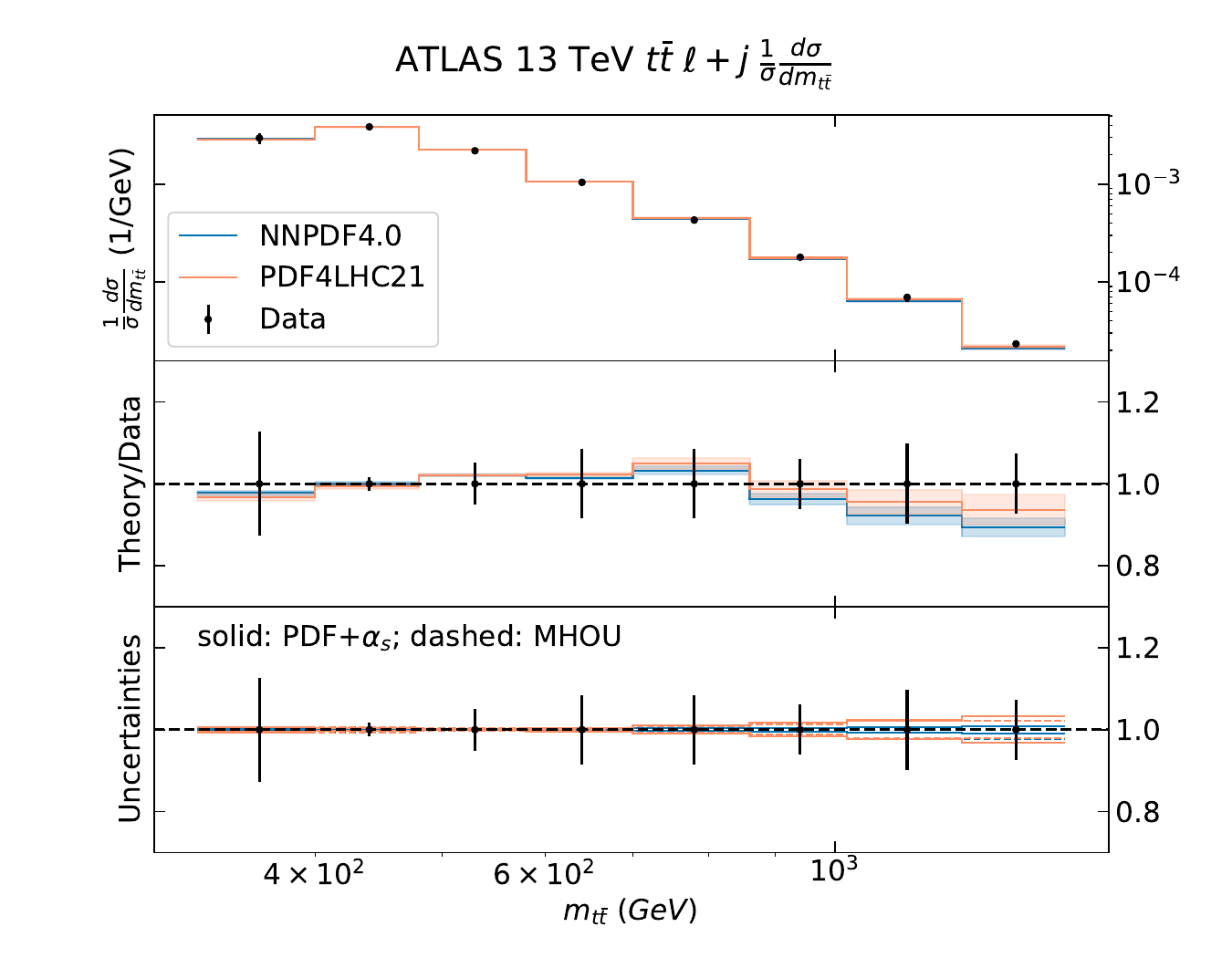}\\
  \includegraphics[width=0.49\textwidth]{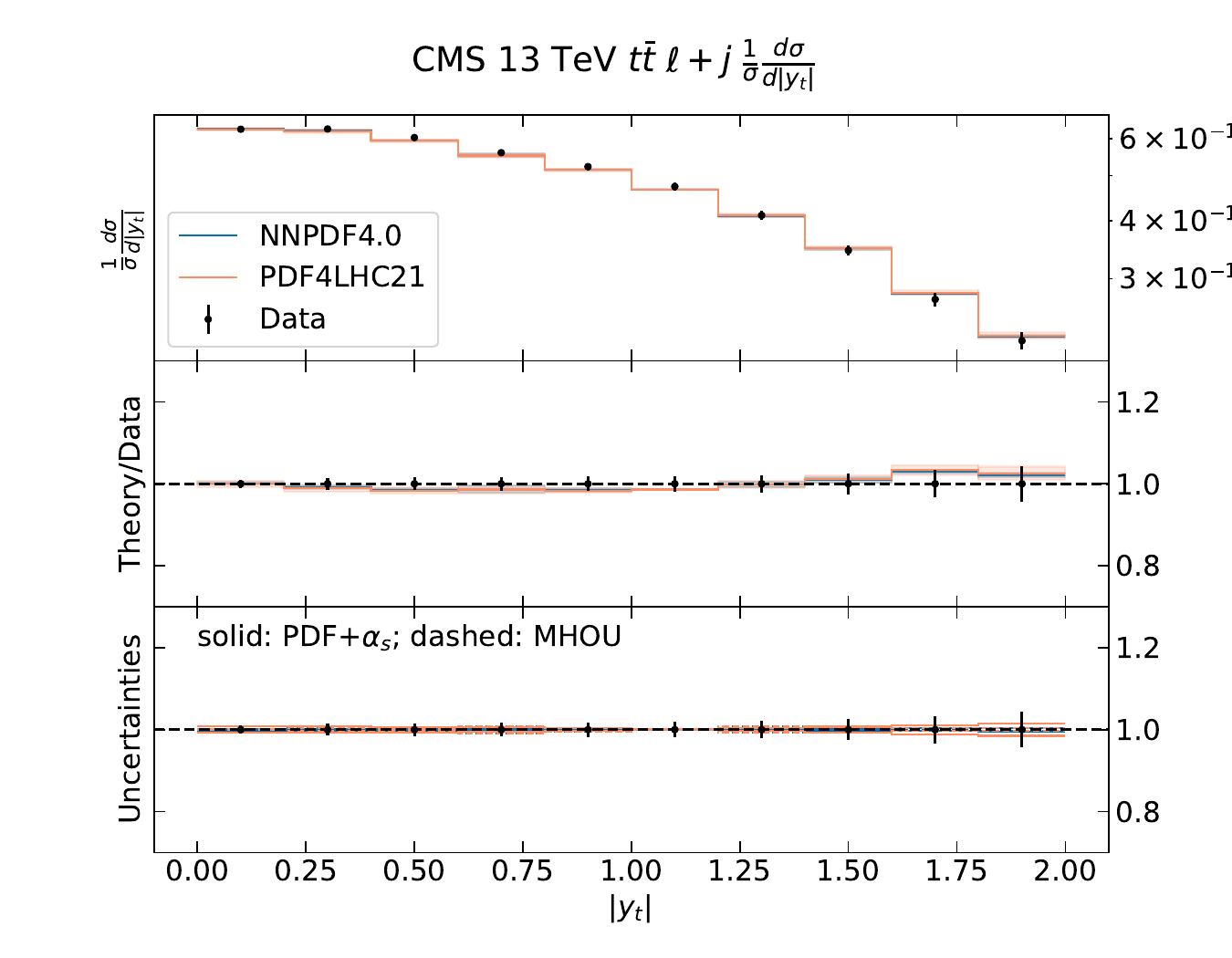}
  \includegraphics[width=0.49\textwidth]{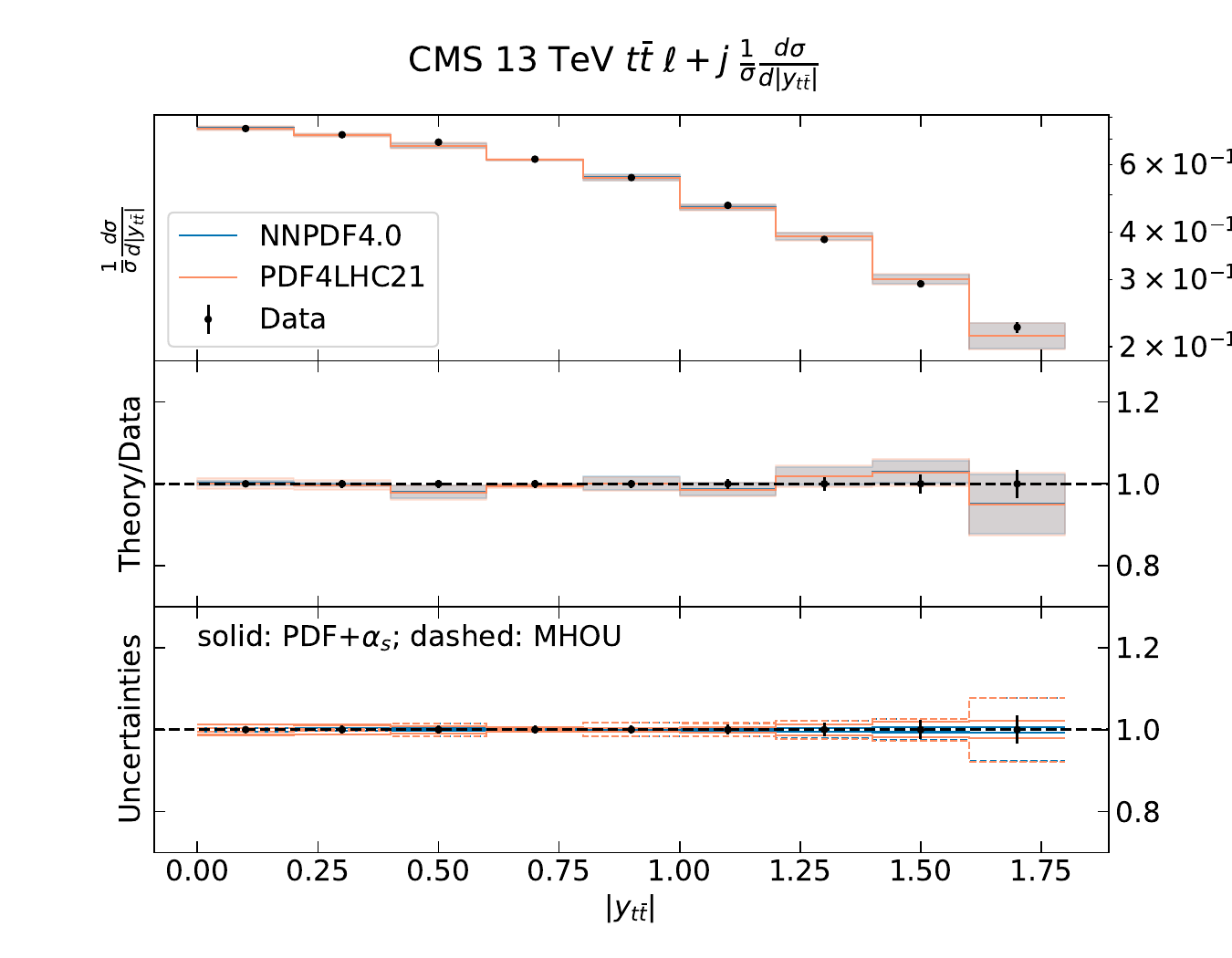}\\
  \includegraphics[width=0.49\textwidth]{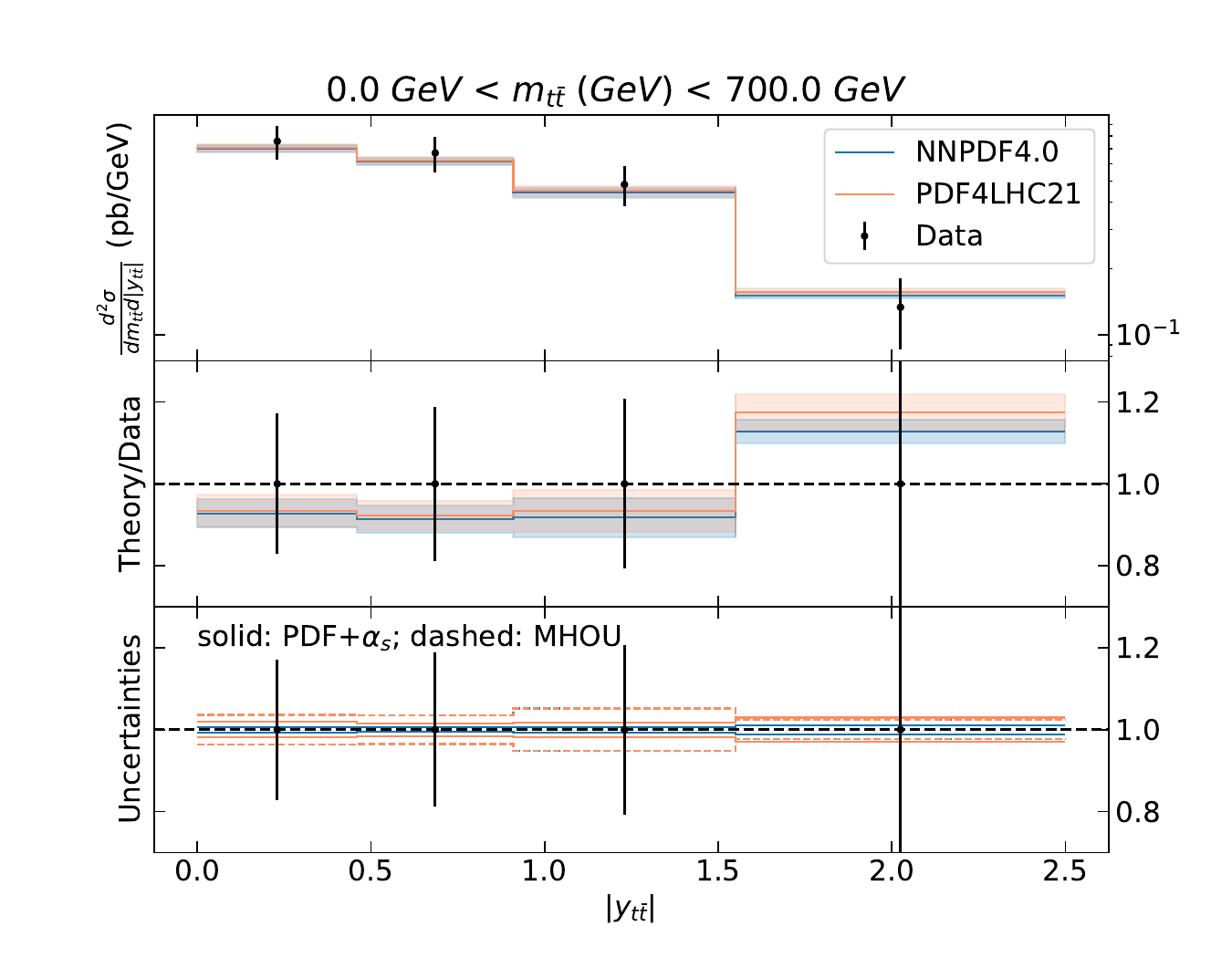}
  \includegraphics[width=0.49\textwidth]{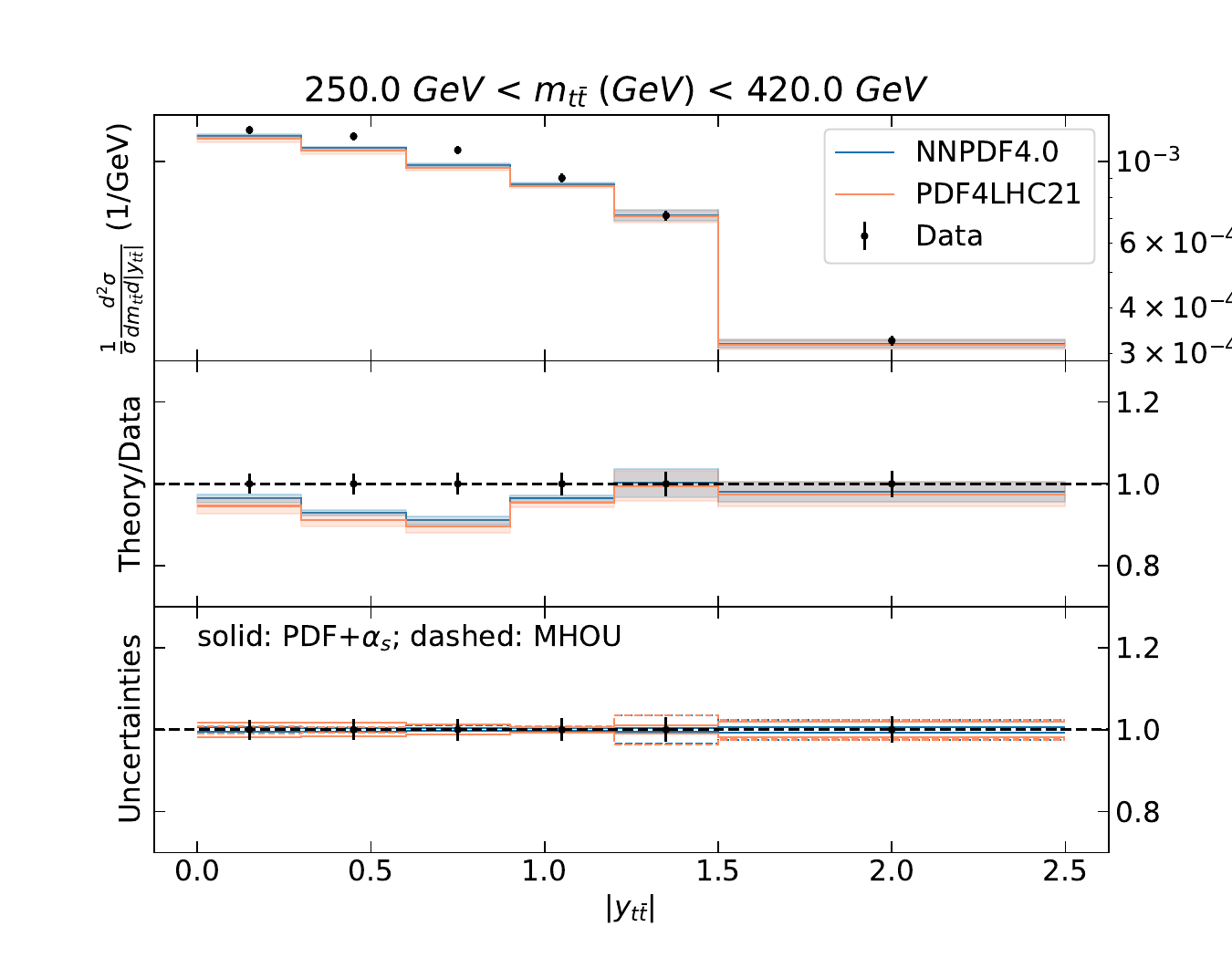}\\
  \caption{Same as Fig.~\ref{fig:DY-datatheory} for the same representative
    subset of top-quark pair measurements of Fig.~\ref{fig:top-ATLAS-2-chi2}.
    In the case of the ATLAS (bottom left) and CMS (bottom right)
    double-differential distributions, only the bin at the lowest invariant
    mass is shown. Additional results are collected in
    Figs.~\ref{fig:datatheory_ttbar_additional_1}--\ref{fig:datatheory_ttbar_additional_2} of Appendix~\ref{app:extra_results}.}
  \label{fig:datatheory-ttbar} 
\end{figure}

We continue by discussing the LHC top-quark pair production measurements
outlined in Sect.~\ref{sec:data}, see also Table~\ref{tab:Z}. The values of
$\chi^2_{\rm exp}$ and $\chi^2_{\rm exp+th}$, computed for each of the PDF sets
summarised in Sect.~\ref{subsec:pdfs}, are reported in
Table~\ref{tab:chi2-ttb}. The experimental covariance matrix
is regularised as explained in Sect.~\ref{subsec:compatibility} for the
following datasets: the ATLAS all-hadronic absolute single-differential
distribution in the invariant mass of the top-quark pair and
double-differential distribution in the invariant mass and absolute rapidity of
the top-quark pair; the ATLAS lepton+jets normalised single-differential
distributions in the invariant mass of the top-quark pair and in the transverse
momentum of the top quark; and the CMS lepton+jets normalised
double-differential distribution in the invariant mass and absolute rapidity of
the top-quark pair. See Appendix~\ref{app:unreg} for the unregularised values
of $\chi^2_{\rm exp+th}$. The breakdown of $\chi^2_{\rm exp+th}$ into
$\chi^2_{\rm exp+mho}$ and $\chi^2_{\rm exp}$ is displayed in
Fig.~\ref{fig:top-ATLAS-2-chi2}, albeit only for a representative subset
of distributions, specifically: the ATLAS lepton+jets normalised cross sections,
single-differential in the transverse momentum of the top quark, $p_T^t$, and
in the invariant mass of the top-quark pair, $m_{t\bar t}$; the CMS
lepton+jets normalised cross sections, single-differential in the absolute
rapidity of the top quark and of the top-quark pair, $|y_t|$ and $|y_{t\bar t}|$;
the ATLAS all-hadronic absolute cross section, double-differential in the
invariant mass and absolute rapidity of the top-quark pair; and the CMS
lepton+jets normalised cross section, double-differential in the invariant
mass and absolute rapidity of the top-quark pair. Histogram plots for the other
datasets are collected in Fig.~\ref{fig:chi2histo_ttbar_additional} of
Appendix~\ref{app:extra_results}. The data-theory
comparison is displayed in Fig.~\ref{fig:datatheory-ttbar} for the same
representative subset of top-quark pair measurements of
Fig.~\ref{fig:top-ATLAS-2-chi2}. In the case of the ATLAS and CMS
double-differential distributions, only the bin at the lowest invariant mass
is shown. Additional results are collected in
Figs.~\ref{fig:datatheory_ttbar_additional_1}-\ref{fig:datatheory_ttbar_additional_2} of Appendix~\ref{app:extra_results}.
Note that, for normalised distributions, we consistently do not display the
last bin, which is linearly dependent from the others by construction. Hence the
number of data points displayed is one unit less than the number of data points
reported in Table~\ref{tab:Z}.

\begin{table}[!t]
  \scriptsize
  \centering
  \renewcommand{\arraystretch}{1.5}
  \begin{tabularx}{\textwidth}{Xrccccccccccc}
Dataset
& \rotatebox{0}{$n_{\rm dat}$}
& \rotatebox{0}{$\sqrt{2/n_{\rm dat}}$}
&
& \rotatebox{80}{ABMP16}
& \rotatebox{80}{CT18}
& \rotatebox{80}{CT18A}
& \rotatebox{80}{CT18Z}
& \rotatebox{80}{MSHT20}
& \rotatebox{80}{NNPDF3.1}
& \rotatebox{80}{NNPDF4.0}
& \rotatebox{80}{PDF4LHC15}
& \rotatebox{80}{PDF4LHC21} \\
\toprule
\multirow{2}{*}{ATLAS 13~TeV $t\bar{t}$ all hadr. $\frac{d\sigma}{dm_{t\bar{t}}}$}
& \multirow{2}{*}{9}
& \multirow{2}{*}{0.47}
& $\chi^2_{\rm exp+th}$
& 0.84 & 0.99 & 0.97 & 0.94 & 0.97 & 0.86 & 0.81 & 0.96 & 0.93 \\
&
&
& $\chi^2_{\rm exp}$
& 0.88 & 1.21 & 1.16 & 1.15 & 1.12 & 0.91 & 0.84 & 1.13 & 1.06 \\
\midrule
\multirow{2}{*}{ATLAS 13~TeV $t\bar{t}$ all hadr. $\frac{1}{\sigma}\frac{d\sigma}{d|y_{t\bar{t}}|}$}
& \multirow{2}{*}{12}
& \multirow{2}{*}{0.41}
& $\chi^2_{\rm exp+th}$
& 0.62 & 0.78 & 0.77 & 0.85 & 0.74 & 0.64 & 0.68 & 0.73 & 0.73 \\
&
&
& $\chi^2_{\rm exp}$
& 0.68 & 0.85 & 0.83 & 0.95 & 0.79 & 0.67 & 0.71 & 0.82 & 0.78 \\ 
\midrule
\multirow{2}{*}{ATLAS 13~TeV $t\bar{t}$ all hadr. $\frac{d^2\sigma}{dm_{t\bar{t}} d|y_{t\bar{t}}|}$}
& \multirow{2}{*}{11}
& \multirow{2}{*}{0.43}
& $\chi^2_{\rm exp+th}$
& 0.92 & 1.38 & 1.39 & 1.42 & 1.48 & 1.12 & 1.22 & 1.22 & 1.39 \\ 
&
&
& $\chi^2_{\rm exp}$
& 1.05 & 2.55 & 2.38 & 2.84 & 2.08 & 1.20 & 1.29 & 2.11 & 2.07 \\ 
\midrule
\multirow{2}{*}{ATLAS 13~TeV $t\bar{t}$ $\ell+j$  $\frac{1}{\sigma}\frac{d\sigma}{dm_{t\bar{t}}}$}
& \multirow{2}{*}{9}
& \multirow{2}{*}{0.47}
& $\chi^2_{\rm exp+th}$
& 1.41 & 1.17 & 1.17 & 1.04 & 1.18 & 1.46 & 1.39 & 1.20 & 1.19 \\ 
&
&
& $\chi^2_{\rm exp}$
& 1.67 & 1.26 & 1.26 & 1.12 & 1.27 & 1.65 & 1.57 & 1.32 & 1.31 \\
\midrule
\multirow{2}{*}{ATLAS 13~TeV $t\bar{t}$ $\ell+j$  $\frac{1}{\sigma}\frac{d\sigma}{dp_T^t}$}
& \multirow{2}{*}{8}
& \multirow{2}{*}{0.50}
& $\chi^2_{\rm exp+th}$
& 0.56 & 0.54 & 0.54 & 0.52 & 0.53 & 0.56 & 0.53 & 0.53 & 0.53 \\
&
&
& $\chi^2_{\rm exp}$
& 0.76 & 0.68 & 0.68 & 0.67 & 0.69 & 0.77 & 0.72 & 0.68 & 0.70 \\
\midrule
\multirow{2}{*}{ATLAS 13~TeV $t\bar{t}$ $\ell+j$ $\frac{d\sigma}{d|y_{t}|}$}
& \multirow{2}{*}{5}
& \multirow{2}{*}{0.63}
& $\chi^2_{\rm exp+th}$
& 1.39 & 1.05 & 1.09 & 0.92 & 1.10 & 1.70 & 1.58 & 1.09 & 1.15 \\
&
&
& $\chi^2_{\rm exp}$
& 1.62 & 1.17 & 1.19 & 1.00 & 1.14 & 1.86 & 1.62 & 1.26 & 1.29 \\
\midrule
\multirow{2}{*}{ATLAS 13~TeV $t\bar{t}$ $\ell+j$ $\frac{1}{\sigma}\frac{d\sigma}{d|y_{t\bar{t}}|}$}
& \multirow{2}{*}{7}
& \multirow{2}{*}{0.53}
& $\chi^2_{\rm exp+th}$
& 0.57 & 0.43 & 0.42 & 0.58 & 0.47 & 0.58 & 0.42 & 0.42 & 0.39 \\
&
&
& $\chi^2_{\rm exp}$
& 0.74 & 0.57 & 0.55 & 0.99 & 0.66 & 0.65 & 0.47 & 0.56 & 0.47 \\
\midrule
\multirow{2}{*}{CMS 13~TeV $t\bar{t}$ $\ell+j$ $\frac{1}{\sigma}\frac{d\sigma}{dm_{t\bar{t}}}$}
& \multirow{2}{*}{15}
& \multirow{2}{*}{0.37}
& $\chi^2_{\rm exp+th}$
& 0.24 & 0.49 & 0.51 & 0.53 & 0.57 & 0.29 & 0.33 & 0.42 & 0.44 \\
&
&
& $\chi^2_{\rm exp}$
& 0.37 & 1.38 & 1.30 & 1.44 & 1.15 & 0.39 & 0.42 & 1.14 & 0.98 \\
\midrule
\multirow{2}{*}{CMS 13~TeV $t\bar{t}$ $\ell+j$ $\frac{1}{\sigma}\frac{d^2\sigma}{dm_{t\bar{t}}d|y_{t\bar{t}}|}$}
& \multirow{2}{*}{35}
& \multirow{2}{*}{0.24}
& $\chi^2_{\rm exp+th}$
& 2.77 & 2.89 & 2.87 & 2.76 & 3.36 & 3.01 & 3.61 & 2.81 & 2.81 \\ 
&
&
& $\chi^2_{\rm exp}$
& 8.37 & 14.2 & 13.7 & 16.6 & 13.1 & 7.31 & 8.14 & 13.1 & 11.6 \\
\midrule 
\multirow{2}{*}{CMS 13~TeV $t\bar{t}$ $\ell+j$ $\frac{1}{\sigma}\frac{d\sigma}{dp_T^t}$ }
& \multirow{2}{*}{16}
& \multirow{2}{*}{0.35}
& $\chi^2_{\rm exp+th}$
& 0.78 & 0.62 & 0.63 & 0.66 & 0.64 & 0.79 & 0.81 & 0.63 & 0.65 \\ 
&
&
& $\chi^2_{\rm exp}$
& 1.31 & 0.68 & 0.70 & 0.69 & 0.72 & 1.24 & 1.17 & 0.74 & 0.78 \\
\midrule
\multirow{2}{*}{CMS 13~TeV $t\bar{t}$ $\ell+j$ $\frac{1}{\sigma}\frac{d\sigma}{d|y_t|}$}
& \multirow{2}{*}{11}
& \multirow{2}{*}{0.43}
& $\chi^2_{\rm exp+th}$
& 1.07 & 1.54 & 1.57 & 1.81 & 1.90 & 1.22 & 1.57 & 1.38 & 1.42 \\ 
&
&
& $\chi^2_{\rm exp}$
& 1.61 & 3.08 & 2.94 & 4.02 & 2.81 & 1.46 & 1.84 & 2.77 & 2.49 \\
\midrule
\multirow{2}{*}{CMS 13~TeV $t\bar{t}$ $\ell+j$ $\frac{1}{\sigma}\frac{d\sigma}{d|y_{t\bar{t}}|}$ }
& \multirow{2}{*}{10}
& \multirow{2}{*}{0.45}
& $\chi^2_{\rm exp+th}$
& 0.94 & 2.01 & 1.89 & 2.16 & 2.44 & 1.76 & 2.71 & 1.53 & 2.00 \\
&
&
& $\chi^2_{\rm exp}$
& 8.65 & 11.0 & 10.7 & 12.9 & 10.4 & 8.06 & 8.72 & 10.6 & 9.82 \\ 
\bottomrule
\end{tabularx}

  \vspace{0.3cm}
  \caption{Same as Table~\ref{tab:chi2-DY} for the ATLAS and CMS
    top-quark pair production measurements at the LHC 13~TeV.}
  \label{tab:chi2-ttb} 
\end{table}

From inspection of Table~\ref{tab:chi2-ttb} and of
Fig.~\ref{fig:top-ATLAS-2-chi2}, we make considerations very similar to those
made for Drell-Yan weak boson production measurements at the LHC 13~TeV.
Namely, that the values of $\chi^2_{\rm exp+th}$, computed with different
input PDFs, are closer to each other than the corresponding values of
$\chi^2_{\rm exp}$, and that the former are generally rather similar across PDF
sets. The only partial exceptions are given by the ATLAS all-hadronic double
differential measurement, for which the values of $\chi^2_{\rm exp+th}$ are away
from one by slightly more than 1$\sigma$ for the MSHT20 and CT18Z sets, and the
ATLAS leptons+jet single-differential measurement in the absolute value of the 
top-quark rapidity, for which the total $\chi^2_{\rm exp+th}$  computed with
NNPDF3.1 is slightly more than 1$\sigma$ away from one.  As far as CMS is
concerned, the leptons+jet double differential distribution is poorly described
by all PDF sets, with the total $\chi^2_{\rm exp+th}$ being away from one by more
than 3$\sigma$ for all PDF sets. Less dramatic, but still significant, is the 
spread of $\chi^2_{\rm exp+th}$ values across PDF sets for the CMS lepton+jets
single-differential distribution in the absolute rapidity of the top-quark and
of the top-quark pair: in the former case, predictions obtained with CT18Z and
MSHT20 are $2\sigma$ away from one, whereas predictions obtained with MSHT20
and NNPDF4.0 are about $3\sigma$ away from one (with NNPDF4.0 performing a
little worse than MSHT20 by a quarter of a sigma).

The breakdown of $\chi^2_{\rm exp+th}$ into its theoretical components depends on
the dataset. The component due to MHO, which is relatively independent from
the PDF set, prevails over the PDF+$\alpha_s$ component in the ATLAS lepton+jets
distributions differential in the transverse momentum of the top quark and in
the invariant mass of the top-quark pair, and in the CMS lepton+jets
distribution differential in the absolute rapidity of the top-quark pair.
The PDF+$\alpha_s$ component prevails in the other datasets, although it
depends on the PDF set: it is generally larger for the CT18 and PDF4LHC21
PDF sets, which are affected by the largest uncertainties,
see Fig.~\ref{fig:lumis}, whereas it is almost immaterial for NNPDF4.0, which
has the smallest PDF uncertainties. 

Overall, the quality of the data description is generally good, being
$\chi^2_{\rm exp+th}\sim 1$ for all the datasets, except in the case of the CMS
normalised single-differential distribution in the absolute rapidity of the
top-quark pair, and double-differential distribution in the absolute
rapidity and invariant mass of the top-quark pair, for which
$\chi^2_{\rm exp+th}\sim 2-3$.
Discrepancies between data and theory that may lead to these results are seen
in Fig.~\ref{fig:datatheory-ttbar}, where experimental data and theoretical
predictions are generally well aligned to each other, within their
uncertainties, except, precisely, for the aforementioned datasets.
Understanding the reason for this behaviour, which is common to most PDF sets,
is left to future investigations. 

\subsection{Single-inclusive jet and di-jet production measurements at the LHC}
\label{subsec:LHCjets}

We now turn to LHC single-inclusive jet and di-jet production measurements
outlined in Sect.~\ref{sec:data}.
The values of $\chi^2_{\rm exp}$ and $\chi^2_{\rm exp+th}$ are reported in
Table~\ref{tab:chi2-jets}. The experimental covariance matrix is regularised
as explained in Sect.~\ref{subsec:compatibility} for all the datasets. 
Without regularisation, the values of $\chi^2_{\rm exp}$ are very poor, as they
are away from one by more than 10-20$\sigma$, independently of the input PDF
set. See Appendix~\ref{app:unreg} for the unregularised values of
$\chi^2_{\rm exp+th}$. Clearly the results that we present here do depend on
regularisation. However, as discussed in Appendix~\ref{app:unreg}, this
dependence does not affect our ability to discriminate how well different PDF
sets describe the data, which is the goal of this work. If, instead, we were
interested to characterise the datasets for inclusion in a PDF determination
(or not), we would consider other ways of decorrelating uncertainties, for
instance by identifying uncertainties that, for experimental reasons, are more
likely to be overcorrelated, see {\it e.g.}~\cite{Harland-Lang:2017ytb}. In this
sense, the main message conveyed by the numbers in Table~\ref{tab:chi2-jets}
is that the single-inclusive jet and di-jet datasets require additional
investigations on the understanding of uncertainty correlations. Only after
accomplishing these investigations, which are beyond the scope of this work,
one may be able to better judge how the various PDF sets comparatively
generalise on them.

\begin{figure}[!t]
  \centering
  \includegraphics[width=0.49\textwidth]{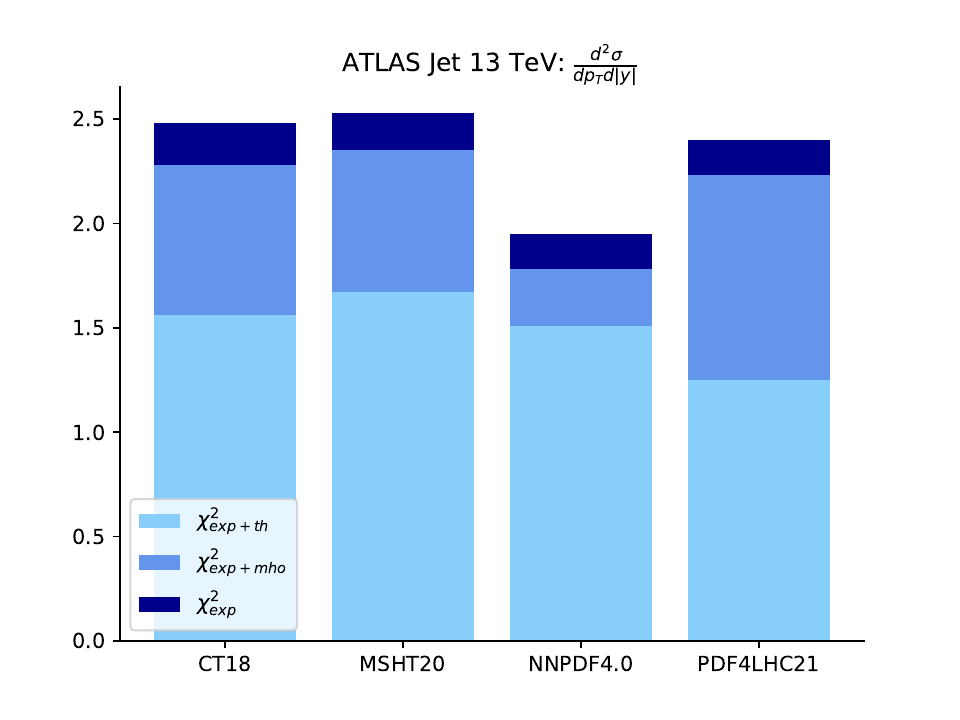}
  \includegraphics[width=0.49\textwidth]{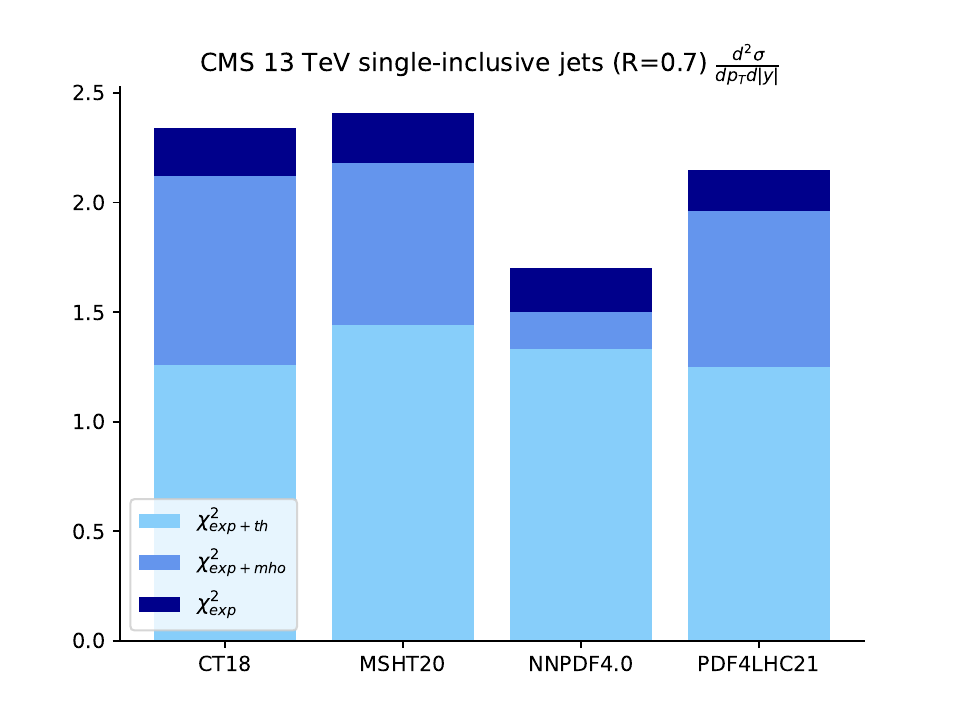}
  \includegraphics[width=0.49\textwidth]{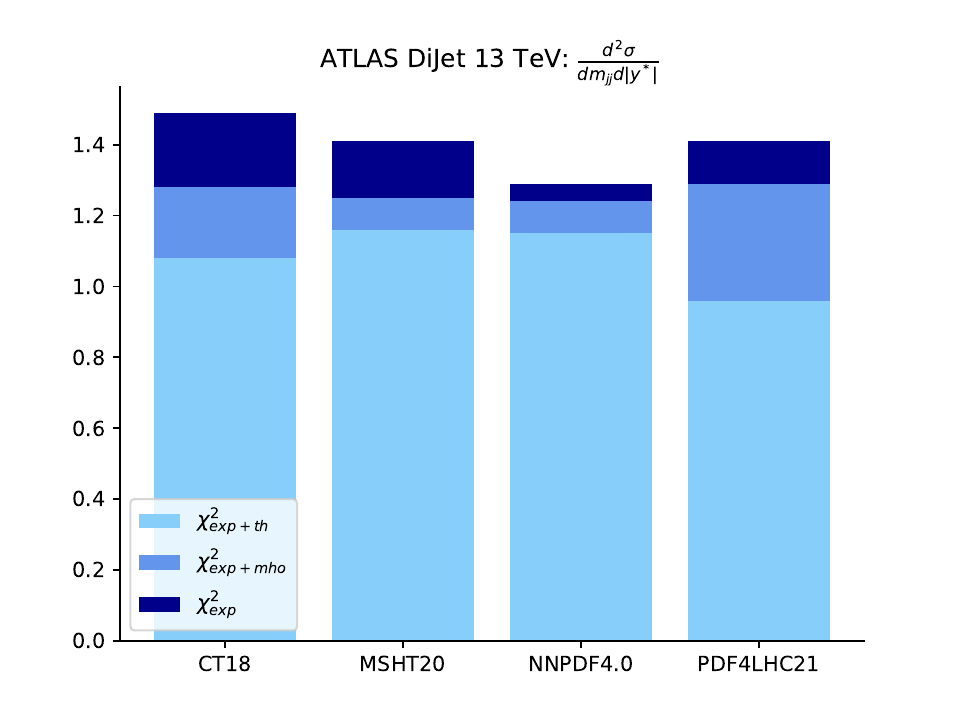}
  \caption{Same as Fig.~\ref{fig:DY-chi2} for the ATLAS and CMS
    single-inclusive (with $n_{\rm dat} = 177$ and $n_{\rm dat} = 78$ respectively, 
    and $\sqrt{2/n_{\rm dat}} = 0.11$ and 0.16 correspondingly) 
    and di-jet production measurements at the LHC 13~TeV ($n_{\rm dat} = 136$, $\sqrt{2/n_{\rm dat}} = 0.12$).}
  \label{fig:jets-chi2} 
\end{figure}

\begin{figure}[!t]
  \includegraphics[width=0.49\textwidth]{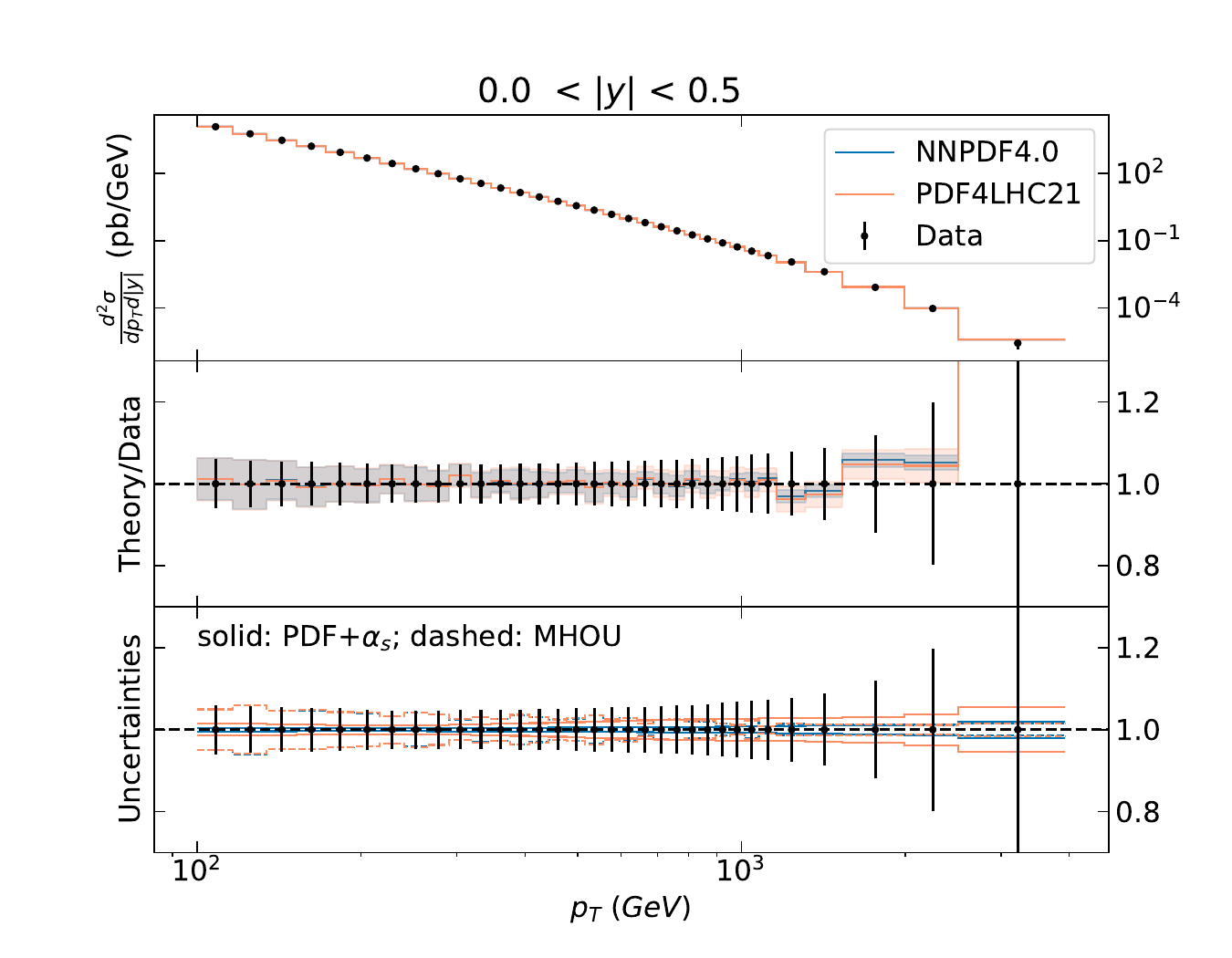}
  \includegraphics[width=0.49\textwidth]{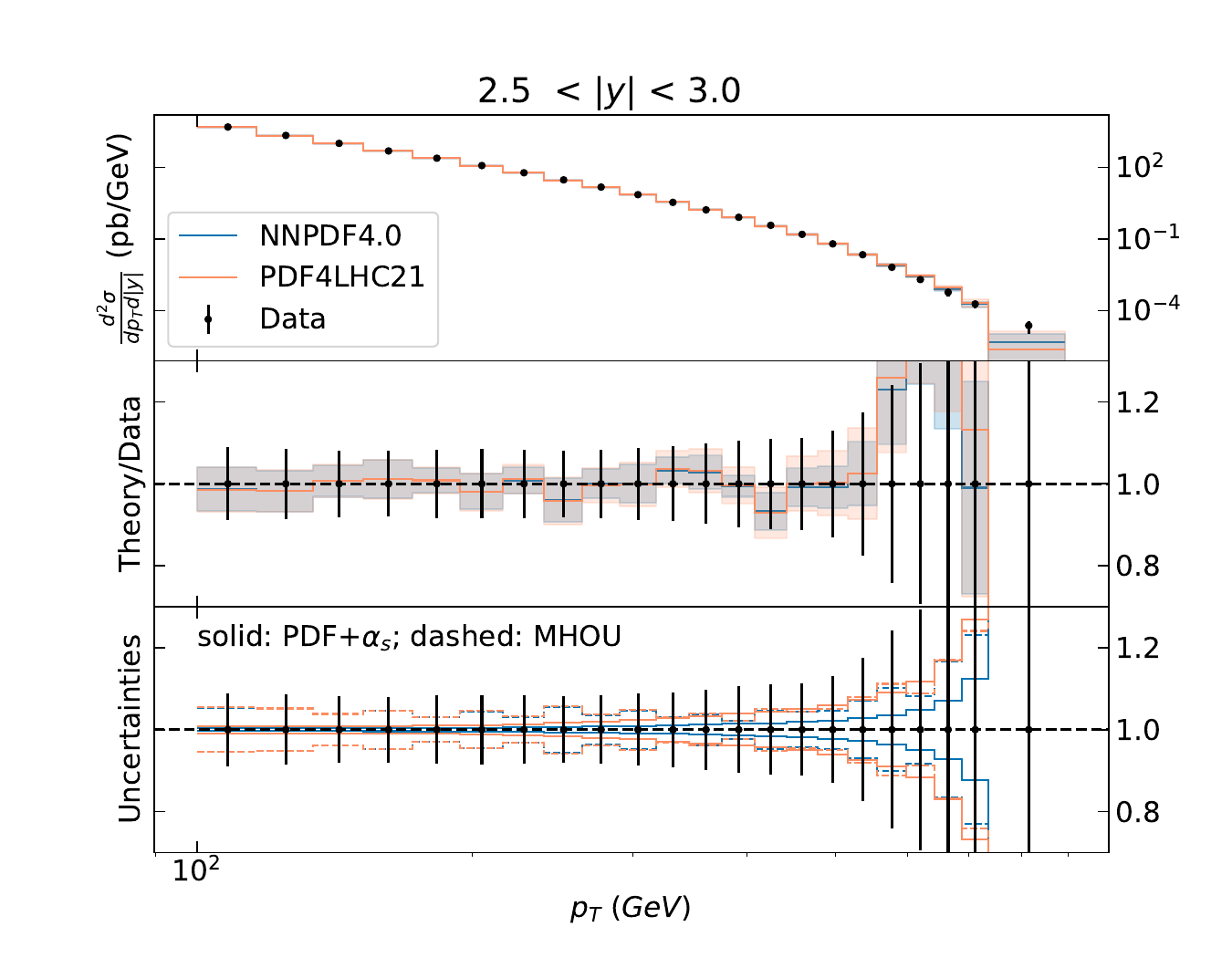}
  \caption{Same as Fig.~\ref{fig:DY-datatheory} for the ATLAS single-inclusive
    jet double differential cross section as a function of the transverse
    momentum of the leading jet, $p_T^j$, for the two outermost bins of the
    absolute value of the jet rapidity, $|y_j|$. The other bins are displayed
    in Fig.~\ref{fig:datatheory_jets_additional_1}
    of Appendix~\ref{app:extra_results}.}
  \label{fig:jets-ATLAS-13tev} 
\end{figure}

\begin{figure}[!t]
  \includegraphics[width=0.49\textwidth]{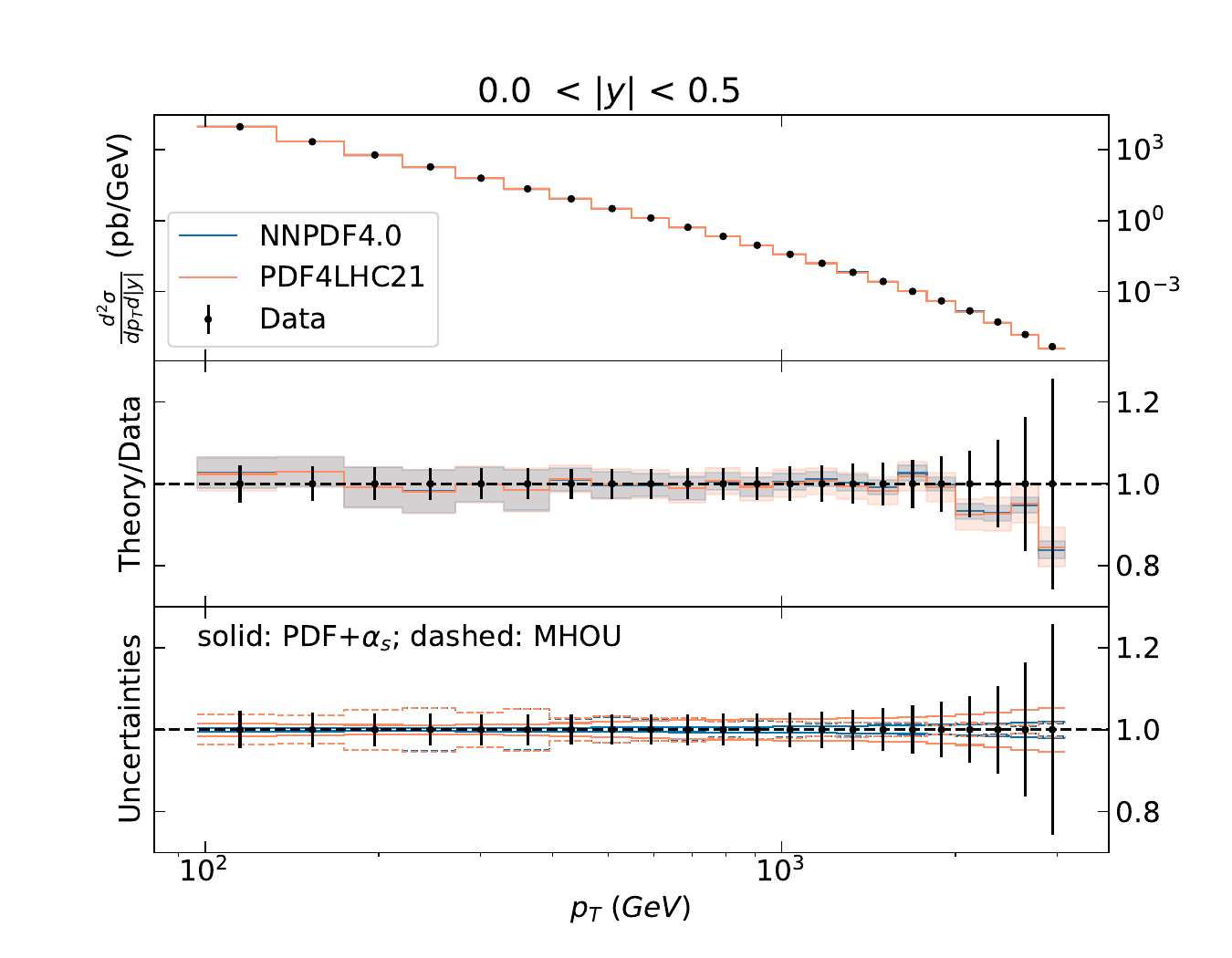}
  \includegraphics[width=0.49\textwidth]{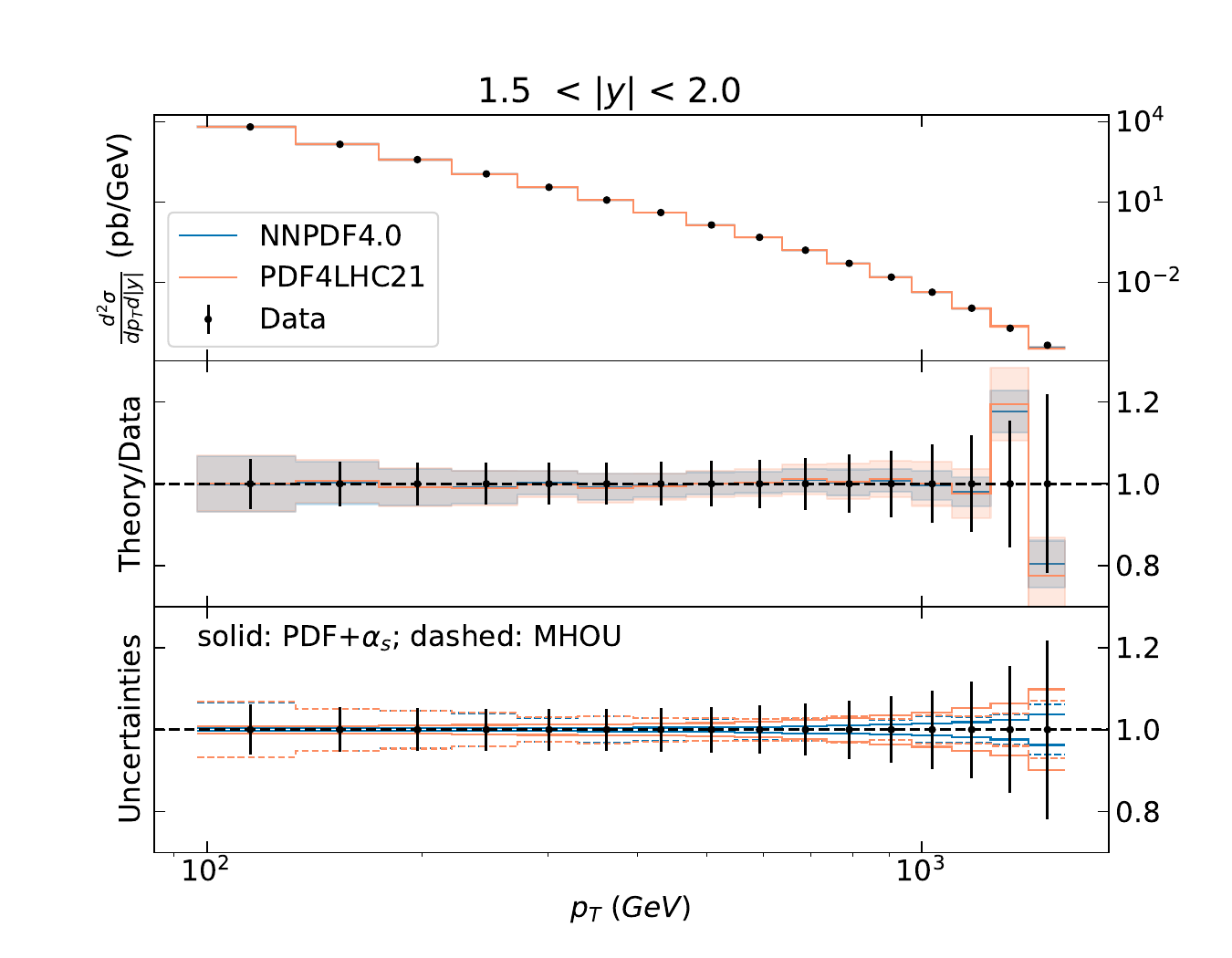}
  \caption{Same as Fig.~\ref{fig:jets-ATLAS-13tev} for the CMS single-inclusive
    jet double differential cross section. The other bins are displayed in
    Fig.~\ref{fig:datatheory_jets_additional_2} of
    Appendix~\ref{app:extra_results}.}
  \label{fig:jets-CMS-13tev-R07} 
\end{figure}

\begin{figure}[!t]
  \includegraphics[width=0.49\textwidth]{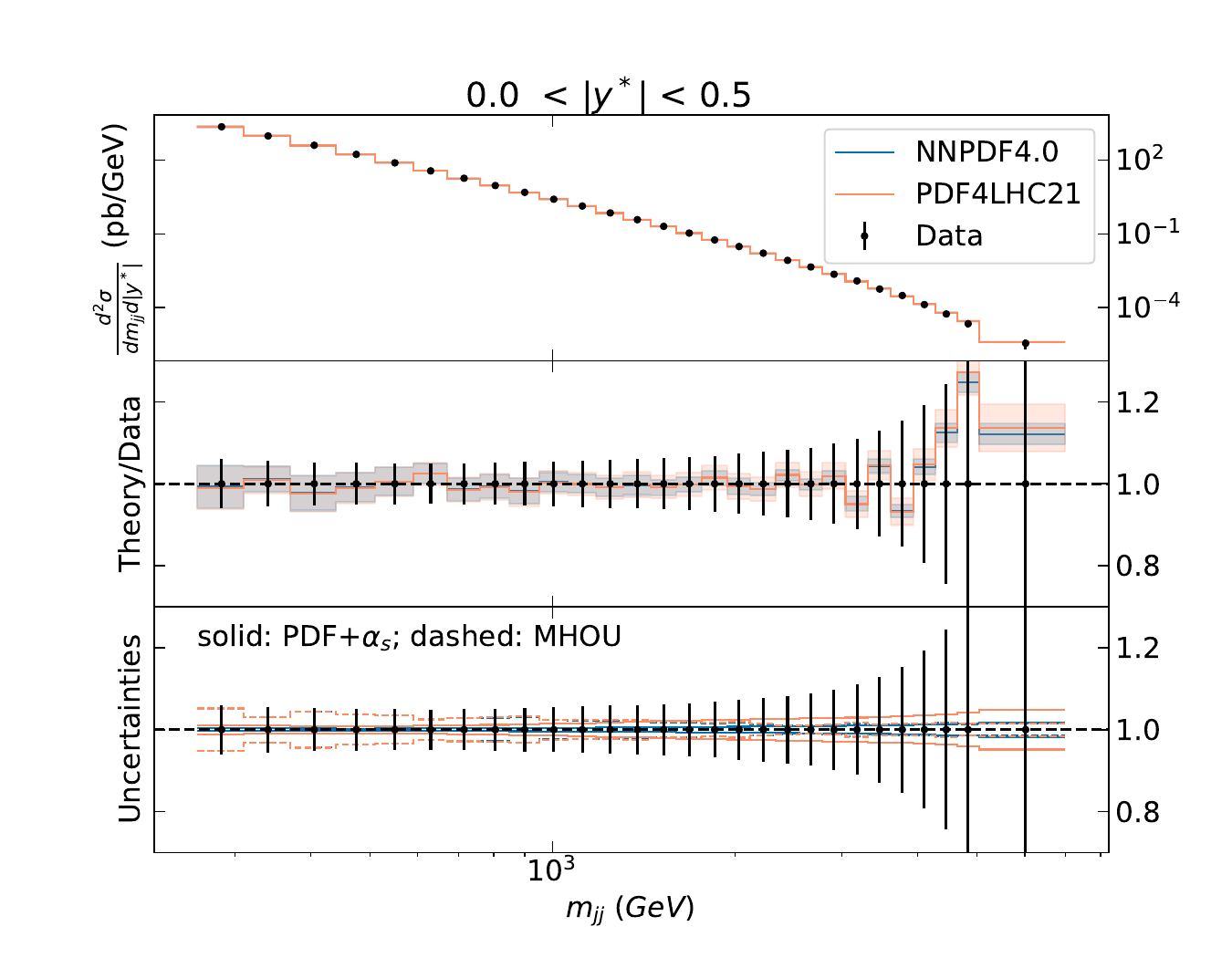}
  \includegraphics[width=0.49\textwidth]{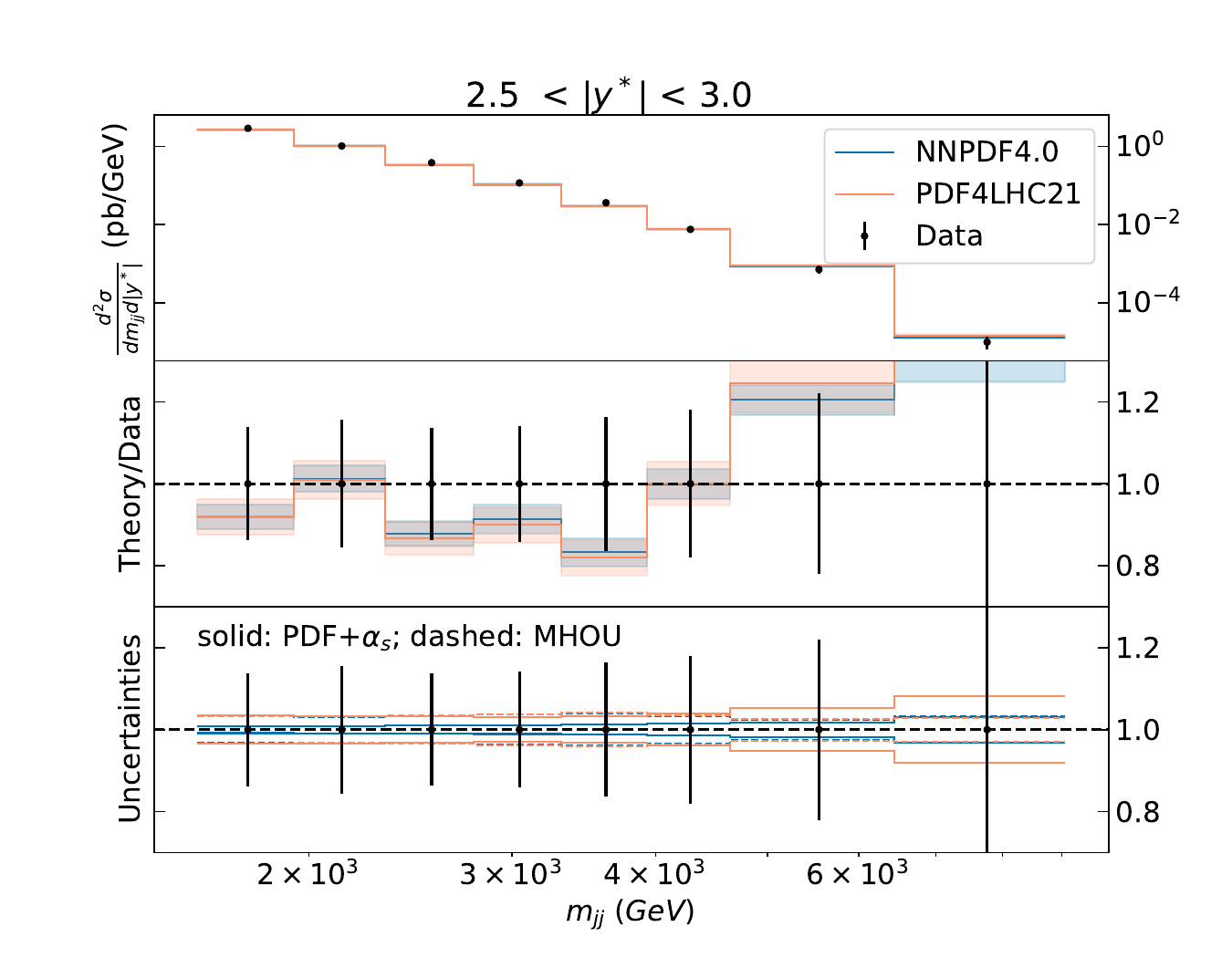}
  \caption{Same as Fig.~\ref{fig:DY-datatheory} for the ATLAS
    di-jet double differential cross section,
    as a function of the invariant mass of the di-jet pair, $m_{jj}$, for the
    two outermost bin in the absolute rapidity separation between the two
    jets. The other bins are displayed in
    Fig.~\ref{fig:datatheory_jets_additional_3} of
    Appendix~\ref{app:extra_results}.}
  \label{fig:dijet-ATLAS-13tev}
\end{figure}

The breakdown of $\chi^2_{\rm exp+th}$ into $\chi^2_{\rm exp+mho}$
and $\chi^2_{\rm exp}$ after regularisation is displayed in
Fig.~\ref{fig:jets-chi2}. 
The data-theory comparison is displayed in Figs.~\ref{fig:jets-ATLAS-13tev},
\ref{fig:jets-CMS-13tev-R07}, and \ref{fig:dijet-ATLAS-13tev}, respectively for
the ATLAS and CMS single-inclusive jet, and for the ATLAS di-jet measurements.
In the first and second (third) cases, we plot the double differential cross
section as a function of the transverse momentum of the leading jet, $p_T^j$
(the invariant mass of the two jets, $m_{jj}$), for the two outermost bins of
the absolute value of the jet rapidity, $|y_j|$ (of the two-jet rapidity
separation $|y^*|$). The other bins are displayed, respectively, in
Figs.~\ref{fig:datatheory_jets_additional_1},
\ref{fig:datatheory_jets_additional_2},
and~\ref{fig:datatheory_jets_additional_3} of
Appendix~\ref{app:extra_results}.

\begin{table}[!t]
  \scriptsize
  \centering
  \renewcommand{\arraystretch}{1.5}
  \begin{tabularx}{\textwidth}{Xrccccccccccc}
Dataset
& \rotatebox{0}{$n_{\rm dat}$}
& \rotatebox{0}{$\sqrt{2/n_{\rm dat}}$}
&
& \rotatebox{80}{ABMP16}
& \rotatebox{80}{CT18}
& \rotatebox{80}{CT18A}
& \rotatebox{80}{CT18Z}
& \rotatebox{80}{MSHT20}
& \rotatebox{80}{NNPDF3.1}
& \rotatebox{80}{NNPDF4.0}
& \rotatebox{80}{PDF4LHC15}
& \rotatebox{80}{PDF4LHC21} \\
\toprule
\multirow{2}{*}{ATLAS 13~TeV incl. jet $\frac{d^2\sigma}{dp_Td|y|}$}
& \multirow{2}{*}{177}
& \multirow{2}{*}{0.11}
& $\chi^2_{\rm exp+th}$
& 1.85 & 1.56 & 1.64 & 1.38 & 1.67 & 1.21 & 1.51 & 1.20 & 1.25 \\
&
&
& $\chi^2_{\rm exp}$
& 2.32 & 2.48 & 2.47 & 2.50 & 2.53 & 2.98 & 1.95 & 3.02 & 2.40 \\
\midrule
\multirow{2}{*}{CMS 13~TeV incl. jet $R_{0.4}$\, $\frac{d^2\sigma}{dp_Td|y|}$}
& \multirow{2}{*}{78}
& \multirow{2}{*}{0.16}
& $\chi^2_{\rm exp+th}$
& 1.64 & 1.58 & 1.60 & 1.52 & 1.64 & 1.47 & 1.50 & 1.48 & 1.43 \\
&
&
& $\chi^2_{\rm exp}$
& 2.05 & 2.29 & 2.25 & 2.26 & 2.23 & 2.21 & 2.02 & 2.30 & 2.18 \\
\midrule
\multirow{2}{*}{ATLAS 13~TeV di-jets $\frac{d^2\sigma}{dm_{jj}d|y^*|}$}
& \multirow{2}{*}{136}
& \multirow{2}{*}{0.12}
& $\chi^2_{\rm exp+th}$
& 1.13 & 1.08 & 1.09 & 1.05 & 1.16 & 1.09 & 1.15 & 1.01 & 0.96 \\
&
&
& $\chi^2_{\rm exp}$
& 1.25 & 1.49 & 1.47 & 1.48 & 1.41 & 1.37 & 1.29 & 1.42 & 1.41 \\
\bottomrule
\end{tabularx}

  \vspace{0.3cm}
  \caption{Same as Table~\ref{tab:chi2-DY} for the ATLAS and CMS
    single-inclusive and di-jet production measurements at the LHC 13~TeV.}
  \label{tab:chi2-jets} 
\end{table}

From inspection of Table~\ref{tab:chi2-jets} and of Fig.~\ref{fig:jets-chi2},
very similar remarks can be drawn for the three considered datasets.
First, when theory errors are not included in the computation of the $\chi^2$,
the NNPDF4.0 PDF set performs better than any of the others, in the sense
that the NNPDF4.0 $\chi^2_{\rm exp}$ is the closest to unity among all PDF
sets, although the $\chi^2_{\rm exp}$ is still away from one by more than
$10\sigma$. Some PDF sets may lead to comparatively larger values of
$\chi^2_{\rm exp}$, such as APMP16, however the statistical significance of these
fluctuations must be seen in units of the $\chi^2$ standard deviation,
as we will further discuss in Sect.~\ref{subsec:global_interpretation}. 
Second, once all the theory errors are included, the values of
$\chi^2_{\rm exp+th}$ become relatively close, irrespective of the input PDF set
used for their computation. However, in the case of single-inclusive jets, they
are all a few units away from one, suggesting that once these datasets are
included in the fit, they might have a significant pull on the gluon PDF. 
The values of $\chi^2_{\rm exp+th}$ being relatively close suggests also that, 
except perhaps for ABMP16, which continues to display
rather large values of $\chi^2_{\rm exp+th}$ even after inclusion of
theoretical uncertainties, it may be difficult to discriminate the quality of
the predicting power of the various PDF sets based solely on these
measurements. 
Third, the relatively homogeneous values of $\chi^2_{\rm exp+th}$ 
occur despite the input PDF sets have very different uncertainties. For
instance, PDF4LHC21 uncertainties are twice the NNPDF4.0 uncertainties, see
Fig.~\ref{fig:lumis}. The breakdown of the theoretical uncertainty into its
various components can be different depending on the PDF set. The MHO
uncertainty remains more or less the same for all PDF sets. Conversely,
the PDF+$\alpha_s$ uncertainty is the smallest for NNPDF4.0.
This is consistent with the fact that NNPDF4.0 PDF uncertainties are typically
the smallest among all the PDF sets considered, see Fig.~\ref{fig:lumis}.
Finally, it is interesting to observe that the balance between the various
components of the theoretical uncertainty depend on the kinematics. From
Figs.~\ref{fig:jets-ATLAS-13tev}-\ref{fig:dijet-ATLAS-13tev}, we see that
the PDF+$\alpha_s$ (MHO) uncertainty dominates at small (large) $p_T^j$ or
$m_{jj}$. 
  
\subsection{Single-inclusive jet and di-jet production measurements at HERA}
\label{subsec:HERAjets}
\begin{figure}[!t]
  \includegraphics[width=0.49\textwidth]{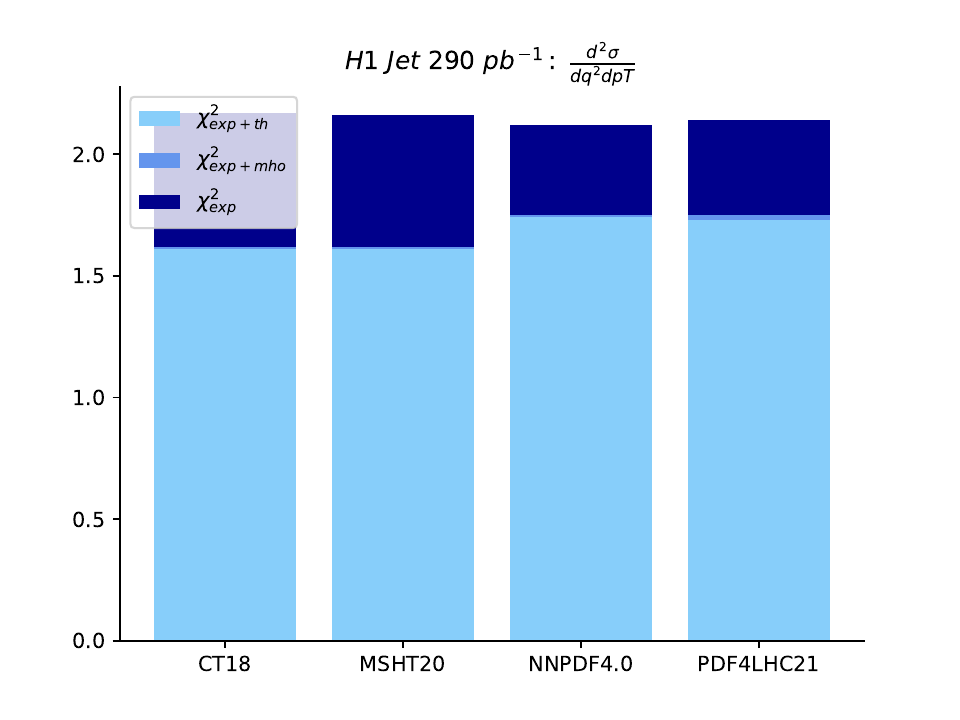}
  \includegraphics[width=0.49\textwidth]{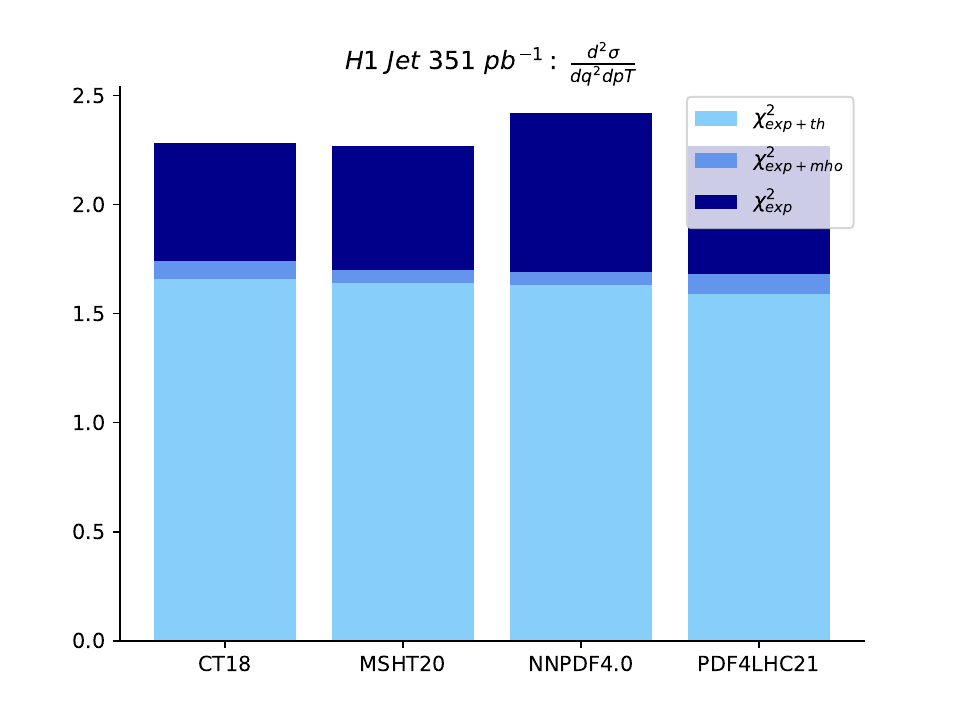}\\
  \includegraphics[width=0.49\textwidth]{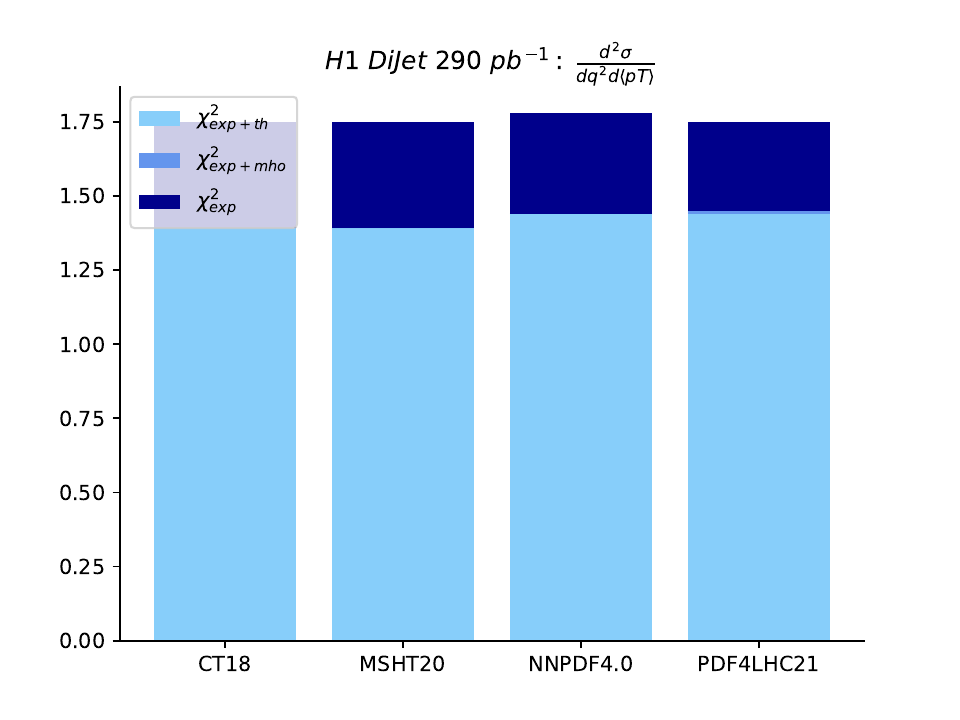}
  \includegraphics[width=0.49\textwidth]{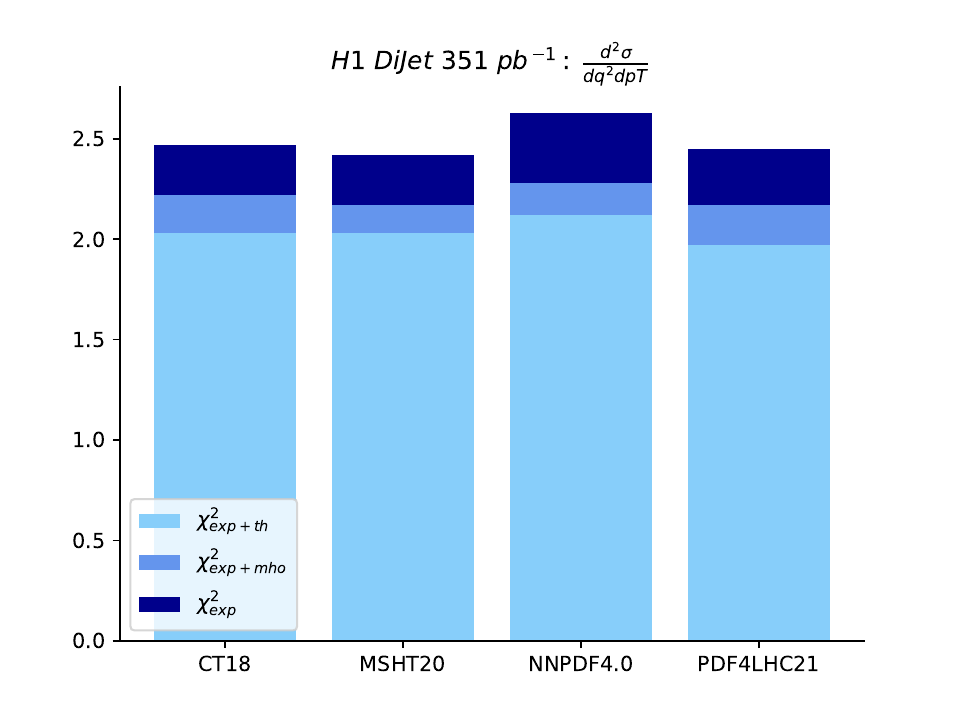}\\  
  \caption{Same as Fig.~\ref{fig:DY-chi2} for the H1 single-inclusive jet (top)
    and di-jet (bottom) datasets. For the measurements on the left plots
    $n_{\rm dat} = 37$ and $\sqrt{2/n_{\rm dat}} = 0.23$, while for the
    measurements on the right plots $n_{\rm dat} = 24$ and
    $\sqrt{2/n_{\rm dat}} = 0.29$.}  
  \label{fig:H1_histos} 
\end{figure}

\begin{figure}[!t]
  \includegraphics[width=0.49\textwidth]{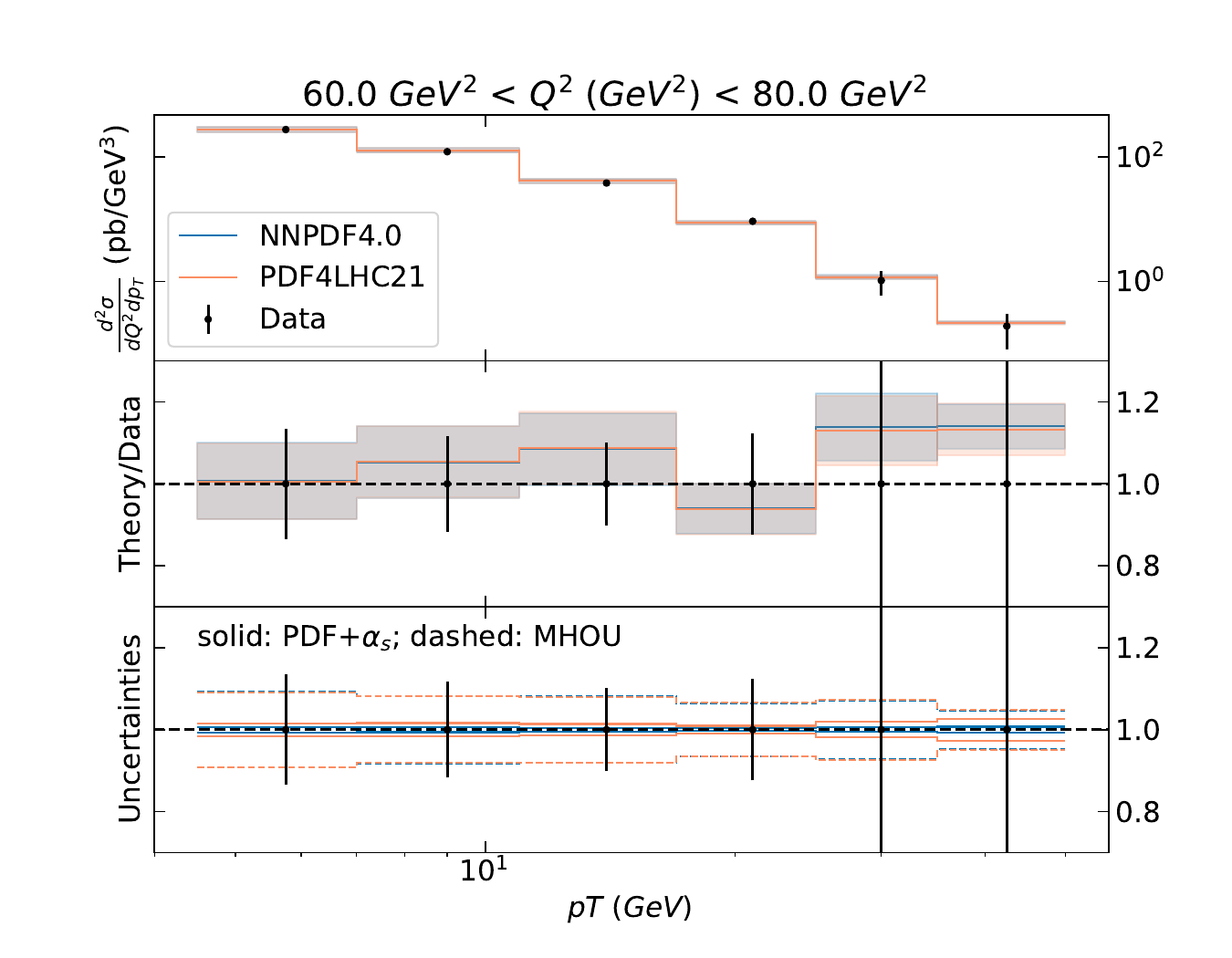}
  \includegraphics[width=0.49\textwidth]{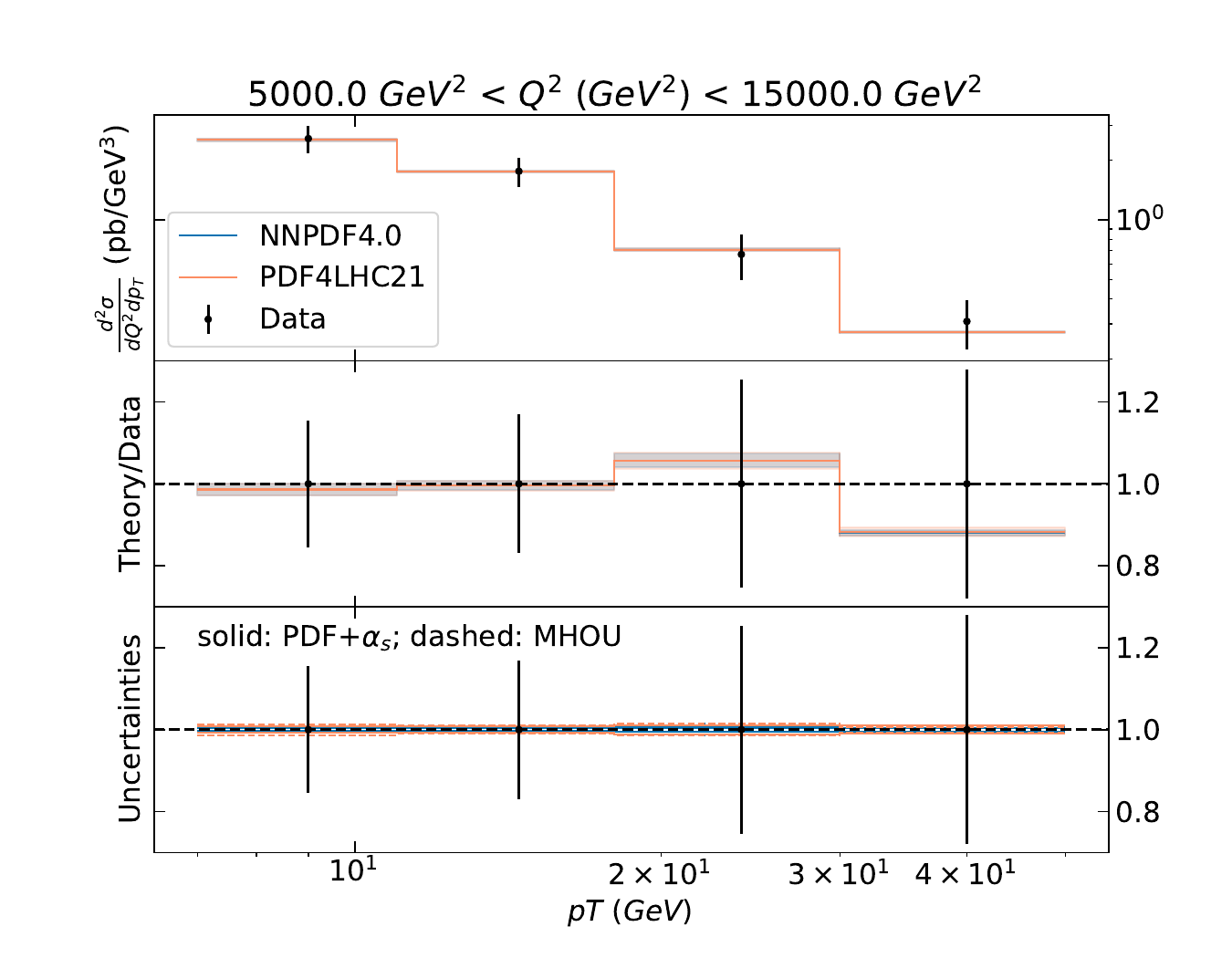}\\
  \includegraphics[width=0.49\textwidth]{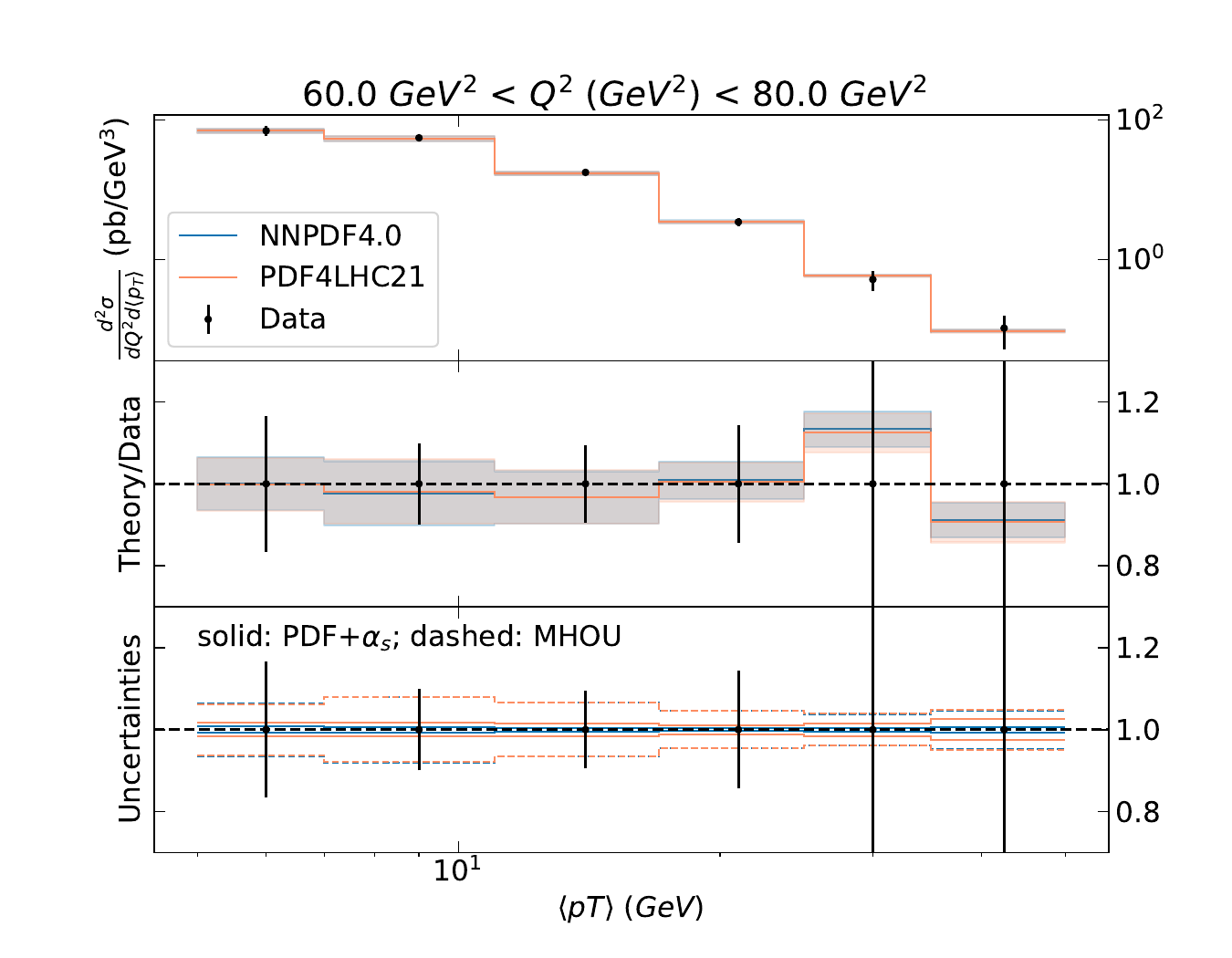}
  \includegraphics[width=0.49\textwidth]{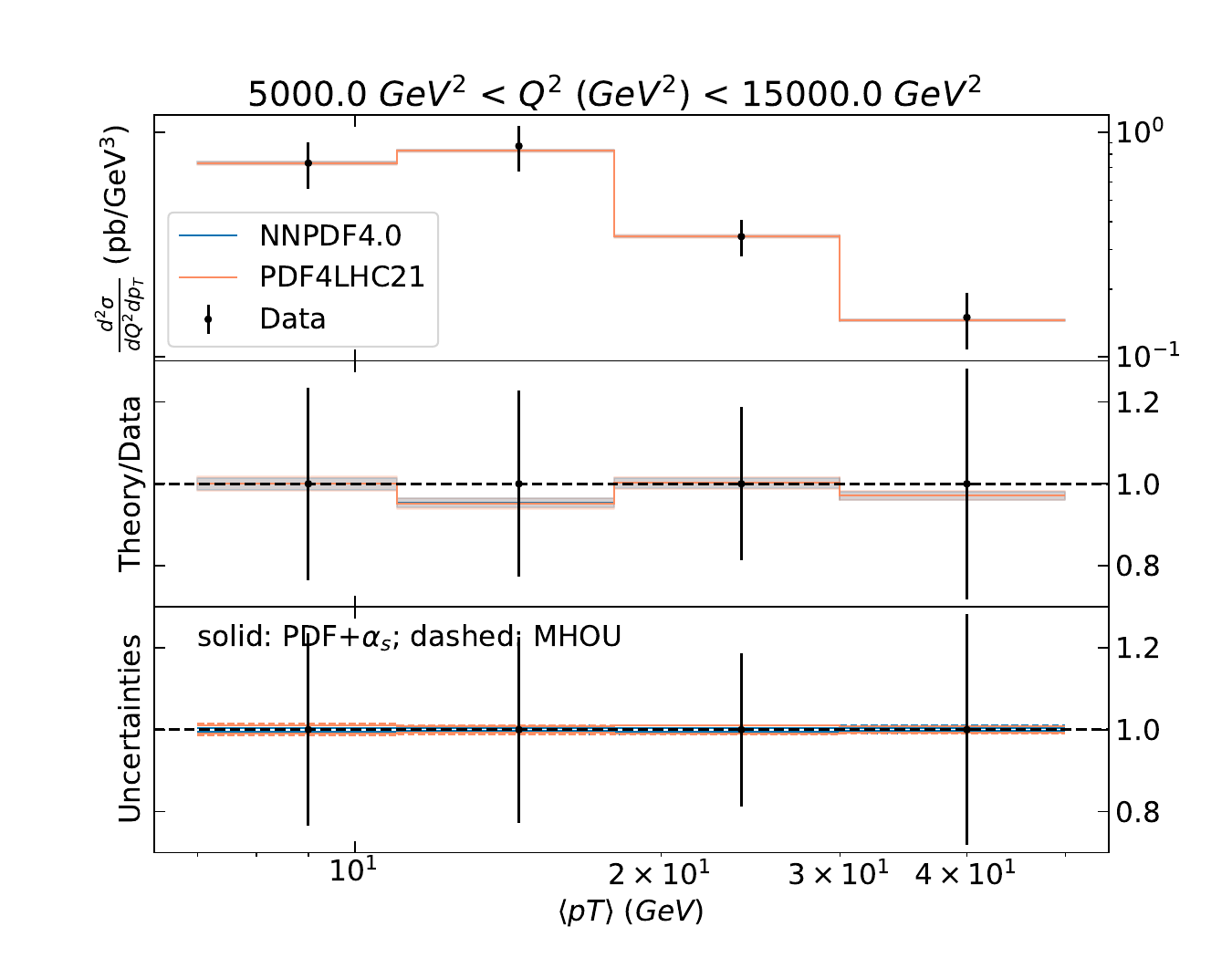}\\
  \caption{Same as Fig.~\ref{fig:DY-datatheory} for the largest $Q^2$ bins
    of the H1 single-inclusive jet (top) and di-jet (bottom) datasets. All the
    other bins are collected in Figs.~\ref{fig:datatheory_DISjets_additional_1}-\ref{fig:datatheory_DISjets_additional_4} of Appendix~\ref{app:extra_results}.}
  \label{fig:H1_datatheory} 
\end{figure}

We finally discuss the HERA single-inclusive jet and di-jet
production measurements outlined in Sect.~\ref{subsec:LHC_jets}.
The values of $\chi^2_{\rm exp}$ and $\chi^2_{\rm exp+th}$ are reported in
Table~\ref{tab:chi2-DISjets}. The experimental covariance matrix of the
H1 low-$Q^2$ single-inclusive jet and di-jet measurements is regularised as
explained in Sect.~\ref{subsec:compatibility}. The unregularised values of
$\chi^2_{\rm exp+th}$ are reported in Appendix~\ref{app:unreg}. The breakdown of
$\chi^2_{\rm exp+th}$ into $\chi^2_{\rm exp+mho}$ and $\chi^2_{\rm exp}$ is displayed
in Fig.~\ref{fig:H1_histos}, albeit only for the H1 data. The data-theory
comparison is displayed in Fig.~\ref{fig:H1_datatheory} for the highest $Q^2$
bin of the H1 single-inclusive jet and di-jet differential cross sections as
a function, respectively, of the transverse momentum of the leading jet and of
the average transverse momentum of the jet pair. Histograms plots for the
ZEUS measurements and data-theory comparison plots for the remaining H1 bins
and for all of the ZEUS bins are collected in
Figs.~\ref{fig:chi2histo_disjets_additional}-\ref{fig:datatheory_DISjets_additional_7} of Appendix~\ref{app:extra_results}.

\begin{table}[!t]
  \scriptsize
  \centering
  \renewcommand{\arraystretch}{1.5}
  \begin{tabularx}{\textwidth}{Xrccccccccccc}
Dataset
& \rotatebox{0}{$n_{\rm dat}$}
& \rotatebox{0}{$\sqrt{2/n_{\rm dat}}$}
&
& \rotatebox{80}{ABMP16}
& \rotatebox{80}{CT18}
& \rotatebox{80}{CT18A}
& \rotatebox{80}{CT18Z}
& \rotatebox{80}{MSHT20}
& \rotatebox{80}{NNPDF3.1}
& \rotatebox{80}{NNPDF4.0}
& \rotatebox{80}{PDF4LHC15}
& \rotatebox{80}{PDF4LHC21} \\
\toprule
\multirow{2}{*}{H1 incl. jet (low $Q^2$) $\frac{d^2\sigma}{dQ^{2} dp_T}$}
& \multirow{2}{*}{37}
& \multirow{2}{*}{0.23}
& $\chi^2_{\rm exp+th}$
& 1.64 & 1.61 & 1.61 & 1.67 & 1.61 & 1.70 & 1.74 & 1.61 & 1.73 \\
&
&
& $\chi^2_{\rm exp}$
& 7.68 & 2.17 & 2.14 & 2.11 & 2.16 & 2.16 & 2.12 & 2.17 & 2.14 \\
\midrule
\multirow{2}{*}{H1 incl. jet (high $Q^2$) $\frac{d^2\sigma}{dQ^{2} dp_T}$}
& \multirow{2}{*}{24}
& \multirow{2}{*}{0.29}
& $\chi^2_{\rm exp+th}$
& 1.62 & 1.66 & 1.62 & 1.63 & 1.64 & 1.49 & 1.63 & 1.58 & 1.59 \\
&
&
& $\chi^2_{\rm exp}$
& 2.40 & 2.28 & 2.20 & 2.18 & 2.27 & 2.43 & 2.42 & 2.33 & 2.27 \\
\midrule
\multirow{2}{*}{ZEUS incl. jet (low lumi.) $\frac{d^2\sigma}{dQ^2dE_{T}}$}
& \multirow{2}{*}{30}
& \multirow{2}{*}{0.26}
& $\chi^2_{\rm exp+th}$
& 0.67 & 0.69 & 0.68 & 0.67 & 0.68 & 0.66 & 0.65 & 0.68 & 0.67 \\ 
&
&
& $\chi^2_{\rm exp}$
& 0.69 & 0.71 & 0.70 & 0.69 & 0.70 & 0.69 & 0.67 & 0.70 & 0.69 \\
\midrule
\multirow{2}{*}{ZEUS incl. jet (high lumi.) $\frac{d^2\sigma}{dQ^2dE_{T}}$}
& \multirow{2}{*}{30}
& \multirow{2}{*}{0.26}
& $\chi^2_{\rm exp+th}$
& 0.77 & 0.77 & 0.77 & 0.76 & 0.78 & 0.77 & 0.76 & 0.77 & 0.77 \\ 
&
&
& $\chi^2_{\rm exp}$
& 0.82 & 0.83 & 0.82 & 0.80 & 0.82 & 0.84 & 0.81 & 0.83 & 0.82 \\ 
\midrule
\multirow{2}{*}{H1 di-jets (low $Q^2$) $\frac{d^2\sigma}{dQ^{2} d \langle p_T \rangle}$}
& \multirow{2}{*}{37}
& \multirow{2}{*}{0.23}
& $\chi^2_{\rm exp+th}$
& 1.37 & 1.39 & 1.38 & 1.37 & 1.39 & 1.42 & 1.44 & 1.36 & 1.44 \\
&
&
& $\chi^2_{\rm exp}$
& 11.0 & 1.75 & 1.73 & 1.68 & 1.75 & 1.82 & 1.78 & 1.77 & 1.75 \\ 
\midrule
\multirow{2}{*}{H1 di-jets (high $Q^2$) $\frac{d^2\sigma}{dQ^{2}\ d \langle p_T \rangle}$}
& \multirow{2}{*}{24}
& \multirow{2}{*}{0.29}
& $\chi^2_{\rm exp+th}$
& 2.21 & 2.03 & 2.00 & 1.95 & 2.03 & 1.84 & 2.12 & 1.94 & 1.97 \\ 
&
&
& $\chi^2_{\rm exp}$
& 2.63 & 2.47 & 2.37 & 2.32 & 2.42 & 2.65 & 2.63 & 2.51 & 2.45 \\ 
\midrule
\multirow{2}{*}{ZEUS di-jets $\frac{d^2\sigma}{dQ^2d\langle E_{T} \rangle}$}
& \multirow{2}{*}{22}
& \multirow{2}{*}{0.30}
& $\chi^2_{\rm exp+th}$
& 0.81 & 0.75 & 0.75 & 0.71 & 0.78 & 0.90 & 0.83 & 0.77 & 0.79 \\ 
&
&
& $\chi^2_{\rm exp}$
& 1.49 & 1.29 & 1.27 & 1.24 & 1.32 & 1.71 & 1.63 & 1.37 & 1.42 \\ 
\bottomrule
\end{tabularx}

  \vspace{0.3cm}
  \caption{Same as Table~\ref{tab:chi2-DY} for HERA single-inclusive jet and
    di-jet data.}
  \label{tab:chi2-DISjets}
\end{table}

From inspection of Table~\ref{tab:chi2-DISjets} and Fig.~\ref{fig:H1_histos},
we observe that the values of $\chi^2_{\rm exp+th}$ and of $\chi^2_{\rm exp}$ are
very similar when different input PDF sets are used. All PDF sets generalise
equally well on these datasets. The largest component of $\chi^2_{\rm exp+th}$
is due to MHO, in a proportion which is roughly the same across PDF sets.
The PDF+$\alpha_s$ component of $\chi^2_{\rm exp+th}$ is almost immaterial (for
the H1 low-$Q^2$ dataset), very small (for the H1 high-$Q^2$ single-inclusive
jet dataset), or as large as the MHO component (for the H1 high-$Q^2$ di-jet
dataset). The quality of the data description is generally very good, with
$\chi^2_{\rm exp+th}\sim 1$ for all the datasets, except for the H1 high-$Q^2$
dataset, in which case $\chi^2_{\rm exp+th}\sim 2$. Investigations into the
reasons for this behaviour, which is consistent throughout PDF sets,  will be
left to future work. For now, we remark that the agreement between experimental
data and the corresponding theoretical predictions, as seen in
Fig.~\ref{fig:H1_datatheory}, is generally good, except
for specific bins that display larger fluctuations between the two.

\subsection{Combined interpretation}
\label{subsec:global_interpretation}

\begin{figure}[!t]
  \centering
  \includegraphics[width=\textwidth]{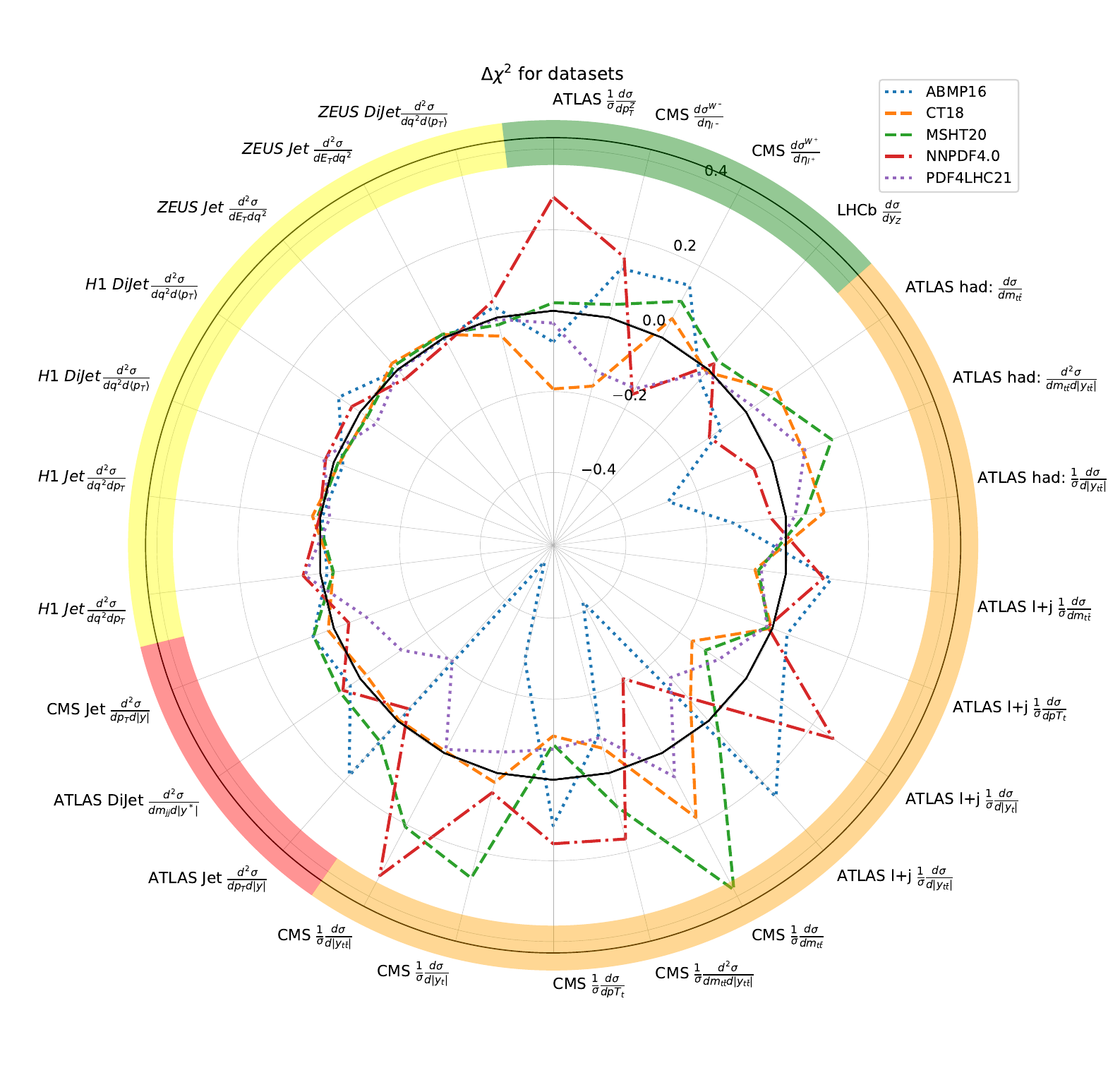}
  \vspace{-1.5cm}
  \caption{The relative change in the total $\chi^2_{\rm exp+th}$ due to the
    change of $i$-th input PDF set, $\Delta \chi^{2(i)}$, with respect to the
    average $\chi^2_{\rm exp+th}$ over PDF sets, see Eq.~\eqref{eq:delta_chi2}.
    The PDF sets considered here are ABMP16, CT18, MSHT20, NNPDF4.0, and
    PDF4LHC21. The datasets are grouped by category: LHC Drell-Yan,
    LHC top-quark pair, LHC single-inclusive jet and di-jet, and HERA
    single-inclusive jet and di-jet production cross sections. The
    circumference corresponding to $\Delta \chi^{2}=0$ is highlighted with a
    solid curve.}
  \label{fig:spider_delta_chi2_pdfas_finalset}
\end{figure}

\begin{figure}[!t]
  \centering
  \includegraphics[width=\textwidth]{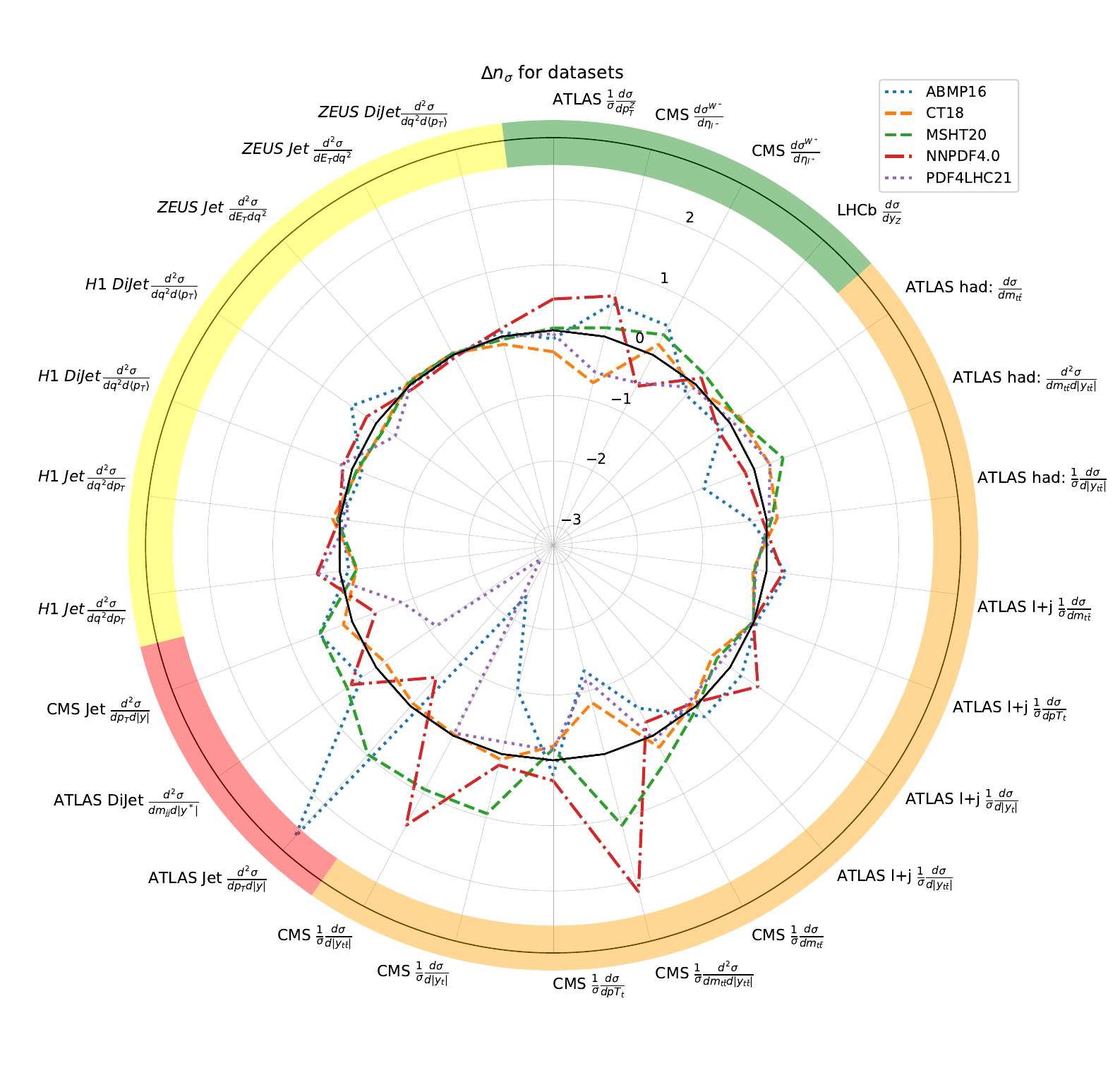}
  \vspace{-1.5cm}
  \caption{Same as Fig.~\ref{fig:spider_delta_chi2_pdfas_finalset} now
  for $\Delta n_{\sigma}$, see Eq.~\eqref{eq:delta_nsigma}.}
  \label{fig:pider_delta_nsigma_pdfas_finalset}
\end{figure}

We now combine the results described in the previous sections to gather the
overall agreement between the considered experimental data and the
corresponding theoretical predictions. To this purpose, in
Fig.~\ref{fig:spider_delta_chi2_pdfas_finalset}, we display $\Delta\chi^{2(i)}$,
the relative change in the total $\chi^2_{\rm exp+th}$ due to the change of input
PDF set with respect to the average $\chi^2_{\rm exp+th}$ over PDF sets,
see Eq.~\eqref{eq:delta_chi2}. The PDF sets considered here are ABMP16, CT18,
MSHT20, NNPDF4.0, and PDF4LHC21. All the datasets listed in
Table~\ref{tab:input_datasets} are considered, except for the 8~TeV ATLAS
Drell-Yan rapidity distribution~\cite{ATLAS:2023lsr}. The reason being that
this dataset, extensively discussed in Sect.~\ref{subsec:ATLAS_8TEV}, is
included in MSHT20 and NNPDF4.0 in the form of an earlier
analysis~\cite{ATLAS:2017rue}, whereas all the other datasets
are not included in any PDF set. Furthermore, all the other data sets are
for the LHC Run II. The datasets are grouped by category:
LHC Drell-Yan, LHC top-quark pair, LHC single-inclusive jet and di-jet, and
HERA single-inclusive jet and di-jet production cross sections.
The circumference corresponding to $\Delta \chi^2=0$ is highlighted with a
solid curve. In Fig.~\ref{fig:pider_delta_nsigma_pdfas_finalset} we display,
in the same format, $\Delta n_\sigma^{(i)}$, the difference between the total
$\chi^2_{\rm exp+th}$ computed with the $i$-th PDF set and the average
$\chi^2_{\rm exp+th}$ over PDF sets, normalised to the standard deviation of the
$\chi^2$ distribution, see Eq.~\eqref{eq:delta_nsigma}.
Figures~\ref{fig:spider_delta_chi2_pdfas_finalset}
and~\ref{fig:pider_delta_nsigma_pdfas_finalset} should be inspected together:
the latter provides an assessment of the statistical significance of
fluctuations from the average $\Delta\chi^2=0$ seen in the former, in units of
the $\chi^2$ standard deviation. Large fluctuations may have low statistical
significance if a dataset has a small number of data points and the other way
around.

On the basis of Figs.~\ref{fig:spider_delta_chi2_pdfas_finalset}
and~\ref{fig:pider_delta_nsigma_pdfas_finalset}, we conclude that the various
classes of datasets are described to a different level of
accuracy. However, whereas the value of $\Delta\chi^2$ displays sizeable
fluctuations depending on the input PDF set, especially in the top-quark pair
and jet sectors, we realise that discrepancies with respect to the average
over PDF sets is almost always within $\Delta n_\sigma=1$. The most relevant
excess occurs with the ABMP16 PDF set in the case of the ATLAS and CMS
single-inclusive jet measurements, and with the NNPDF4.0 PDF set in the case
of the CMS double-differential and rapidity-differential top-quark pair
measurements. In these cases, the excess is between one and three sigma.
This fact may be explained by assuming that these measurements disfavor
the softer (harder) large-$x$ gluon of ABMP16 (NNPDF4.0). We also note an
anomalous deficiency, close $\Delta n_\sigma=3$, with the PDF4LHC21 PDF set in
the case of the ATLAS single-inclusive jet measurements, and with the ABMP16
PDF set in the case of the CMS rapidity-differential top-quark pair measurement.
We therefore conclude that, whereas HERA jet and LHC Drell-Yan measurements may
not be able to discriminate between PDF sets, LHC jet and top-quark pair
measurements may help put stronger constraints on PDFs, especially those
datasets for which the largest fluctuations among different PDF sets are
observed in terms of $\Delta n_\sigma$.

As for the general trend displayed by individual PDF sets, on top of the
aforementioned dataset-specific considerations, we remark two interesting
facts. First, the NNPDF4.0 PDF set, despite displaying the smallest
uncertainties among all the PDF sets considered in this work (see
Fig.~\ref{fig:lumis}), provides a
description of the data which is overall not worse than the others (with the
aforementioned few exceptions). We therefore conclude that theoretical
predictions obtained with NNPDF4.0 are overall as accurate as those
obtained with the other PDF sets, despite their smaller PDF uncertainties, 
once experimental, MHO, and $\alpha_s$ uncertainties are taken into account,
and the regularisation of the covariance matrix is applied when needed.
This is true on average on the examined datasets: as discussed, there
are cases in which NNPDF4.0 performs better than other PDF sets, others in
which it performs worse than these, and others in which the experimental
precision of the datasets cannot discriminate among different PDF sets.
Second, the PDF4LHC21 PDF set generally
displays the value of $\Delta\chi^2$ and $\Delta n_\sigma$ closest to zero among
all the PDF sets considered in this work. This fact is unsurprising, given that
PDF4LHC21 is the unweighted average of the CT18, MSHT20 and NNPDF3.1 PDF
sets. Deviations from the mean $\Delta\chi^2=0$ and $\Delta n_\sigma=0$,
obtained with individual PDF sets, cancel out by construction. In this sense,
PDF4LHC21 is a conservative PDF set, as already illustrated
in~\cite{PDF4LHCWorkingGroup:2022cjn}, although it remains the least precise.

\section{Summary and outlook}
\label{sec:summary}

In this paper we have compared theoretical predictions, computed at NNLO
accuracy in perturbative QCD using different input PDF sets, with a wide
array of experimental measurements, typically not yet included in PDF
determination. Specifically, we have considered differential cross sections
measured at the LHC, for Drell-Yan gauge boson, top-quark pair,
single-inclusive jet and di-jet production, and at HERA, for
single-inclusive jet and di-jet production. We have considered the most widely
used PDF sets in LHC experimental analyses, namely, ABMP16, CT18 (and its
variants), MSHT20, NNPDF3.1, NNPDF4.0, PDF4LHC15, and PDF4LHC21. We have
accounted for all the relevant sources of experimental and theoretical
uncertainties, in particular due to PDFs, $\alpha_s$, and MHOUs.

The aim of our work has been twofold. First, to test the predictive power of
different PDF sets, by assessing the goodness with which they describe
the datasets not included in their determination. Second, to quantify the
various sources of uncertainty that enter theoretical predictions, specifically
PDF, $\alpha_s$, and MHO uncertainties. These two objectives are becoming
increasingly relevant given the ever higher precision of LHC experiments to
determine SM parameters, such as the strong coupling $\alpha_s(m_Z)$,
the $W$-boson mass $m_W$, and the effective lepton mixing angle
$\sin^2\theta^\ell_{\rm eff}$. This precision is now comparable to, if not better
than, that obtained at LEP. This outstanding result requires a careful estimate
of all of the sources of uncertainties that accompany it, in particular the PDF
uncertainty, which is often dominant in LHC measurements.

In this work we have considered the data-theory agreement between predictions
obtained with a broad range of input PDF sets and experimental data,
but we have not included any of the examined datasets in a PDF determination.
Only the simultaneous inclusion of a given subset of statistically independent
datasets considered in this work in future PDF determinations 
will allow us to determine how the resulting PDFs can adapt to the data, and 
possibly improve the description of the data, as well as the precision of PDFs. 

The main outcome of our investigations is summarised in the overview plots
presented in Sect.~\ref{subsec:global_interpretation}. We have found that the
CT18, MSHT20, NNPDF4.0, and PDF4LHC21 PDF sets provide a comparable description
of all of the datasets considered in this work, once all sources of theoretical
uncertainty are taken into account. We have therefore concluded that all PDF
sets have a similar predictive power and generalise similarly well to unseen
data. Incorporating PDF, $\alpha_s$ and MHO
uncertainties is crucial to reach this conclusion. These outcomes may seem
counter-intuitive given that individual PDF sets differ among each other for
their central values and uncertainties, by an amount that is not always
encompassed by the latter. Within this general picture, the NNPDF4.0 and
PDF4LHC21 sets represent opposite cases. On the one hand, the NNPDF4.0 set has
by far the smallest uncertainties of all PDF sets, hence it is the most precise.
In spite of this fact, it describes the examined data, on average, as well as
the other PDF sets. On the other hand, the PDF4LHC21 set has some of the
largest uncertainties of all PDF sets, hence it is the least precise. This is
by construction, given that it is the combination of three different PDF sets.
However, the fact that it describes the data as well as the other PDF sets
means that it does not need to be as accurate as these. 

The only exception to this overall trend is represented by the ATLAS 8~TeV
inclusive measurement of the $Z$ rapidity distributions extrapolated to the
full phase space, which underlies the recent $\alpha_s(m_Z)$ extraction from the
companion measurement differential in the transverse momentum of the $Z$ boson.
In this case we have found that, despite the excellent agreement of NNPDF4.0
theoretical predictions with the central values of the experimental data, the
peculiar slope in rapidity, combined with the dominance of the luminosity
normalisation uncertainty, leads to a poor $\chi^2$. The $\chi^2$ is instead
better for other PDF sets because of their larger PDF uncertainty. This may
therefore be a case in which the accuracy of the NNPDF4.0 set does not match
its very high precision. We have investigated whether this is truly the case.
We have found that using variants of the NNPDF4.0 set that incorporate MHOU
and/or aN$^3$LO corrections improves the $\chi^2$ only marginally. We have also
observed that the NNPDF4.0 set and any of its variants provide an excellent
description of the earlier ATLAS measurement of the $Z$ boson rapidity
distributions based on the same collider events. We have finally determined
that the $\chi^2$ of this dataset can be lowered only if it is included, and
overweighted, in a fit, at the price of a slight deterioration
in the description of the other datasets. We have therefore concluded that
there are residual tensions between this dataset and the other datasets
in NNPDF4.0.

Two important by-products of this work have been the computation of fast
interpolation grids, accurate to NNLO, and the implementation of the
experimental information, in the NNPDF format, for all the considered datasets.
These facts will allow us to streamline their inclusion in future
NNPDF releases. The fast interpolation framework has two major advantages.
First, we will abandon the use of $K$-factors to account for NNLO corrections
in partonic matrix elements. Second, we will be able to readily vary the
renormalisation and factorisation scales in the computation of theoretical
predictions and determine MHOU.

To conclude, as the LHC experiments finalise the Run II legacy measurements,
start to release datasets based on the Run III luminosity, and prepare for
the HL-LHC era, our analysis demonstrates the importance of testing the
predictive power of PDFs on a broad set of high-precision measurements with
state-of-the-art theoretical predictions, which must crucially include all
possible sources of theoretical uncertainty. The methodology laid out in
this work can be applied to any upcoming and future LHC measurements that may
eventually provide a clear guidance concerning which PDF sets are preferred
by the experimental data.

\subsection*{Acknowledgments}

We thank our colleagues of the NNPDF Collaboration for discussions on the
results presented in this work, in particular C.~Schwan for his work
on maintaining and supporting the {\sc PineaAPPL} library. We thank the authors
of {\sc NNLOJET} and {\sc MATRIX}, in particular A.~Huss, S.~Devoto, and
S.~Kallweit, for providing access to preliminary versions of their codes
interfaced to {\sc PineaAPPL}. We would like to thank the experimentalists who have 
helped us interpreting and implementing their measurements, in particular Stefano Camarda, 
Marco Cipriani, Josh Bendavid, Regina Demina and Otto Hindrichs. 
E.R.N. is supported by the Italian Ministry of University and Research (MUR)
through the ``Rita Levi-Montalcini'' Program. M.U. and M.N.C. are supported by the European
Research Council under the European Union's Horizon 2020 research and
innovation Programme (grant agreement n.950246), and partially by the STFC
consolidated grant ST/T000694/1 and ST/X000664/1. J.R. is partially supported
by NWO, the Dutch Research Council. J.R. and T.R.R. are supported by an ASDI
grant from the Netherlands eScience Center (NLeSC). R.S. is supported by the
UK Science and Technology Facility Council (STFC)
consolidated grants ST/T000600/1 and ST/X000494/1.

\appendix

\section{Impact of the experimental covariance matrix regularisation}
\label{app:unreg}

As discussed in Sect.~\ref{subsec:compatibility}, the experimental covariance
matrix of all the datasets with $Z_{\mathcal{L}}>4$ is regularised by means of
the procedure laid out in~\cite{Kassabov:2022pps}. The values of
$\chi^2_{\rm exp}$ and $\chi^2_{\rm exp+th}$ reported in Sect.~\ref{sec:results}
are computed accordingly. In this appendix, we recompute the values of
$\chi^2_{\rm exp+th}$ with the original, unregularised experimental covariance
matrix. These values, called $\chi^2_{\rm exp+th,orig}$, are compared to
the $\chi^2_{\rm exp+th}$ values of Sect.~\ref{sec:results} in
Table~\ref{tab:chi2-unreg}. We also substantiate what the largest changes in
covariances and correlations are upon applying our regularisation procedure.
To this purpose, we report in Table~\ref{tab:regularisation}, for each of the
regularised data sets, the maximum relative difference of the variances
$\Delta\sigma_r$ (in percent) and the maximum absolute difference of the
correlation $|\Delta\rho|$ computed between the corresponding unregularised
and regularised matrices.

\begin{table}[!t]
  \scriptsize
  \centering
  \renewcommand{\arraystretch}{1.5}
  \begin{tabularx}{\textwidth}{Xlccccccccc}
    Dataset
    &
    & \rotatebox{80}{ABMP16}
    & \rotatebox{80}{CT18}
    & \rotatebox{80}{CT18A}
    & \rotatebox{80}{CT18Z}
    & \rotatebox{80}{MSHT20}
    & \rotatebox{80}{NNPDF3.1}
    & \rotatebox{80}{NNPDF4.0}
    & \rotatebox{80}{PDF4LHC15}
    & \rotatebox{80}{PDF4LHC21} \\
    \toprule
    \multirow{2}{*}{CMS 13~TeV $W^+$ $\frac{d\sigma}{d|\eta|}$}
    & $\chi^2_{\rm exp+th}$
    & 1.31 & 1.20 & 1.11 & 1.05 & 1.26 & 0.85 & 0.96 & 1.15 & 0.98 \\
    & $\chi^2_{\rm exp+th,orig}$
    & 14.6 & 10.9 & 10.8 & 10.8 & 12.2 & 10.2 & 11.2 & 10.7 & 10.5 \\ 
    \midrule
    \multirow{2}{*}{CMS 13~TeV $W^-$ $\frac{d\sigma}{d|\eta|}$}
    & $\chi^2_{\rm exp+th}$
    & 1.56 & 1.15 & 1.11 & 1.10 & 1.43 & 1.12 & 1.60 & 1.14 & 1.20 \\ 
    & $\chi^2_{\rm exp+th,orig}$
    & 13.9 & 8.22 & 8.32 & 8.51 & 10.4 & 8.26 & 11.5 & 8.40 & 8.59 \\ 
    \midrule
    \multirow{2}{*}{ATLAS 13~TeV $t\bar{t}$ all hadr. $\frac{d\sigma}{dm_{t\bar{t}}}$}
    & $\chi^2_{\rm exp+th}$
    & 0.84 & 0.99 & 0.97 & 0.94 & 0.97 & 0.86 & 0.81 & 0.96 & 0.93 \\ 
    & $\chi^2_{\rm exp+th,orig}$
    & 0.92 & 1.07 & 1.05 & 1.02 & 1.05 & 0.94 & 0.87 & 1.04 & 1.01 \\
    \midrule
    \multirow{2}{*}{ATLAS 13~TeV $t\bar{t}$ all hadr. $\frac{d^2\sigma}{dm_{t\bar{t}} d|y_{t\bar{t}}|}$}
    & $\chi^2_{\rm exp+th}$
    & 0.93 & 1.38 & 1.39 & 1.42 & 1.48 & 1.12 & 1.22 & 1.22 & 1.39 \\  
    & $\chi^2_{\rm exp+th,orig}$
    & 0.97 & 1.44 & 1.44 & 1.46 & 1.53 & 1.17 & 1.28 & 1.26 & 1.45 \\
    \midrule
    \multirow{2}{*}{ATLAS 13~TeV $t\bar{t}$ $\ell+j$  $\frac{1}{\sigma}\frac{d\sigma}{dm_{t\bar{t}}}$}
    & $\chi^2_{\rm exp+th}$
    & 1.41 & 1.17 & 1.17 & 1.04 & 1.18 & 1.46 & 1.39 & 1.20 & 1.19 \\ 
    & $\chi^2_{\rm exp+th,orig}$
    & 1.45 & 1.24 & 1.23 & 1.10 & 1.24 & 1.48 & 1.41 & 1.26 & 1.24 \\ 
    \midrule
    \multirow{2}{*}{ATLAS 13~TeV $t\bar{t}$ $\ell+j$  $\frac{1}{\sigma}\frac{d\sigma}{dp_T^t}$} & $\chi^2_{\rm exp+th}$
    & 0.56 & 0.54 & 0.54 & 0.52 & 0.53 & 0.56 & 0.53 & 0.53 & 0.53 \\ 
    & $\chi^2_{\rm exp+th,orig}$
    & 0.80 & 0.83 & 0.83 & 0.81 & 0.82 & 0.81 & 0.78 & 0.81 & 0.81 \\ 
    \midrule
    \multirow{2}{*}{CMS 13~TeV $t\bar{t}$ $\ell+j$ $\frac{1}{\sigma}\frac{d^2\sigma}{dm_{t\bar{t}}d|y_{t\bar{t}}|}$}
    & $\chi^2_{\rm exp+th}$
    & 2.77 & 2.89 & 2.87 & 2.76 & 3.36 & 3.01 & 3.61 & 2.81 & 2.81 \\
    & $\chi^2_{\rm exp+th,orig}$
    & 3.16 & 3.29 & 3.27 & 3.27 & 3.80 & 3.47 & 4.19 & 3.21 & 3.26 \\ 
    \midrule
    \multirow{2}{*}{ATLAS 13~TeV single-inclusive jets $\frac{d^2\sigma}{dp_Td|y|}$}
    & $\chi^2_{\rm exp+th}$
    & 1.85 & 1.56 & 1.64 & 1.38 & 1.67 & 1.21 & 1.51 & 1.20 & 1.25 \\
    & $\chi^2_{\rm exp+th,orig}$
    & 3.25 & 2.77 & 2.90 & 2.49 & 2.86 & 2.17 & 2.80 & 2.28 & 2.38 \\
    \midrule
    \multirow{2}{*}{CMS 13~TeV single-inclusive jets ($R=0.4$) $\frac{d^2\sigma}{dp_Td|y|}$}
    & $\chi^2_{\rm exp+th}$
    & 1.64 & 1.58 & 1.60 & 1.52 & 1.64 & 1.47 & 1.50 & 1.48 & 1.43 \\
    & $\chi^2_{\rm exp+th,orig}$
    & 3.04 & 2.68 & 2.69 & 2.63 & 2.94 & 2.48 & 2.70 & 2.51 & 2.62 \\
    \midrule
    \multirow{2}{*}{ATLAS 13~TeV di-jets $\frac{d^2\sigma}{dm_{jj}d|y^*|}$}
    & $\chi^2_{\rm exp+th}$
    & 1.13 & 1.08 & 1.09 & 1.05 & 1.16 & 1.09 & 1.15 & 1.01 & 0.96 \\ 
    & $\chi^2_{\rm exp+th,orig}$
    & 1.73 & 1.56 & 1.58 & 1.52 & 1.72 & 1.53 & 1.70 & 1.46 & 1.41 \\ 
    \midrule
    \multirow{2}{*}{H1 single-inclusive-jets (low $Q^2$) $\frac{d^2\sigma}{dQ^{2} dp_T}$}
    & $\chi^2_{\rm exp+th}$
    & 1.64 & 1.61 & 1.61 & 1.67 & 1.61 & 1.70 & 1.74 & 1.61 & 1.73 \\ 
    & $\chi^2_{\rm exp+th,orig}$
    & 2.20 & 2.19 & 2.19 & 2.27 & 2.20 & 2.30 & 2.34 & 2.17 & 2.33 \\ 
    \midrule
    \multirow{2}{*}{H1 di-jets (low $Q^2$) $\frac{d^2\sigma}{dQ^{2} d \langle p_T \rangle}$}
    & $\chi^2_{\rm exp+th}$
    & 1.37 & 1.39 & 1.38 & 1.37 & 1.39 & 1.42 & 1.44 & 1.36 & 1.44 \\ 
    & $\chi^2_{\rm exp+th,orig}$
    & 2.28 & 2.29 & 2.29 & 2.21 & 2.31 & 2.32 & 2.34 & 2.27 & 2.34 \\ 
    \bottomrule
    \end{tabularx}

  \vspace{0.3cm}
  \caption{A comparison of the values of $\chi^2_{\rm exp+th}$,
    computed in Sect.~\ref{sec:results} by regularizing the experimental
    covariance matrix with the procedure of~\cite{Kassabov:2022pps}, to the
    corresponding values $\chi^2_{\rm exp+th,orig}$, computed with the original,
    unregularised covariance matrix.}
  \label{tab:chi2-unreg}
\end{table}

\begin{table}[!t]
  \scriptsize
  \centering
  \renewcommand{\arraystretch}{1.5}
  \begin{tabularx}{\textwidth}{Xcccccccccccc}
& \rotatebox{85}{CMS 13~TeV $W^+$ $\frac{d\sigma}{d|\eta|}$}
& \rotatebox{85}{CMS 13~TeV $W^-$ $\frac{d\sigma}{d|\eta|}$}
& \rotatebox{85}{ATLAS 13~TeV $t\bar{t}$ all hadr. $\frac{d\sigma}{dm_{t\bar{t}}}$}
& \rotatebox{85}{ATLAS 13~TeV $t\bar{t}$ all hadr. $\frac{d^2\sigma}{dm_{t\bar{t}} d|y_{t\bar{t}}|}$}
& \rotatebox{85}{ATLAS 13~TeV $t\bar{t}$ $\ell+j$  $\frac{1}{\sigma}\frac{d\sigma}{dm_{t\bar{t}}}$}
& \rotatebox{85}{ATLAS 13~TeV $t\bar{t}$ $\ell+j$  $\frac{1}{\sigma}\frac{d\sigma}{dp_T^t}$}
& \rotatebox{85}{CMS 13~TeV $t\bar{t}$ $\ell+j$ $\frac{1}{\sigma}\frac{d^2\sigma}{dm_{t\bar{t}}d|y_{t\bar{t}}|}$}
& \rotatebox{85}{ATLAS 13~TeV single-inclusive jets $\frac{d^2\sigma}{dp_Td|y|}$}
& \rotatebox{85}{CMS 13~TeV single-inclusive jets ($R=0.4$) $\frac{d^2\sigma}{dp_Td|y|}$}
& \rotatebox{85}{ATLAS 13~TeV di-jets $\frac{d^2\sigma}{dm_{jj}d|y^*|}$}
& \rotatebox{85}{H1 single-inclusive-jets (low $Q^2$) $\frac{d^2\sigma}{dQ^{2} dp_T}$}
& \rotatebox{85}{H1 di-jets (low $Q^2$) $\frac{d^2\sigma}{dQ^{2} d \langle p_T \rangle}$}\\
\toprule
$\Delta\sigma_r$
& 5.48 & 5.45 & 2.54 & 1.37 & 3.06 & 2.71 & 4.07 & 5.54 & 5.41 & 4.91 & 2.76 & 3.73 \\
$|\Delta\rho|$
& 0.06 & 0.06 & 0.03 & 0.02 & 0.04 & 0.04 & 0.06 & 0.06 & 0.06 & 0.06 & 0.03 & 0.04 \\
\bottomrule
\end{tabularx}

  \vspace{0.3cm}
  \caption{The maximum relative difference of the variances $\Delta\sigma_r$
    (in percent) and the maximum absolute difference of the correlation
    $|\Delta\rho|$ computed between the corresponding unregularised
    and regularised matrices for the regularised datasets outlined in
    Table~\ref{tab:chi2-unreg}.}
  \label{tab:regularisation}
\end{table}

From Table~\ref{tab:chi2-unreg}, we see that the effect of regularisation
on the $\chi^2$ depends on the dataset. For some of these, the effect is
huge, {\it e.g.}~for the CMS Drell-Yan measurement or for the ATLAS and CMS
single-inclusive jet and di-jet measurements. Specifically, it amounts to a
reduction of the $\chi^2$ of more than $20\sigma$ for the first and of about
$7\sigma$ for the latter. For others, the effect is small,
{\it e.g.}~for the ATLAS and CMS top-quark pair measurements or for the H1
single-inclusive jet measurements. This is unsurprising, given that the first
datasets have the largest value of $Z_{\mathcal{L}}$ among all the datasets
selected in Table~\ref{tab:Z}. One may think that the regularisation procedure
is significantly modifying uncertainties and correlations. However, as we can
see from Table~\ref{tab:regularisation}, changes are relatively mild: the
largest relative change of uncertainties is of order 5\%, whereas the
largest absolute change of correlation is of order 0.06. The first of this
changes results in an effective increase of the diagonal elements of the
covariance matrix. The increase is however moderate: typical LHC uncertainties
are around a few percent, so the actual increase in uncertainty is of the
order of a few permil. The second of this changes results in an effective
decrease of correlations by about 6\%. We consider this decrease to also be
moderate, and at the same time we appreciate that it may be experimentally
very challenging to quantify correlations with a precision of 6\%.

It may seem counter-intuitive that a relatively small change in the covariance
matrix leads to a variation of several standard deviations in the $\chi^2$. We
refer the reader to~\cite{Kassabov:2022pps} for a mathematical demonstration of
this fact. Here we shall note that the degree of knowledge of experimental
correlations related to a $1\sigma$ variation of the $\chi^2$ depends on the
size of the uncertainties. The smaller the uncertainty, the
higher the required degree of precision. Roughly speaking, as one can see
from Fig.~3 of~\cite{Kassabov:2022pps}, for a 1\% (2.5\%) uncertainty,
correlations ought to be known with a precision of 2\% (12\%) in order to be
within a variation of the $\chi^2$ of one standard deviation. It is therefore
unsurprising that the largest improvements in the value of the $\chi^2$ occur
for the most precise datasets, which are affected by percent-level (if not
sub-percent-level) uncertainties. That being said, we reiterate the fact that
the regularisation procedure applies to the covariance matrix as a whole, hence
it does not discriminate across different uncertainites that could have a
different physical meaning. If we assessed a dataset for inclusion in a PDF
determination, understanding which uncertainties are responsible for the bad
conditioning of the covariance matrix would be mandatory. However, we consider
that all of these observations do not affect our ability to comparatively judge
the performance of different PDF sets at describing the data. In our view, the
regularisation procedure does not alter the relative pattern of $\chi^2$
among different PDF sets and datasets. Therefore, the conclusions of
Sect.~\ref{sec:results} continue to hold.

\section{Additional results}
\label{app:extra_results}

In this appendix, we collect additional results, complementing those presented
in Sect.~\ref{sec:results}, for the breakdown of $\chi^2_{\rm exp+th}$ into its
$\chi^2_{\rm exp+mho}$ and $\chi^2_{\rm exp}$ components, and for the data-theory
comparisons. The additional results refer to the following categories of
measurements.

\begin{description}

\item[Top-quark pair production.] Figure~\ref{fig:chi2histo_ttbar_additional}
  displays the breakdown of $\chi^2_{\rm exp+th}$ into its $\chi^2_{\rm exp+mho}$
  and $\chi^2_{\rm exp}$ components for the datasets not displayed in
  Fig.~\ref{fig:top-ATLAS-2-chi2}, namely: the ATLAS all-hadronic absolute
  differential distribution in the invariant mass of the top-quark pair;
  the ATLAS all-hadronic normalised differential distribution in the absolute
  rapidity of the top-quark pair; the ATLAS lepton+jets normalised
  differential distributions in the absolute rapidity of the top quark and of
  the top-quark pair; and the CMS lepton+jets normalised differential
  distributions in the transverse momentum of the top quark and of the
  invariant mass of the top-quark pair.
  Figure~\ref{fig:datatheory_ttbar_additional_1} displays the data-theory
  comparison for the top-quark pair single-differential distributions not
  displayed in Fig.~\ref{fig:datatheory-ttbar}, namely: the ATLAS all-hadronic
  normalised distribution differential in the absolute value of the top-quark
  pair rapidity; the ATLAS all-hadronic absolute distribution differential in
  the invariant mass of the top-quark pair; the ATLAS lepton+jets normalised
  distributions differential in the absolute rapidity of the top quark and of
  the top-quark pair; and the CMS lepton+jets normalised distributions
  differential in the transverse momentum of the top quark and in the
  invariant mass of the top-quark pair.
  Figure~\ref{fig:DY-datatheory} displays the data-theory comparison for the
  top-quark pair bins of the ATLAS and CMS double-differential distributions
  not displayed in Fig.~\ref{fig:datatheory-ttbar}.

  \begin{figure}[!t]
    \centering
    \includegraphics[width=0.49\textwidth]{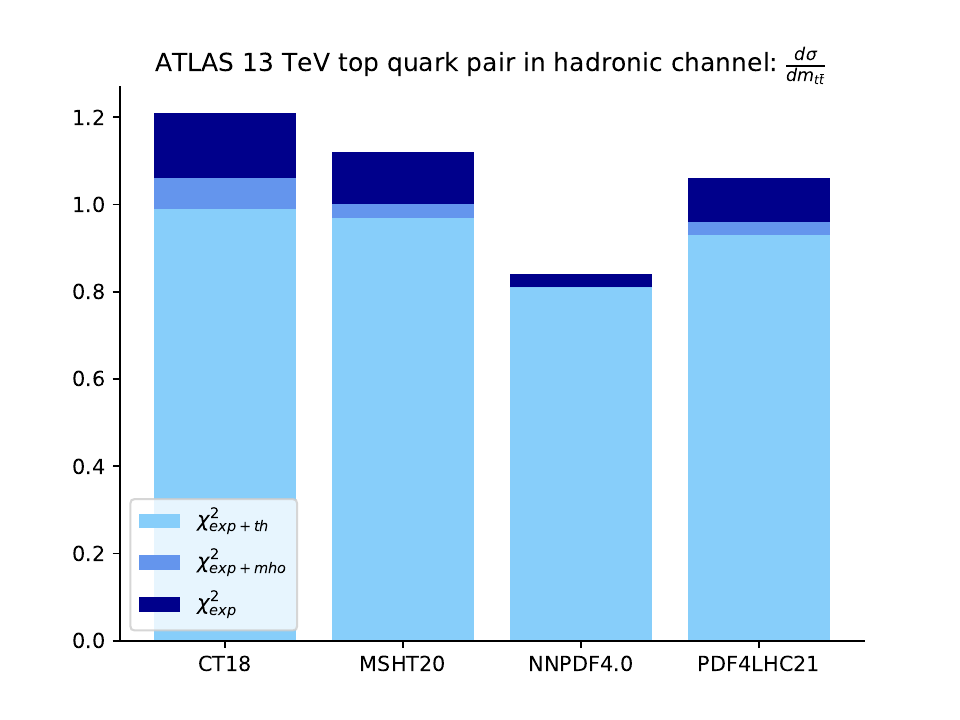}
    \includegraphics[width=0.49\textwidth]{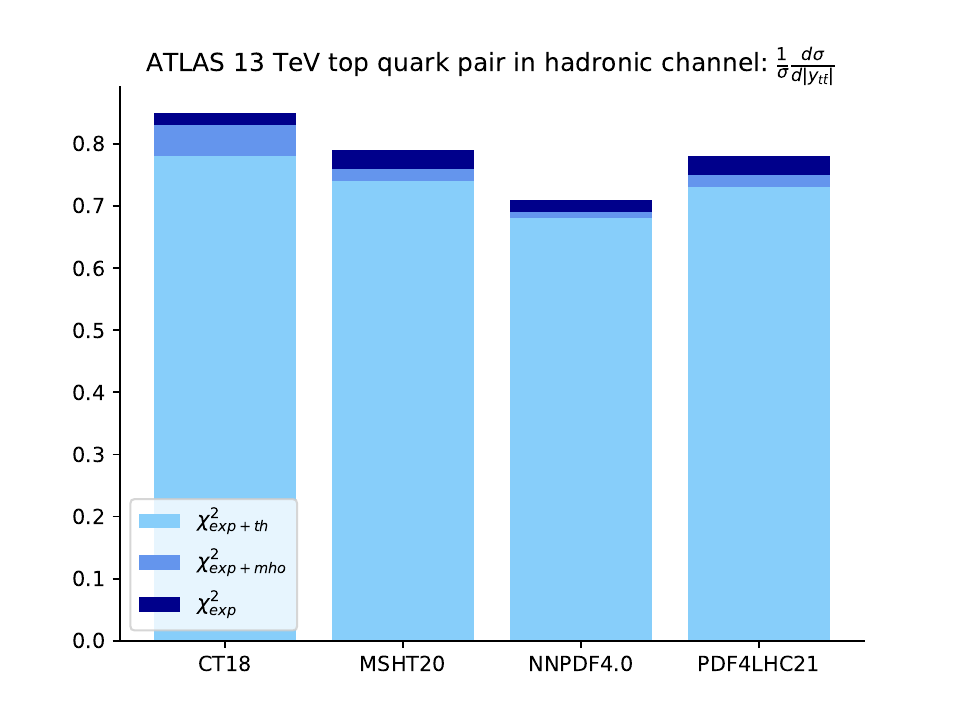}\\
    \includegraphics[width=0.49\textwidth]{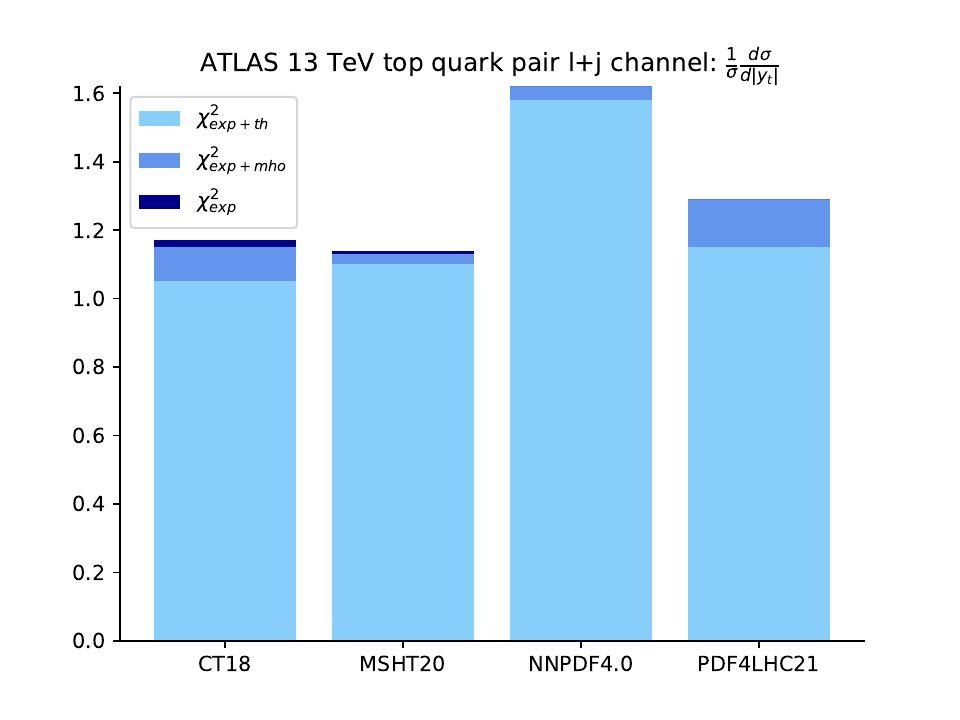}
    \includegraphics[width=0.49\textwidth]{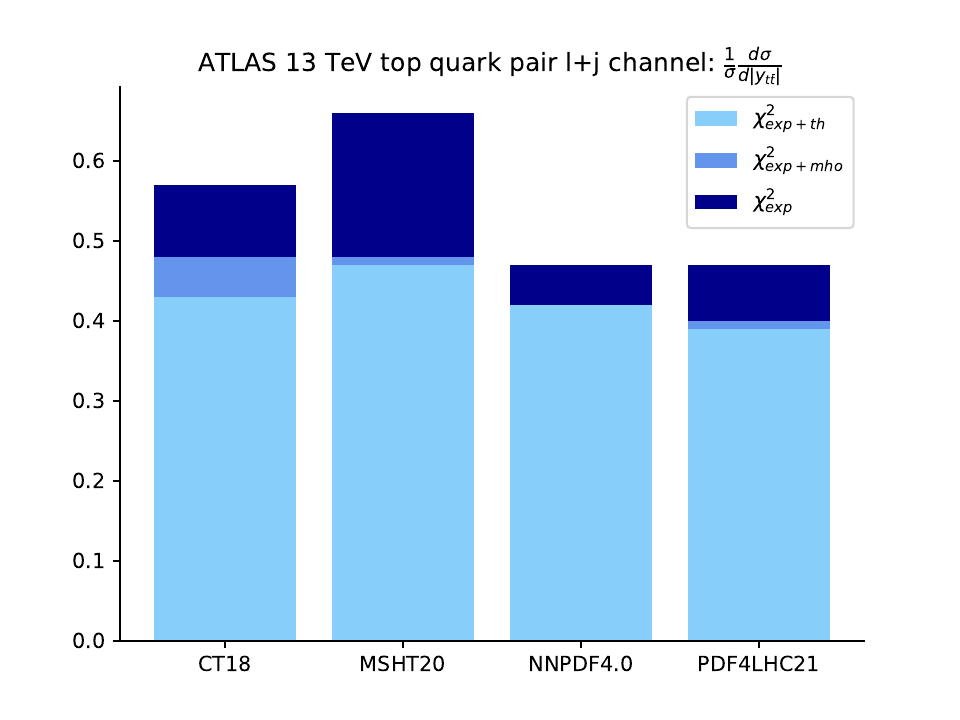}\\
    \includegraphics[width=0.49\textwidth]{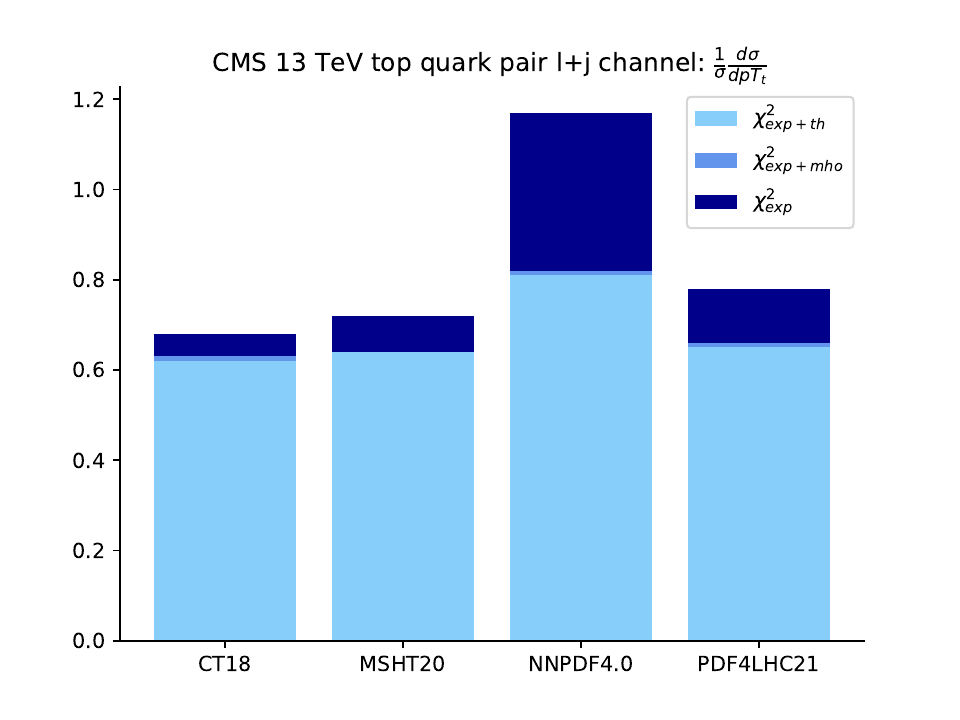}
    \includegraphics[width=0.49\textwidth]{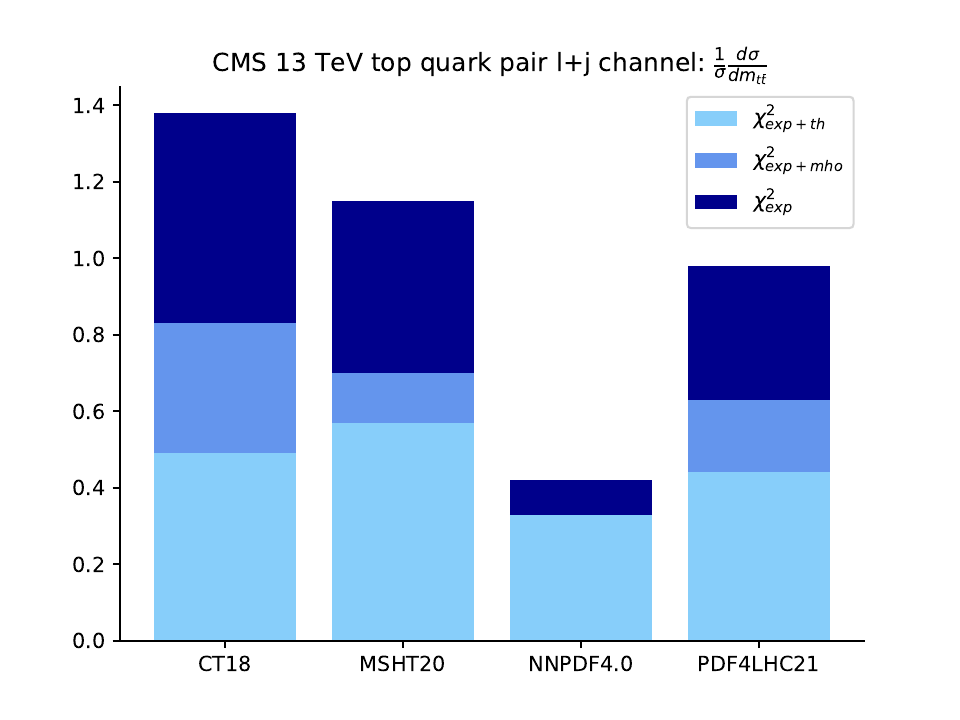}\\
    \caption{Same as Fig.~\ref{fig:top-ATLAS-2-chi2} for the ATLAS and CMS
      datasets not shown there. From top to bottom, left to right: the ATLAS
      all-hadronic absolute differential distribution in the invariant mass of
      the top-quark pair ($n_{\rm dat} = 9$, $\sqrt{2/n_{\rm dat}} = 0.47$); the ATLAS all-hadronic normalised differential
      distribution in the absolute rapidity of the top-quark pair ($n_{\rm dat} = 12$, $\sqrt{2/n_{\rm dat}} = 0.41$); the ATLAS
      lepton+jets normalised differential distributions in the absolute rapidity
      of the top quark ($n_{\rm dat} = 5$, $\sqrt{2/n_{\rm dat}} = 0.63$) 
      and of the top-quark pair ($n_{\rm dat} = 7$, $\sqrt{2/n_{\rm dat}} = 0.53$); and the CMS lepton+jets
      normalised differential distributions in the transverse momentum of the
      top quark ($n_{\rm dat} = 16$, $\sqrt{2/n_{\rm dat}} = 0.35$) and of the invariant mass of the top-quark pair
      ($n_{\rm dat} = 15$, $\sqrt{2/n_{\rm dat}} = 0.37$).}
    \label{fig:chi2histo_ttbar_additional}
  \end{figure}

  \begin{figure}[!t]
    \centering
    \includegraphics[width=0.49\textwidth]{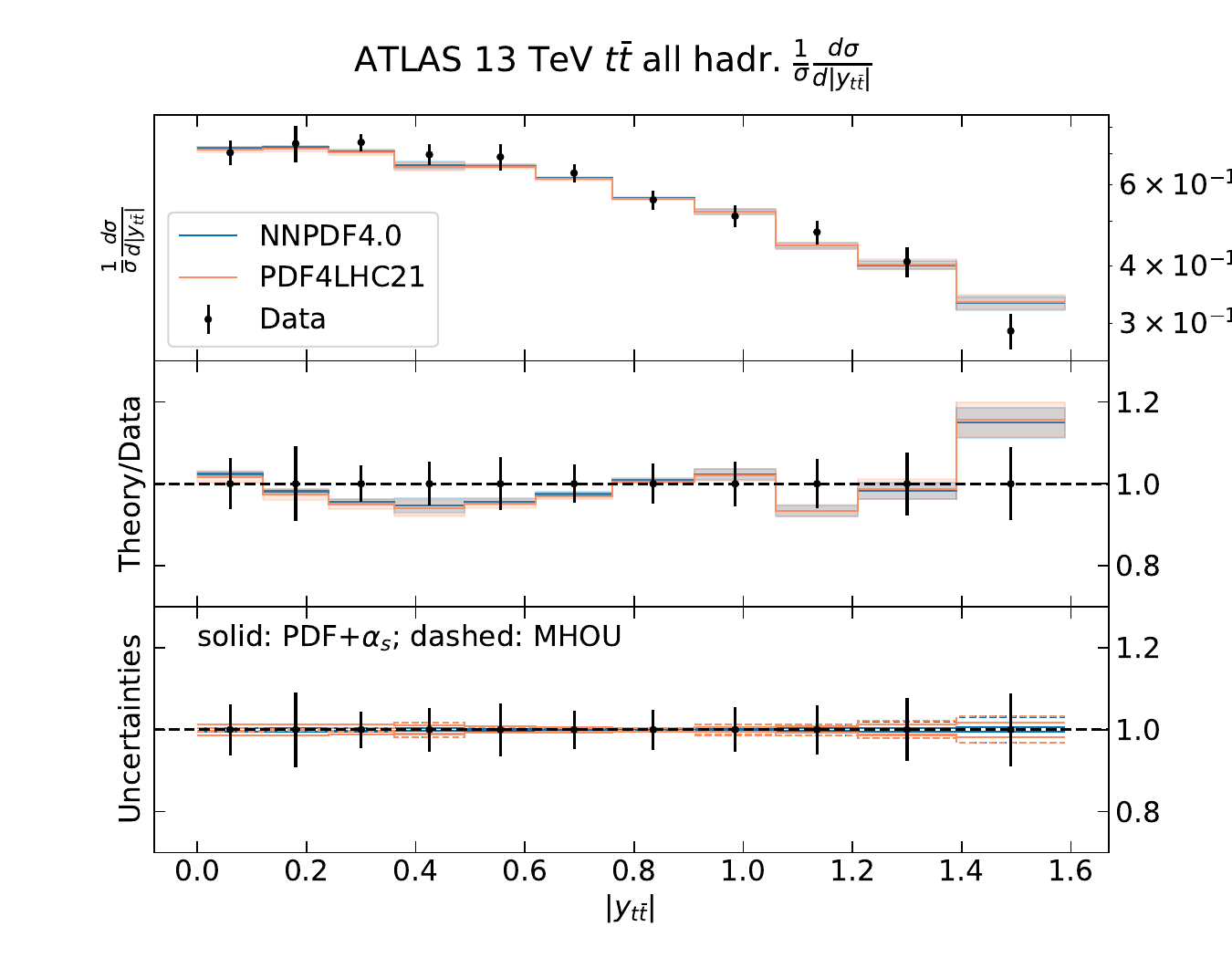}
    \includegraphics[width=0.49\textwidth]{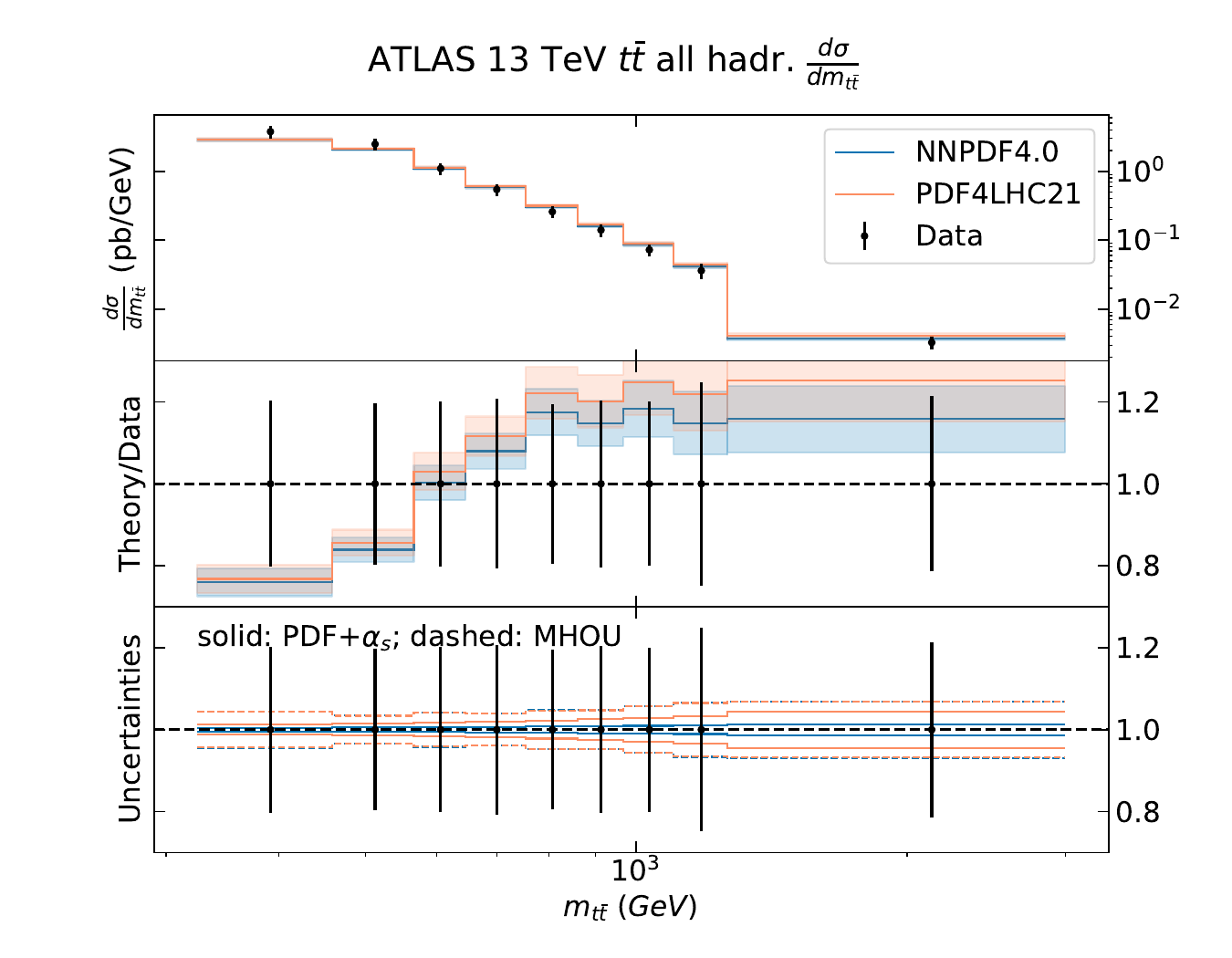}
    \includegraphics[width=0.49\textwidth]{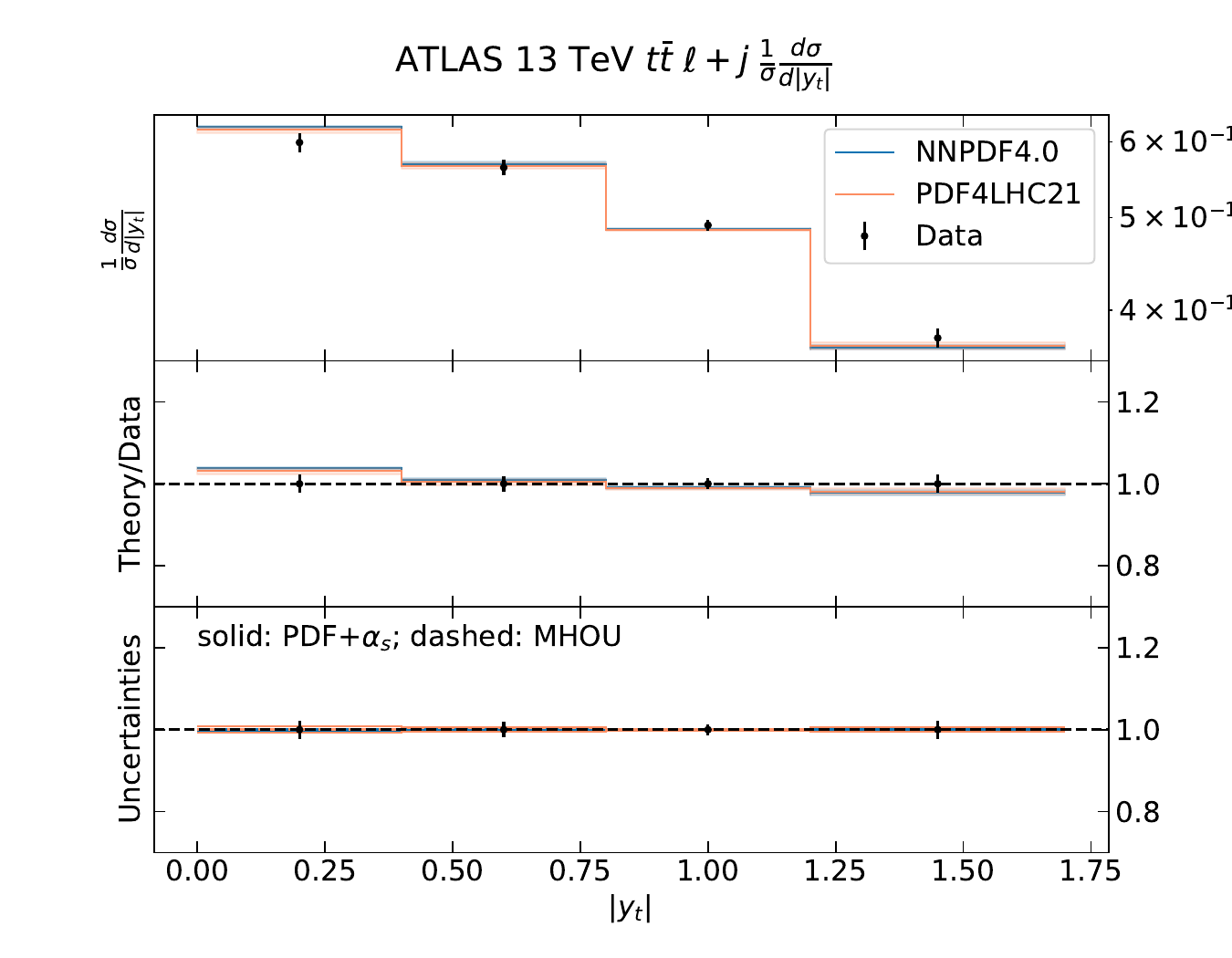}
    \includegraphics[width=0.49\textwidth]{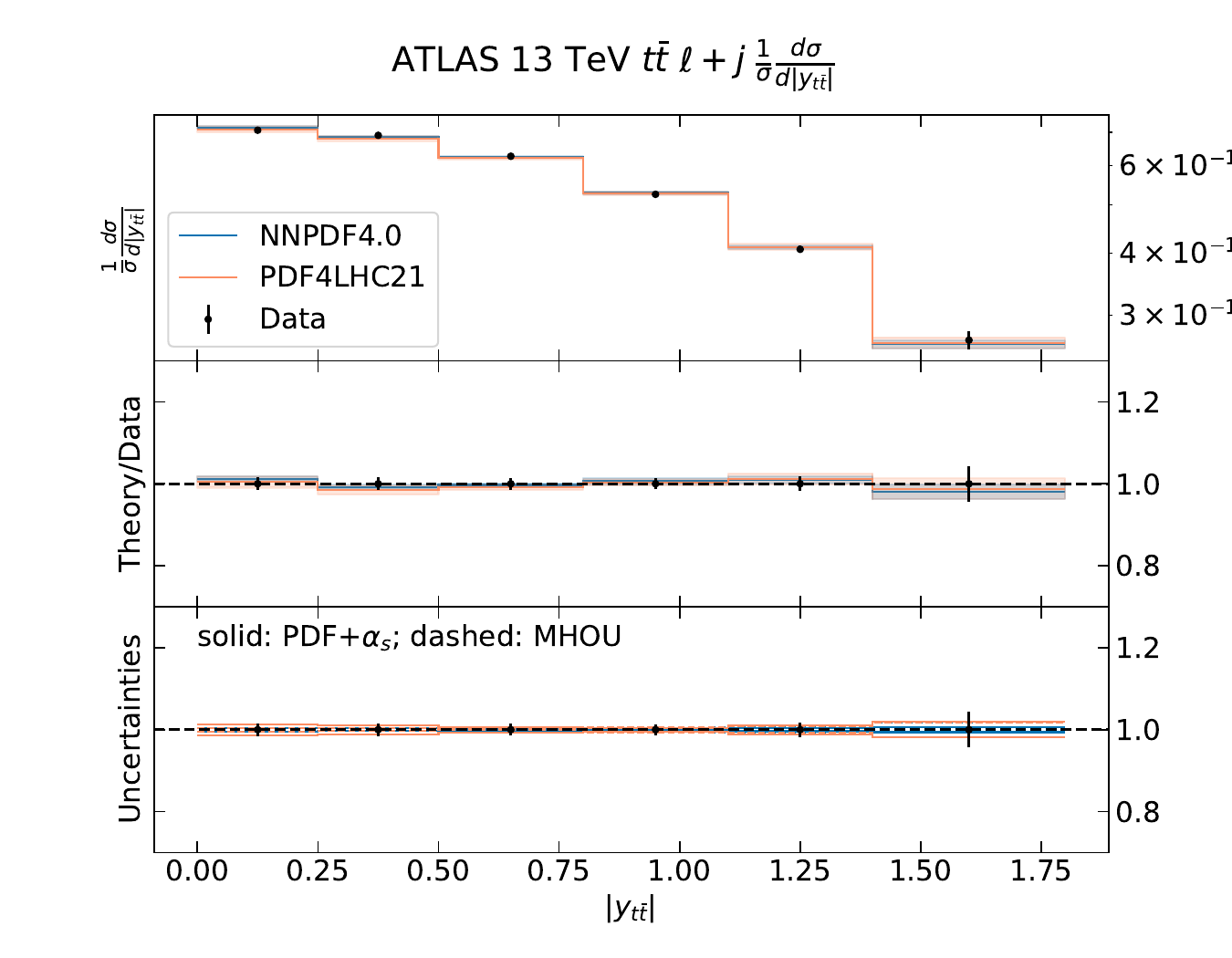}\\
    \includegraphics[width=0.49\textwidth]{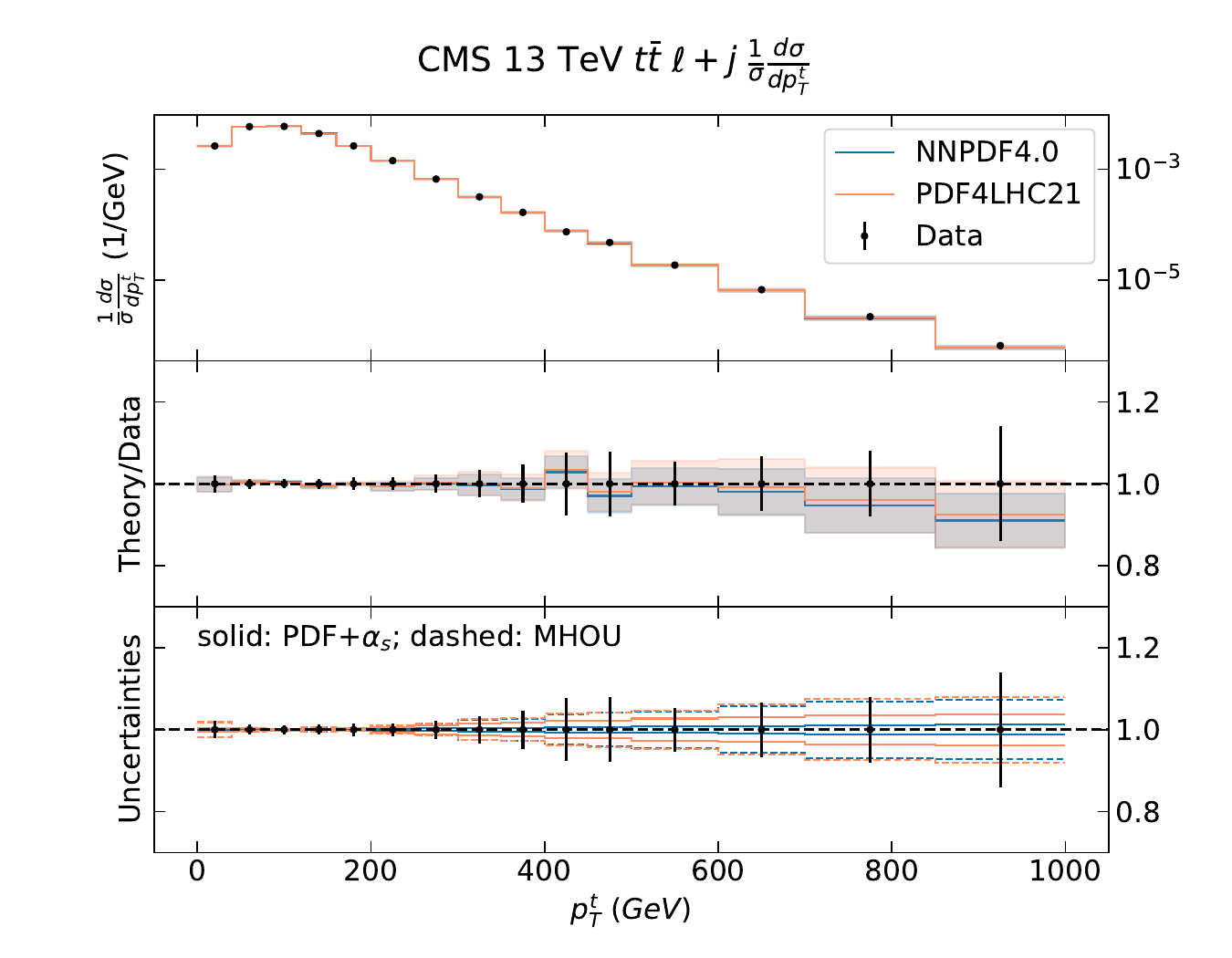}
    \includegraphics[width=0.49\textwidth]{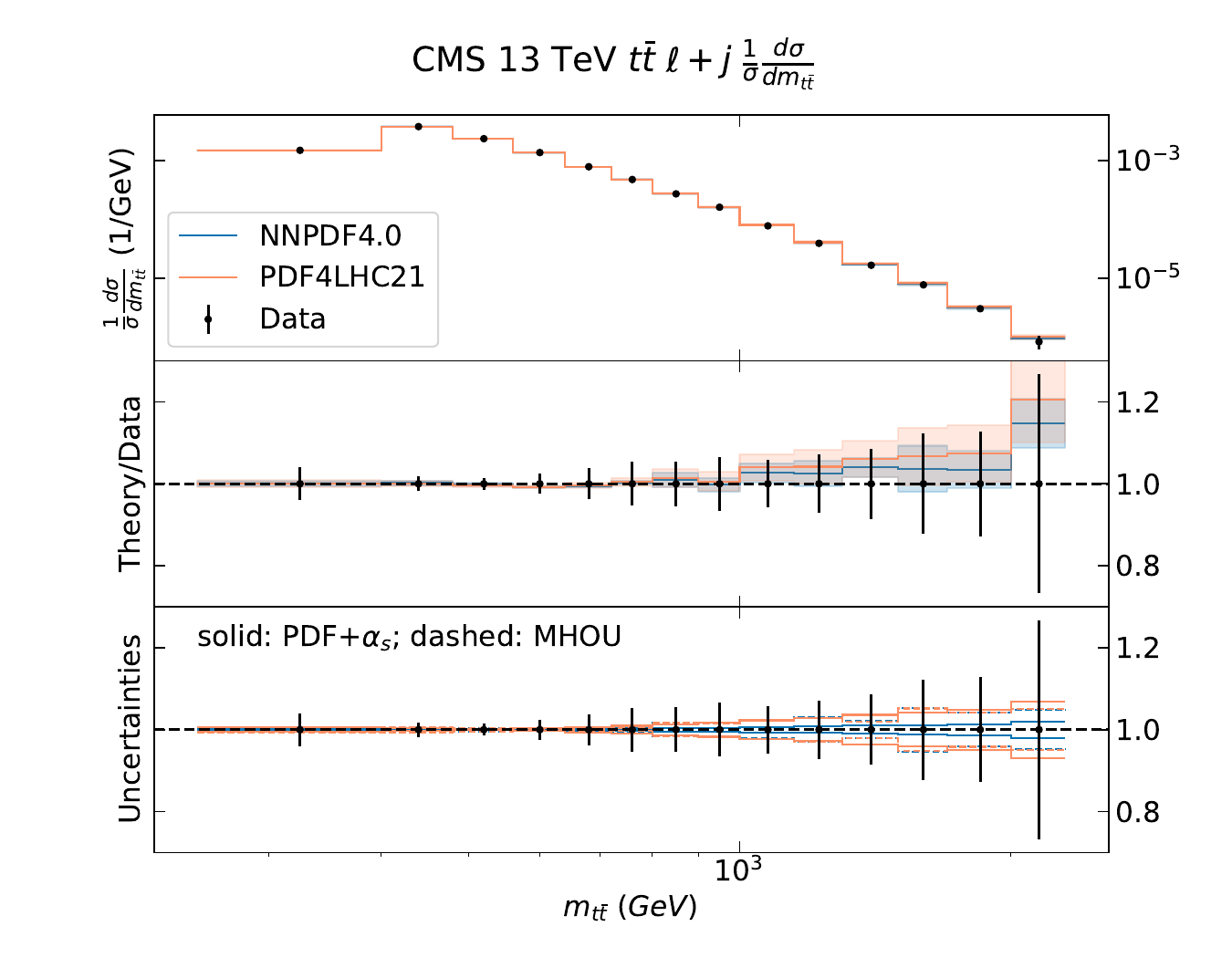}\\
    \caption{Same as Fig.~\ref{fig:DY-datatheory} for the top-quark pair
      single-differential distributions not displayed in
      Fig.~\ref{fig:datatheory-ttbar}, namely: the ATLAS all-hadronic
      normalised distribution differential in the absolute value of the
      top-quark pair rapidity; the ATLAS all-hadronic absolute distribution
      differential in the invariant mass of the top-quark pair; the ATLAS
      lepton+jets normalised distributions differential in the absolute
      rapidity of the top quark and of the top-quark pair; and the CMS
      lepton+jets normalised distributions differential in the transverse
      momentum of the top quark and in the invariant mass of the top-quark
      pair.}
    \label{fig:datatheory_ttbar_additional_1}
  \end{figure}

  \begin{figure}[!t]
    \includegraphics[width=0.49\textwidth]{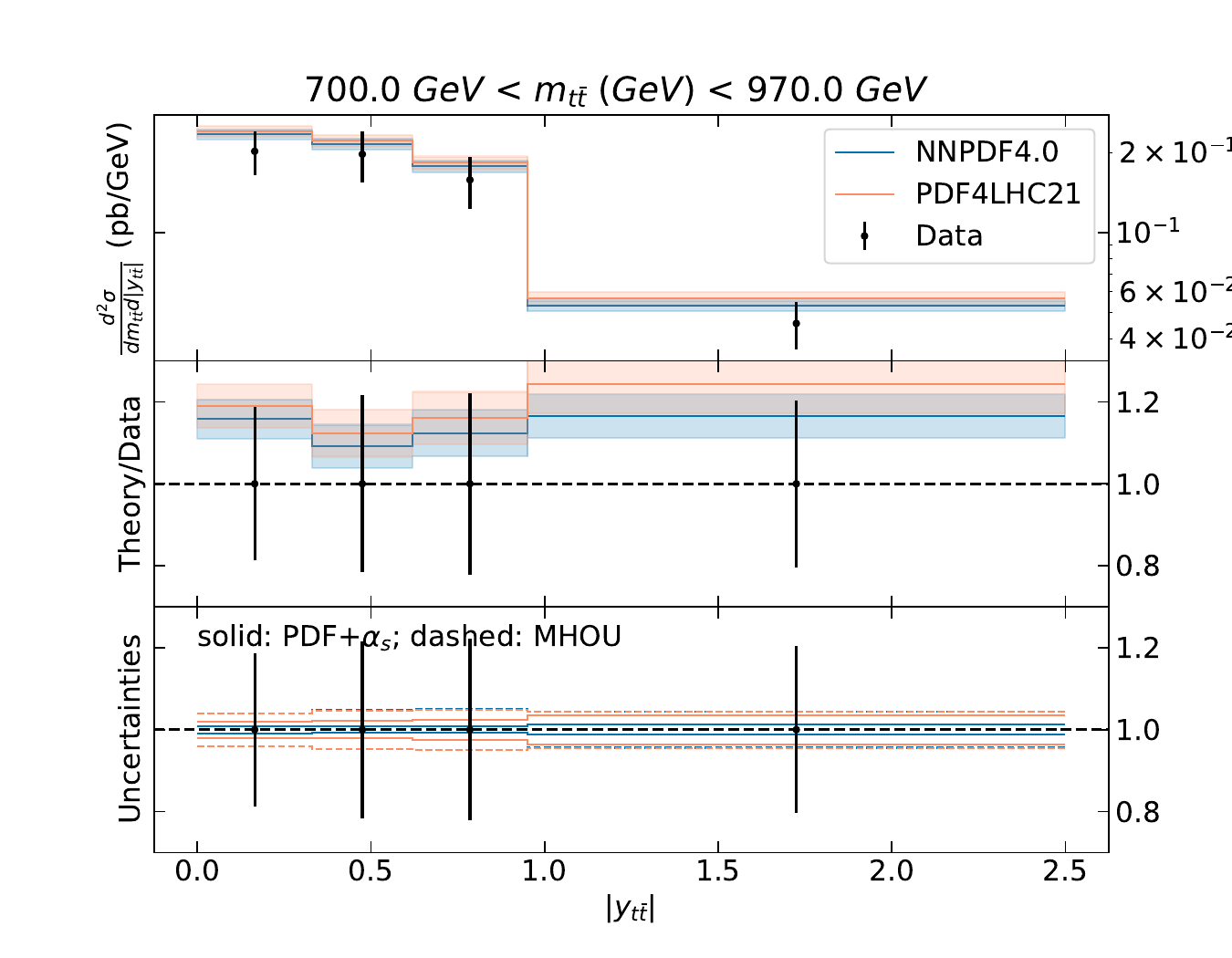}
    \includegraphics[width=0.49\textwidth]{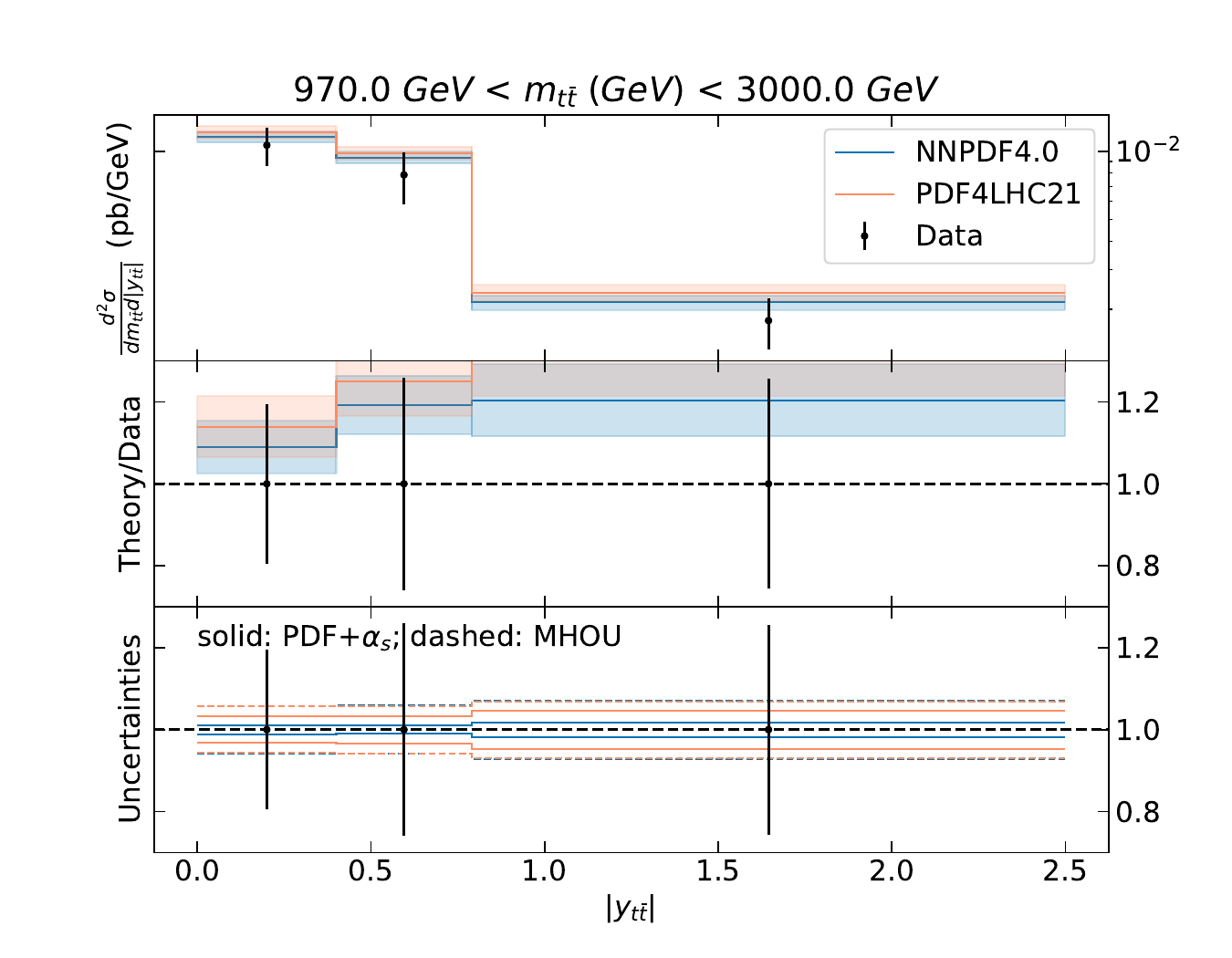}\\
    \includegraphics[width=0.49\textwidth]{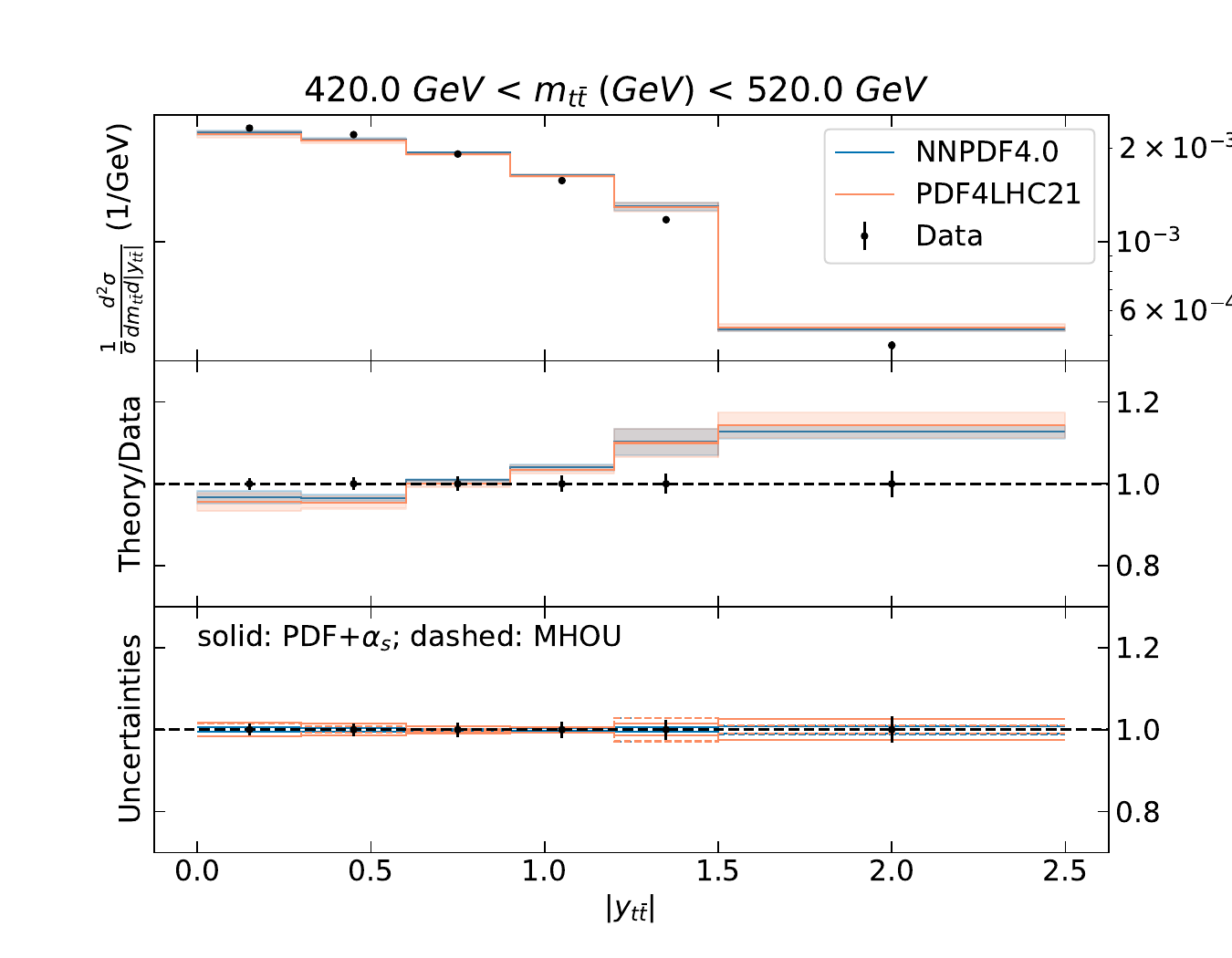}
    \includegraphics[width=0.49\textwidth]{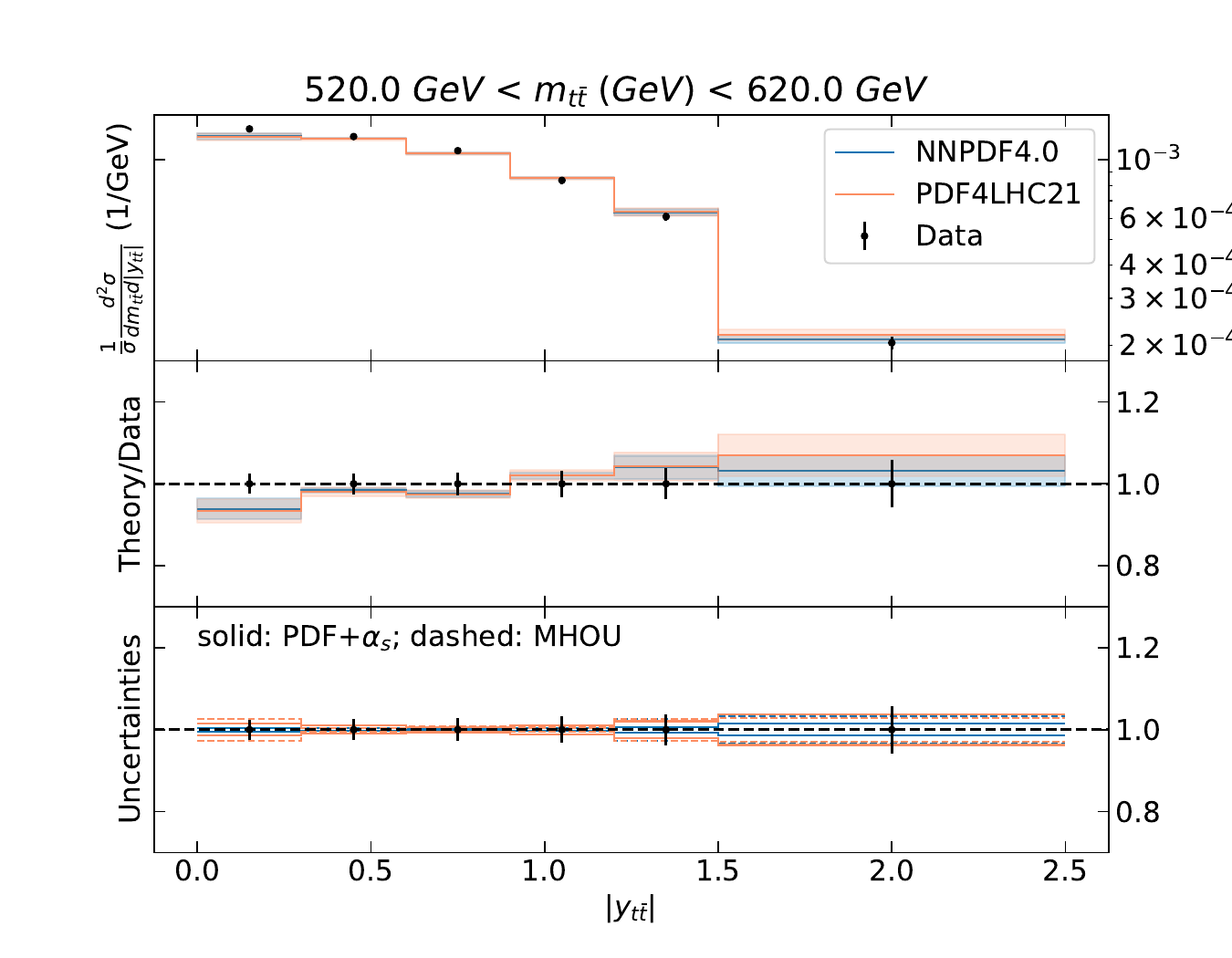}\\
    \includegraphics[width=0.49\textwidth]{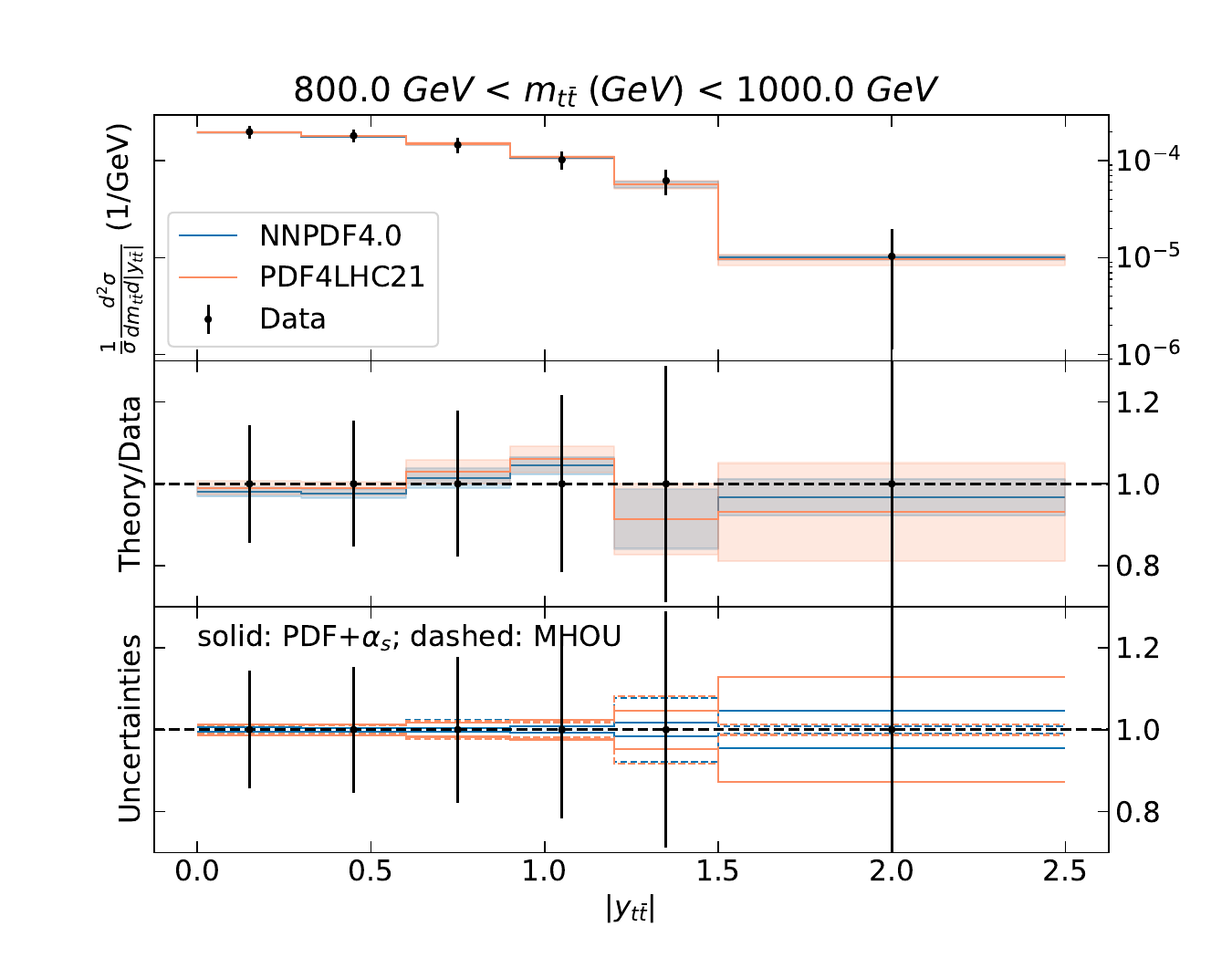}
    \includegraphics[width=0.49\textwidth]{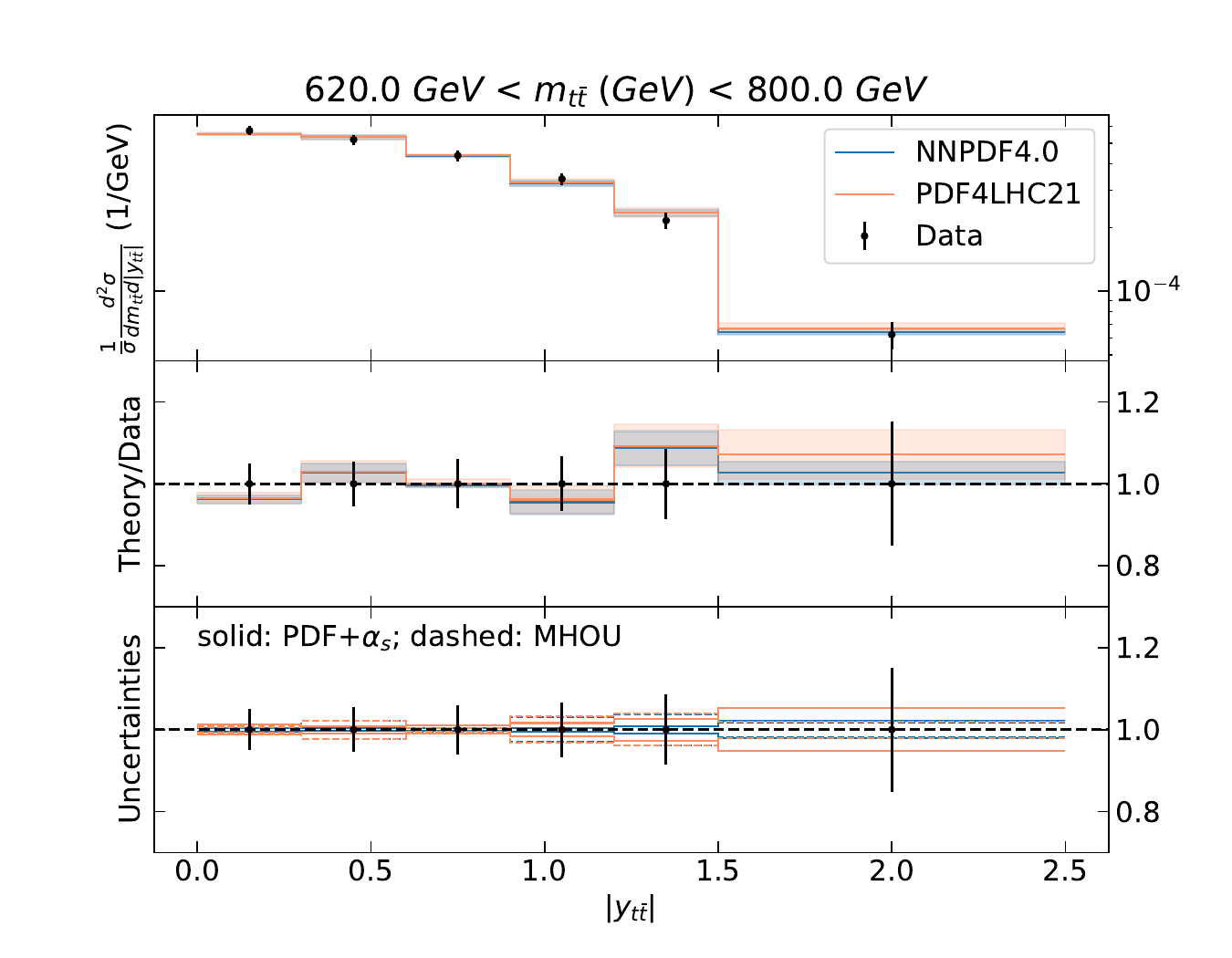}\\
    \caption{Same as Fig.~\ref{fig:DY-datatheory} for the top-quark pair bins
      of the double-differential distributions not displayed in
      Fig.~\ref{fig:datatheory-ttbar}: the first row corresponds to the ATLAS
      measurement in the all-hadronic final state; the second and third rows
      correspond to the CMS measurement in the lepton+jets final state.}
    \label{fig:datatheory_ttbar_additional_2} 
  \end{figure}

\item[Single-inclusive jet and di-jet production at the LHC.]
  Figures~\ref{fig:datatheory_jets_additional_1},
  \ref{fig:datatheory_jets_additional_2},
  and~\ref{fig:datatheory_jets_additional_3}
  display the data-theory comparison for the remaining rapidity bins not
  shown in Figs.~\ref{fig:jets-ATLAS-13tev}, \ref{fig:jets-CMS-13tev-R07},
  and~\ref{fig:dijet-ATLAS-13tev}, respectively.
  Figure~\ref{fig:datatheory_jets_additional_1} corresponds to the ATLAS
  single-inclusive jet measurement; Fig.~\ref{fig:datatheory_jets_additional_2}
  corresponds to the CMS single-inclusive jet measurement; and
  Fig.~\ref{fig:datatheory_jets_additional_2} corresponds to the ATLAS di-jet
  measurement.

  \begin{figure}[!t]
    \includegraphics[width=0.49\textwidth]{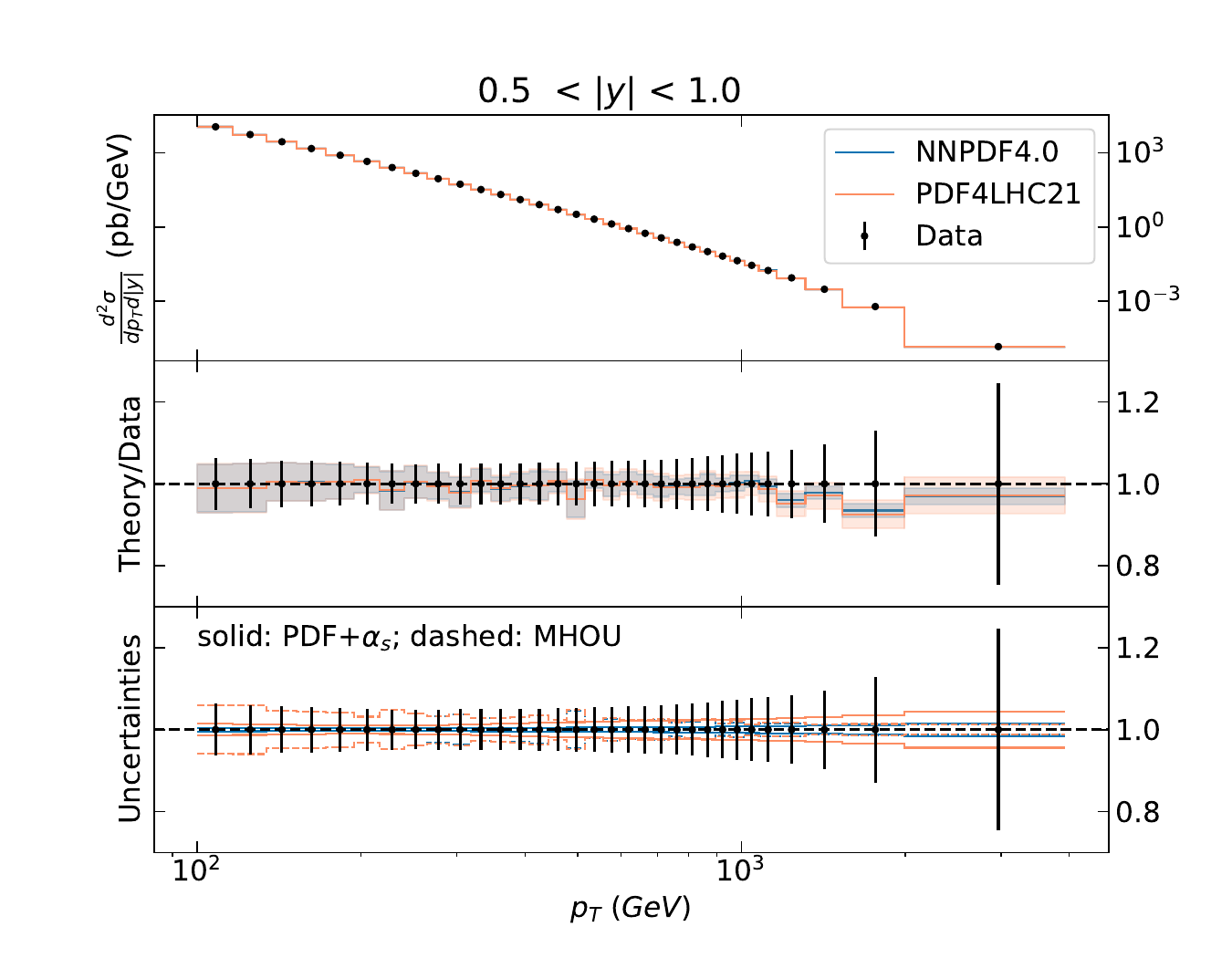}
    \includegraphics[width=0.49\textwidth]{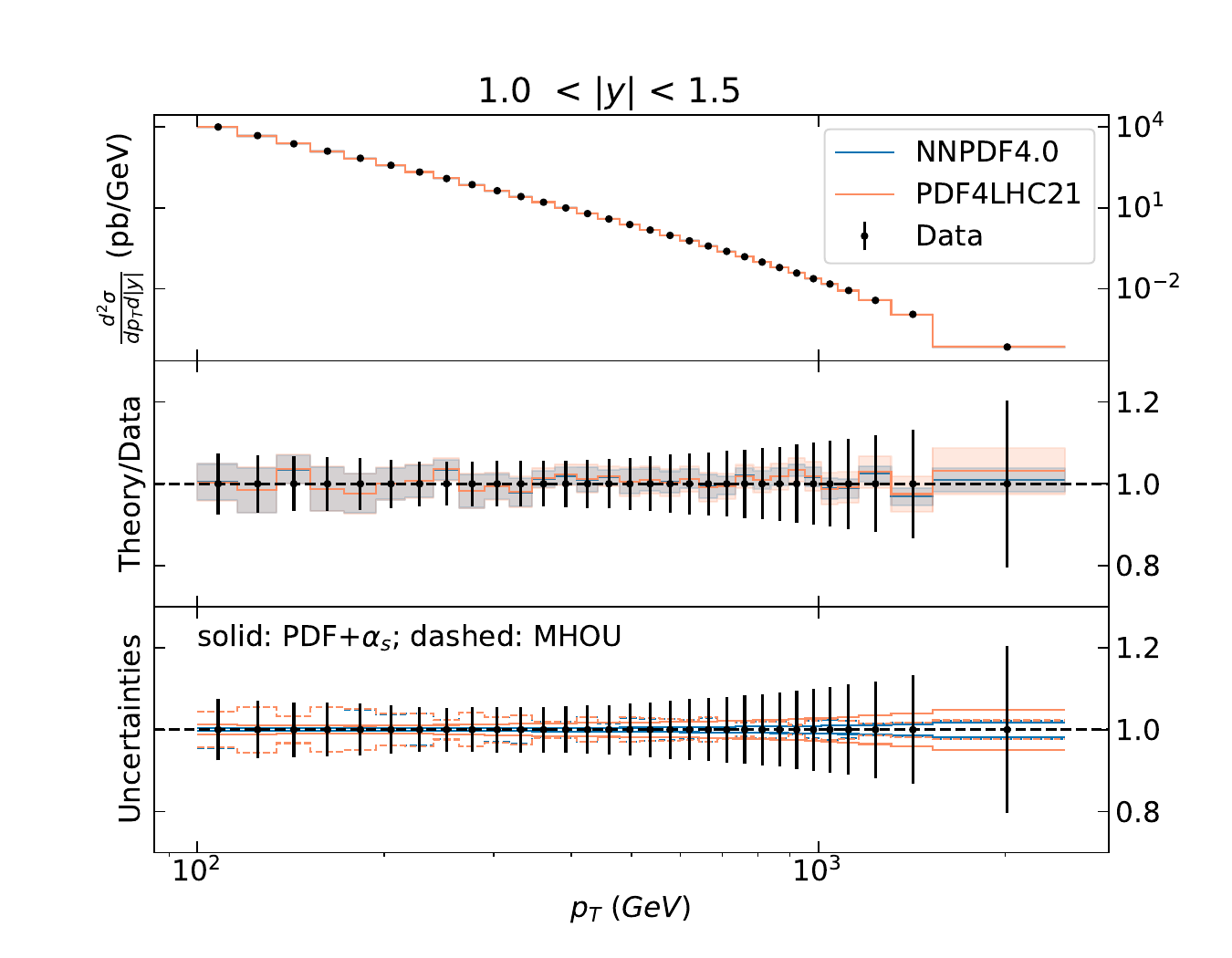}\\
    \includegraphics[width=0.49\textwidth]{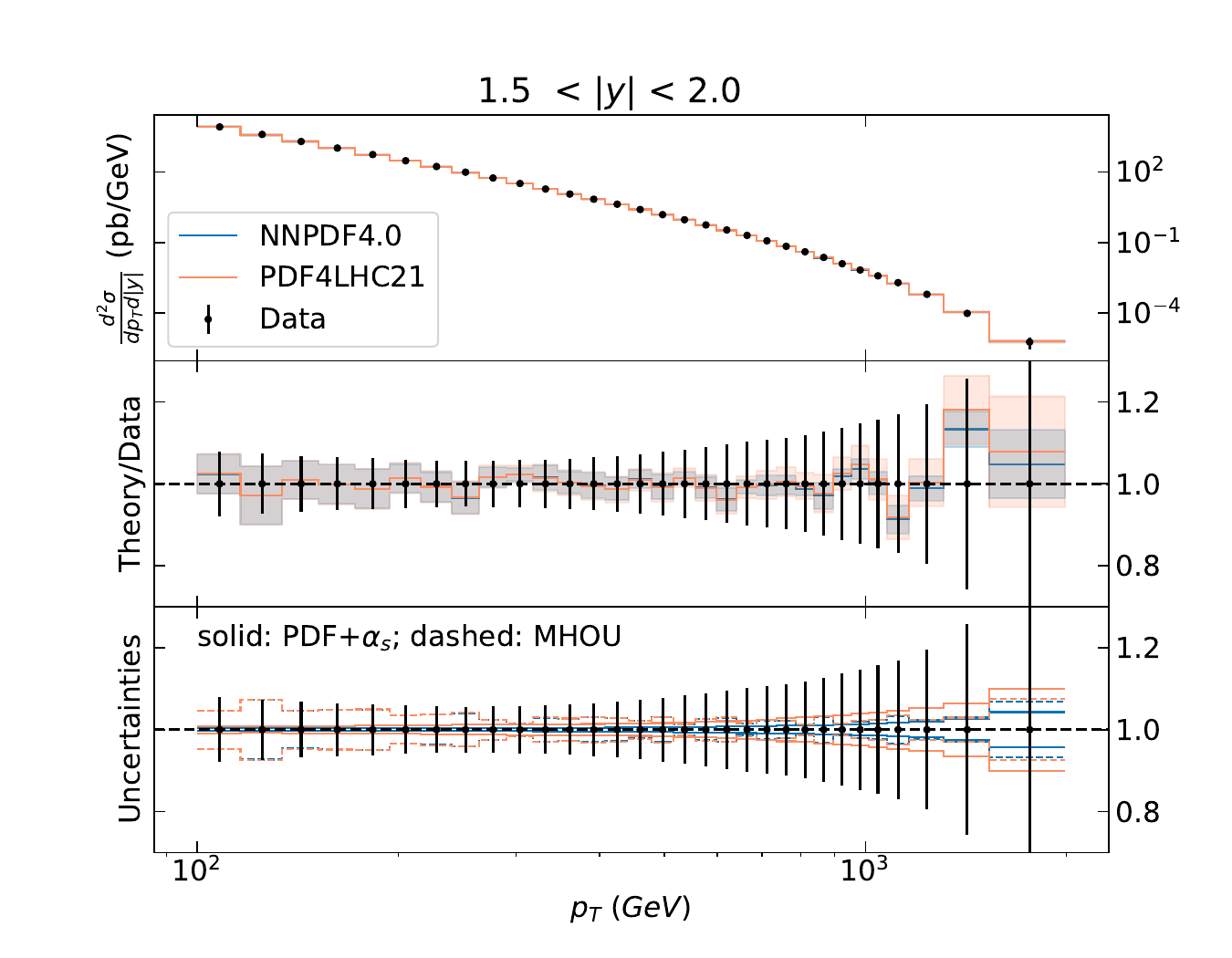}
    \includegraphics[width=0.49\textwidth]{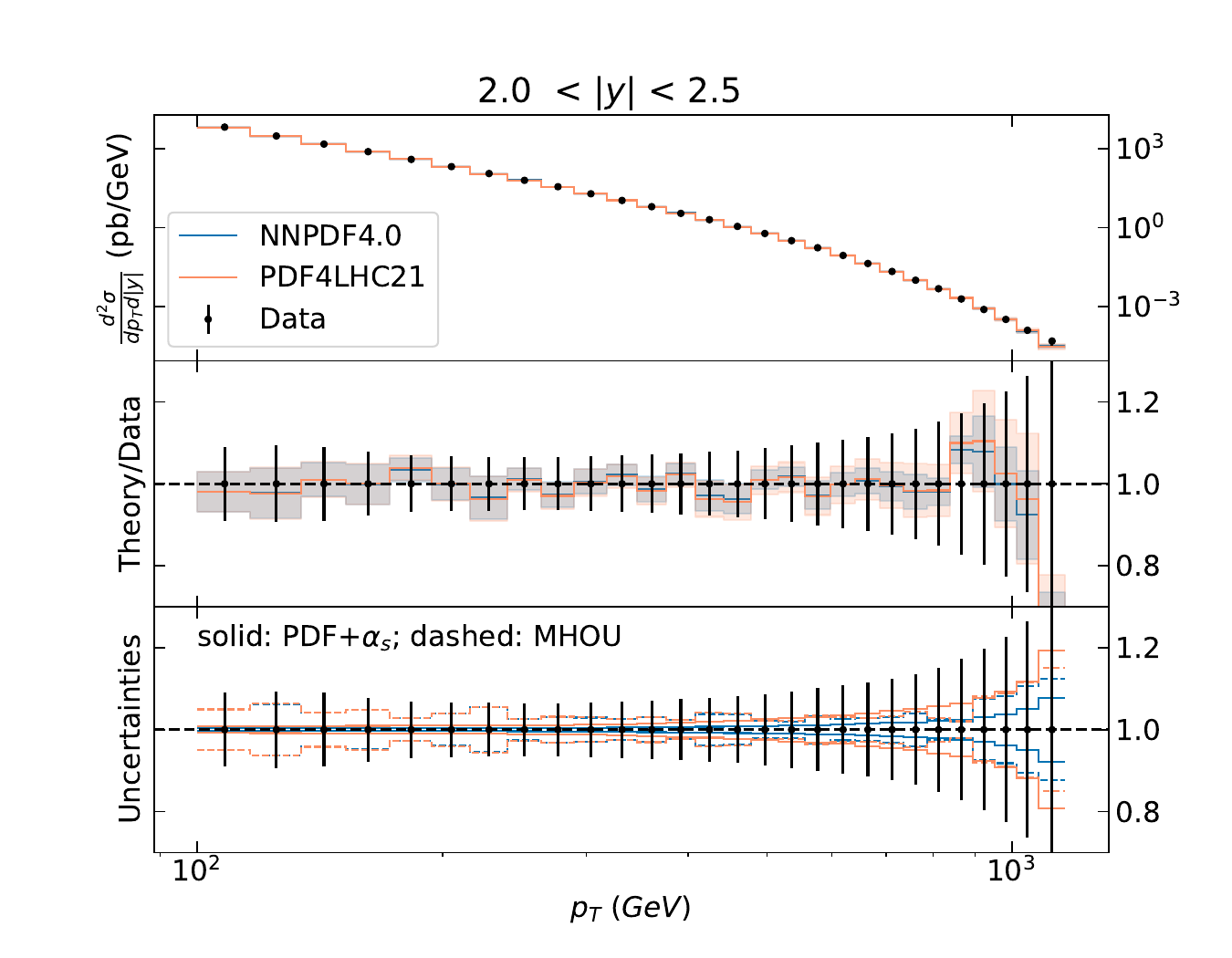}\\
    \caption{Same as Fig.~\ref{fig:jets-ATLAS-13tev} for the intermediate
      rapidity bins.}
    \label{fig:datatheory_jets_additional_1}
  \end{figure}
  
  \begin{figure}[!t]
    \includegraphics[width=0.49\textwidth]{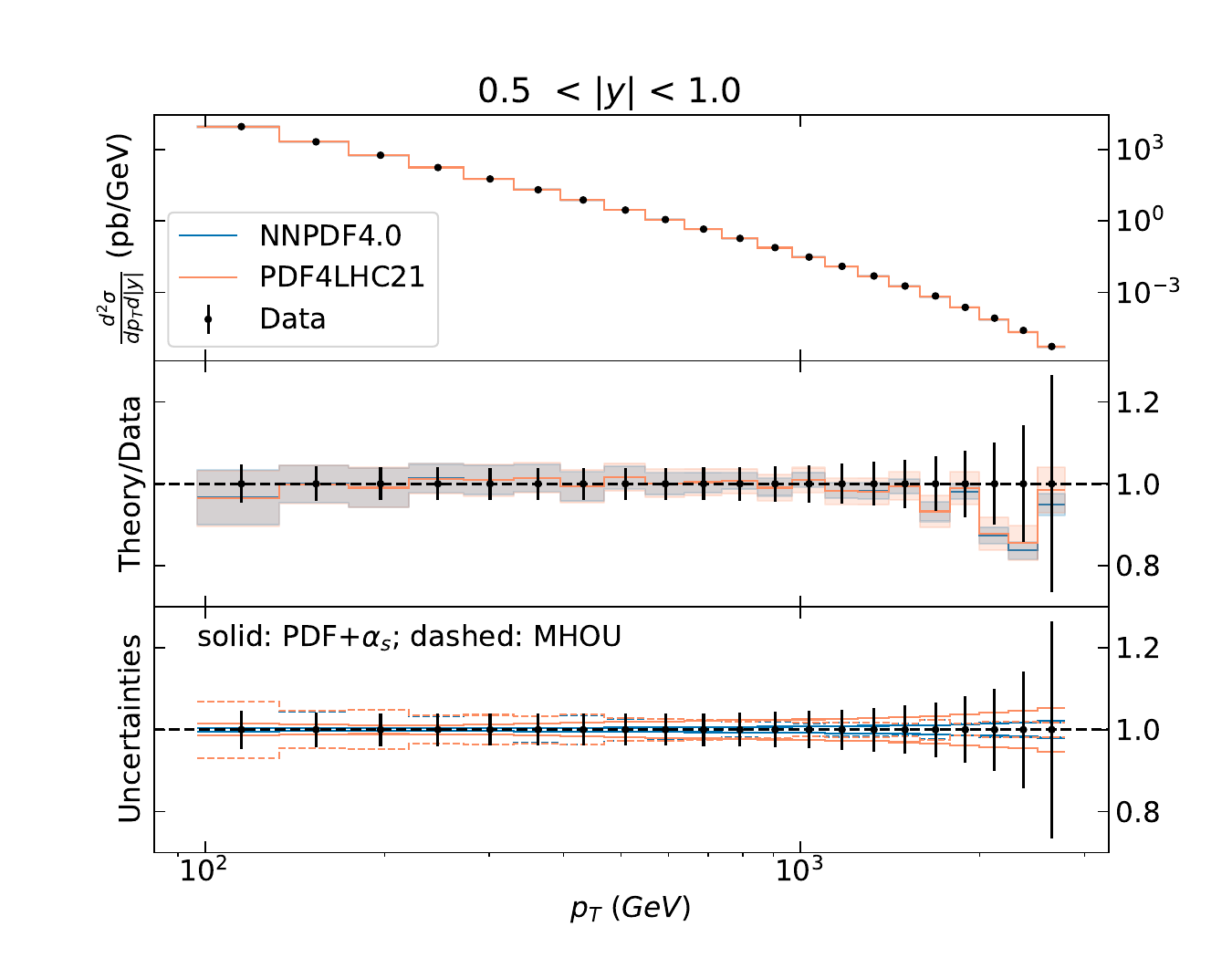}
    \includegraphics[width=0.49\textwidth]{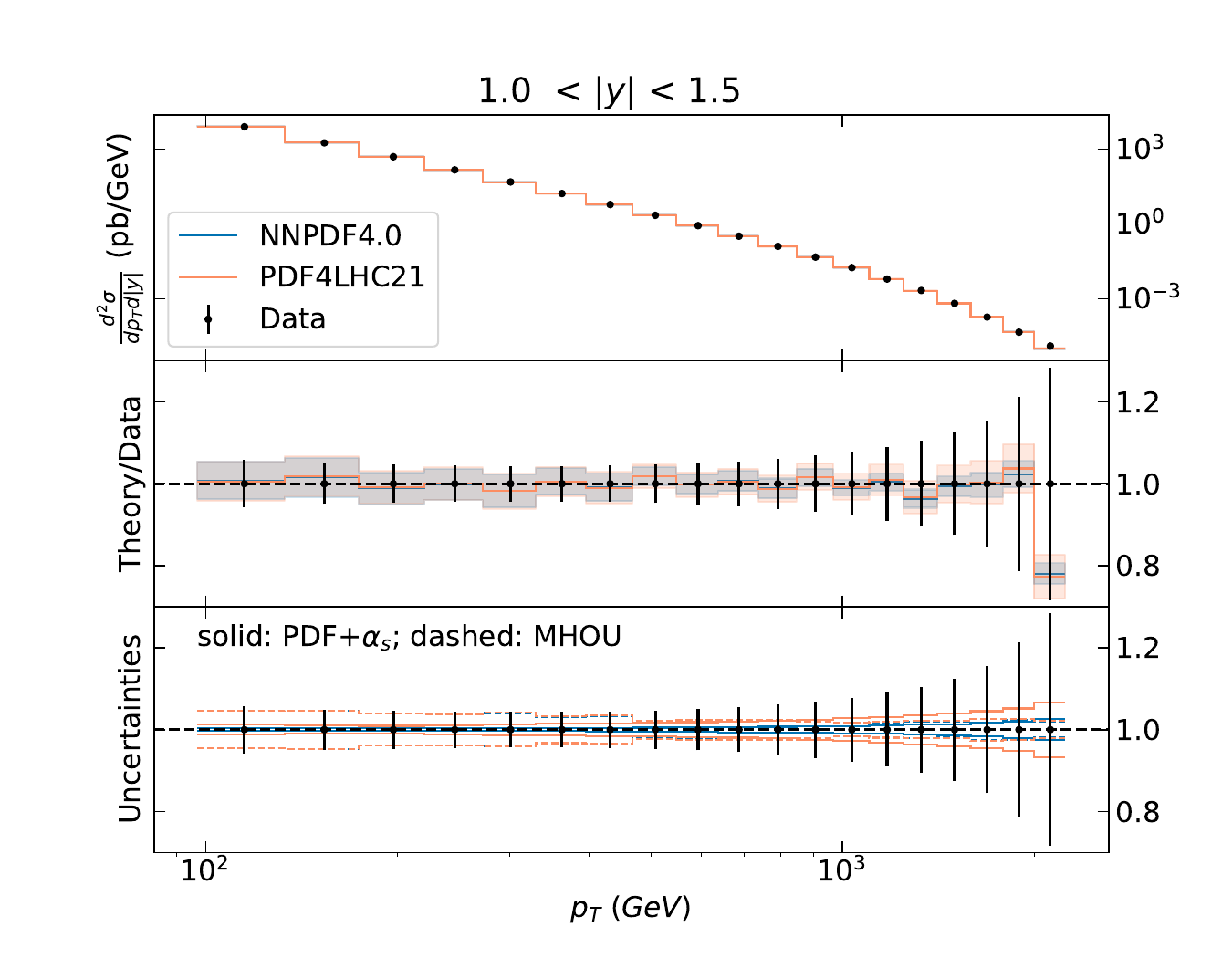}
    \caption{Same as Fig.~\ref{fig:jets-CMS-13tev-R07} for the intermediate
      rapidity bins.}
    \label{fig:datatheory_jets_additional_2} 
  \end{figure}
  
  \begin{figure}[!t]
    \includegraphics[width=0.49\textwidth]{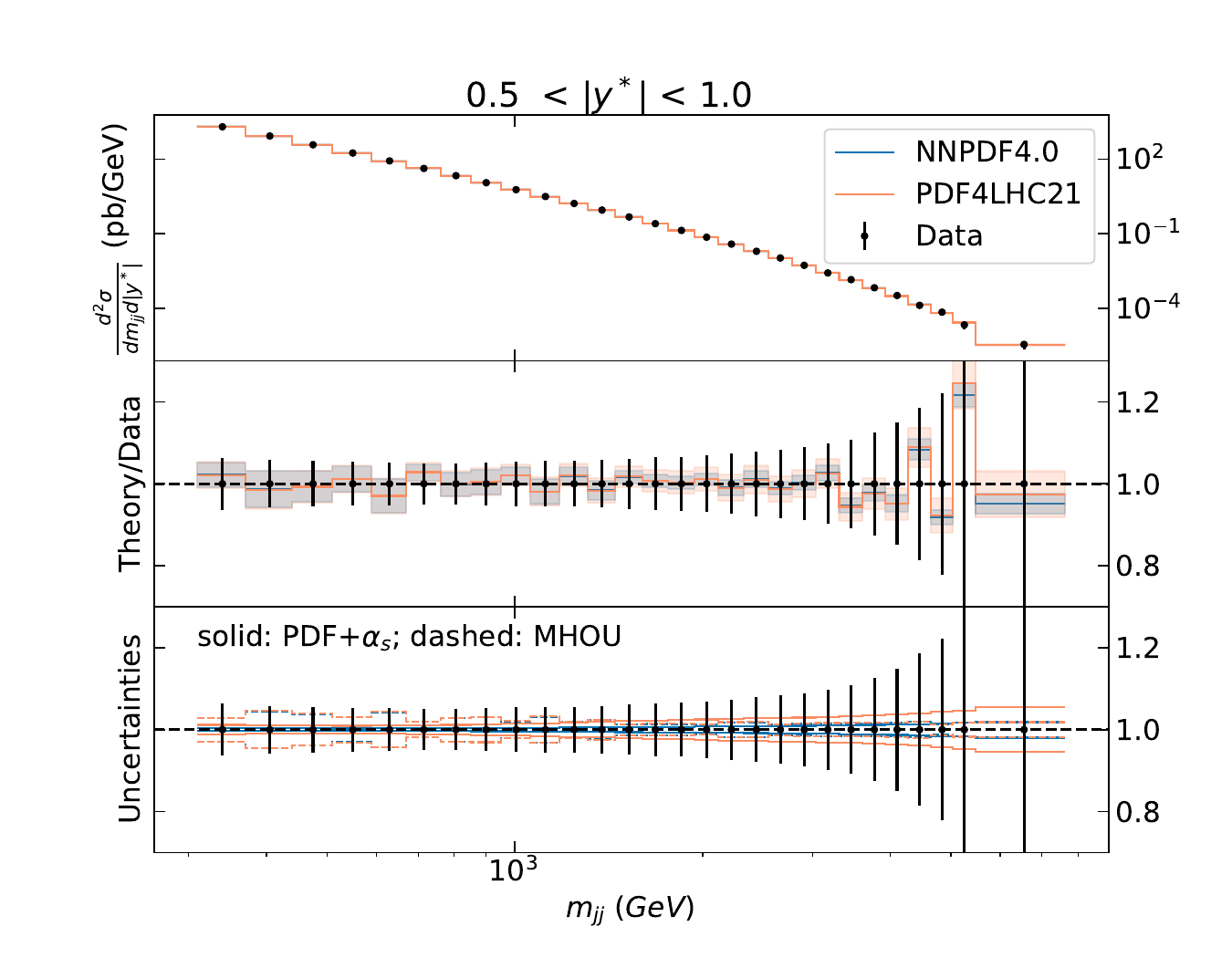}
    \includegraphics[width=0.49\textwidth]{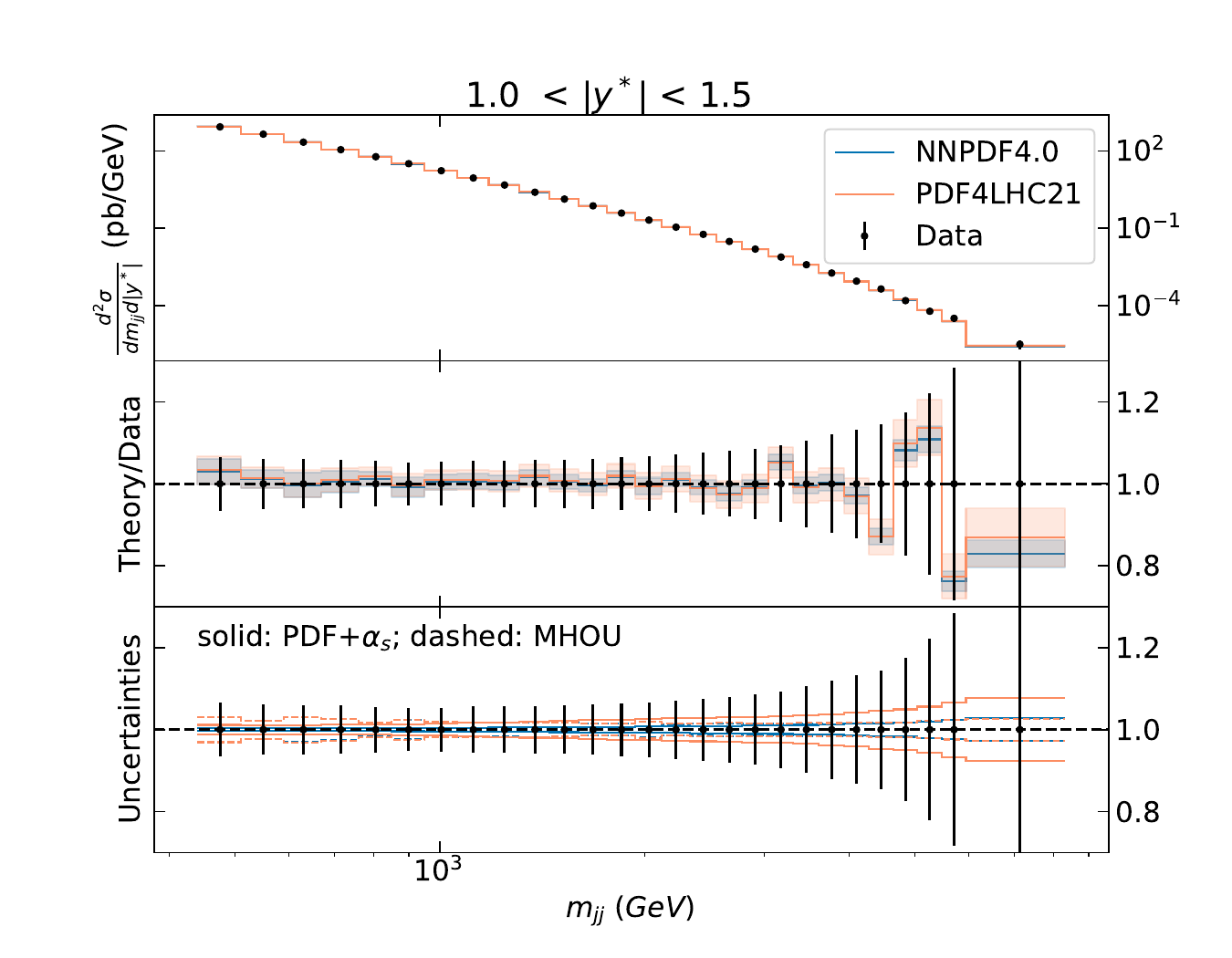}\\
    \includegraphics[width=0.49\textwidth]{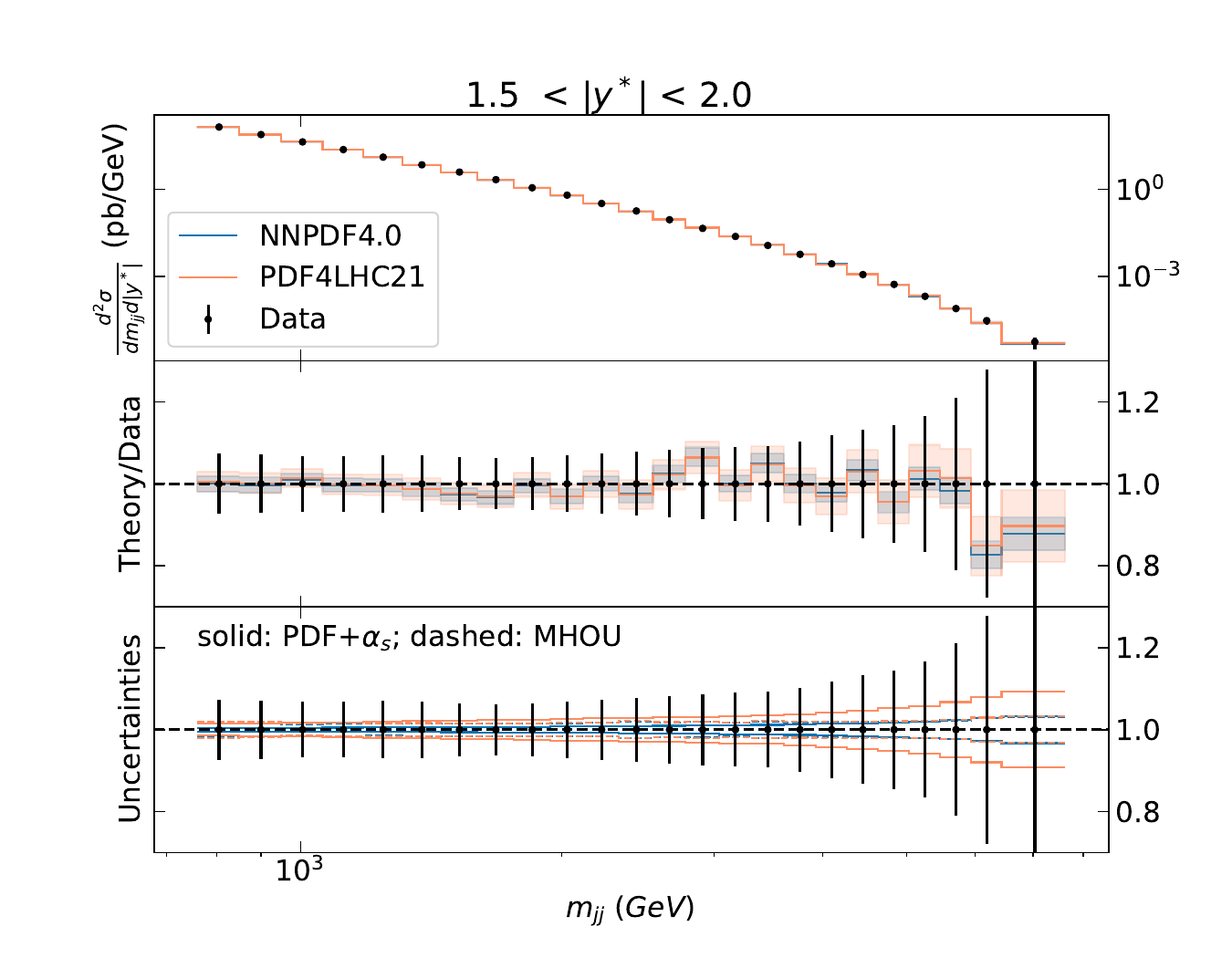}
    \includegraphics[width=0.49\textwidth]{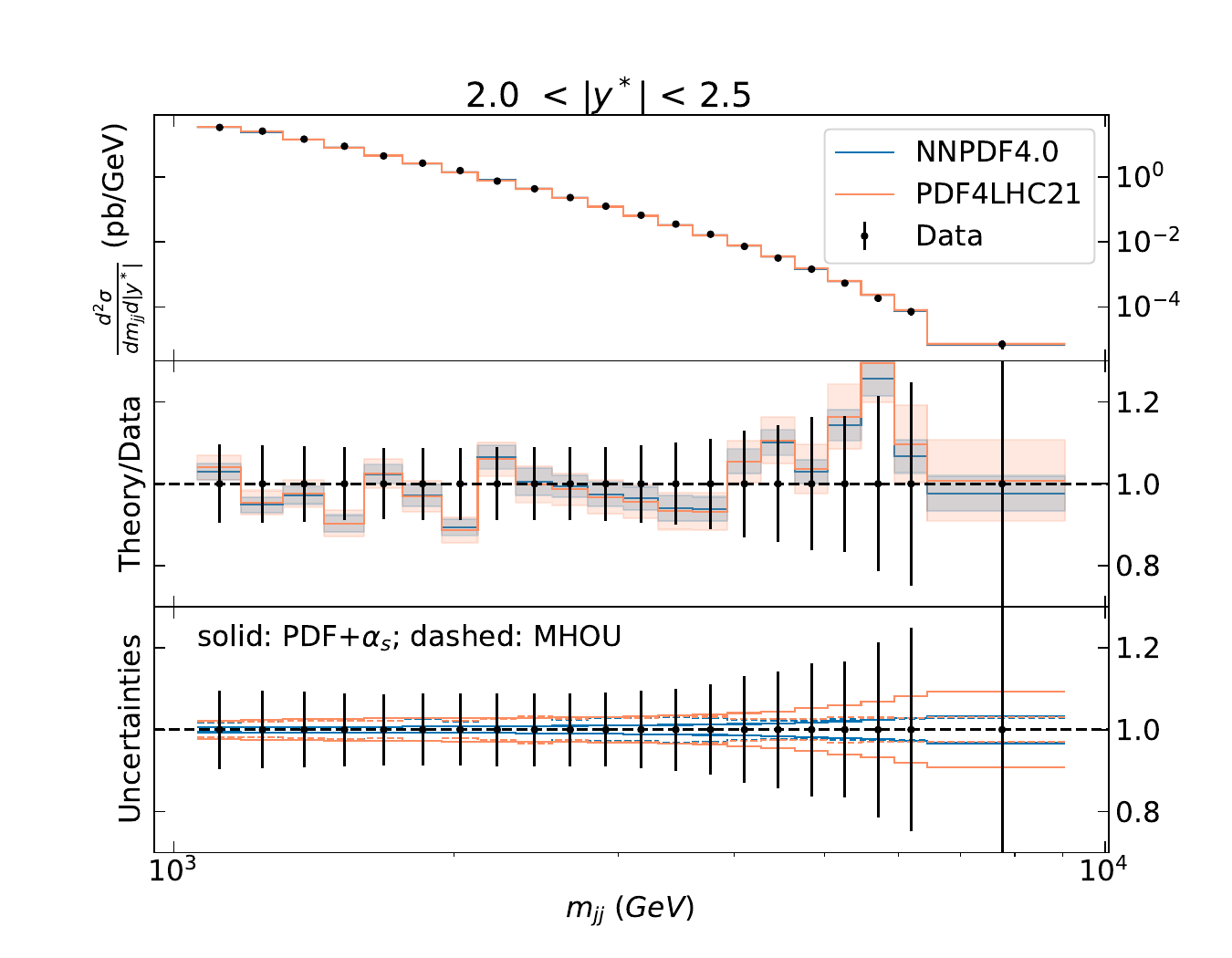}
    \caption{Same as Fig.~\ref{fig:dijet-ATLAS-13tev} for the intermediate
      rapidity bins.}
    \label{fig:datatheory_jets_additional_3} 
  \end{figure}

\item[Single-inclusive jet and di-jet production at HERA.]
  Figure~\ref{fig:chi2histo_disjets_additional} displays the breakdown of
  $\chi^2_{\rm exp+th}$ into its $\chi^2_{\rm exp+mho}$ and $\chi^2_{\rm exp}$
  components for the ZEUS single-inclusive jet and di-jet measurements
  outlined in Table~\ref{tab:Z}.
  Figures~\ref{fig:datatheory_DISjets_additional_1}-\ref{fig:datatheory_DISjets_additional_4} display the data-theory comparison for the H1 $Q^2$ bins not
  displayed in Fig.~\ref{fig:H1_datatheory}, respectively, for the low-$Q^2$
  single-inclusive jet and di-jet measurements, and for the high-$Q^2$
  single-inclusive jet and di-jet measurements.
  Figures~\ref{fig:datatheory_DISjets_additional_5}-\ref{fig:datatheory_DISjets_additional_7} display the data-theory comparison, respectively, for the
  ZEUS low-luminosity single-inclusive jet, high-luminosity single-inclusive
  jet, and di-jet measurements.

  \begin{figure}[!t]
    \centering
    \includegraphics[width=0.49\textwidth]{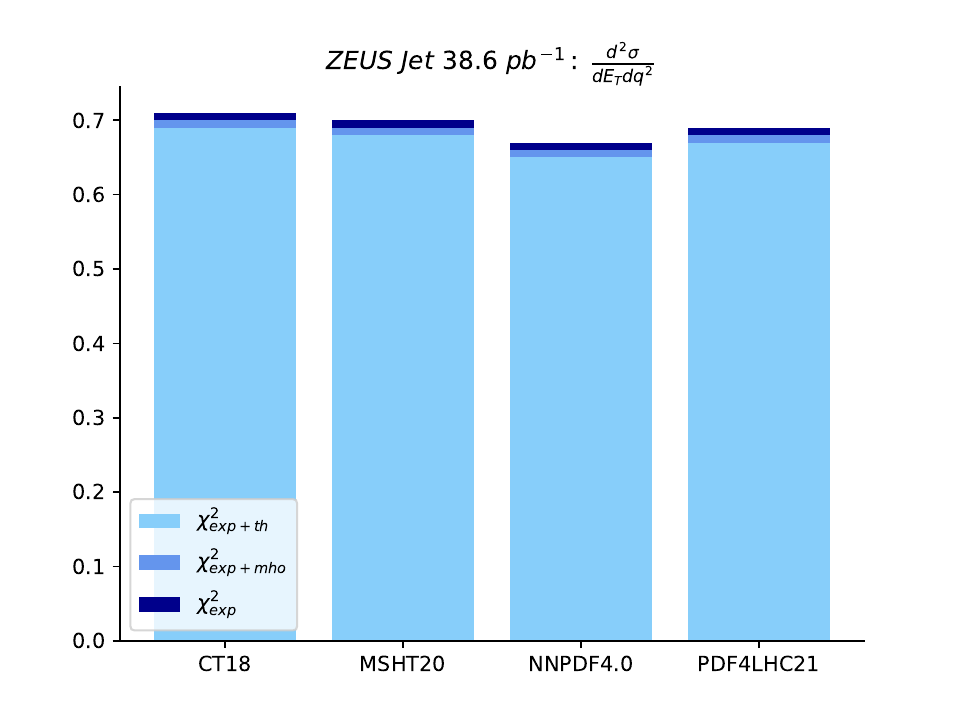}
    \includegraphics[width=0.49\textwidth]{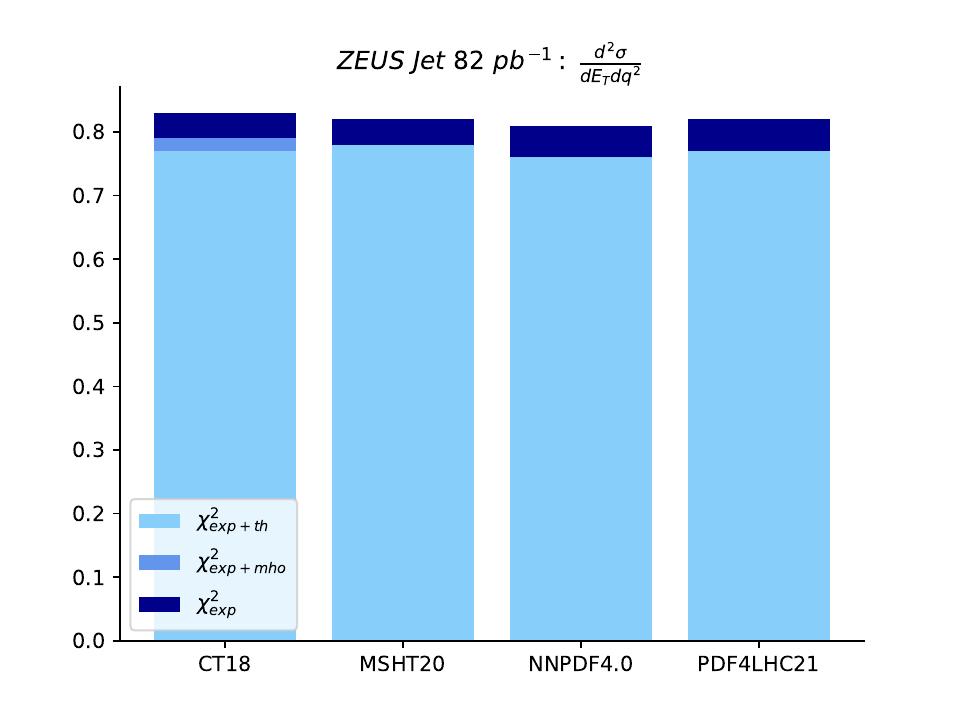}
    \includegraphics[width=0.49\textwidth]{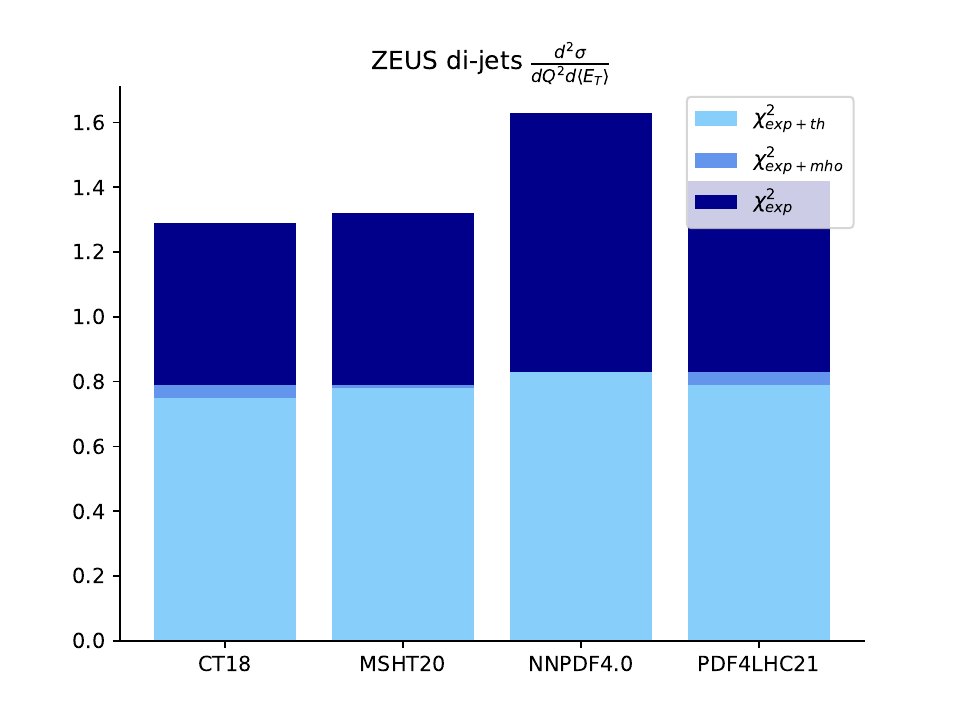}
    \caption{Same as Fig.~\ref{fig:DY-chi2} for the ZEUS single-inclusive jet dataset 
    comprising $n_{\rm dat} = 30$ data points (and $\sqrt{2/n_{\rm dat}} = 0.26$)
      (top) and di-jet (bottom) dataset comprising $n_{\rm dat} = 22$ data points (and $\sqrt{2/n_{\rm dat}} = 0.30$).}
    \label{fig:chi2histo_disjets_additional}
  \end{figure}

  \begin{figure}[!t]
    \centering
    \includegraphics[width=0.49\textwidth]{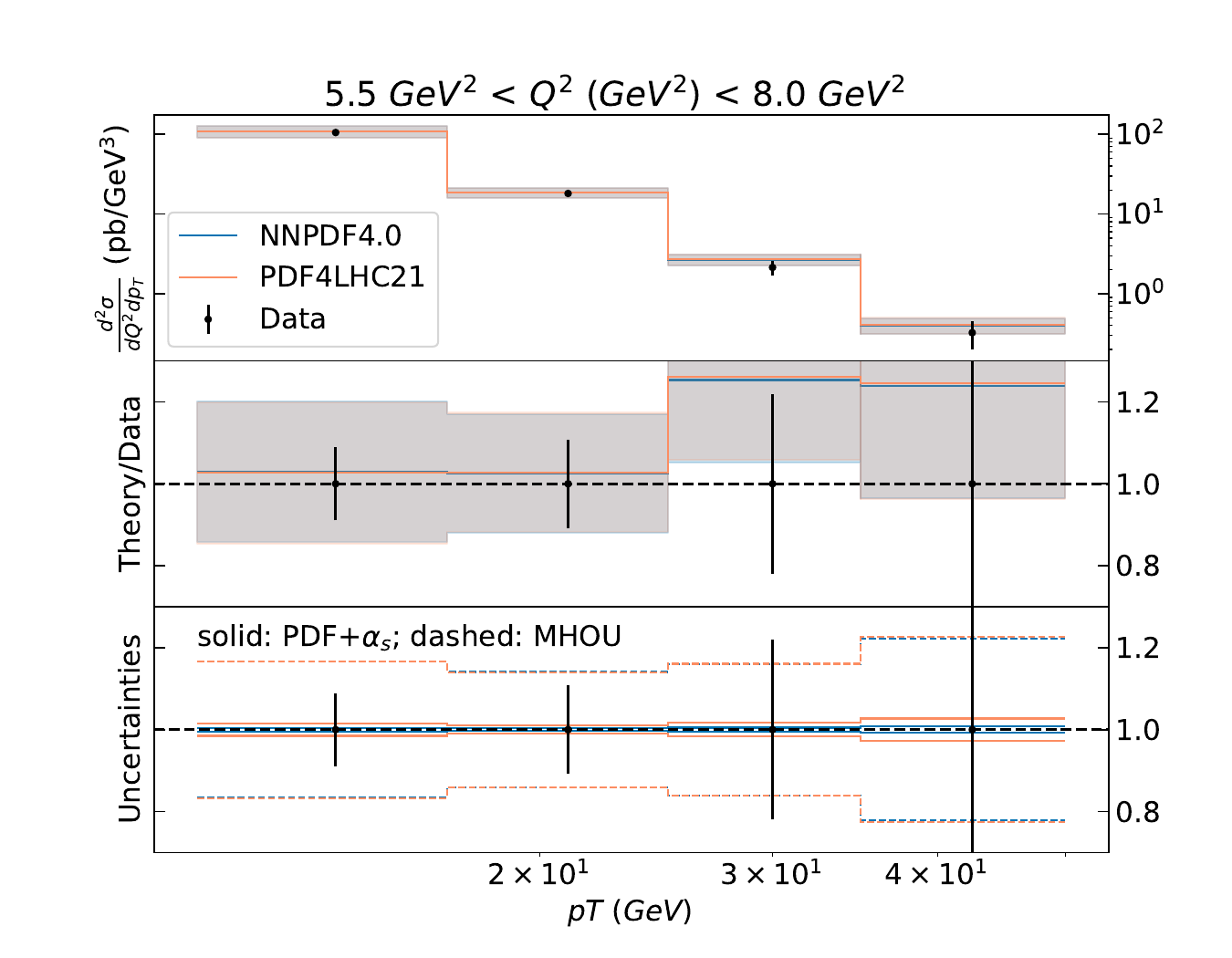}
    \includegraphics[width=0.49\textwidth]{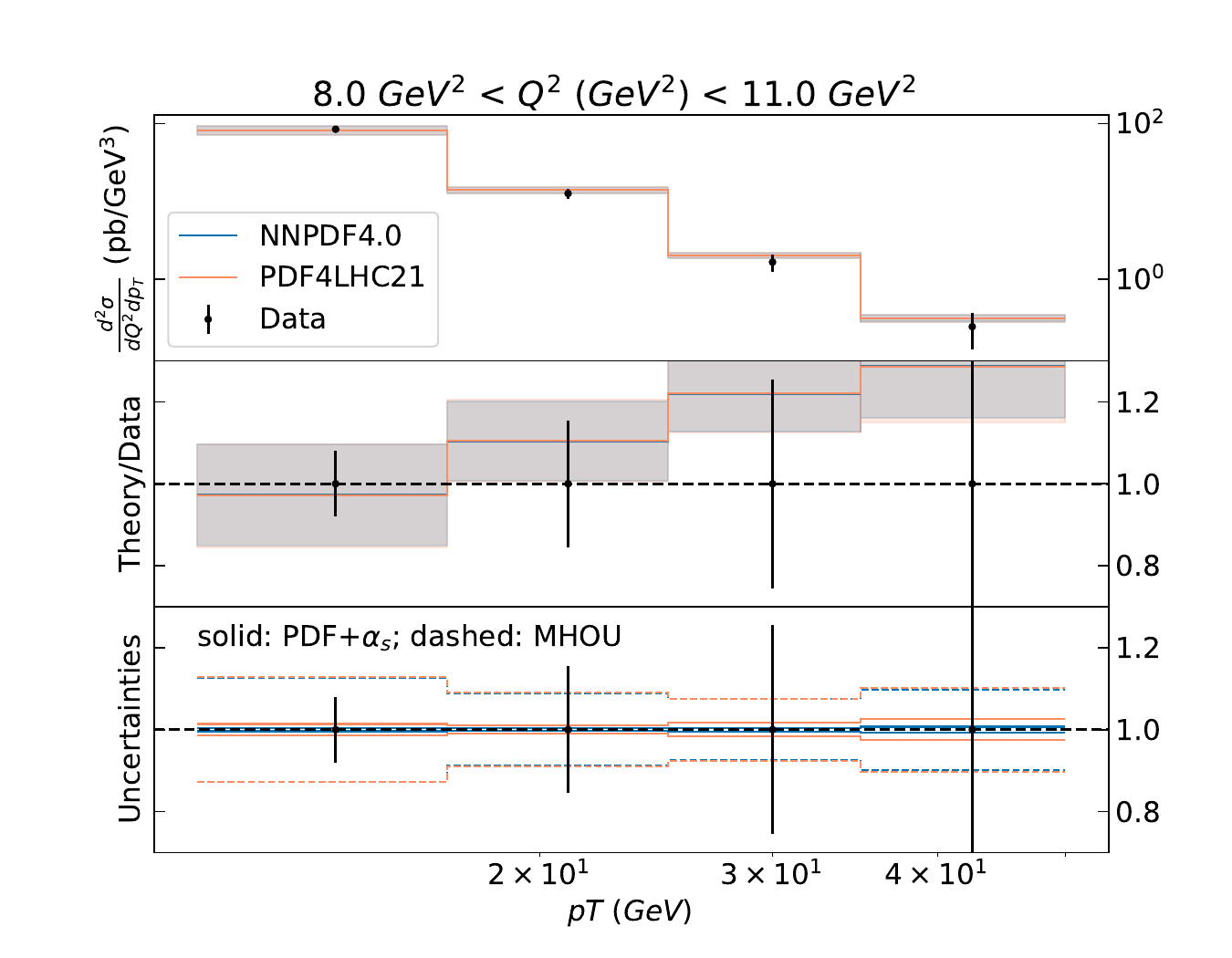}\\
    \includegraphics[width=0.49\textwidth]{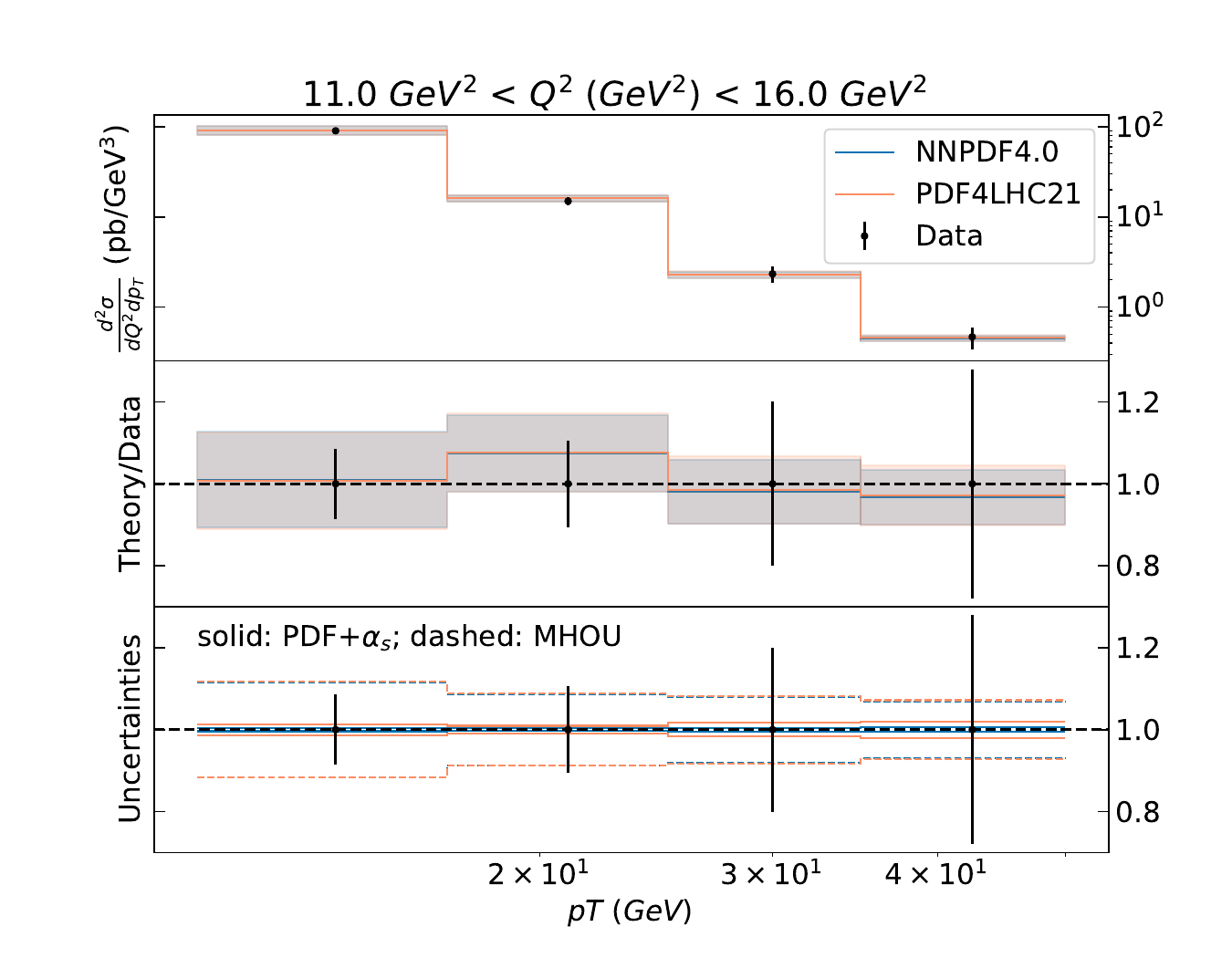} 
    \includegraphics[width=0.49\textwidth]{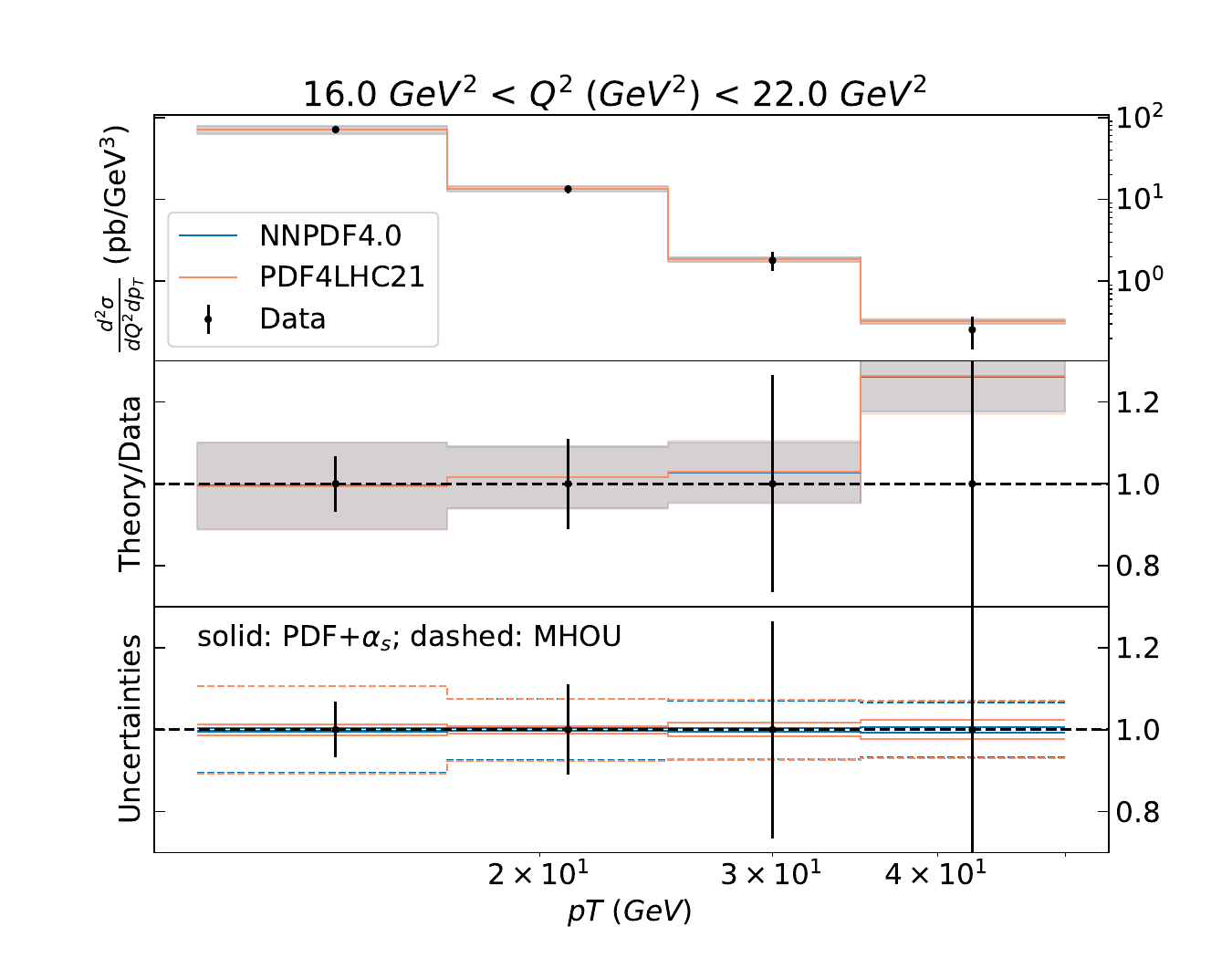}\\
    \includegraphics[width=0.49\textwidth]{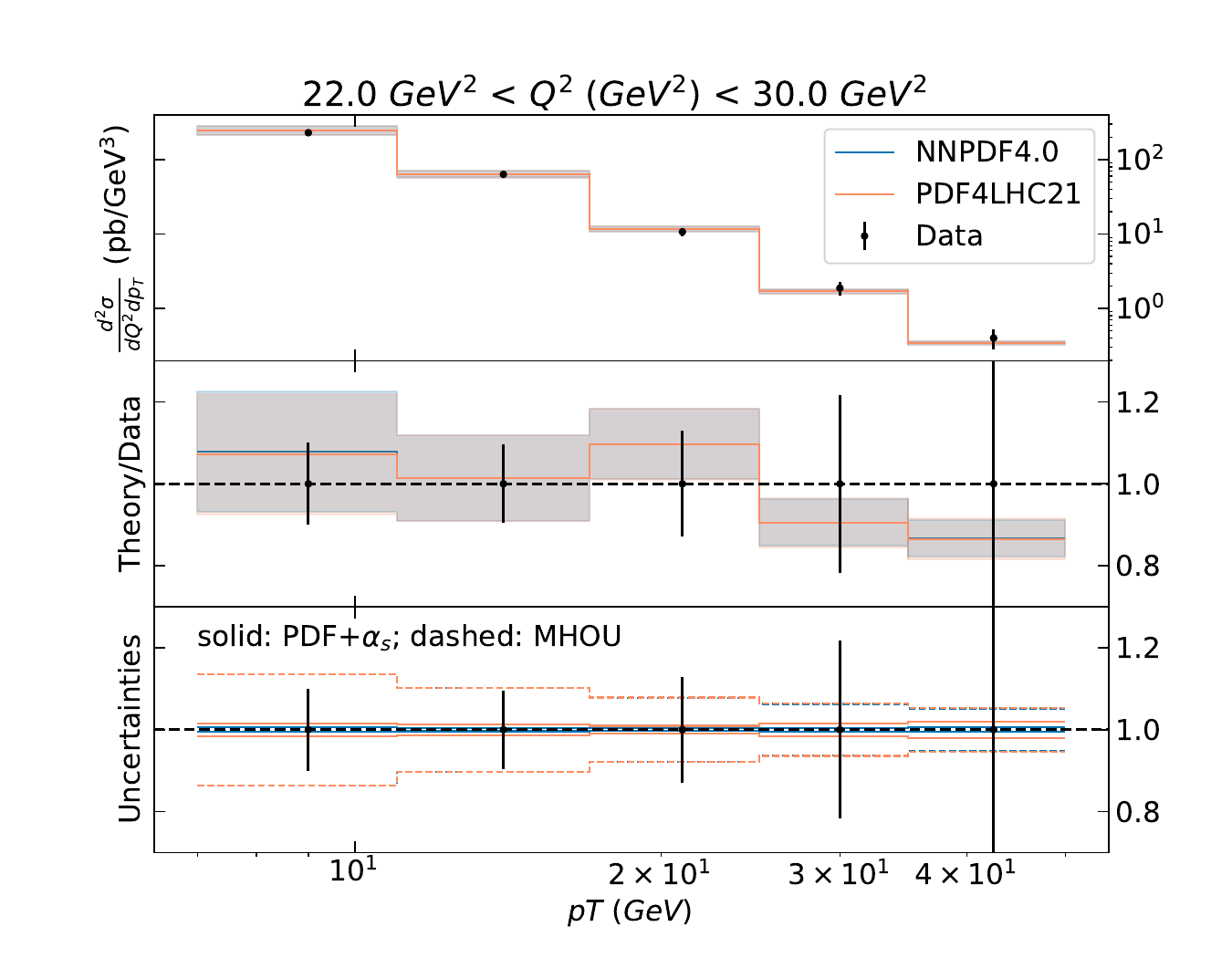}
    \includegraphics[width=0.49\textwidth]{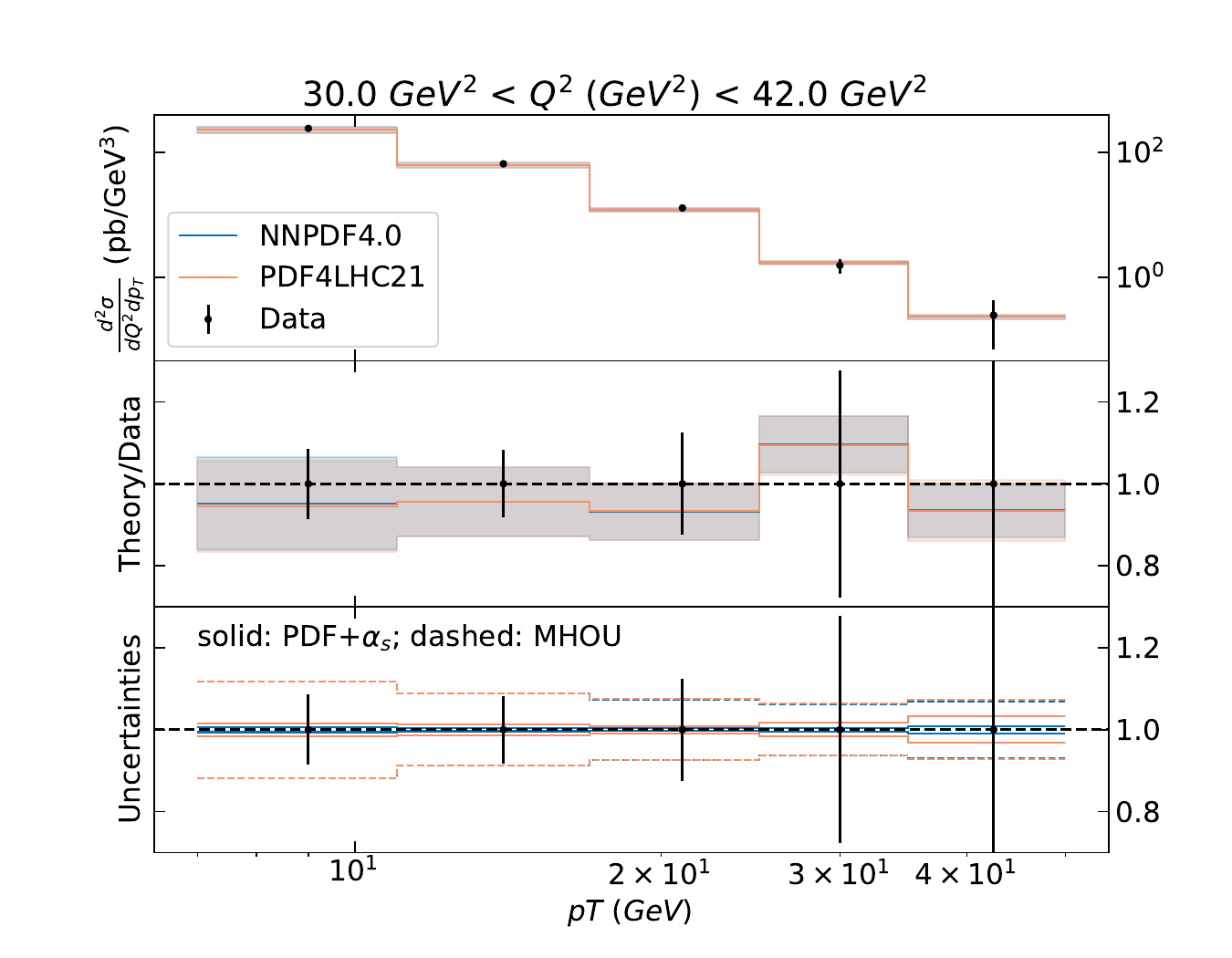}\\
    \caption{Same as Fig.~\ref{fig:H1_datatheory} for the bins of the H1
      low-$Q^2$ single-inclusive jet measurement not displayed in
      Fig.~\ref{fig:H1_datatheory}.}
    \label{fig:datatheory_DISjets_additional_1}
  \end{figure}

  \begin{figure}[!t]
    \centering
    \includegraphics[width=0.49\textwidth]{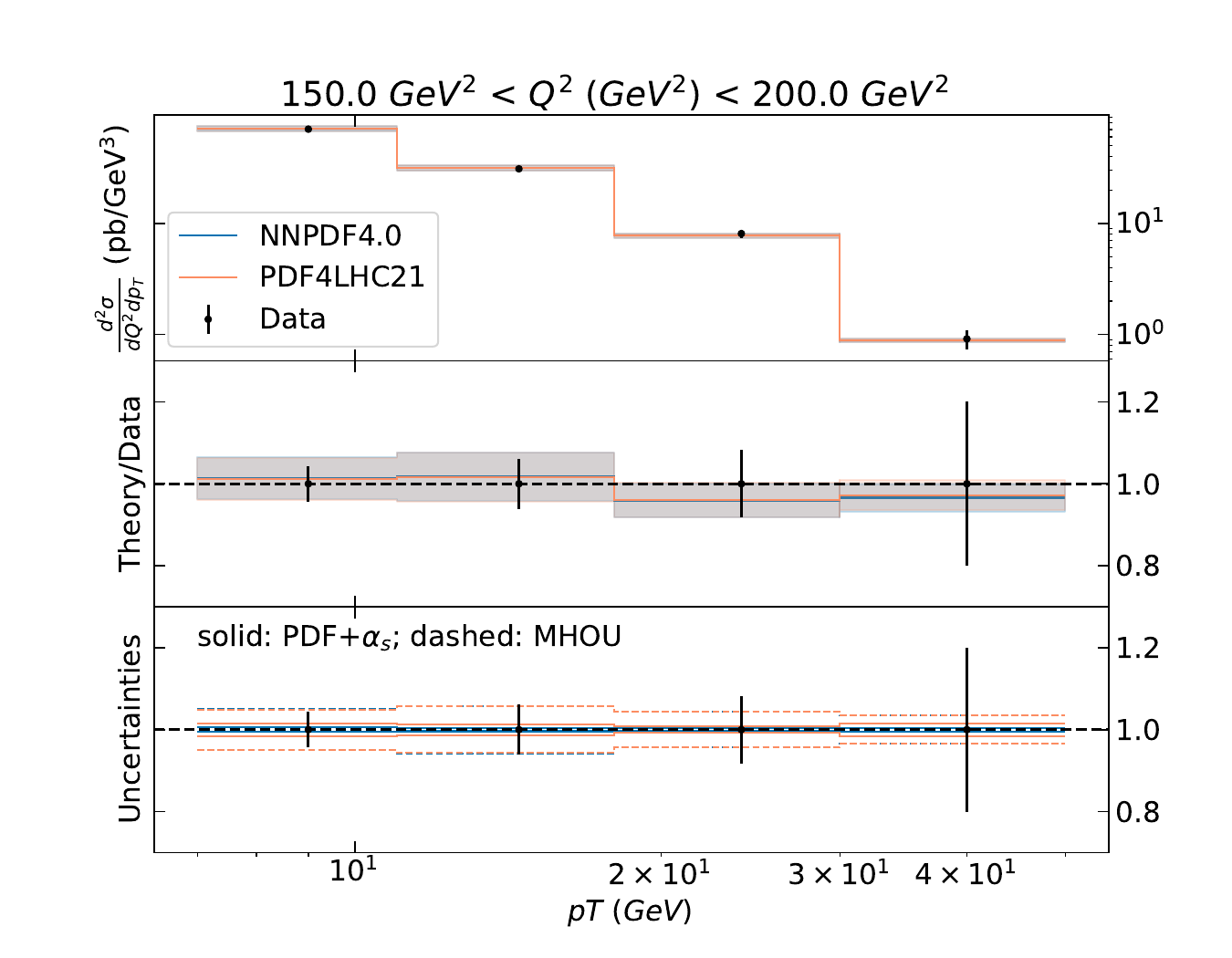}
    \includegraphics[width=0.49\textwidth]{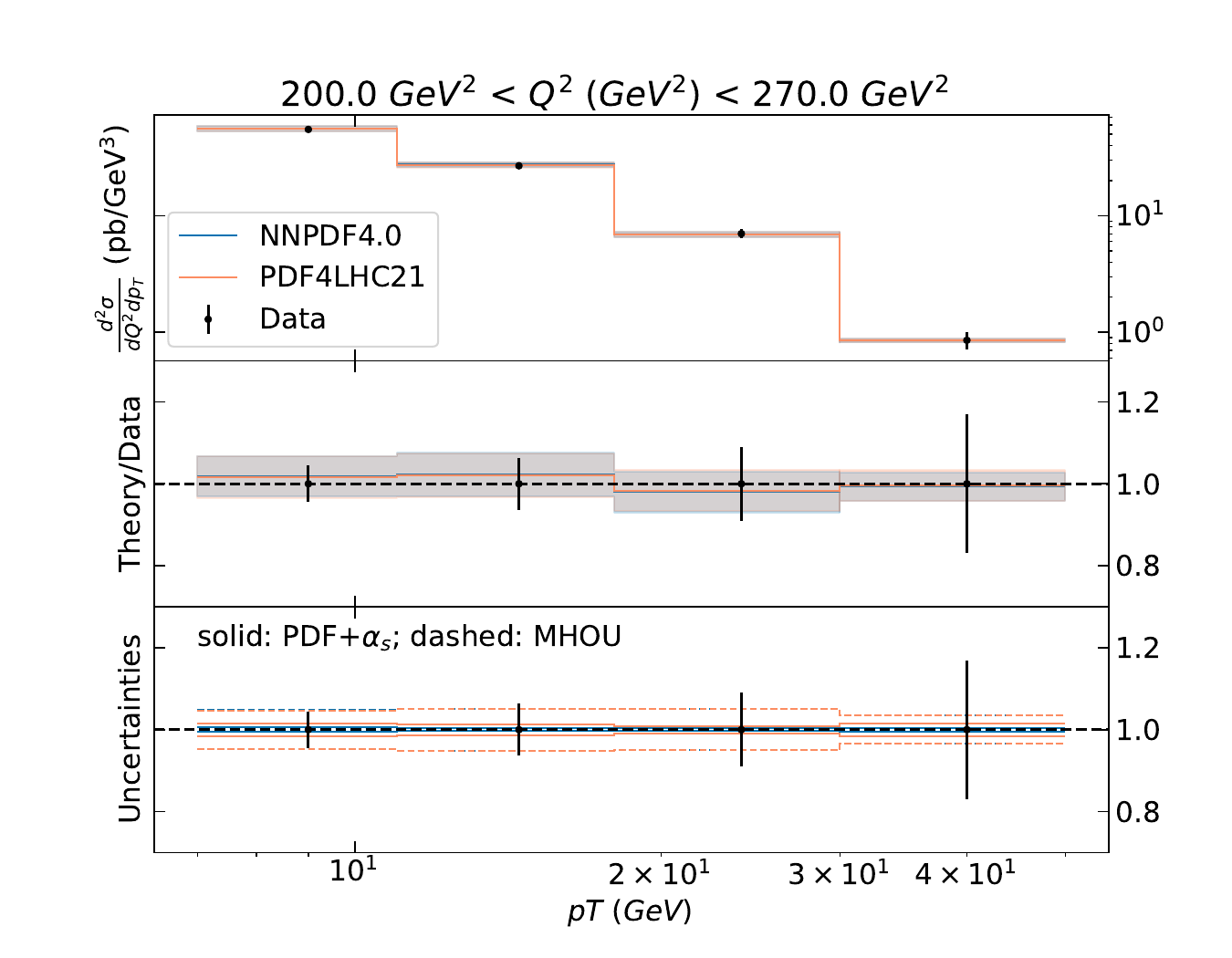}\\
    \includegraphics[width=0.49\textwidth]{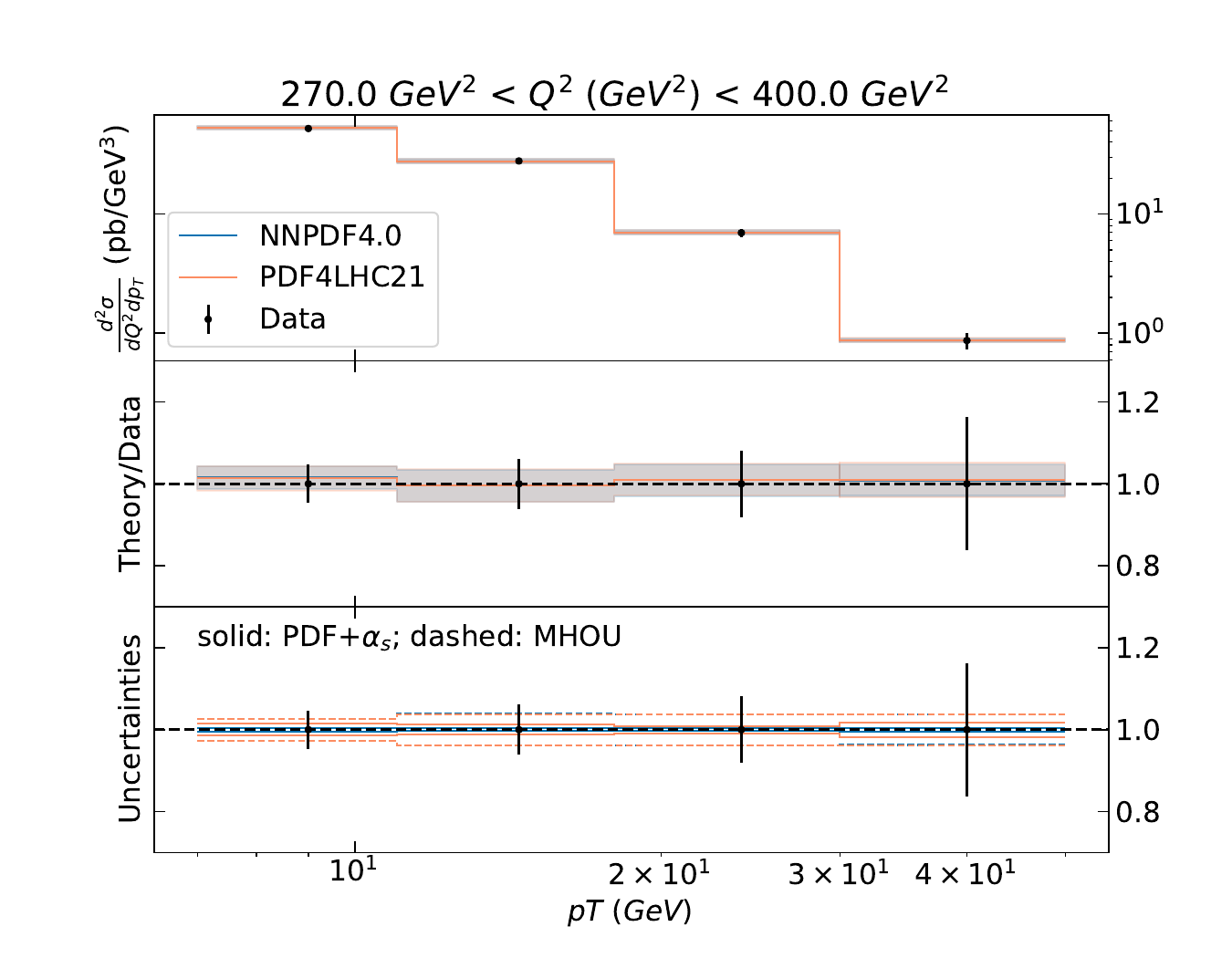} 
    \includegraphics[width=0.49\textwidth]{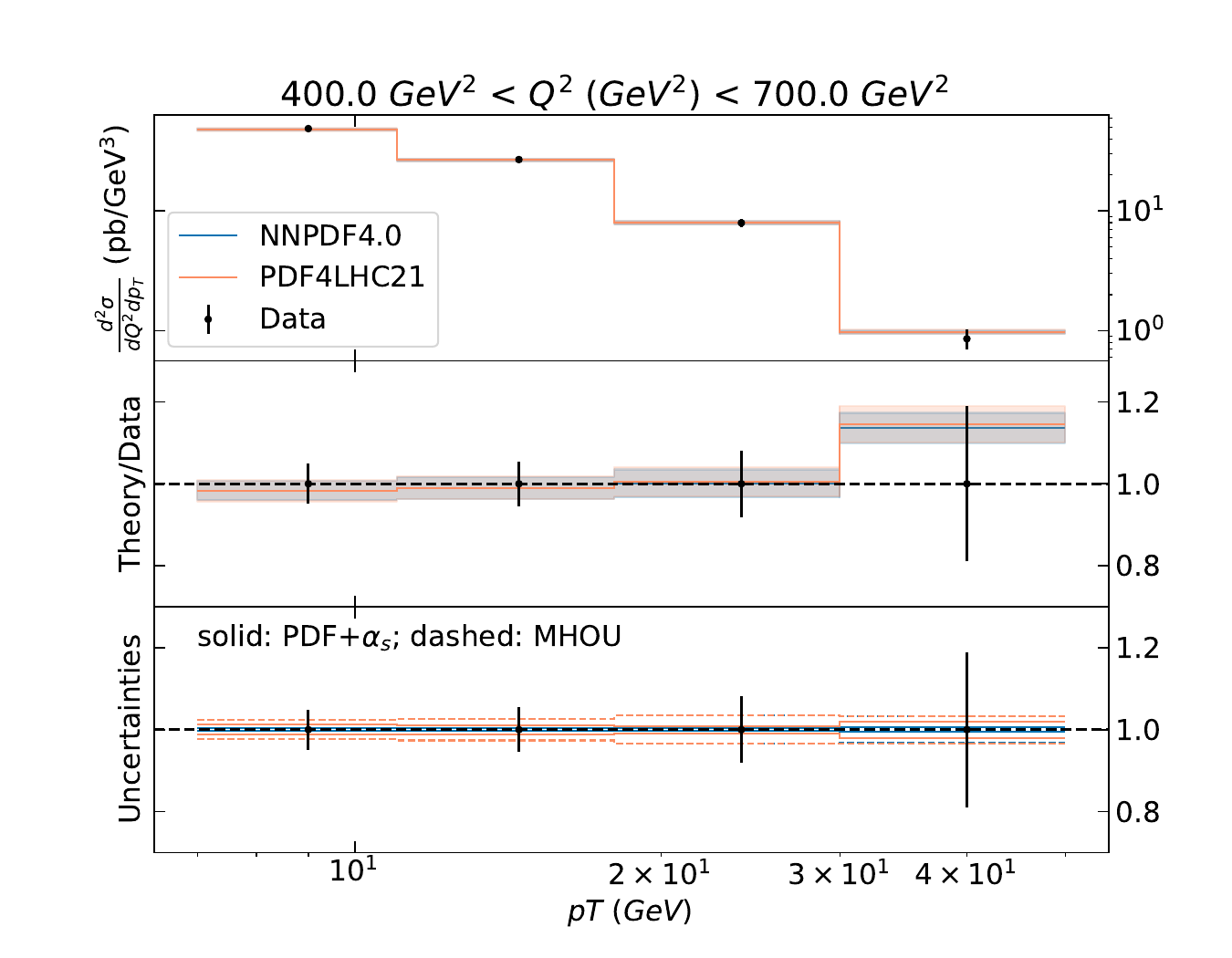}\\
    \includegraphics[width=0.49\textwidth]{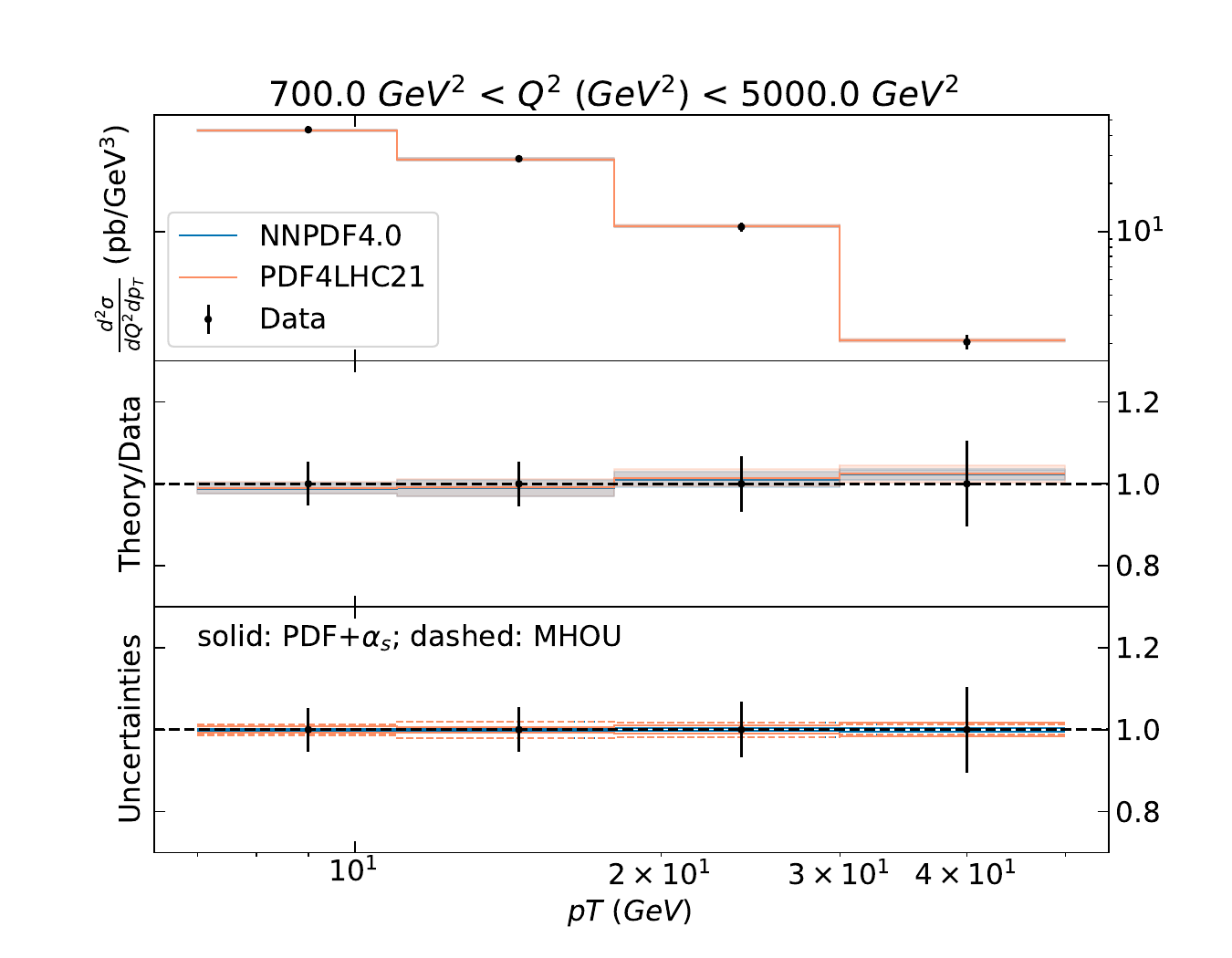}
    \caption{Same as Fig.~\ref{fig:H1_datatheory} for the bins of the H1
      high-$Q^2$ single-inclusive jet measurement not displayed in
      Fig.~\ref{fig:H1_datatheory}.}
    \label{fig:datatheory_DISjets_additional_2}
  \end{figure}

  \begin{figure}[!t]
    \centering
    \includegraphics[width=0.49\textwidth]{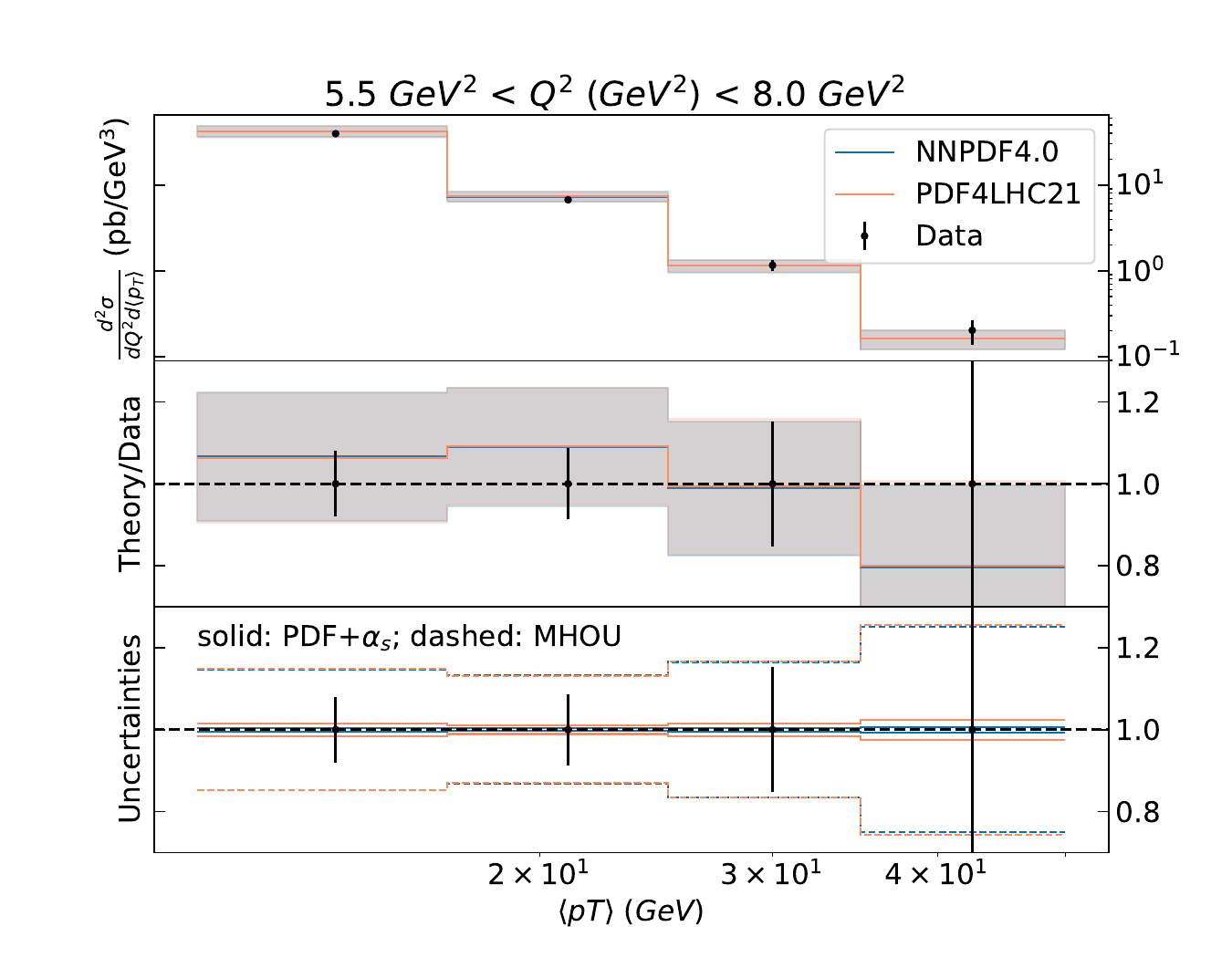}
    \includegraphics[width=0.49\textwidth]{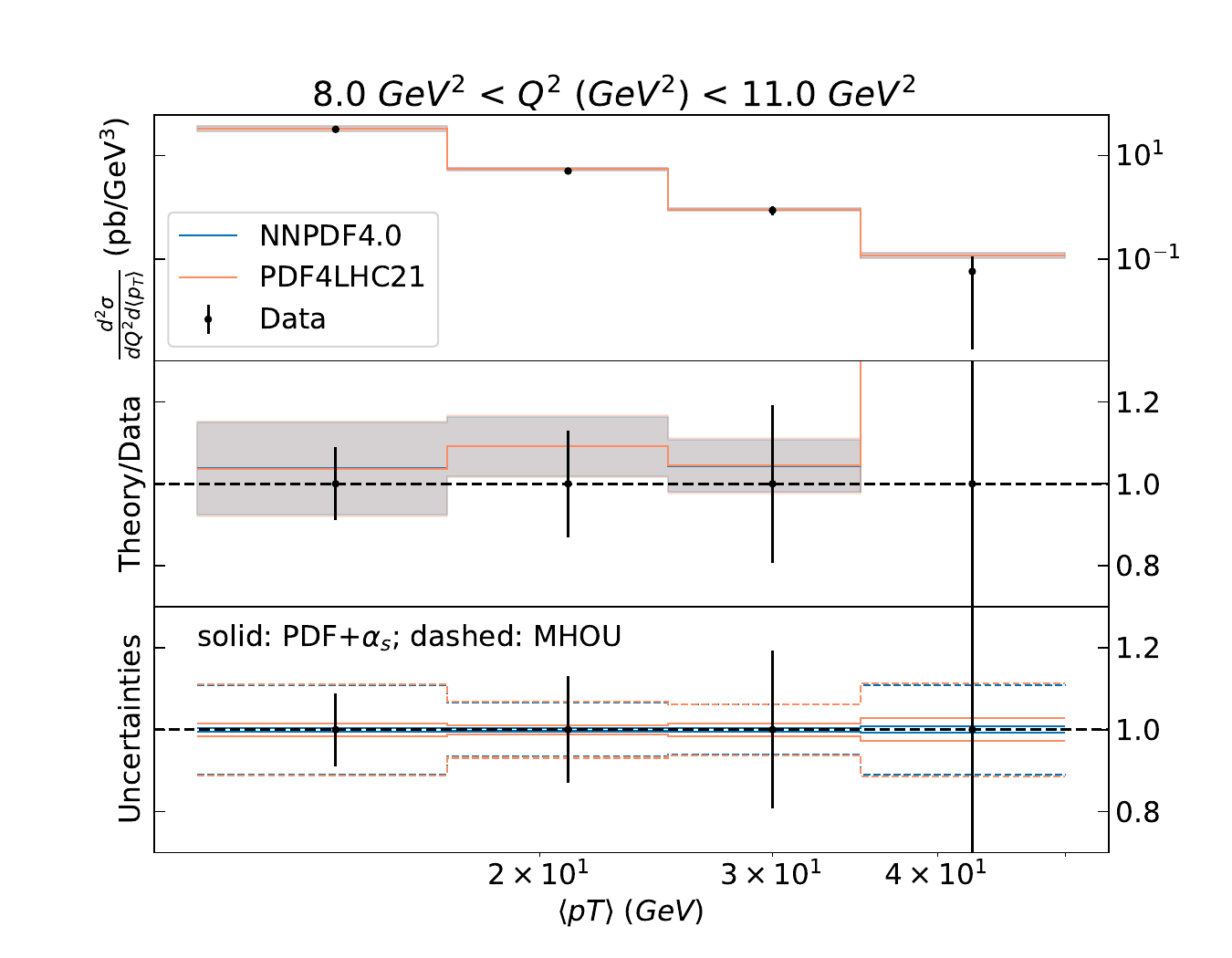}\\
    \includegraphics[width=0.49\textwidth]{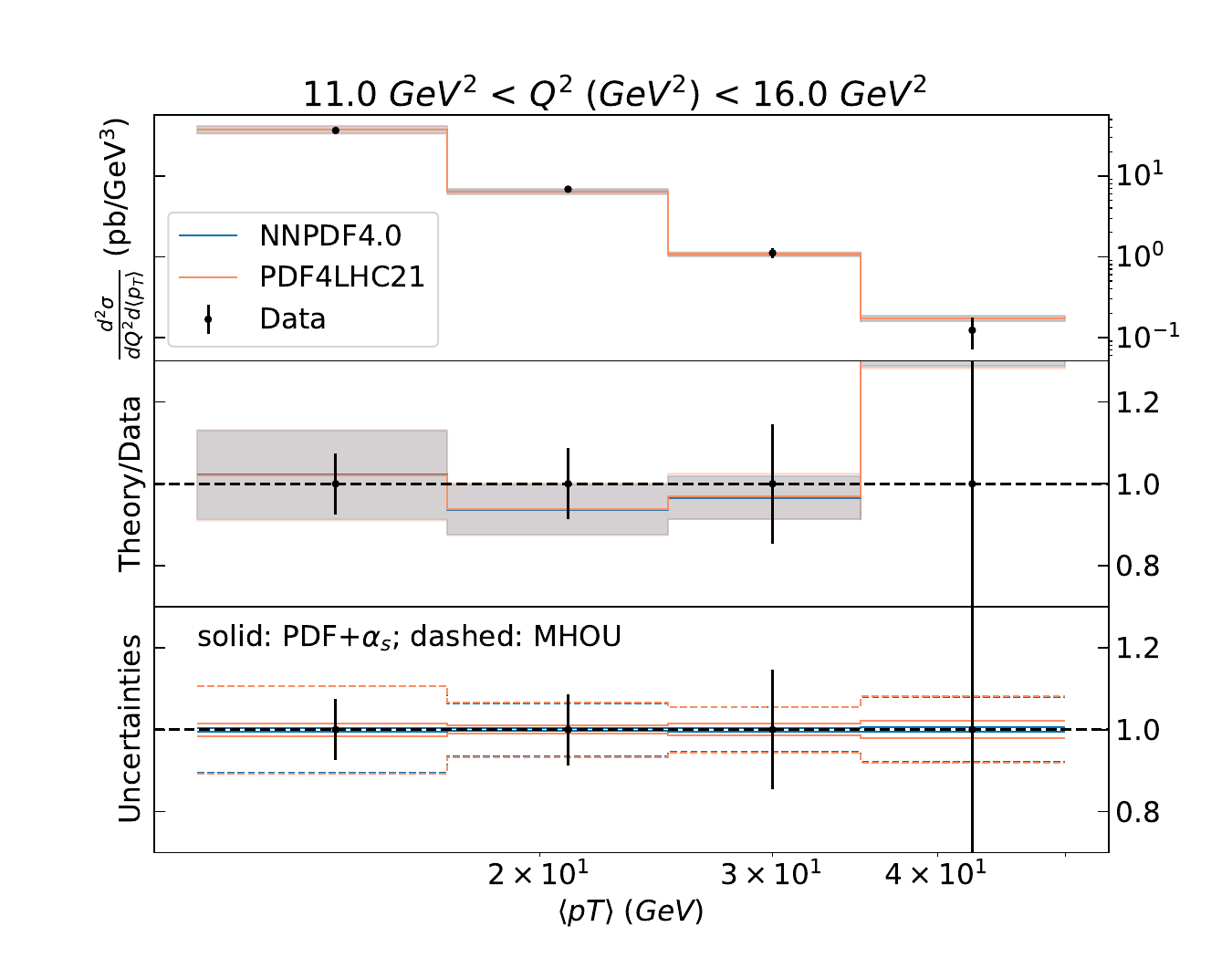} 
    \includegraphics[width=0.49\textwidth]{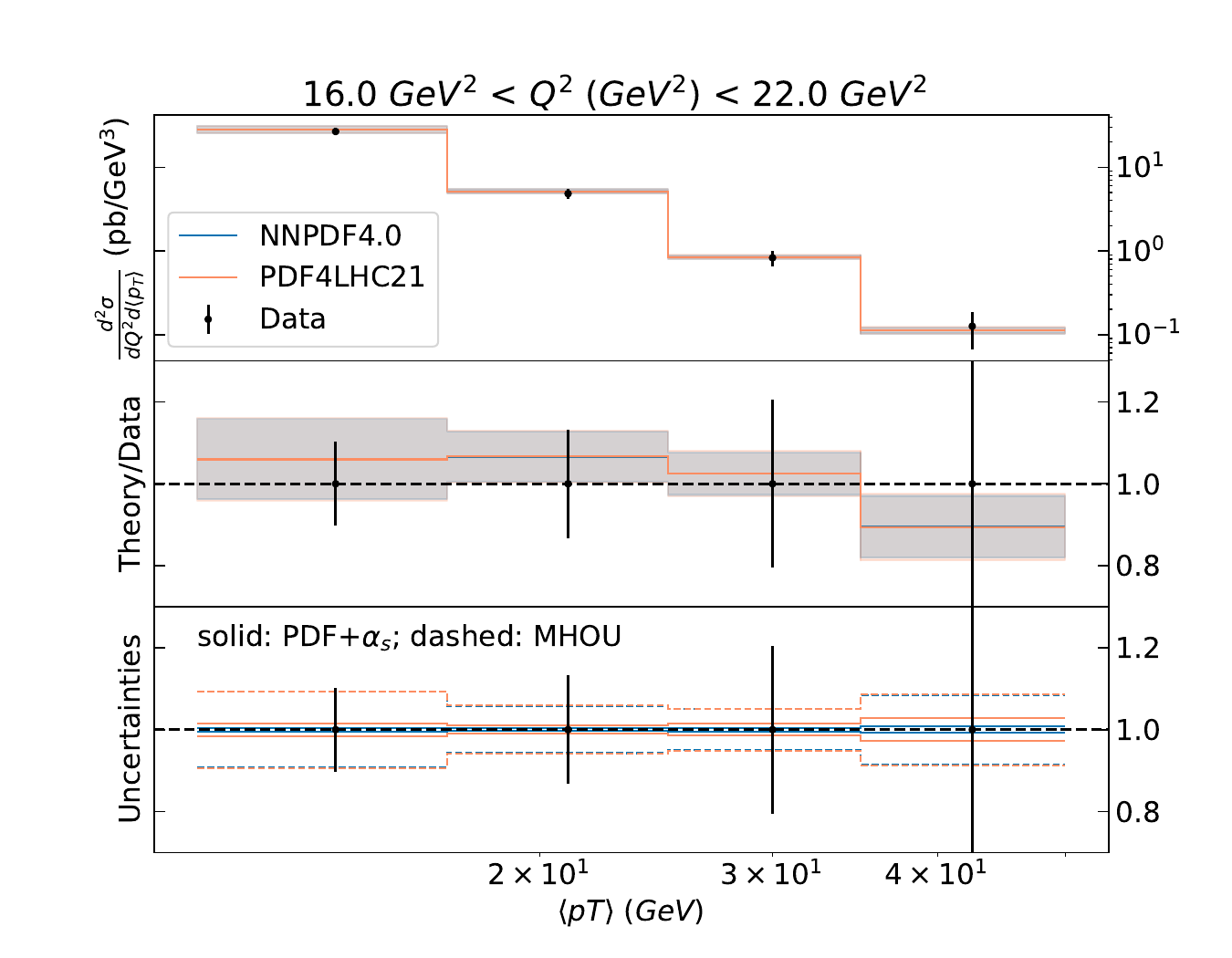}\\
    \includegraphics[width=0.49\textwidth]{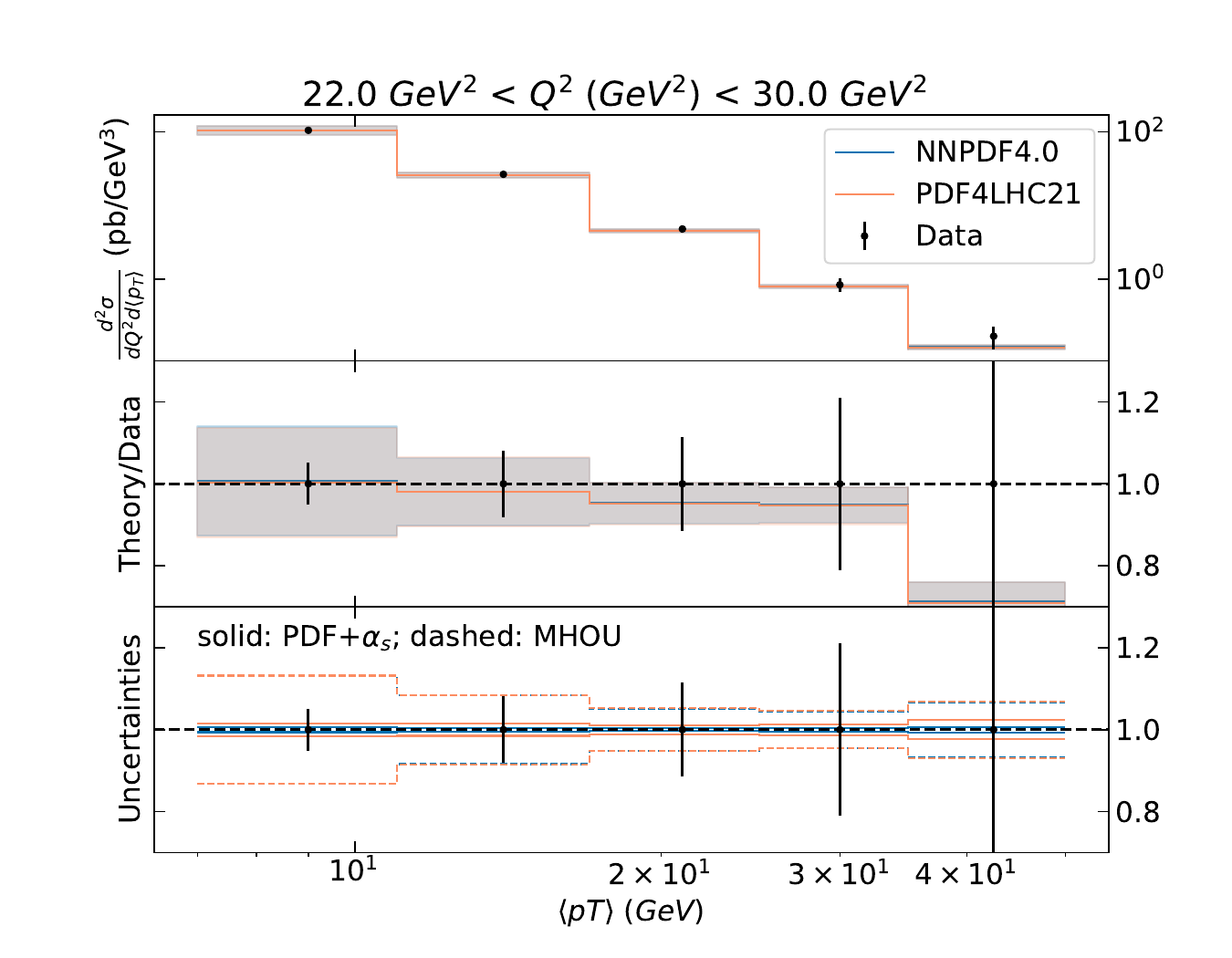}
    \includegraphics[width=0.49\textwidth]{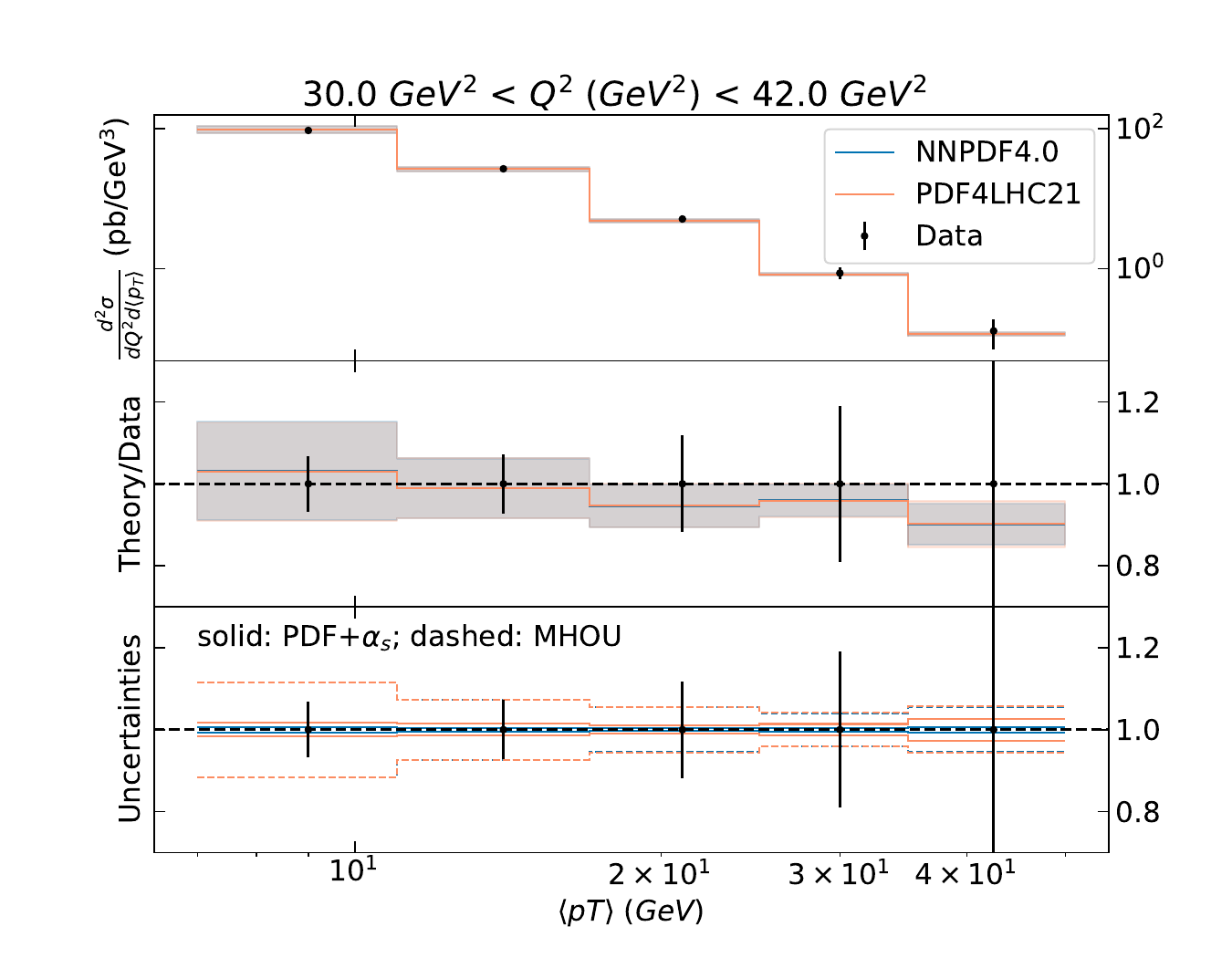}\\
    \caption{Same as Fig.~\ref{fig:H1_datatheory} for the bins of the H1
      low-$Q^2$ di-jet measurement not displayed in
      Fig.~\ref{fig:H1_datatheory}.}
    \label{fig:datatheory_DISjets_additional_3}
  \end{figure}
  
  \begin{figure}[!t]
    \centering
    \includegraphics[width=0.49\textwidth]{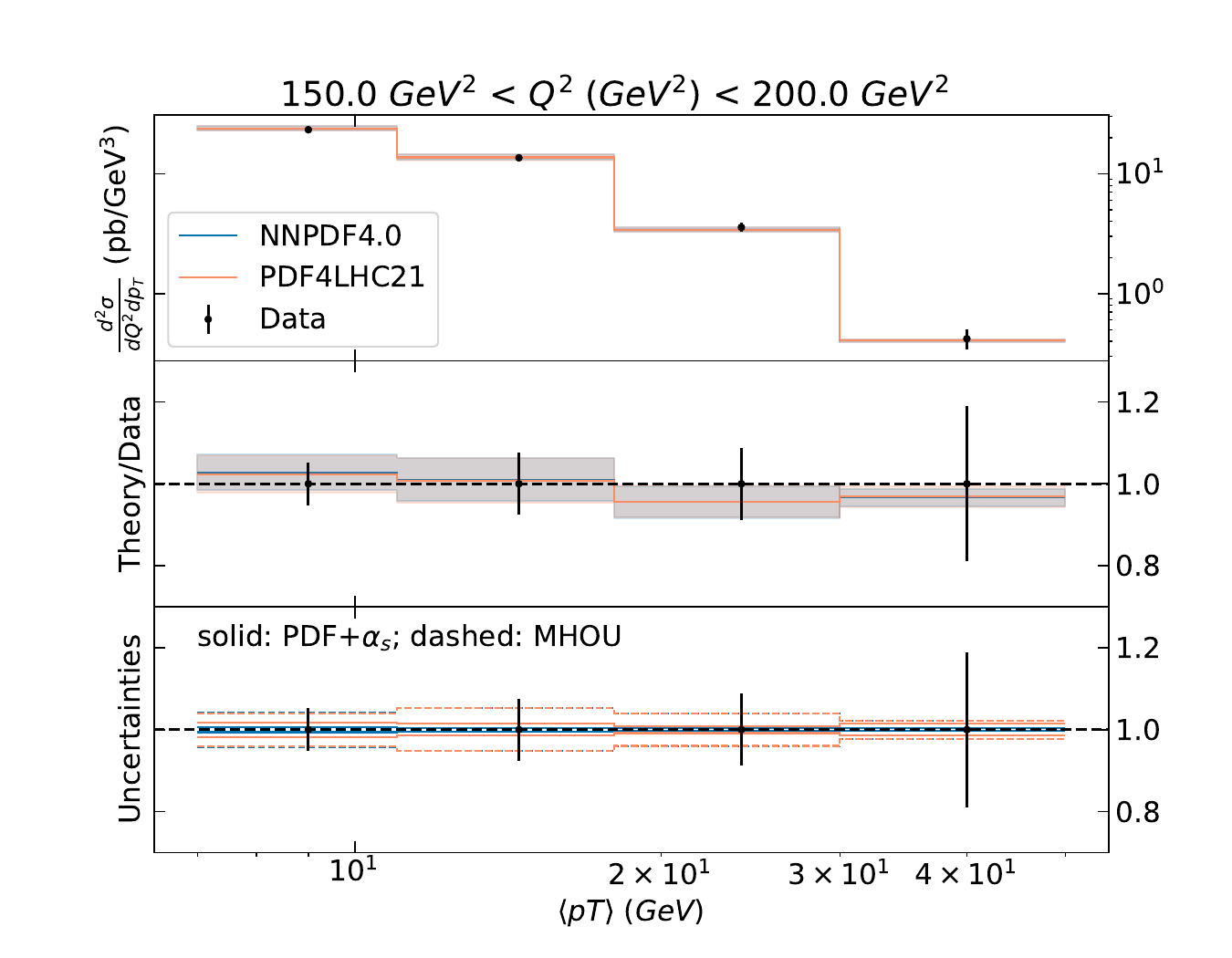}
    \includegraphics[width=0.49\textwidth]{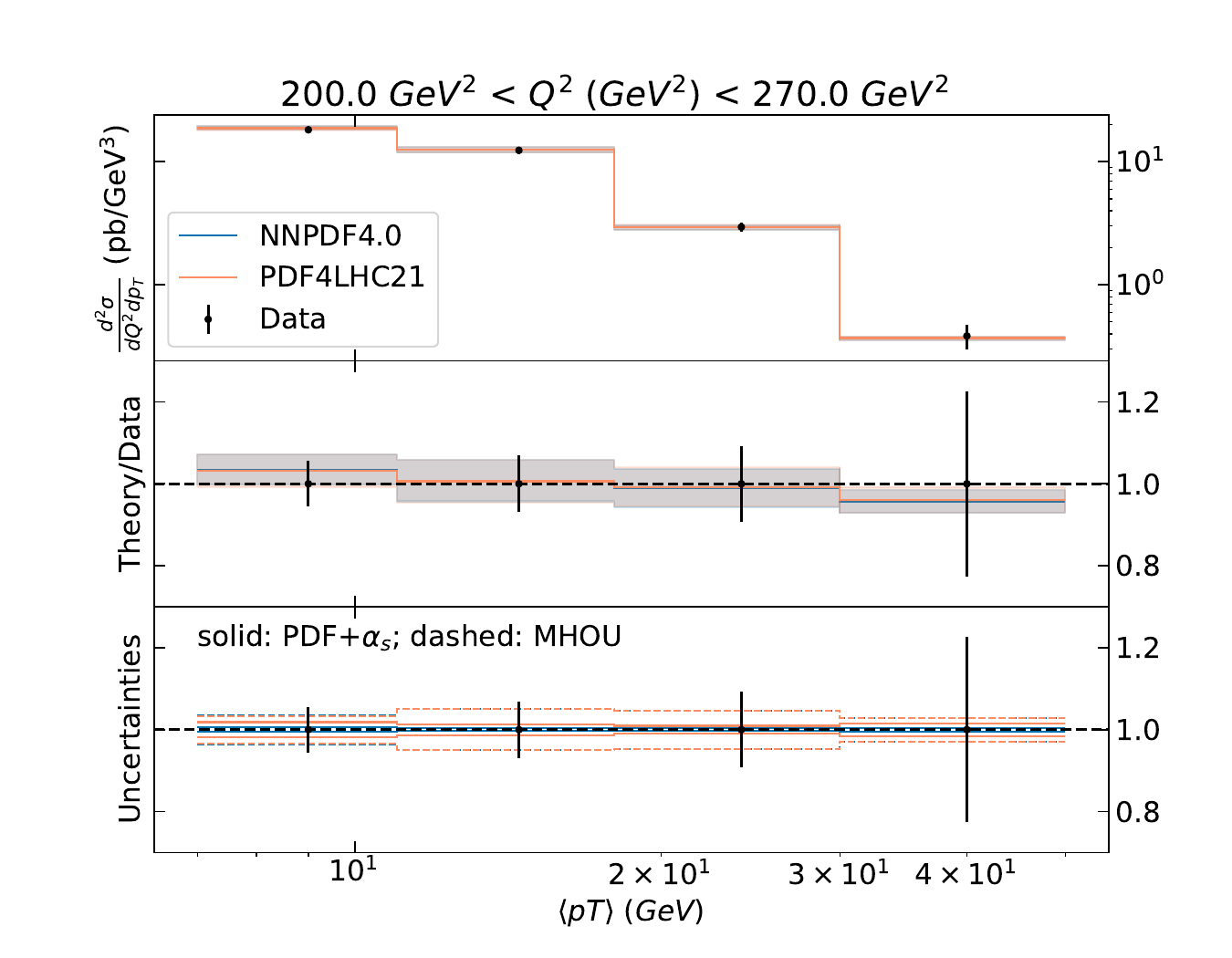}\\
    \includegraphics[width=0.49\textwidth]{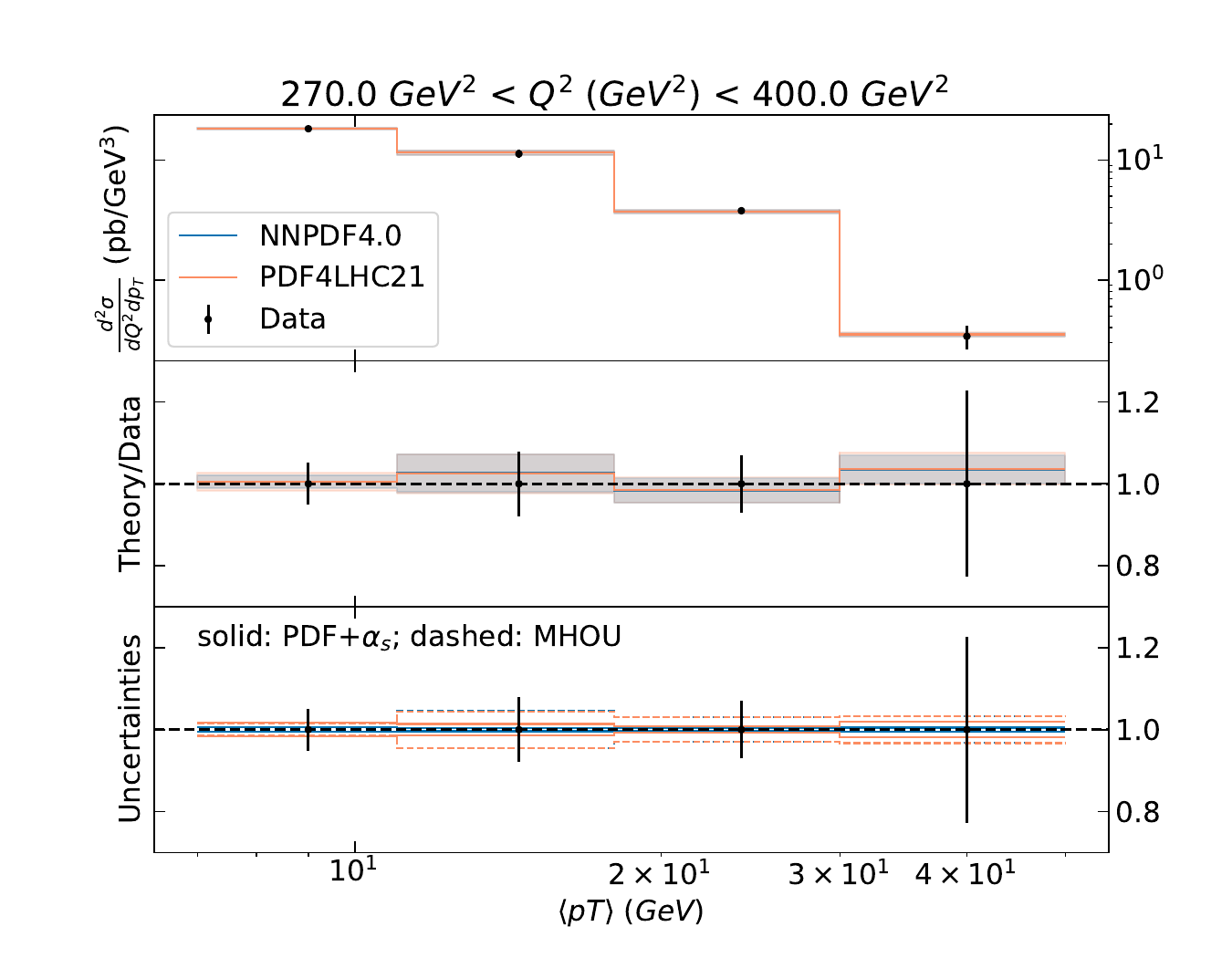} 
    \includegraphics[width=0.49\textwidth]{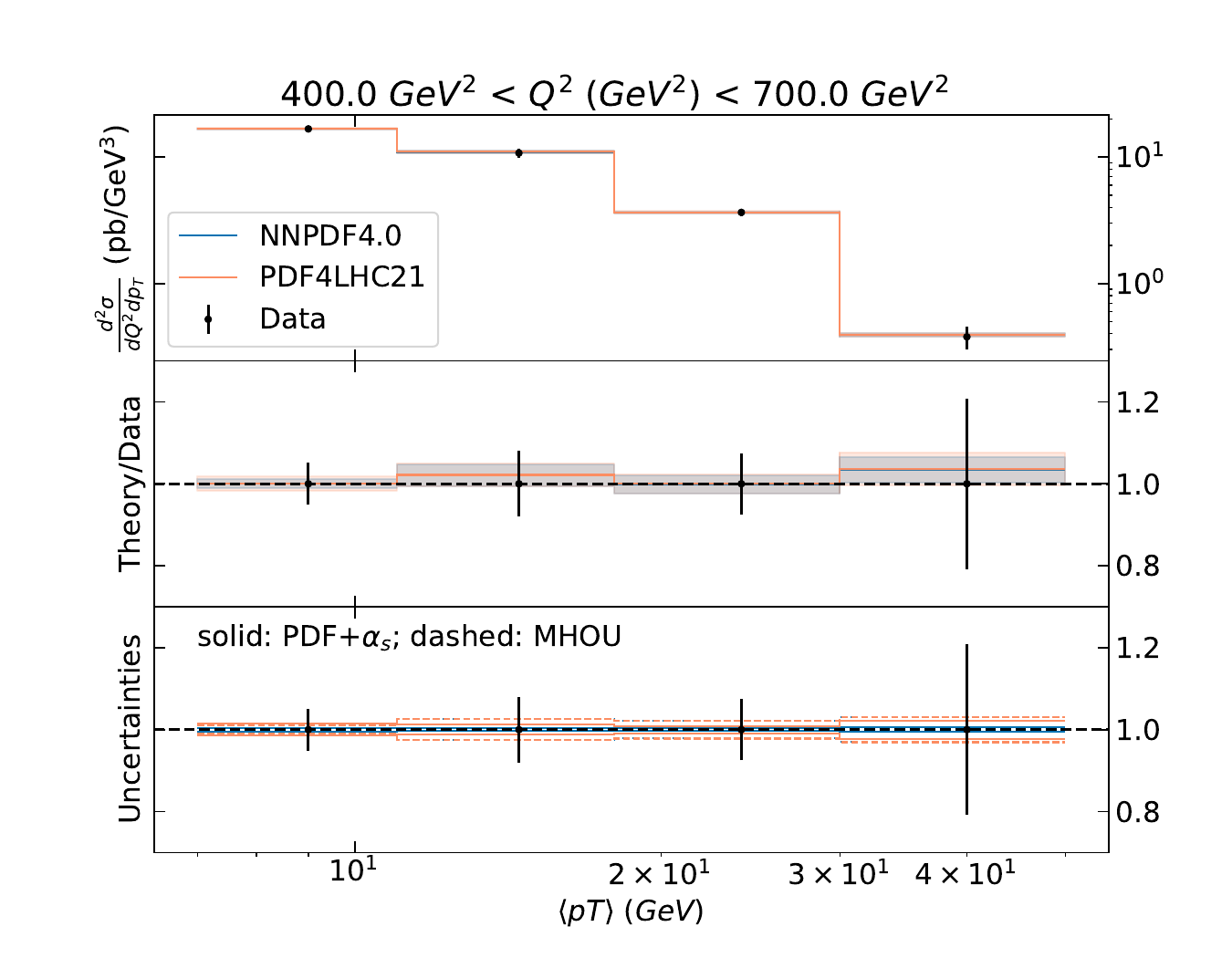}\\
    \includegraphics[width=0.49\textwidth]{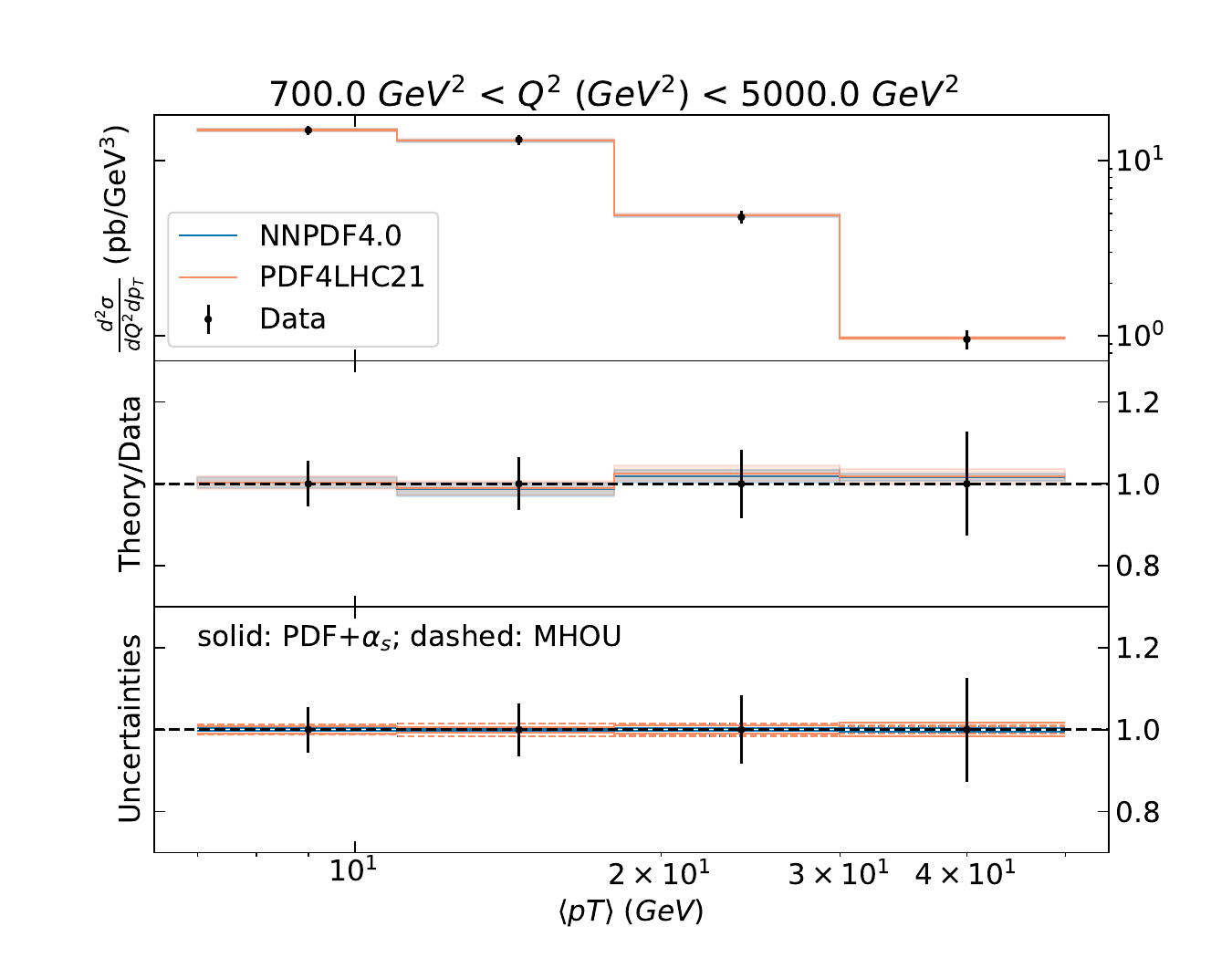}
    \caption{Same as Fig.~\ref{fig:H1_datatheory} for the bins of the H1
      high-$Q^2$ di-jet measurement not displayed in
      Fig.~\ref{fig:H1_datatheory}.}
    \label{fig:datatheory_DISjets_additional_4}
  \end{figure}

  \begin{figure}[!t]
    \centering
    \includegraphics[width=0.49\textwidth]{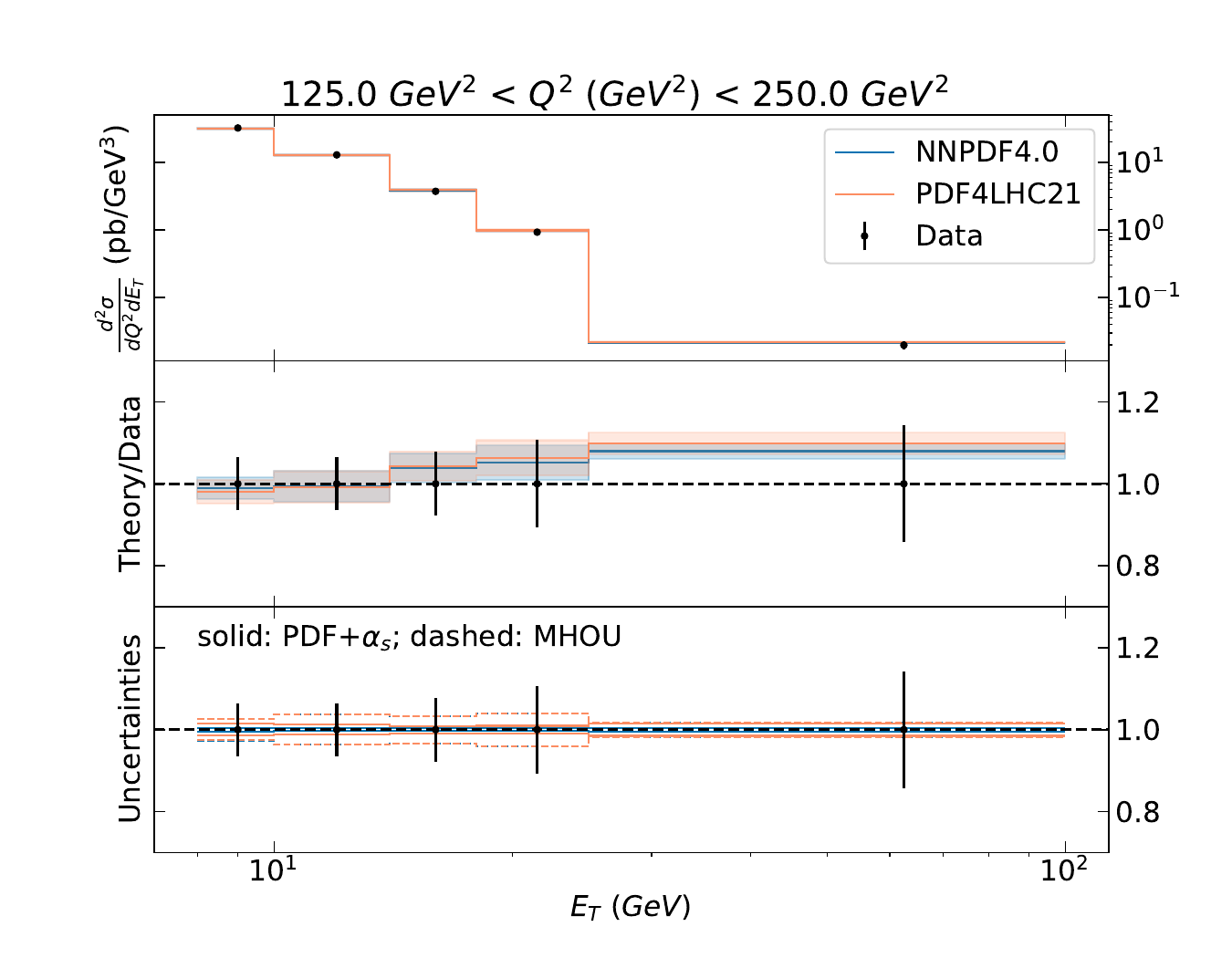}
    \includegraphics[width=0.49\textwidth]{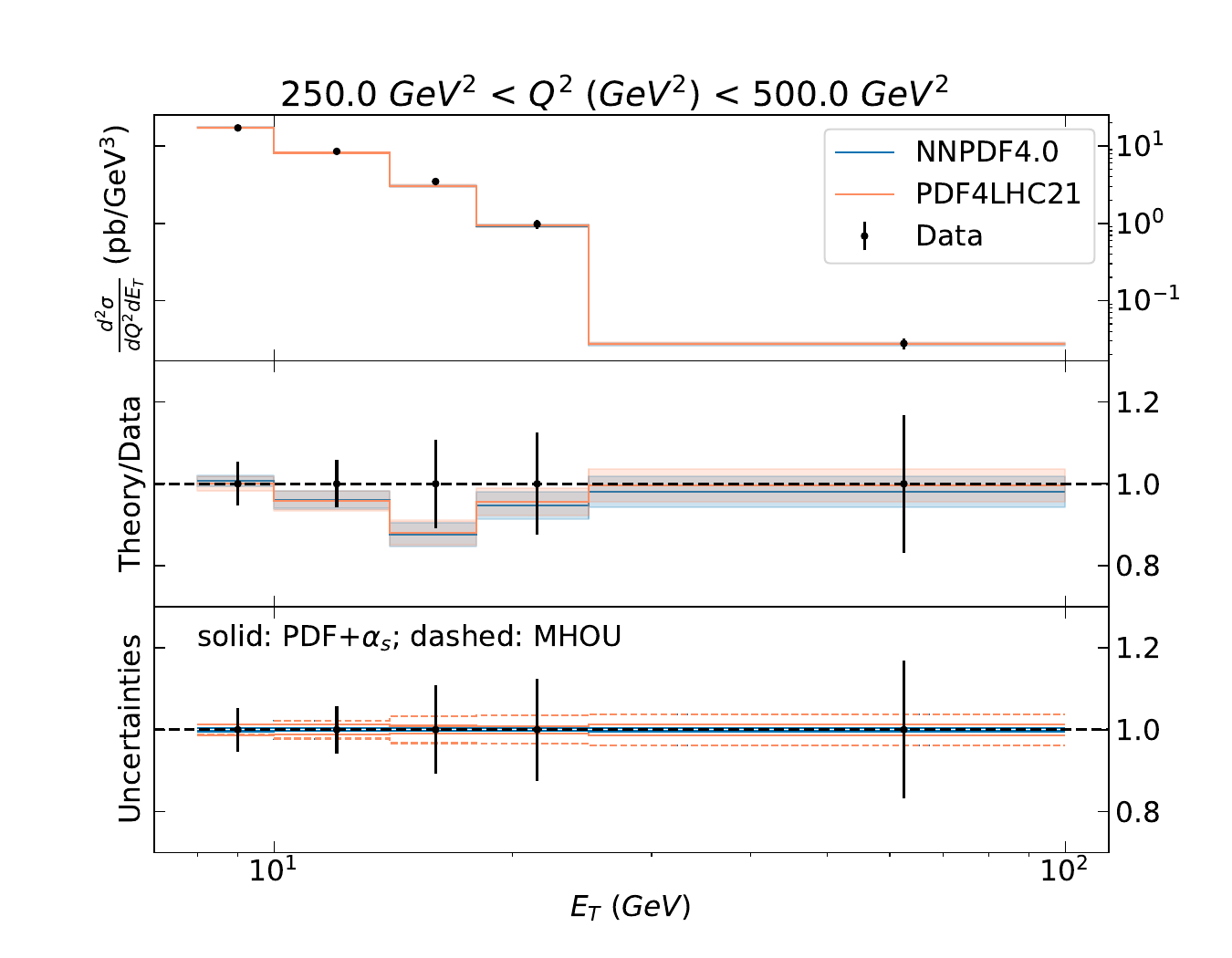}\\
    \includegraphics[width=0.49\textwidth]{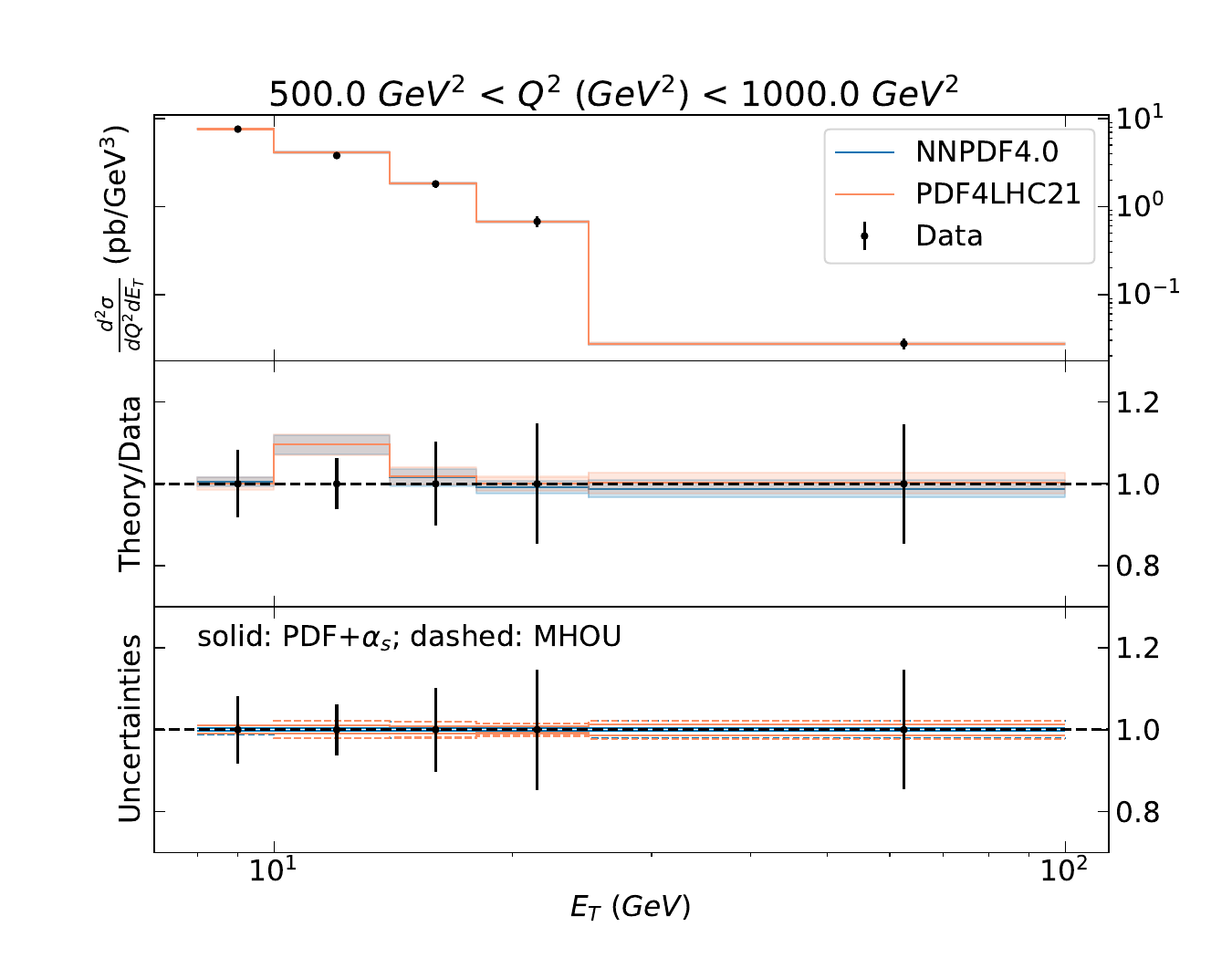} 
    \includegraphics[width=0.49\textwidth]{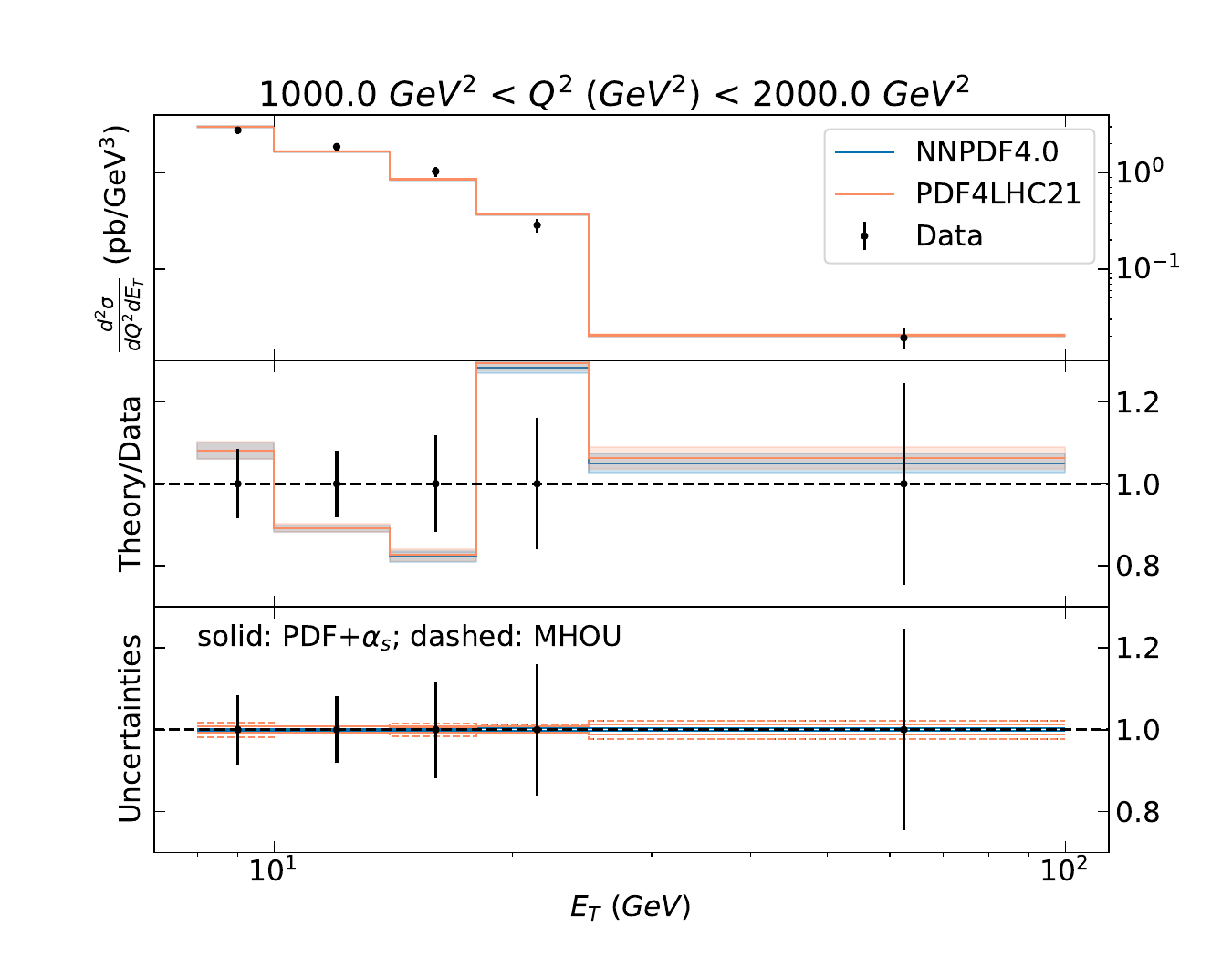}\\
    \includegraphics[width=0.49\textwidth]{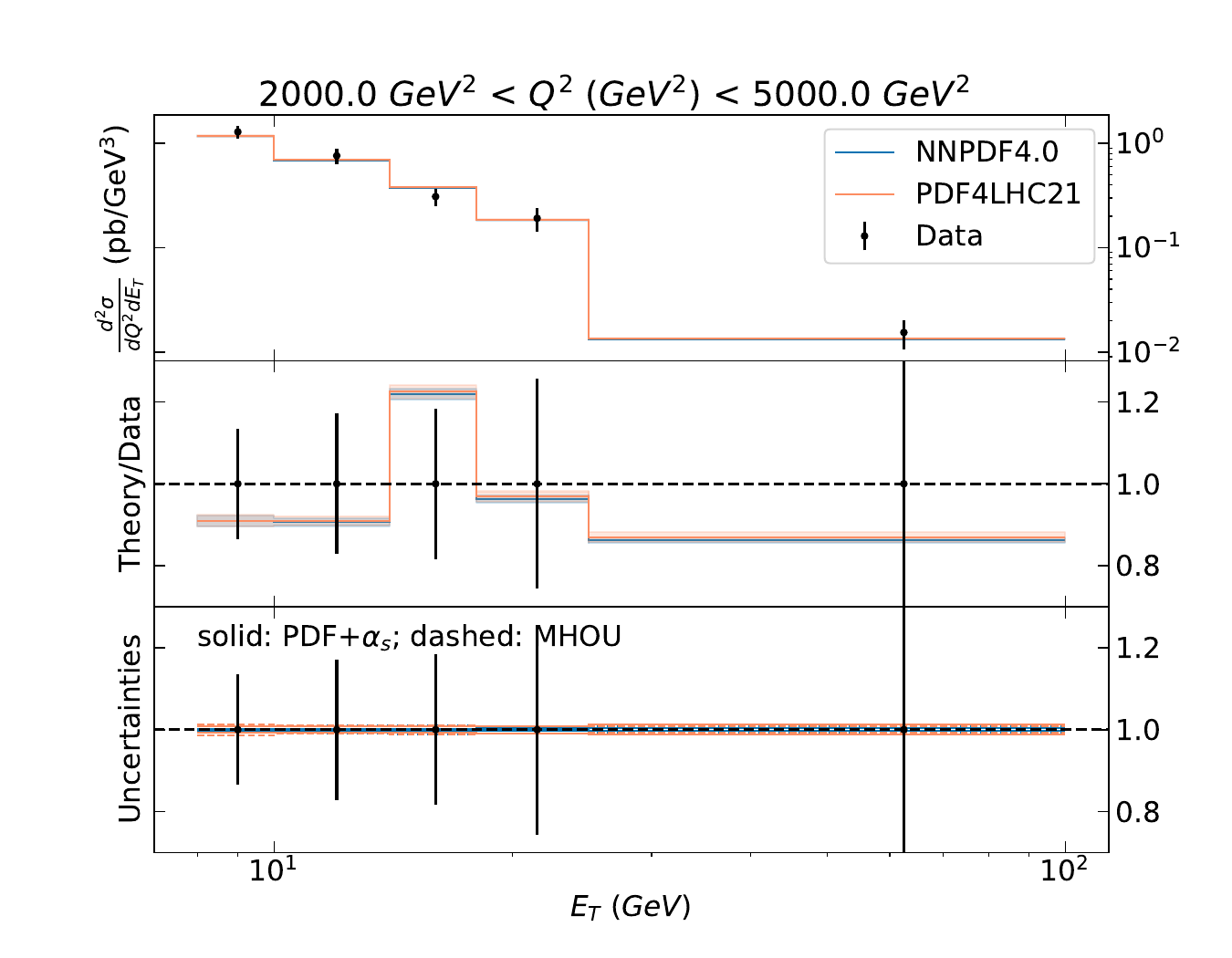}
    \includegraphics[width=0.49\textwidth]{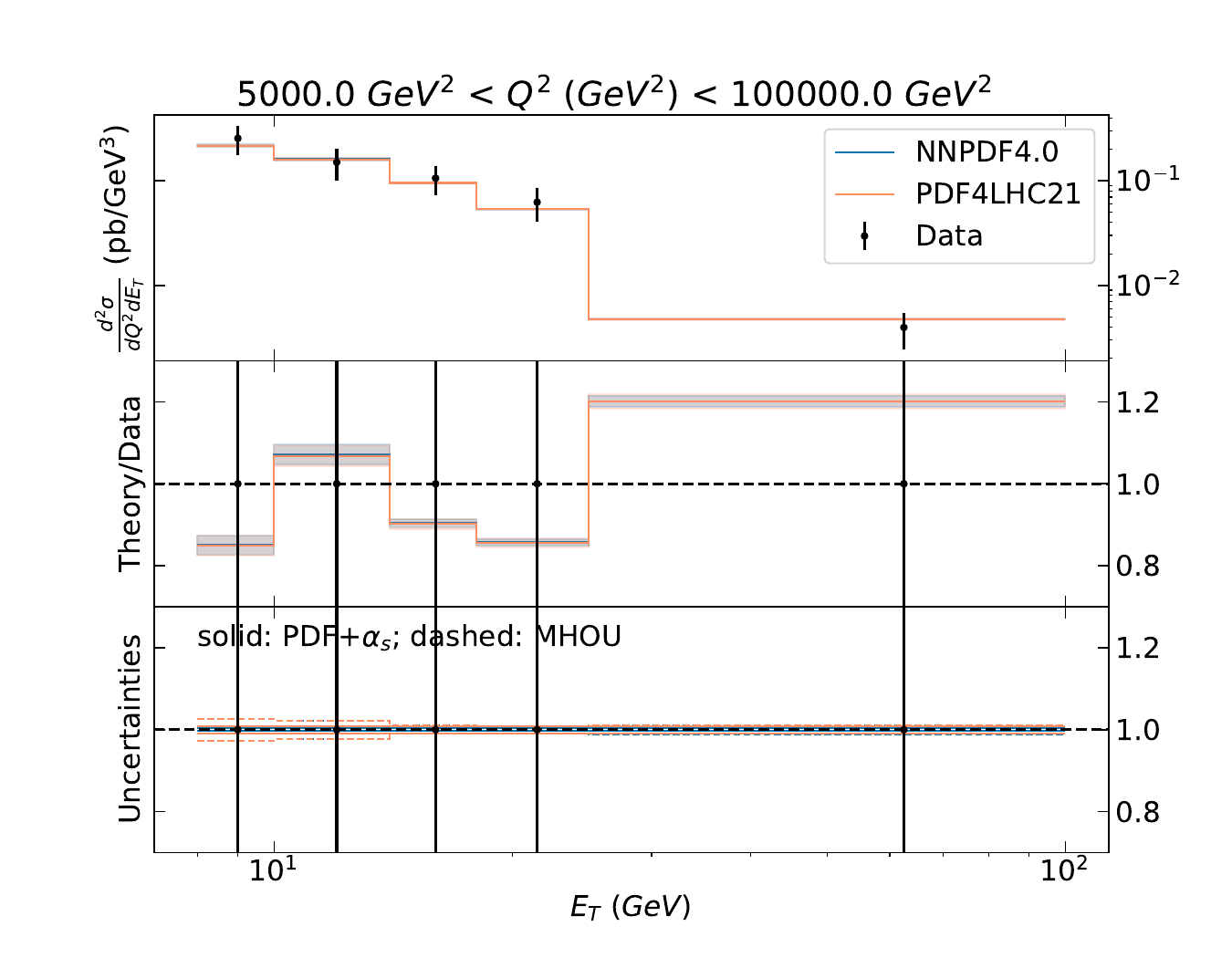}\\ 
    \caption{Same as Fig.~\ref{fig:DY-datatheory} for the ZEUS low-luminosity
      single-inclusive jet production measurement.}
    \label{fig:datatheory_DISjets_additional_5}
  \end{figure}
  
  \begin{figure}[!t]
    \centering
    \includegraphics[width=0.49\textwidth]{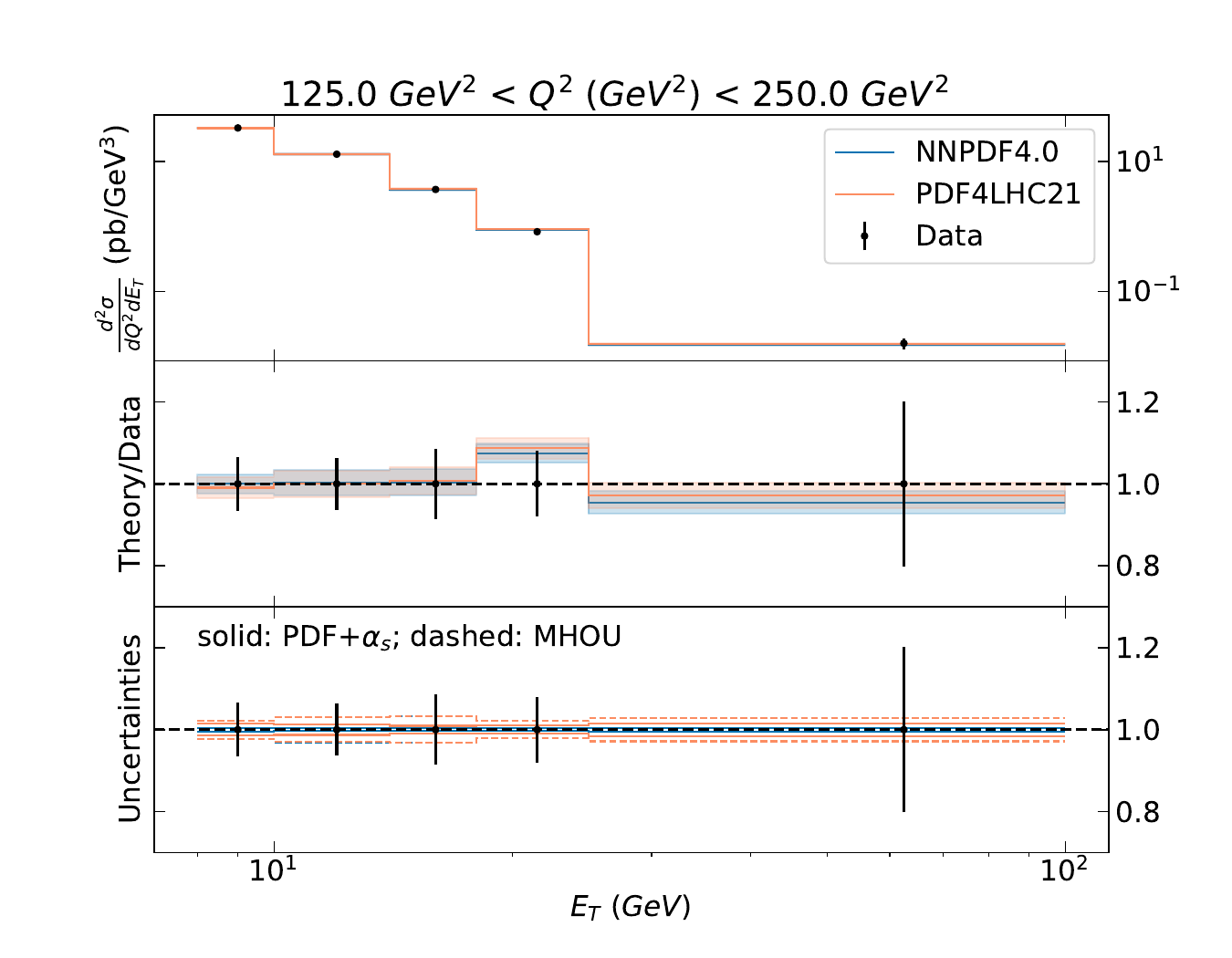}
    \includegraphics[width=0.49\textwidth]{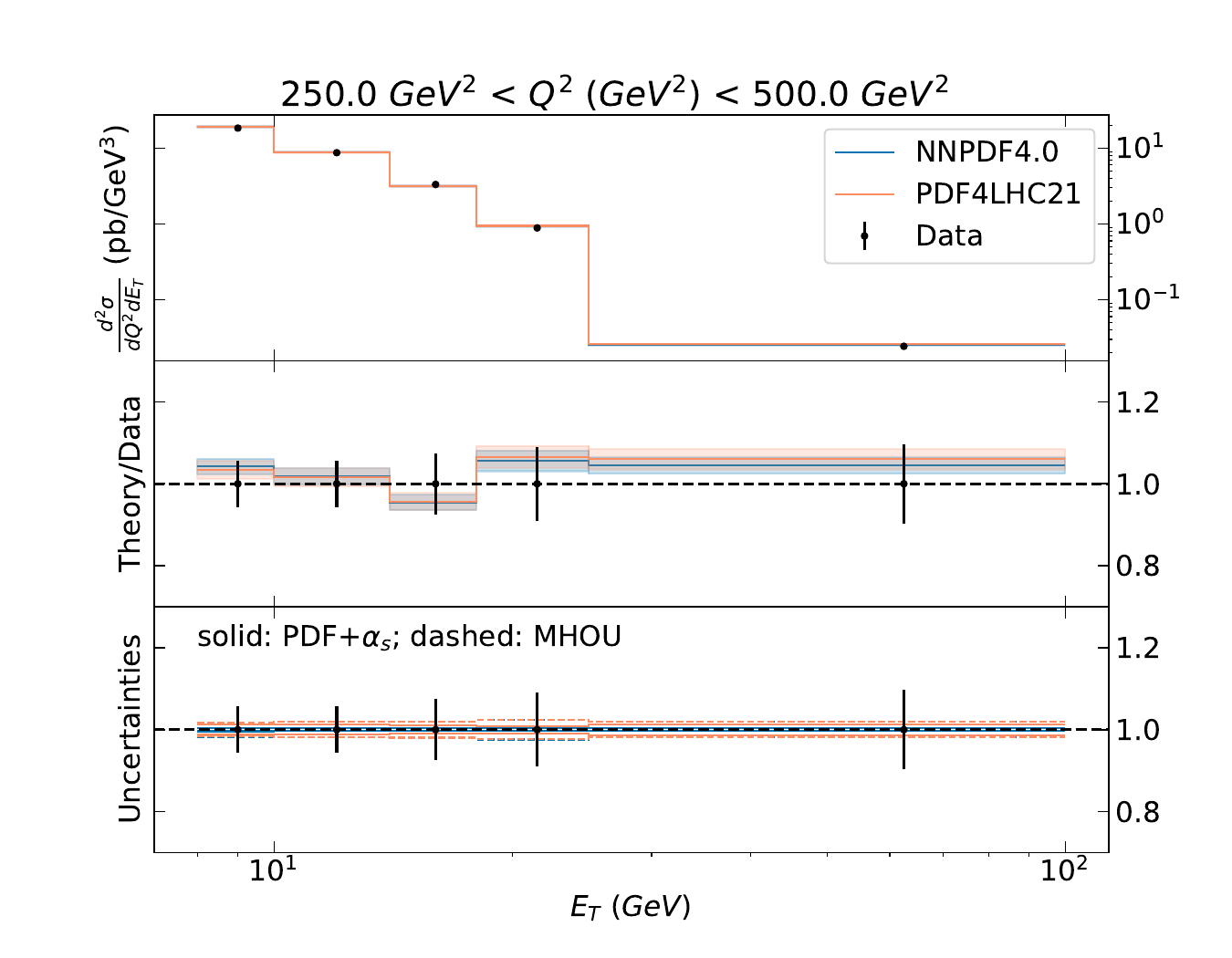}\\
    \includegraphics[width=0.49\textwidth]{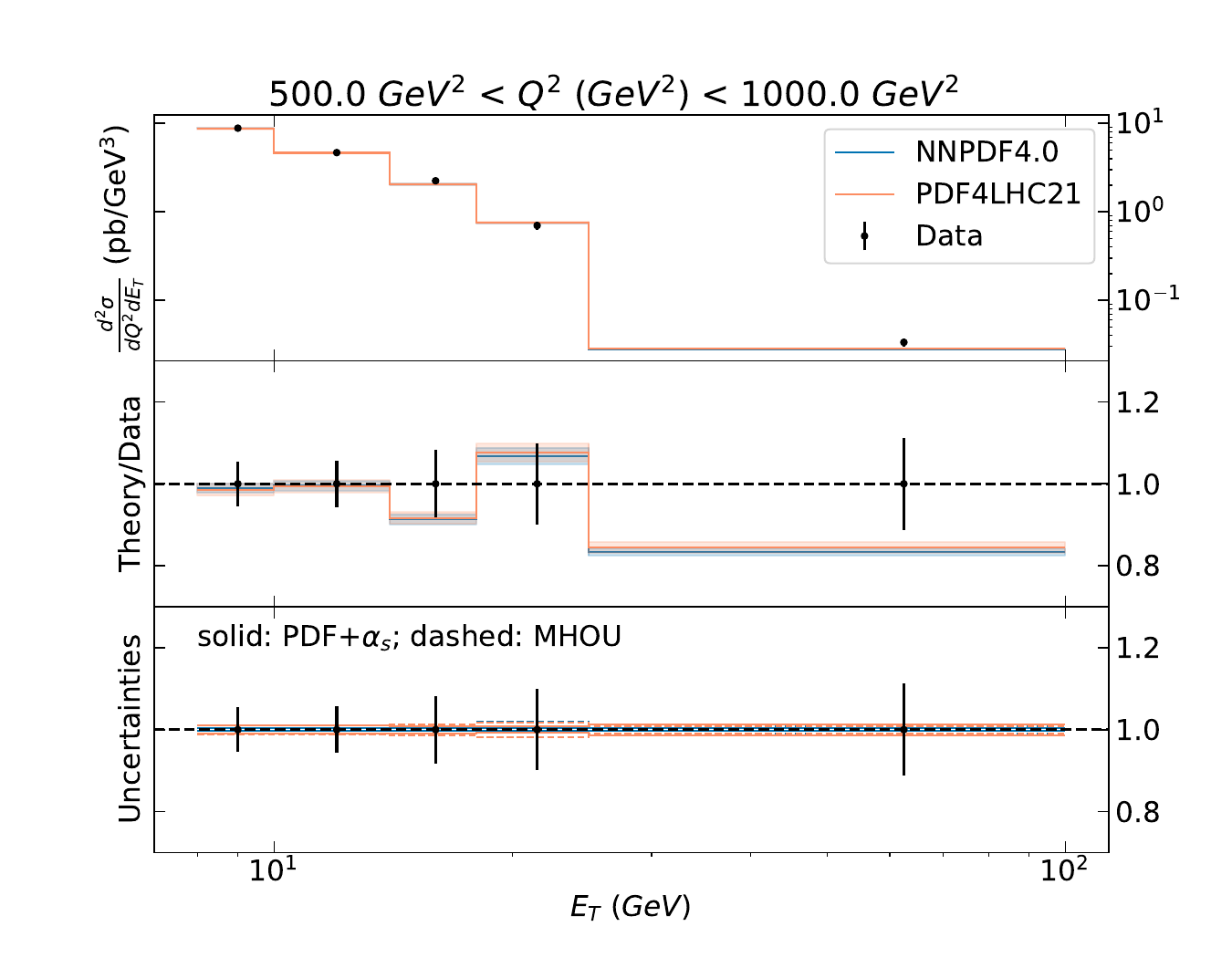} 
    \includegraphics[width=0.49\textwidth]{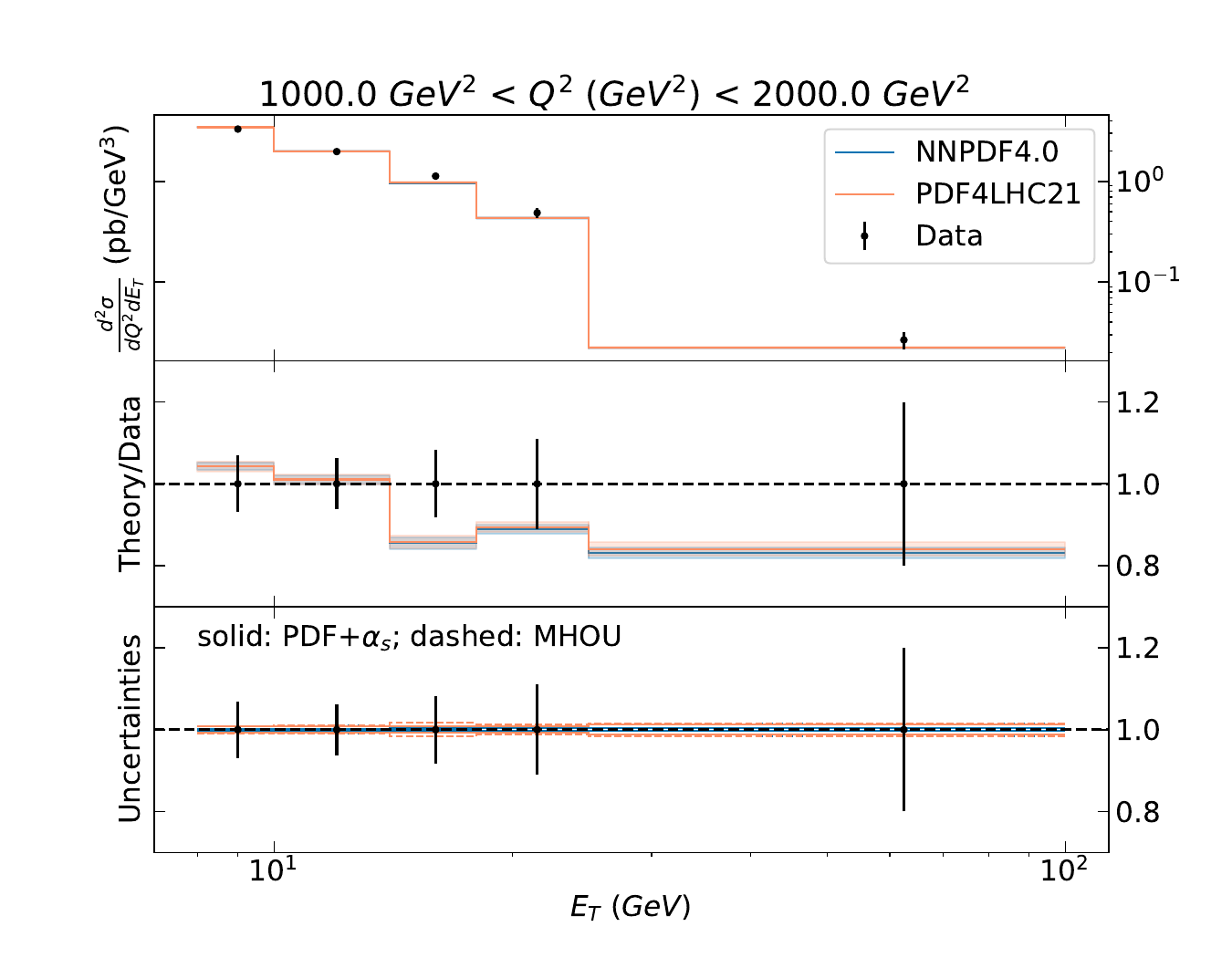}\\
    \includegraphics[width=0.49\textwidth]{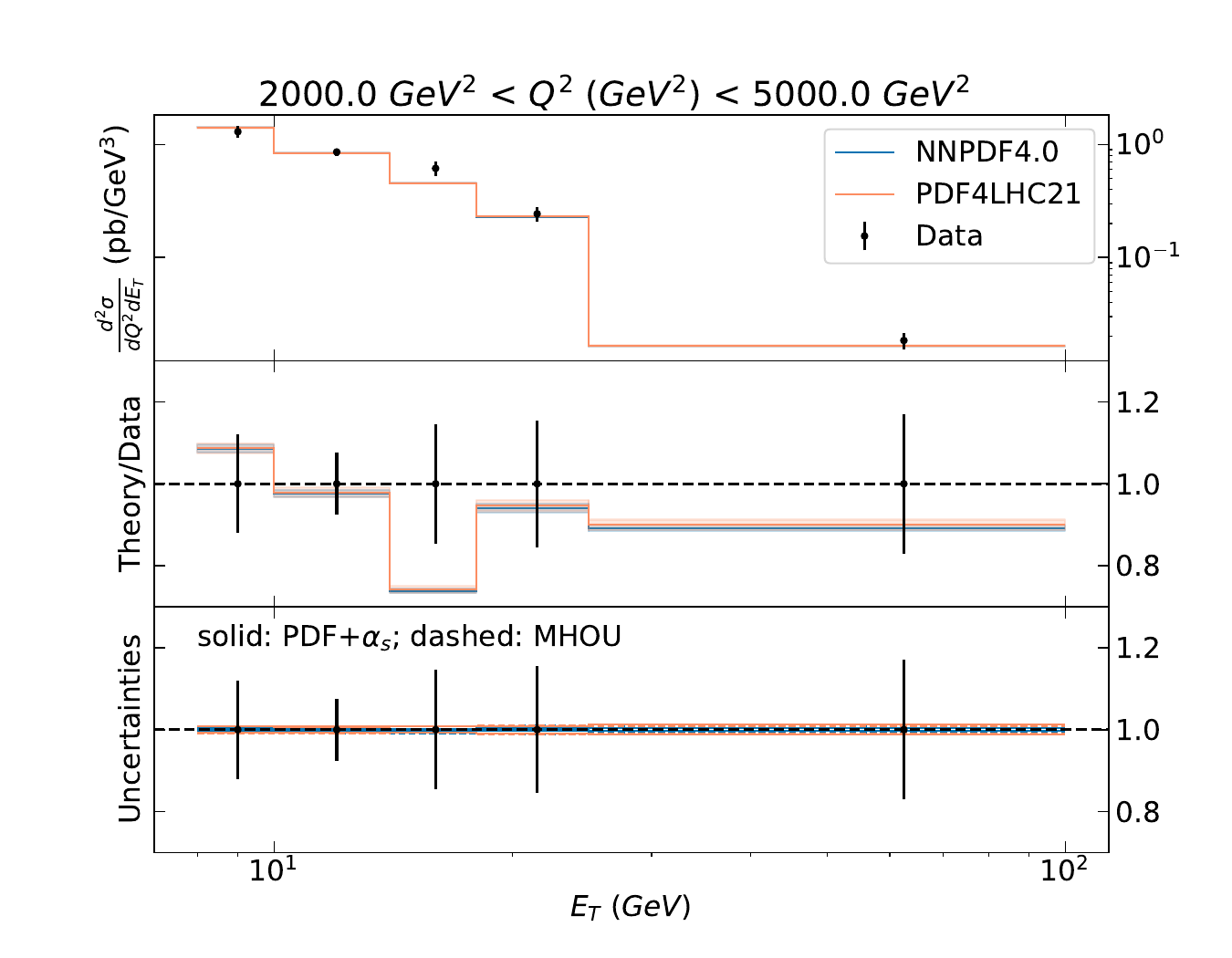}
    \includegraphics[width=0.49\textwidth]{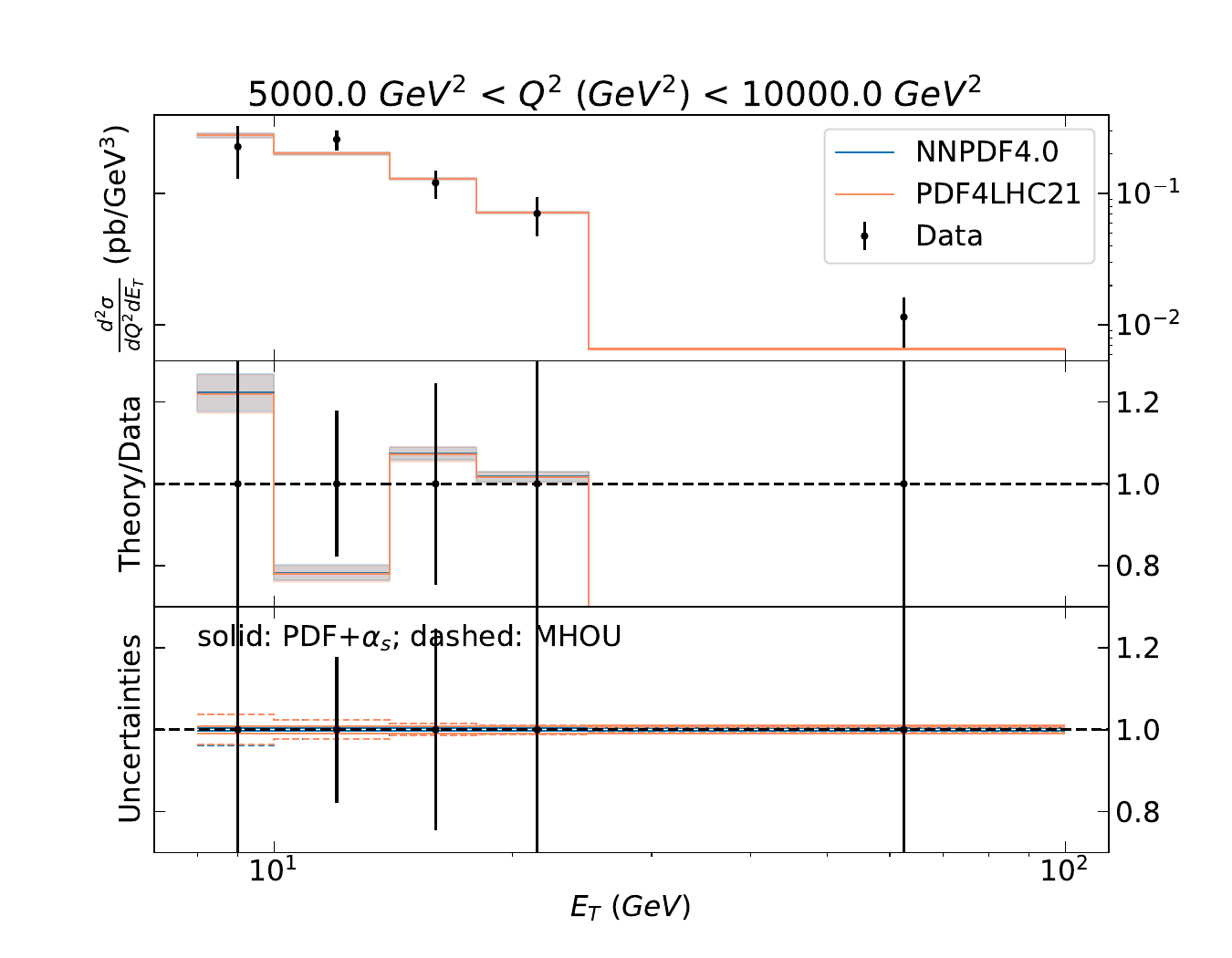}\\ 
    \caption{Same as Fig.~\ref{fig:DY-datatheory} for the ZEUS high-luminosity
      single-inclusive jet production measurement.}
    \label{fig:datatheory_DISjets_additional_6}
  \end{figure}

  \begin{figure}[!t]
    \centering
    \includegraphics[width=0.49\textwidth]{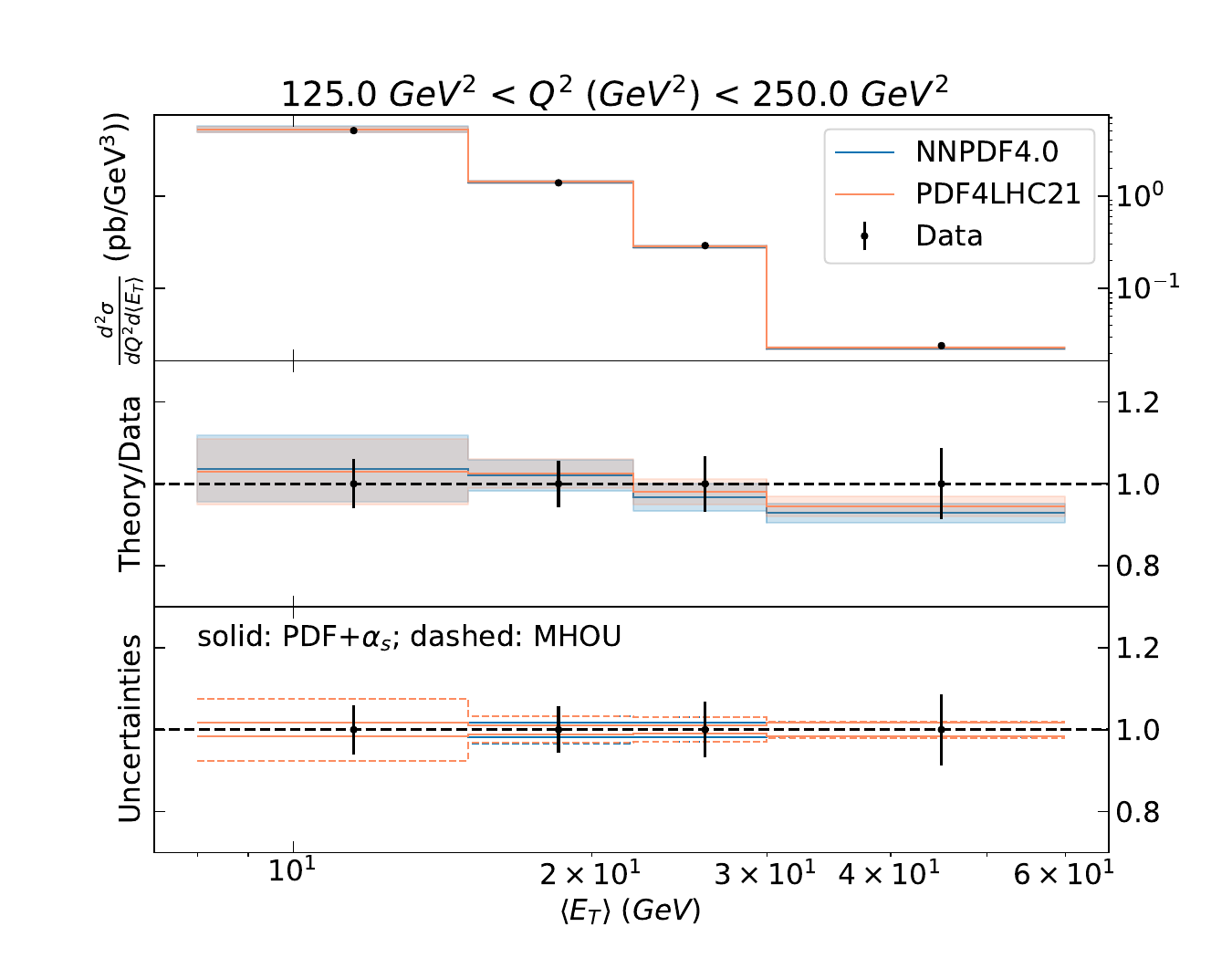}
    \includegraphics[width=0.49\textwidth]{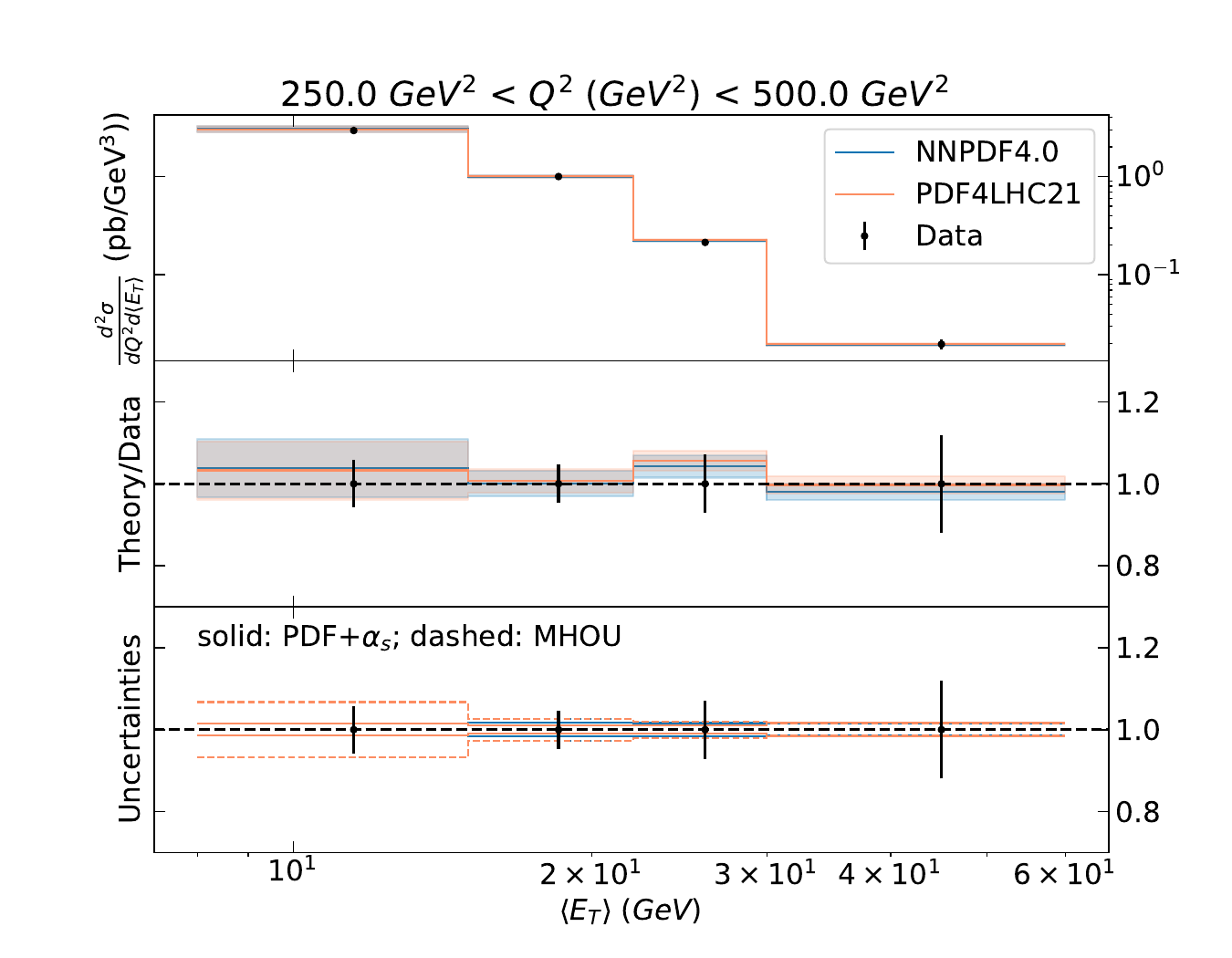}\\
    \includegraphics[width=0.49\textwidth]{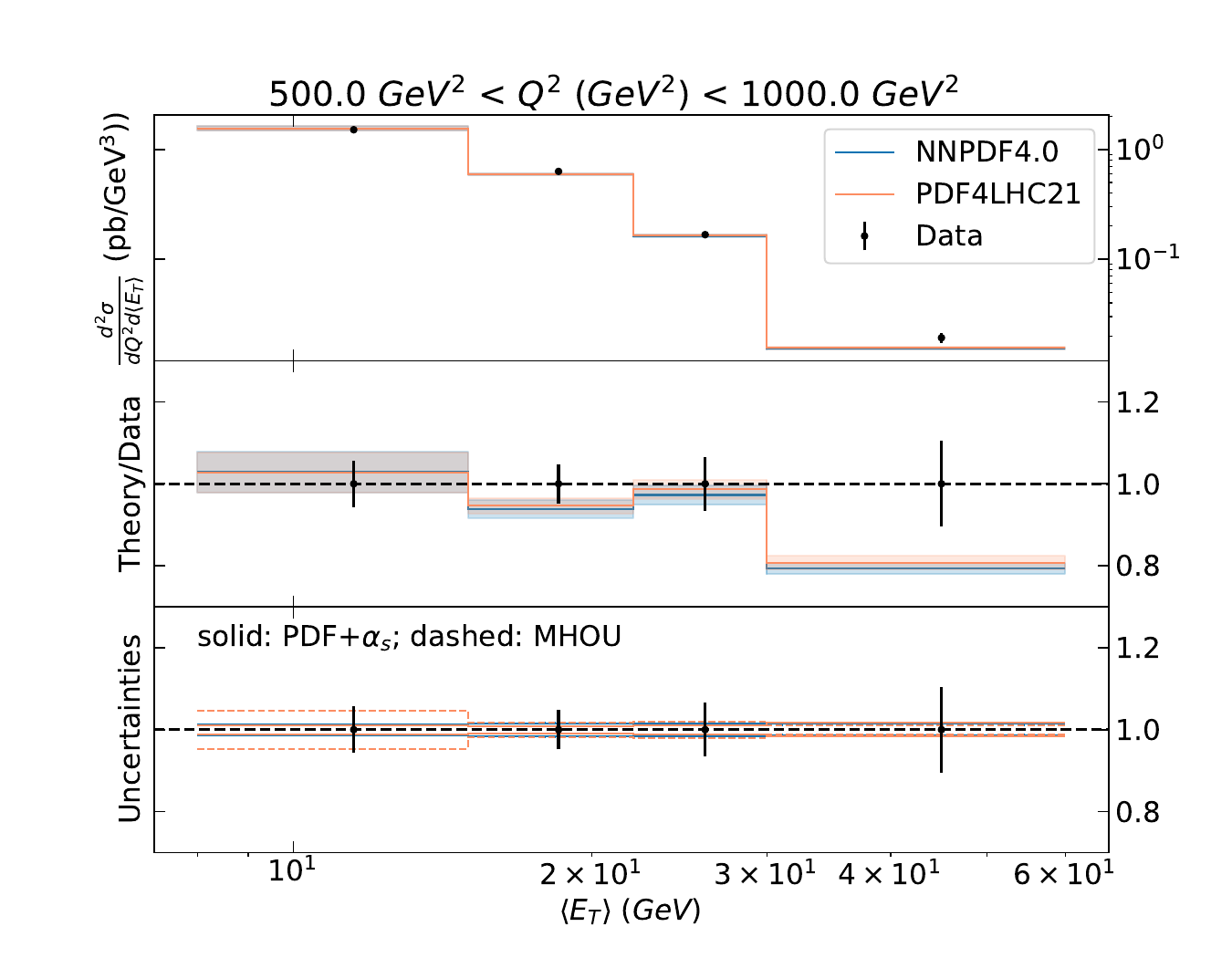} 
    \includegraphics[width=0.49\textwidth]{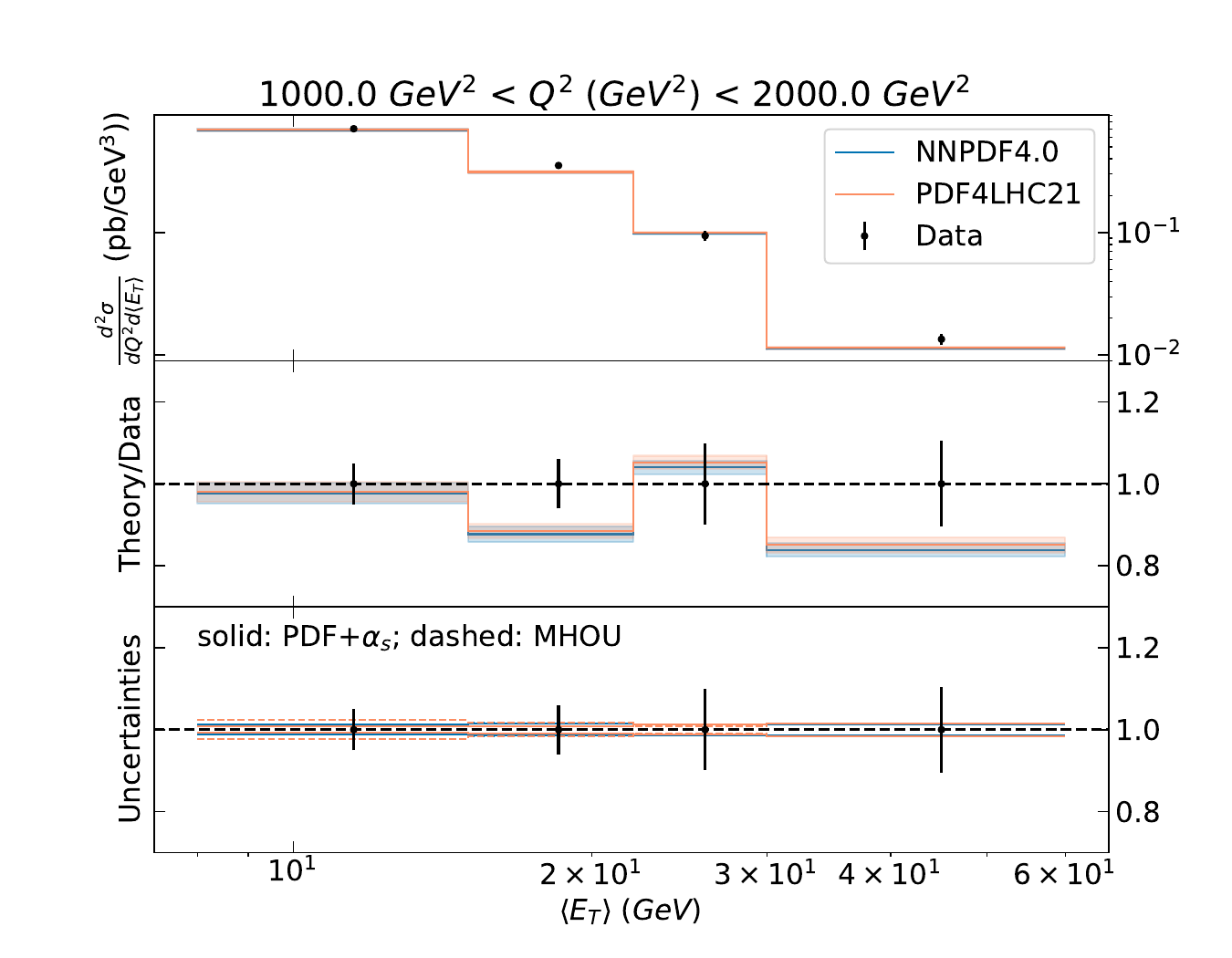}\\
    \includegraphics[width=0.49\textwidth]{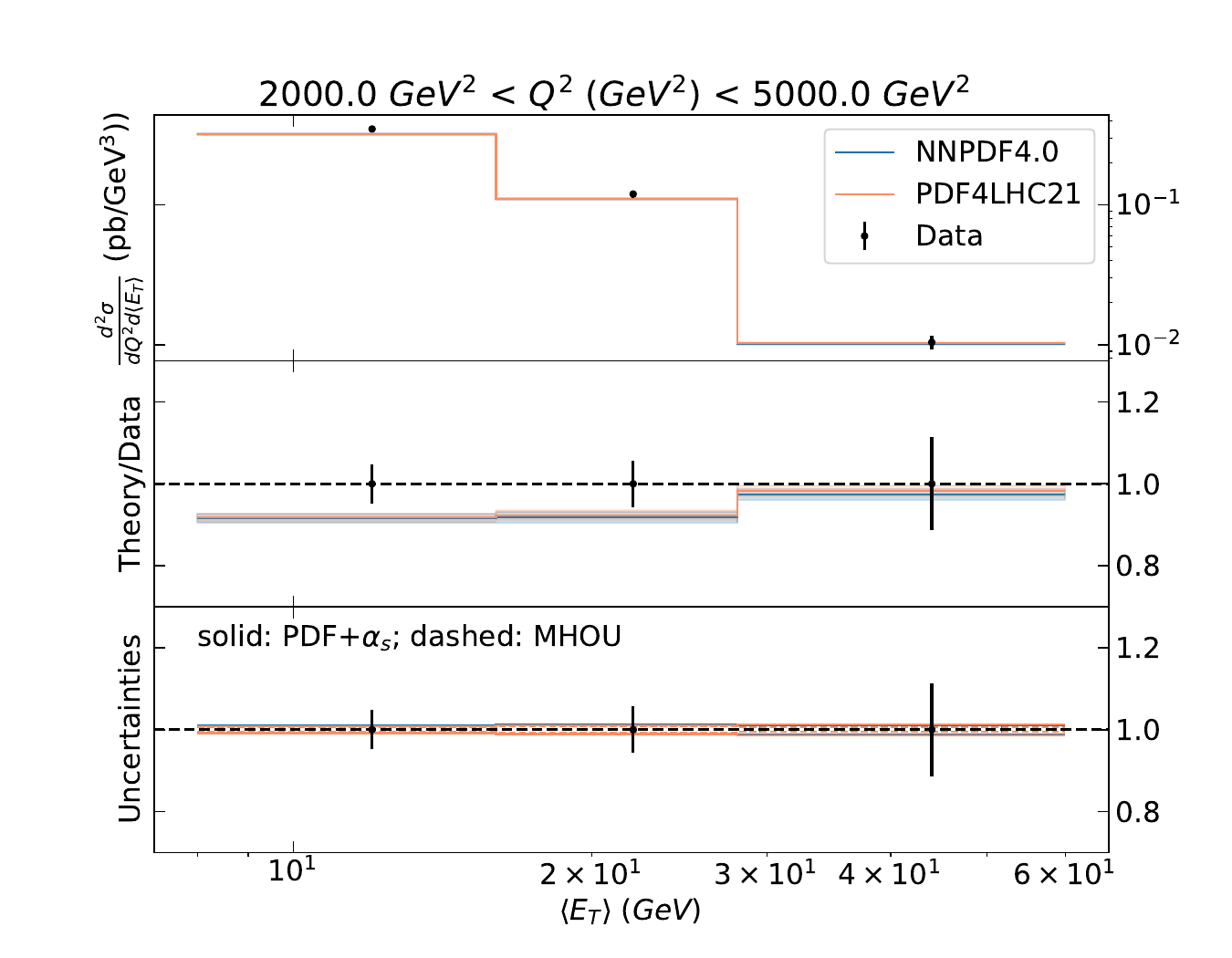}
    \includegraphics[width=0.49\textwidth]{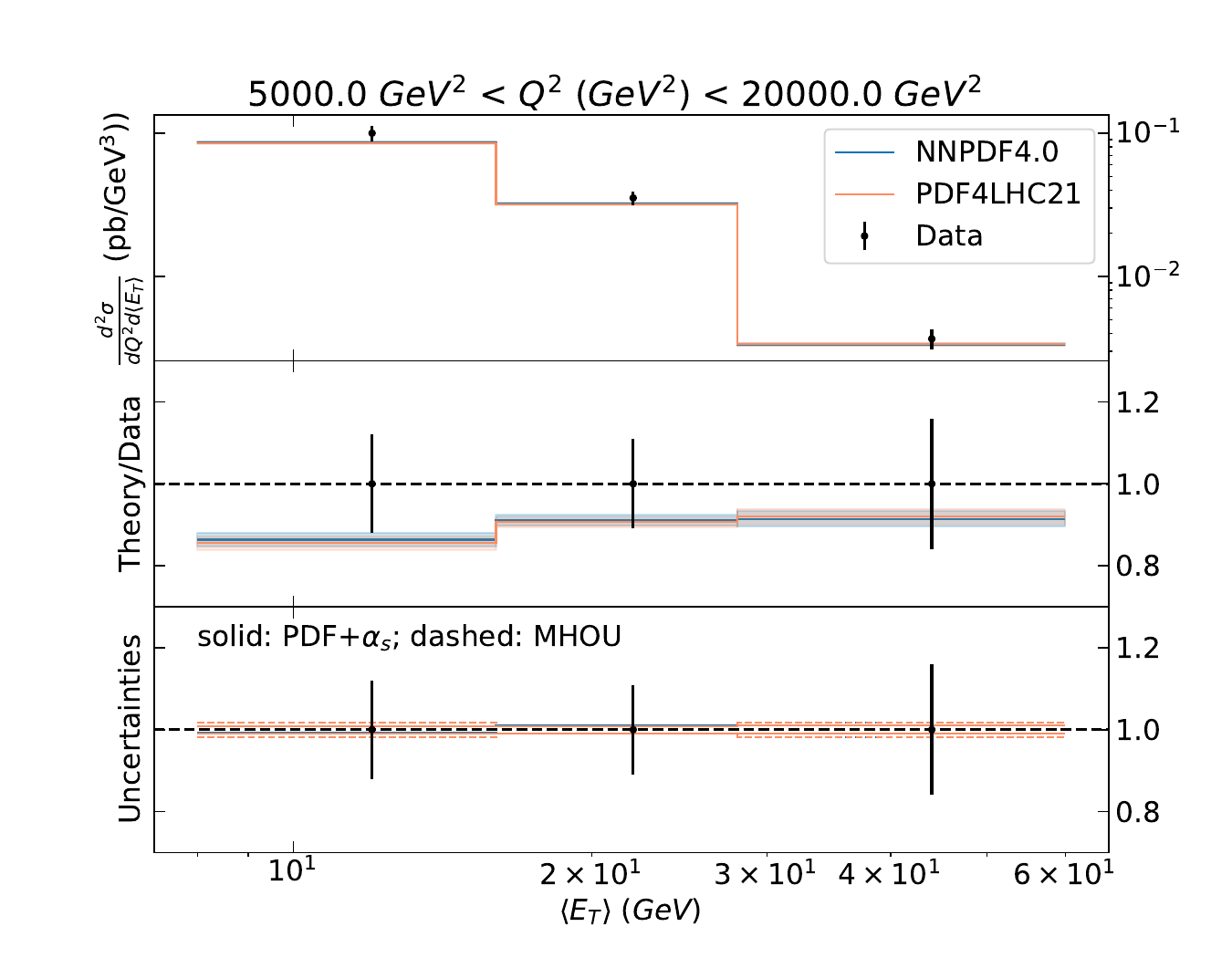}\\ 
    \caption{Same as Fig.~\ref{fig:DY-datatheory} for the ZEUS
      di-jet production measurement.}
    \label{fig:datatheory_DISjets_additional_7}
  \end{figure}

\end{description}

\clearpage

\providecommand{\href}[2]{#2}\begingroup\raggedright\endgroup

\end{document}